\documentclass[a4paper,twoside,openany,12pt]{memoir}

\usepackage{amsmath}
\usepackage{amsfonts}
\usepackage{amssymb}
\usepackage{bigfoot}
\usepackage{bm}
\usepackage[]{caption}
\usepackage{color}
\definecolor{mygreen}{rgb}{0,0.6,0}
\usepackage{graphicx}
\usepackage{indentfirst}
\usepackage[utf8]{inputenc}
\usepackage[
  safeinputenc,
  style=phys,
  backend=biber,
  sorting=none,
  minnames=10,
  maxnames=10,
  biblabel=brackets,
  pageranges=false,
]{biblatex}
\AtEveryBibitem{\clearlist{language}}
\renewbibmacro{in:}{}
\usepackage[ttscale=.875]{libertine}
\usepackage{listings}
\usepackage{oxfordthesis}
\usepackage{paralist}
\usepackage{physics}
\OnehalfSpacing
\usepackage{subcaption}
\usepackage[disable]{todonotes}
\usepackage{upgreek}

\thetitle{Subcritical Turbulence in the Mega Ampere Spherical
          Tokamak}
\theauthor{Ferdinand van Wyk}
\degreedate{Michaelmas 2016}
\degree{Doctor of Philosophy}
\college{Mansfield College}

\usepackage{xr}
\usepackage{xr-hyper}
\usepackage[
	pdftitle={\oxfthetitle},
	pdfauthor={\oxftheauthor},
	pdfsubject={Thesis for the Degree of \oxfdegree, \oxfdegreedate},
	pdfborder=0,
	bookmarks=true,
	bookmarksnumbered=true,
	bookmarksopen=true,
	bookmarksopenlevel=1,
	plainpages=false,
	pdfpagelabels=true,
	colorlinks=false,
    citecolor=blue
]{hyperref}
\usepackage{memhfixc}

\DeclareMathOperator{\sgn}{sgn}

\newcommand{\exb}{$\vb*{E} \times \vb*{B}$ }

\newcommand{\ensav}[2]{\left<{#1}\right>_{#2}}
\newcommand{\vareps}{\varepsilon}
\newcommand{\figref}[1]{Figure~\ref{fig:#1}}
\newcommand{\figsref}[2]{Figures~\ref{fig:#1} and~\ref{fig:#2}}

\newcommand{\Figref}[1]{Figure~\ref{fig:#1}}
\newcommand{\Figsref}[2]{Figures~\ref{fig:#1} and~\ref{fig:#2}}

\newcommand{\pd}[2]{\ensuremath{ \frac{\partial #1} {\partial #2} } }

\begin{document}

\titlepage

\frontmatter

\begin{abstract}
  The transport of heat out of tokamak plasmas by turbulence is the dominant
  mechanism limiting the performance of fusion reactors. Turbulence can be
  driven by the ion temperature gradient (ITG) and suppressed by toroidal
  equilibrium scale sheared flows. Numerical simulations attempting to
  understand, and ultimately reduce, turbulence are crucial for guiding the
  design and optimisation of future reactors.

  In this thesis, we investigate ion-scale turbulence by means of local
  gyrokinetic simulations in the outer core of the Mega Ampere Spherical
  Tokamak (MAST). We perform a parameter scan in the values of the ITG and the
  flow shear. We show that nonlinear simulations reproduce the experimental ion
  heat flux and that the experimentally measured values of the ITG and the flow
  shear lie close to the turbulence threshold. We demonstrate that the system
  is subcritical in the presence of flow shear, i.e., the system is formally
  stable to small perturbations, but transitions to a turbulent state given a
  large enough initial perturbation. We propose a scenario for the transition
  to subcritical turbulence previously unreported in tokamak plasmas: close to
  the threshold, the plasma is dominated by a low number of coherent long-lived
  structures; as the system is taken away from the threshold into the more
  unstable regime, the number of these structures increases until they fill the
  domain and a more conventional turbulence emerges.

  We make quantitative comparisons of correlation properties between our
  simulations and experimental measurements of ion-scale density fluctuations
  from the MAST BES diagnostic. We apply a synthetic diagnostic to our
  simulation data and find reasonable agreement of the correlation properties
  of the simulated and experimental turbulence, most notably of the correlation
  time, for which significant discrepancies were found in previous numerical
  studies of MAST turbulence. We show that the properties of turbulence are
  essentially functions of the distance to threshold, as quantified by the ion
  heat flux.  We find that turbulence close to the threshold is strongly
  affected by flow shear, whereas far from threshold, the turbulence resembles
  a conventional ITG-driven, zonal-flow damped regime.
\end{abstract}

\begin{acknowledgements}

  I would like to thank my supervisors Prof. Alex Schekochihin, Dr. Edmund
  Highcock, and Dr. Colin Roach for their support and guidance throughout my
  DPhil, and for instilling in me the importance of taking care of the details.
  As result of their guidance, I have become better scientist, writer, and
  programmer.

  I benefitted over the years from the advice and expertise of
  senior members of the extended Oxford and Culham plasma physics groups: Ian
  Abel, Anthony Field, Felix Parra, Bill Dorland, Paul Dellar, Michael
  Barnes, and Nuno Loureiro who were always willing to help.  I also greatly
  appreciated the company of my fellow plasma physics graduate students who
  made it a much easier journey: Michael Fox, Justin Ball, Greg Colyer,
  Alessandro Geraldini, Joseph Parker, and last but not least, George Wilkie,
  whose humour made every conference a more enjoyable experience.

  There are many other fellow graduate students from Mansfield College and
  elsewhere, who made the DPhil journey with me either in person, or in spirit,
  and whose support was greatly appreciated: Alex Rowell, Nicole Miranda, Chris
  Birkl, James Fagan, Katia Damer, and Kevin O'Keeffe. Particular thanks goes
  to Cherese Thakur, Daniel Lordan, and Phelim Bradley who went through all the
  ups and downs of graduate life with me.

  From my time at University College Cork, I would like to thank Dr. Paddy
  McCarthy for helping me reach my goal of researching fusion energy and Prof.
  Frank Peters for his engaging lectures and encouragement.

  Finally, this DPhil would not have been possible without the support of my
  family. My mom and her partner who have allowed me to follow my interests and
  to reach my potential, but above all my brother Adriaan whose limitless good
  humour and high spirits have always been a source of optimism.

\end{acknowledgements}

\tableofcontents
\listoffigures

\mainmatter

\listoftodos
\chapter{Introduction}

\section{Nuclear fusion}

  Nuclear fusion is the process that powers the stars. Confined by the
  gravitational force and heated to very high temperatures, hydrogen isotopes
  can collide and fuse to form helium and release large amounts of energy. When
  it comes to harnessing this power for use on Earth, the most promising fusion
  reaction is between deuterium and tritium isotopes of hydrogen, which
  produces a 3.5~MeV helium nucleus and a 14.6~MeV neutron. Utilising this
  reaction for the purposes of electricity generation has been the goal of
  fusion scientists since the idea was first proposed in the 1950s.

\begin{figure}[t]
    \centering
    \includegraphics[width=0.7\linewidth]{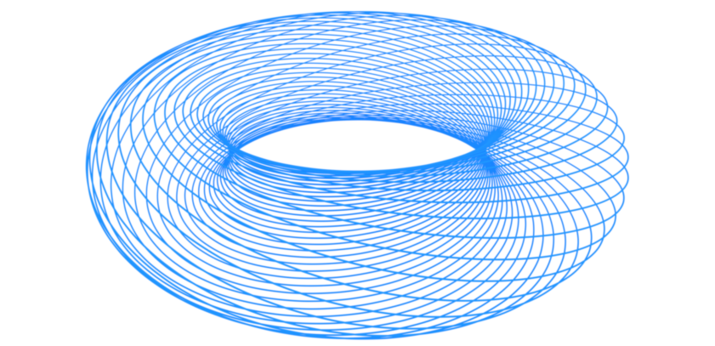}
    \caption[Toroidal configuration of magnetic field lines]{
      Toroidal configuration of magnetic field lines used to confine fusion
      plasmas.
    }
    \label{fig:cbc_field_lines}
  \end{figure}

  The tokamak has emerged as the most promising concept for confining this
  reaction by using a toroidal configuration of magnetic field lines
  (see~\figref{cbc_field_lines}). At the temperatures
  required for fusion to occur, deuterium and tritium become fully ionised and
  the gas becomes a plasma.  In the presence of a magnetic field, these charged
  particles are forced to gyrate about the magnetic field lines in a plane
  perpendicular to the field lines and although they can freely stream along
  them, they remain confined. This is because in the toroidal configuration,
  magnetic field lines lie on a single surface and so provide no direct
  route out of the plasma.  This is crucial given that no material one could
  feasibly build a fusion reactor out of, can withstand direct contact with the
  extremely high temperature fusion plasma. This also
  necessarily means that large pressure gradients are set up between the hot
  core, where fusion reactions take place, and the relatively cool edge near
  the reactor walls. It is these gradients that give rise to a physical process
  that has hindered the realisation of fusion energy since the first attempts
  to build reactors large enough to produce electricity: turbulence.

\section{Turbulence}

  Even in the ideal confinement scenario described above, there are still
  processes by which plasma particles can escape. These processes include
  collisions with other particles in addition to
  particle drifts due the presence of an electric field, magnetic field line
  curvature, and magnetic field gradients.  The transport of particles,
  momentum, and heat out of the plasma due to these processes is known as
  neoclassical transport~\cite{Hinton1976, Hinton1985, Catto1987, Sugama1997,
  Helander2005}.  While important, neoclassical transport alone would not
  significantly hinder the viability of a well-designed fusion reactor.
  Instead, it is turbulence that presents a much greater challenge to fusion
  power as an energy source~\cite{Coppi1967,Catto1973,Cowley1991, Connor1994,
  Jenko2000a,Dorland2000,Dannert2005,Tynan2009}. In the presence of gradients
  of density, flow, or temperature, small perturbations to the plasma state can
  grow exponentially, and eventually interact with each other, leading to a
  turbulent state.  This turbulent state gives rise to enhanced radial
  transport of particles, momentum, and heat, which can significantly exceed
  neoclassical estimates~\cite{Hawryluk1998,Lazarus1996,Akers2003,Mantica2009}.
  This presents a challenge for sustaining the temperatures and densities
  necessary in the core for fusion. Thus, reducing or eliminating turbulence
  completely would be the most effective means of achieving improved fusion
  power.

  Experimental, numerical, and theoretical studies have
  shown that turbulent transport is strongly dependent on the ion temperature
  gradient (ITG)~\cite{Baker2001,Tardini2002,Mantica2009,Mantica2011,Ghim2014,
  Horton1980,Waltz1988,Kotschenreuther1995a,Dimits1996,
  Highcock2010,Barnes2011a}. Additionally, the electron temperature and density
  gradients, which give rise to the electron temperature gradient (ETG)
  mode~\cite{Dorland2000,Jenko2000a} and the trapped electron mode
  (TEM)~\cite{Dannert2005}, can also drive turbulence. In this work, we will
  focus on turbulence driven by the ITG, which is a source of free energy and
  drives the well-known ITG instability~\cite{Coppi1967, Cowley1991}, in
  combination with turbulence driven by the TEM, which also drives
  turbulence at ion scales.  It is well-established that modest increases in
  the ITG lead to large increases in ion heat flux, so-called ``stiff
  transport'' (see~\cite{Mantica2009} for a recent experimental study). The
  phenomenon of stiff transport is an important consideration in the design of
  fusion reactors. In order to maximise the temperature in the core (to
  increase fusion power) we want to maximise the temperature gradient between
  the core and the relatively cool edge, where technological constraints (e.g.
  material strain due to heat deposition, melting due to high temperatures,
  etc.) must be taken into account. However, enhanced ITG-driven transport
  (which reduces the ITG) would set an upper bound on the ITG and, hence, on
  the core temperature that we can achieve. That is, unless some process
  exists that can reduce or eliminate turbulence, driven by the ITG
  instability, without reducing the ITG itself. Fortunately, it has been shown
  that such a process exists in the form of sheared flows perpendicular to the
  magnetic field lines.

\section{Sheared flows and subcritical turbulence}

  It has been shown experimentally that toroidal rotation, or more
  specifically the differential rotation between surfaces of constant magnetic
  field, can lead to a
  reduction or even complete suppression of
  turbulence~\cite{Ritz1990,Burrell1997,Mantica2009, Mantica2011,Field2011}.
  Toroidal differential rotation can be driven by the neutral beam injection
  (NBI) system present in most fusion experiments~\cite{Field2011}. The NBI
  system injects deuterium atoms at high energy to heat the plasma and
  simultaneously generates a toroidal flow in the plasma. This gives rise to a
  sheared flow (since the NBI system deposits most of its momentum and heat
  at the core of the plasma) with components both parallel and perpendicular to
  the direction of the magnetic field. Perpendicular flow shear has been shown
  to reduce, or even eliminate, turbulence, while parallel flow shear has been
  shown to drive a linear instability~\cite{Catto1973} (the parallel-velocity
  gradient (PVG) instability), which can increase the level of turbulence.
  This effect has been confirmed in many numerical studies~\cite{Waltz1994,
  Waltz1997, Kinsey2005, Camenen2009, Roach2009, Barnes2011a,Highcock2010}.
  However, it was shown that large flow shears and temperature gradients are
  required before the destabilising effect of the parallel flow shear is strong
  enough to overcome the stabilising effect of the perpendicular flow
  shear~\cite{Highcock2010,Barnes2011a}. For this reason, PVG-driven turbulence
  is not expected to play a large role in the experimentally relevant plasmas
  we will consider in this work, given the modest levels of the ITG and flow
  shear.  To summarise, we see that there is a competition in fusion plasmas
  between the destabilising effects of the ITG and PVG instabilities, and the
  stabilising effect of the perpendicular flow shear.

  Perpendicular flow shear can reduce turbulence levels in two ways: by
  stabilising the linear instabilities that amplify small perturbations, and by
  shearing apart eddies that characterise the turbulent state. It has
  been shown that perpendicular flow shear can, in fact, render the plasma
  completely linearly stable. However, there may still be substantial transient
  growth of perturbations and, given large enough initial perturbations, this
  transient growth can still lead to a saturated nonlinear state -- a
  phenomenon known as ``subcritical'' turbulence~\cite{Newton2010,Highcock2010,
  Highcock2011,Barnes2011a, Schekochihin2012, Landreman2015}.  This is a
  well-known phenomenon in neutral fluid systems, such as Couette and
  Poiseuille flows, where, though they are linearly stable, finite
  perturbations can nonetheless lead to a turbulent
  state~\cite{Reynolds1883, Salwen1980, Trefethen1993, Kerswell2005, Avila2011,
  Barkley2015}. Understanding the transition to a turbulent state in
  subcritical systems is a long-standing challenge in neutral fluids and, more
  recently, in fusion plasmas, where dramatically improved confinement is
  possible in the absence of turbulence. However, there is currently very
  little known about the transition to subcritical turbulence in
  fusion-relevant plasmas -- an issue we address in this thesis.

\section{Comparisons between simulations and experimental measurements}

  At the temperatures and densities found in fusion experiments, such as MAST,
  it can be shown that the conditions for a fluid description are rarely
  satisfied and that a kinetic description must be used
  (see~\cite{Schekochihin2007} for a recent discussion).
  Gyrokinetics~\cite{Frieman1982, Sugama1998,Abel2013} has emerged as the most
  appropriate first-principles description in the context of plasma turbulence
  in the core of tokamaks -- the focus of this thesis. The nonlinear gyrokinetic
  equation is derived via an asymptotic expansion of the Fokker-Planck
  equation. In general, it can only be solved numerically, and a number of
  codes have been developed for this purpose, for example,
  GS2~\cite{Kotschenreuther1995,Dorland2000} (the code used in this work),
  GENE~\cite{Jenko2000a,Gorler2011}, and GYRO~\cite{Candy2003}.  There has been
  a concerted effort to include in these codes a large number of physical
  effects relevant to experimental plasmas, such as realistic magnetic-surface
  geometries, arbitrary numbers of kinetic species, realistic Fokker-Planck
  collision operators, and so on. This has allowed the simulation of turbulence
  in fusion plasmas with sufficient realism to be compared quantitatively to
  experimental measurements. These ``local'' codes, such as GS2, take as input
  the values and first derivatives of equilibrium quantities at a
  particular radial location, and predict a host of quantities that could
  theoretically be measured by an experimental diagnostic, for example, the
  flux of particles, momentum, and heat, or density, flow, and temperature
  fluctuations.

  In conjunction with increasingly realistic modelling, more sophisticated
  diagnostic techniques have been designed, which aid in our understanding of
  the conditions inside the reactor and allow us to make comparisons with
  modelling results. Initial comparisons between simulations and experiments
  were limited to averaged quantities such as the transport of particles,
  momentum, and heat. More recently, diagnostics that measure fluctuating
  quantities have been developed: beam emission spectroscopy (BES) that
  measures ion-scale density fluctuations~\cite{McKee2003,Field2009,Field2012,
  Smith2010}; Doppler reflectometry that measures density fluctuation at scales
  intermediate to ion and electron scales, rotation velocity of turbulent
  structures, and the radial electron field~\cite{Hennequin2006,Hillesheim2012,
  Hillesheim2015}; scattering diagnostics that measure electron scale density
  fluctuations~\cite{Mazzucato2008}; and correlation electron cyclotron
  emission (CECE) diagnostics~\cite{White2008b} that measure electron
  temperature fluctuations. Measurements of fluctuating quantities allow
  more extensive quantitative comparisons between experiment and simulations.
  However, meaningful comparisons are only possible via the use of ``synthetic
  diagnostics'' that take account of the measurement characteristics of the
  particular diagnostic and modify the simulation output
  accordingly~\cite{White2008b, Holland2009, Shafer2012, Ghim2012,
  Field2014,Fox2016}.

  In this work, we will focus on measurements from the BES system on
  MAST~\cite{Field2009,Field2012}. The BES diagnostic
  infers ion-scale turbulent density fluctuations from D$_\alpha$ emission (the
  emission of light resulting from the dominant transition of ionised
  deuterium), which is generated as a result of the injection of neutral
  particles by the NBI system. The BES diagnostic takes measurements in a
  two-dimensional radial-poloidal plane. In the case of an ITG- or TEM-unstable
  plasma, the characteristic turbulence length scale in the direction
  perpendicular to the magnetic field is of the order of the ion
  gyroradius~\cite{Barnes2011}: $l_\perp \sim \rho_i$, and it is these
  turbulent structures that BES is designed to measure. Such two-dimensional
  measurements provide insight into the structure of turbulence, and they have
  allowed turbulence to be visualised for the first time.  From the BES
  measurements, it is possible to estimate the turbulence correlation time
  $\tau_c$ via the cross-correlation time delay (CCTD)
  method~\cite{Durst1992,Ghim2012,Fox2016}, the radial and poloidal correlation lengths
  $l_R$ and $l_Z$, and the relative density-fluctuation field $\delta
  n_i/n_i$~\cite{Ghim2013,Fox2016}. A recent experimental
  study~\cite{Field2014} used the BES diagnostic to measure turbulent density
  fluctuations in the outer core of a MAST L-mode plasma and compared with
  global gyrokinetic simulations. While there was some agreement at mid-radius,
  serious discrepancies remained at outer radii, where ITG turbulence may not
  be fully suppressed by flow shear, in predictions of turbulence
  characteristics, such as the ion heat flux and turbulence correlation time.
  In this work, we will study turbulence in the outer-core region of the MAST
  discharge in Ref.~\cite{Field2014} using high-resolution local gyrokinetic
  simulations.

  In simulating experimentally-relevant plasmas using gyrokinetic
  codes, we aim to achieve the following. First, we want to better understand
  the physical mechanisms that most affect influence turbulence and its
  associated enhanced transport. Specifically, how do turbulence characteristics
  (such as transport, spatial scales, time scales, etc.) change in the outer
  core of MAST with the ITG and the flow shear? Secondly, in light of newly
  available experimental data from the MAST BES diagnostic~\cite{Field2014}, do
  the turbulence characteristics found in local gyrokinetic GS2 simulations
  agree with experimental BES measurements within the experimental
  uncertainties of the ITG and flow shear? Such quantitative comparisons with
  experimental results are essential in developing confidence in our
  theoretical models and numerical implementations. In understanding the
  properties of turbulence, we ultimately aim to guide the optimisation and
  design of future experiments and fusion reactors to mitigate or eliminate the
  causes of turbulence.

\section{Thesis outline}
  The rest of this thesis is organised as follows. In
  Chapter~\ref{sec:exp_setup}, we give an overview of MAST, the MAST BES
  diagnostic, and discuss the discharge we will be considering in this work. In
  Chapter~\ref{sec:gk_modelling}, we give an overview of gyrokinetics and
  the GS2 code that we use to solve the system of gyrokinetic equations in an
  axisymmetric torus. We discuss the toroidal geometry that is appropriate to
  tokamaks and the relevant approximations in this setting that are used to
  derive the gyrokinetic equation. We discuss details of the numerical
  implementation of GS2 pertinent to our study, such as the extraction of
  geometric information from experimental output, the calculation of collision
  frequencies, and the implementation of flow shear and hyperviscosity.
  Finally, we detail the numerical setup for our study, including the extent of
  our parameter scan, the physics we have included, the approximations we have
  made, the numerical resolutions we have used (along with a justification for
  choosing them), and lastly a comprehensive table of parameters extracted from
  the experiment required to run a numerical study.

  The main results of this work are split into two parts. In
  Chapter~\ref{sec:nl}, we will study, numerically, the effect on turbulence in
  the outer core of MAST, when the ITG and perpendicular flow shear are
  changed.  We will show that turbulent
  transport is stiff with respect to changes in the ITG, but also that the
  perpendicular flow shear is effective at suppressing turbulence.
  Performing an extensive parameter scan in these two equilibrium parameters,
  we map out the turbulence threshold (the line separating regions of enhanced
  turbulent transport and neoclassical transport) and show that the
  experimental level of ion heat flux corresponds to values of the ITG and
  flow shear
  close to the turbulence threshold.  We discover that the system is
  subcritical and that large initial perturbations are required to ignite
  turbulence, a phenomenon not previously observed for experimentally-relevant
  plasmas.  Furthermore, we discover that the near-threshold state is one
  dominated by long-lived, coherent structures, which exist against a
  background of much smaller fluctuations. We argue that these structures are a
  direct consequence of the subcritical nature of the system, which
  concentrates plasma into these structures as a means of maintaining the
  minimum amplitude below which fluctuations would be quenched.  Sufficiently
  far from the turbulence threshold in parameter space, we recover a more
  conventional turbulent state consisting of many strongly interacting eddies
  simultaneously being sheared apart by the perpendicular flow shear.  The
  number and amplitude of the above structures are shown to be functions of the
  distance from the turbulence threshold in the parameter space of ITG and flow
  shear -- both increasing as the ITG is increased or as the flow shear is
  decreased.  In this way, we identify three distinct regions of parameter
  space: the region of no turbulence (where transport would be neoclassical); a
  marginally unstable, intermediate state between the non-turbulent and
  fully turbulent states, characterised by long-lived, coherent structures; and a
  conventional chaotic, turbulent state far from the turbulence threshold.

  In Chapter~\ref{sec:struc_of_turb}, we make direct comparisons
  with experimental measurements from the BES. We review the existing methods
  for performing a correlation analysis of BES measurements and discuss the
  differences in applying such an analysis to our simulations. Additional
  analyses are performed, such as calculating the parallel correlation length
  -- something not currently experimentally measured. We then proceed to
  present two types of correlation analysis of our simulations: with and
  without a synthetic diagnostic. We show that there is reasonable agreement with
  experimental measurements in the case of analysis with the synthetic
  diagnostic. However, radial correlation lengths predicted by GS2 are shown to
  be below the resolution threshold of the BES diagnostic (an issue discussed in
  detail in Ref.~\cite{Fox2016}). Our analysis without the synthetic
  diagnostic shows that the synthetic diagnostic has a measurable effect on
  several turbulence characteristics, including the poloidal correlation length
  and the fluctuation amplitude, consistent with work in Ref.~\cite{Fox2016}.
  Finally, we present the correlation properties as functions of the ion
  heat flux and show that the structure of the turbulence in our
  simulations is effectively only a function of this parameter, which measures
  the distance to the turbulence threshold.

  Our discussion and conclusions are presented in Chapter~\ref{sec:conclusion},
  along with suggestions for future work.

\chapter{MAST experimental configuration}
\label{sec:exp_setup}

\section{The Mega Ampere Spherical Tokamak}
\label{sec:mast}

  MAST~\cite{Darke1994,Morris2012} is a medium-sized, low-aspect ratio
  spherical tokamak. Along with the National Spherical Torus Experiment
  Upgrade (NSTX-U)\cite{Ono2000, Menard2012} in Princeton, USA, it is one of
  the leading spherical tokamaks: a novel reactor design that is under active
  research as an alternative to conventional high-aspect ratio
  reactors~\cite{Peng2000}, such as the Joint European Torus (JET). Spherical
  tokamaks offer a number of potential advantages over conventional tokamaks
  that could make them suitable as fusion reactors~\cite{Sykes1999,Peng2000}:
  \begin{inparaenum}[(i)]
    \item lower cost due to compact design;
    \item higher plasma $\beta$ (ratio of plasma pressure to magnetic
      pressure), as a result of more efficient confinement;
    \item superconducting magnets are not strictly needed due to already high
      plasma $\beta$;
    \item in the case of MAST, high rotation and resulting sheared flows can
      suppress turbulence.
  \end{inparaenum}
  The energy confinement of spherical tokamaks has been shown to be comparable to
  conventional tokamaks~\cite{Counsell2002} and promisingly, spherical tokamaks
  show more favourable energy confinement scalings with experimental
  parameters~\cite{Kaye2007,Valovic2009}.

\begin{figure}[t]
  \begin{minipage}[t]{\linewidth}
  \begin{minipage}{0.47\linewidth}
    \centering
    \includegraphics[width=\linewidth]{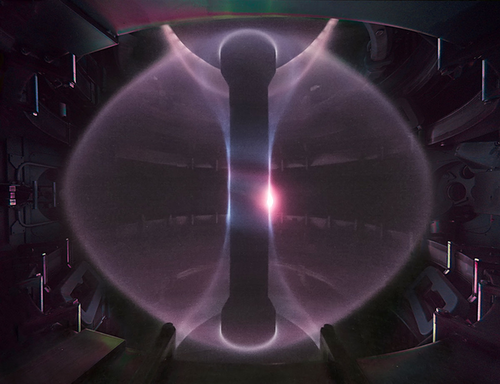}
    \captionof{figure}[Image of MAST plasma]{
      Image of the MAST tokamak in operation highlighting the compact D-shaped
      geometry aided by a narrow central magnet column. The bright spot is
      the location at which deuterium fuel is pumped into the plasma and is
      ionised.
    }
    \label{fig:mast}
  \end{minipage}
  \hfill
  \begin{minipage}{0.47\linewidth}
    \vspace{-50pt}
    \centering
    {\renewcommand{\arraystretch}{1.4}%
    \begin{tabular}{r c}
      \toprule
      Major radius $R$ & $\approx 0.9$~m \\
      Minor radius $a$ & $\approx 0.6$~m \\
      Aspect ratio $A = R/a$ & $\approx 1.5$ \\
      Plasma current $I_p$ & $1.3$ MA \\
      Magnetic field $B$ & $0.5$ T \\
      Pulse duration & $0.5$ s \\
      Power injected & $ 3.8$~MW \\
      \bottomrule
    \end{tabular}}
    \captionof{table}{Experimental parameters for the MAST experiment.}
    \label{tab:mast_params}
  \end{minipage}
  \end{minipage}
  \end{figure}

  \Figref{mast} shows an image of a typical MAST
  plasma\footnote{\url{http://www.opendata.ccfe.ac.uk/mast/}}
  and Table~\ref{tab:mast_params} gives some important parameters of the MAST
  device\cite{Morris2012}.
  MAST is equipped with two NBI systems directed tangential to the flux
  surfaces that heat the plasma, with injected power up to $3.8$~MW. The NBI
  system also gives rise to toroidal rotation and differential toroidal
  rotation, which will be the subject of our investigation. MAST is one of the
  more well-diagnosed tokamaks in operation, making it an ideal experiment to
  test theoretical predictions against. We detail the range of diagnostics that
  have allowed us to perform our numerical transport study in
  Section~\ref{sec:exp_profiles} and the review the BES system in
  Section~\ref{sec:bes} with which we compared our simulation results.

\section{Equilibrium profiles}
\label{sec:exp_profiles}

\subsection{MAST discharge \#27274}
  In this work, we will focus on the MAST discharge \#27274, which forms part of a
  set of three nominally identical experiments (i.e., identical profiles and
  equilibria) previously reported in Ref.~\cite{Field2014}, differing only in
  the radial viewing location of the BES system. The three discharges are
  \#27272, \#27268, and \#27274, wherein the centre of the BES was located at
  $1.05$~m, $1.2$~m, and $1.35$~m, respectively. Each discharge produced an
  L-mode plasma with strong toroidal rotation and, hence, flow shear
  perpendicular and parallel to the magnetic field~\cite{Field2014}. The MAST
  BES diagnostic~\cite{Field2009,Field2012} observes an area of approximately
  $16\times8$~cm$^2$ in the radial and poloidal directions, respectively,
  corresponding to approximately one third of the minor radius of the plasma.
  Therefore, the combination of the above discharges provided a complete radial
  profile of BES measurements on the outboard side of the plasma.

  Previous investigations of MAST turbulence for similar
  configurations~\cite{Roach2009, Field2011}, found that ion-scale turbulence is
  suppressed in the core region by flow shear. However, flow shear is weaker in
  the outer-core region,
  which may still be unstable to ITG modes, making it possibly to study
  ion-scale turbulence. Turbulence is also driven partly by trapped electron
  modes (TEMs) and the electron temperature gradient (ETG). In this work, we
  will restrict our attention to the time window $t=0.250\pm0.002$~s and the
  radial location $r = D/2a = 0.8~(\equiv r_0)$ of \#27274, where $D$ is the
  diameter of the flux surface and $a$ is the half diameter of the last closed
  flux surface (LCFS), both measured at the height of the magnetic axis.
  Importantly, there is no large-scale and disruptive magnetohydrodynamic (MHD)
  activity at this time and radial location~\cite{Field2014}; as such activity
  would interfere with the quality of BES measurements. The normalized radial
  location $r=0.8$ corresponds to a major radius of approximately $1.32$~m and,
  therefore, falls within the viewing area covered by discharge \#27274 [see
  \figref{flux_surfaces}].

\subsection{A note on radial grids}

  We use $r=D/2a$ as the definition of the radial location because it corresponds
  to the radial coordinate used by the Miller specification of flux-surface
  geometry~\cite{Miller1998} (see Section~\ref{sec:miller}). In terms of other
  commonly used radial coordinates, $r = 0.8$
  corresponds to $\rho_\mathrm{tor} =
  \sqrt{\psi_\mathrm{tor}/\psi_\mathrm{tor,LCFS}} = 0.7$ where
  \begin{equation}
    \psi_\mathrm{tor} = {\qty(\frac{1}{2\pi})}^2 \int_0^V dV \vb*{B} \cdot \nabla \phi
    \label{psi_tor}
  \end{equation}
  is the toroidal magnetic flux, $V$ is the volume enclosed by the flux surface,
  $\vb*{B}$ is the magnetic field, $\phi$ is the toroidal angle, and
  $\psi_\mathrm{tor,LCFS}$ is the toroidal flux enclosed by the last closed
  flux surface [see \figref{flux_surfaces}]. In terms of the poloidal magnetic
  flux, $\rho_\mathrm{pol} = \sqrt{\psi_\mathrm{pol}/\psi_\mathrm{pol,LCFS}} =
  0.87$, where
  \begin{equation}
    \psi \equiv \psi_\mathrm{pol} = {\qty(\frac{1}{2\pi})}^2 \int_0^V dV \vb*{B} \cdot \nabla \theta
    \label{psi_pol}
  \end{equation}
   is the poloidal magnetic flux, $\theta$ is the poloidal angle, and
   $\psi_\mathrm{pol,LCFS}$ is the poloidal flux enclosed by the LCFS.

\subsection{MAST profile diagnostics}
\label{sec:mast_diagnostics}
  \begin{figure}[t]
    \centering
    \begin{subfigure}[t]{0.55\linewidth}
      \includegraphics[width=\linewidth]{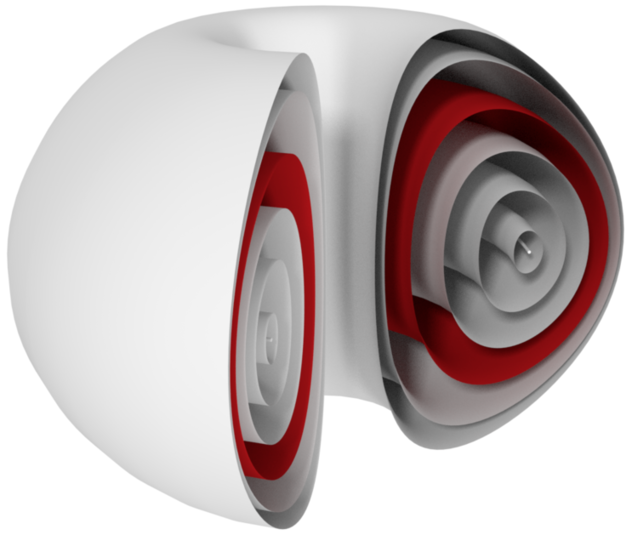}
      \caption{}
      \label{fig:nested_flux_surfaces}
    \end{subfigure}
    \begin{subfigure}[t]{0.37\linewidth}
      \includegraphics[width=\linewidth]{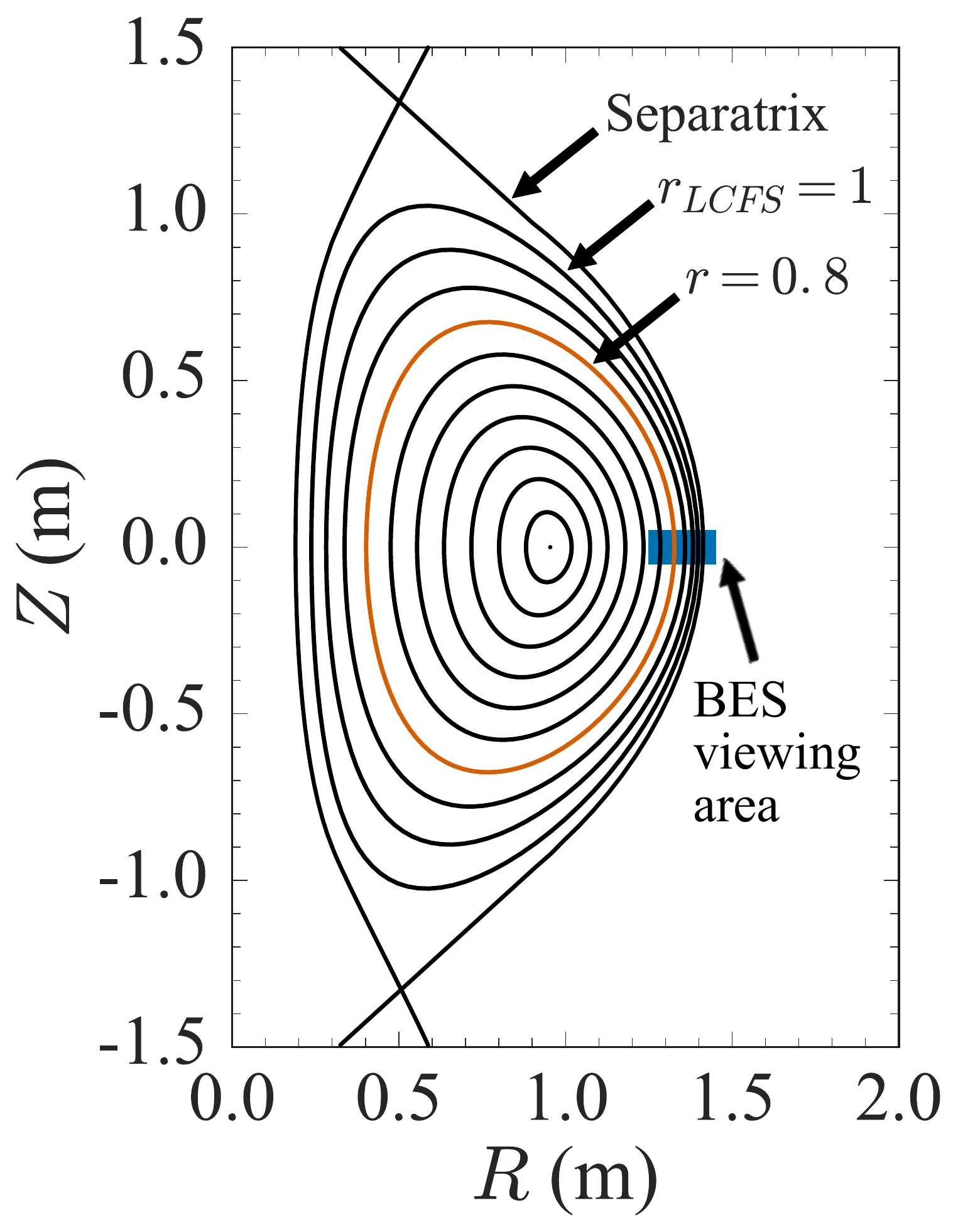}
      \caption{}
      \label{fig:flux_surfaces}
    \end{subfigure}
    \caption[MAST flux surfaces]{\subref*{fig:nested_flux_surfaces} A
      three-dimensional view of the nested flux surfaces.
      \subref*{fig:flux_surfaces} The poloidal cross-section of the magnetic
      geometry along with the LCFS and the separatrix, which separates closed
      field lines from open ones. The flux surface of interest is at $r = 0.8$,
      shown in red. It was chosen so that this surface intersects the BES
      measurement plane for discharge \#27274.  The blue shaded region
      indicates the location of the BES diagnostic.
    }
  \end{figure}

  MAST has a range of high-quality diagnostics, which allow us to extract the
  equilibrium parameters required to conduct a numerical transport study. The ion
  temperature, $T_i$, and toroidal flow velocity, $u_\phi = R \omega$, where
  $\omega$ is the toroidal angular rotation frequency, were obtained from
  charge-exchange-recombination spectroscopy (CXRS) measurements of C$^{+6}$
  impurity ions with a spatial resolution of $\sim 1$~cm~\cite{Conway2006}. The
  electron density, $n_e$, and temperature, $T_e$, were obtained from a
  Thomson-scattering (TS) diagnostic~\cite{Scannell2010} with resolution comparable
  to the CXRS system. These measured profiles were mapped onto flux-surface
  coordinates by the pre-processing code $MC^\mathit{3}$ using a
  motional-Stark-effect-(MSE)-constrained EFIT equilibrium~\cite{Lao1985}.
  These equilibrium profiles served as input to the transport analysis
  code TRANSP\footnote{\url{http://w3.pppl.gov/transp/}}~\cite{Hawryluk1981},
  which calculates the transport coefficients of particles, momentum, and heat.
  \Figref{nested_flux_surfaces} shows a three-dimensional view of the
  axisymmetric nested flux surfaces and \figref{flux_surfaces} shows the
  poloidal cross-section of the flux surfaces extracted from an EFIT
  equilibrium. The $r=0.8$ surface is highlighted in both plots. The
  measurement window of the BES diagnostic for discharge \#27274 is also shown
  in \figref{flux_surfaces}.
  The chosen flux surface at $r = 0.8$ intersects the measurement window
  at the outboard midplane, allowing direct comparisons between our numerical
  predictions of turbulence and experimental measurements.

\subsection{Equilibrium profiles}

  The important experimental quantities needed to conduct a numerical
  study are the radial profiles of $T_i$, $T_e$, $n_i$ (the ion density), $n_e$, and
  $\omega$. MAST does not take direct measurements of $n_i$, but we assume that
  it is equal to $n_e$, as measured by the TS diagnostic, due
  to quasineutrality. As explained in Section~\ref{sec:gk_theory}, it is
  assumed in the local formulation of gyrokinetics that only the physical
  quantities (and their first derivatives) at the location of the flux tube
  determine the characteristics of the turbulence. Therefore, to conduct a
  numerical study of turbulence we need only the equilibrium values and their
  first derivatives (or for some quantities their gradient length scales) at
  $r=0.8$ to simulate turbulence at that radius. The appropriate (normalised)
  gradient length scales of $T_i$, $T_e$, and $n_e$, and flow shear (gradient
  of $\omega$) are
  \begin{align}
      \label{ti_prime}
      \frac{1}{L_{Ti}} &= - \dv{\ln T_i}{r} \equiv \kappa_T, \\
      \label{te_prime}
      \frac{1}{L_{Te}} &= - \dv{\ln T_e}{r}, \\
      \label{ne_prime}
      \frac{1}{L_{ne}} &= - \dv{\ln n_e}{r}, \\
      \label{flow_shear}
      \gamma_E &= \frac{r_0}{q_0} \dv{\omega}{r} \frac{a}{v_{\mathrm{th}i}},
  \end{align}
  where $q(\psi) = \pdv*{\psi_\mathrm{tor}}{\psi_{\mathrm{pol}}}$ is the safety
  factor and $q_0$ is the value at $r_0$, $v_{\mathrm{th}i} = \sqrt{2T_i/m_i}$
  is the ion thermal velocity, and $m_i$ is the mass of the ion species
  (deuterium).  In a tokamak, the safety factor is approximately
  $q(\psi) \sim (r/R)(B_\phi/B_\theta)$, where $B$ is the magnetic field, $B_\theta
  = |\nabla \psi|/R$ is the poloidal component of $B$, and $R$ is the major
  radius at the location of the flux surface at the outboard midplane.
  The flow shear parameter $\gamma_E$ can be interpreted as the
  (non-dimensionalised) shear of the component of the toroidal shear
  perpendicular to the local magnetic field. The sign of $\gamma_E$ is
  determined in Section~\ref{sec:sign_omega}, given that $\omega$ can be
  positive or negative depending on the sign convention used.

\begin{figure}
    \centering
    \begin{subfigure}[t]{0.35\textwidth}
      \includegraphics[width=\textwidth]{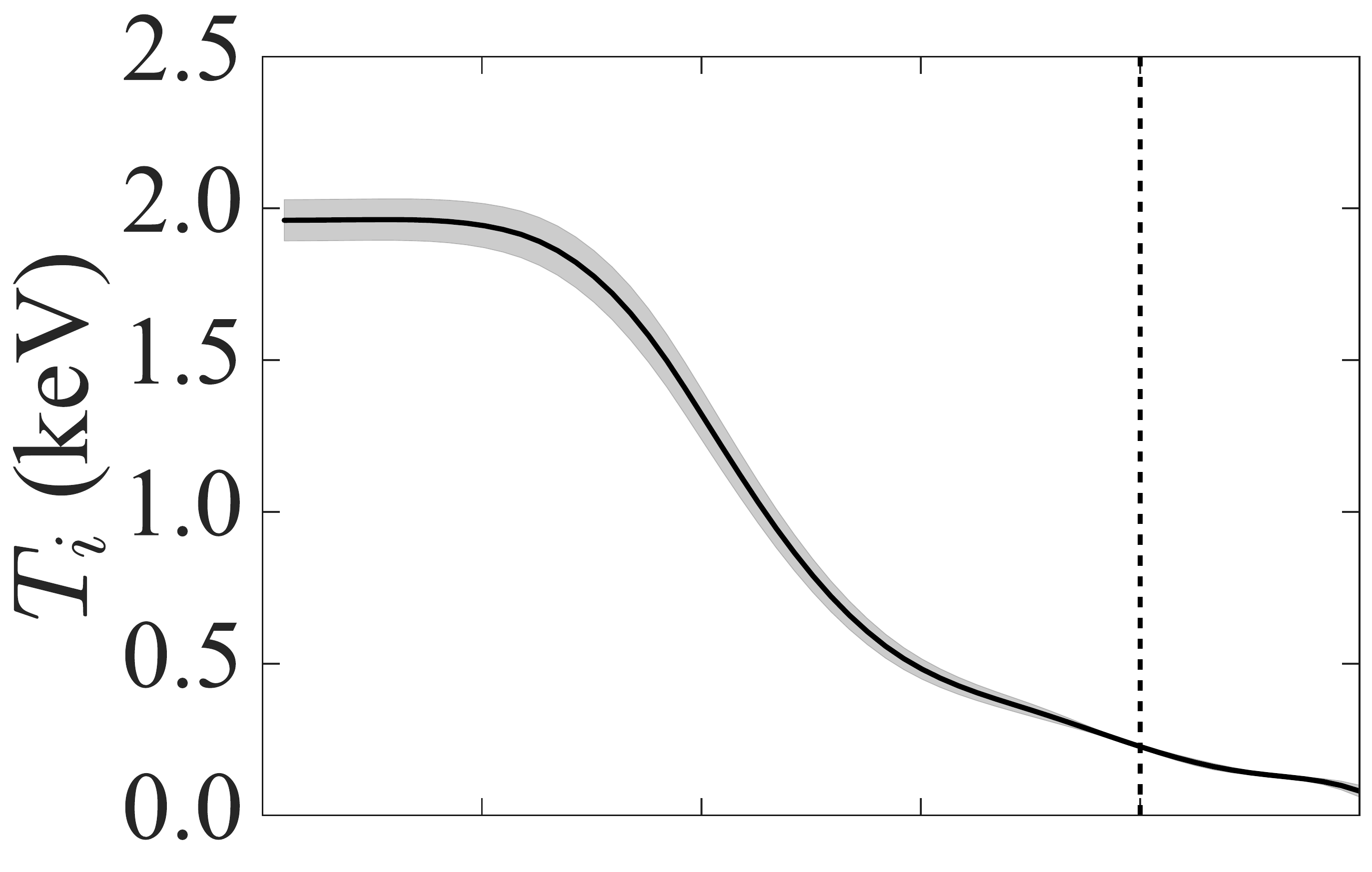}
      \caption{}
      \label{fig:ti}
    \end{subfigure}
    \begin{subfigure}[t]{0.35\textwidth}
      \includegraphics[width=\textwidth]{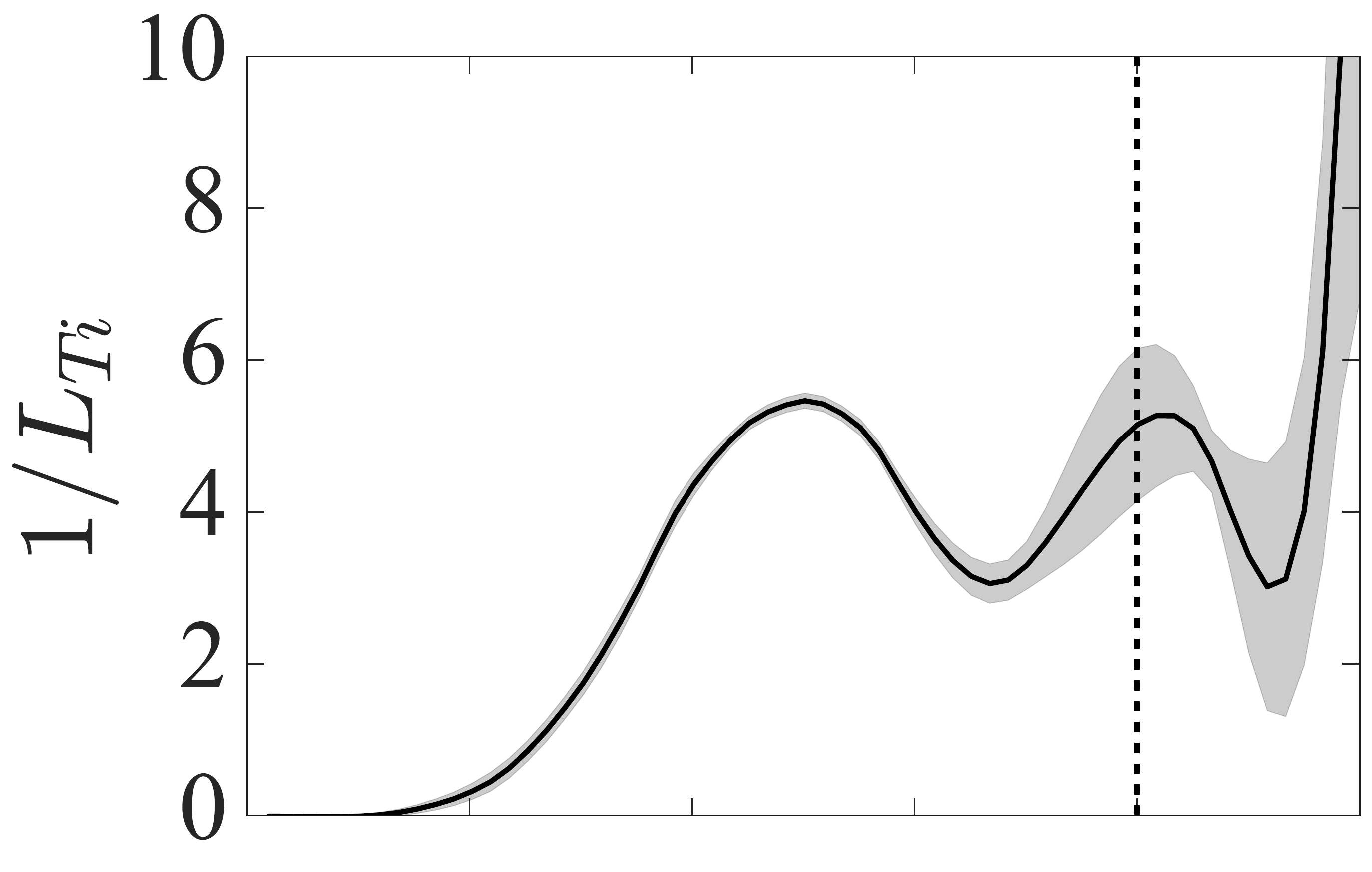}
      \caption{}
      \label{fig:ti_prime}
    \end{subfigure}
    \begin{subfigure}[t]{0.35\textwidth}
      \includegraphics[width=\textwidth]{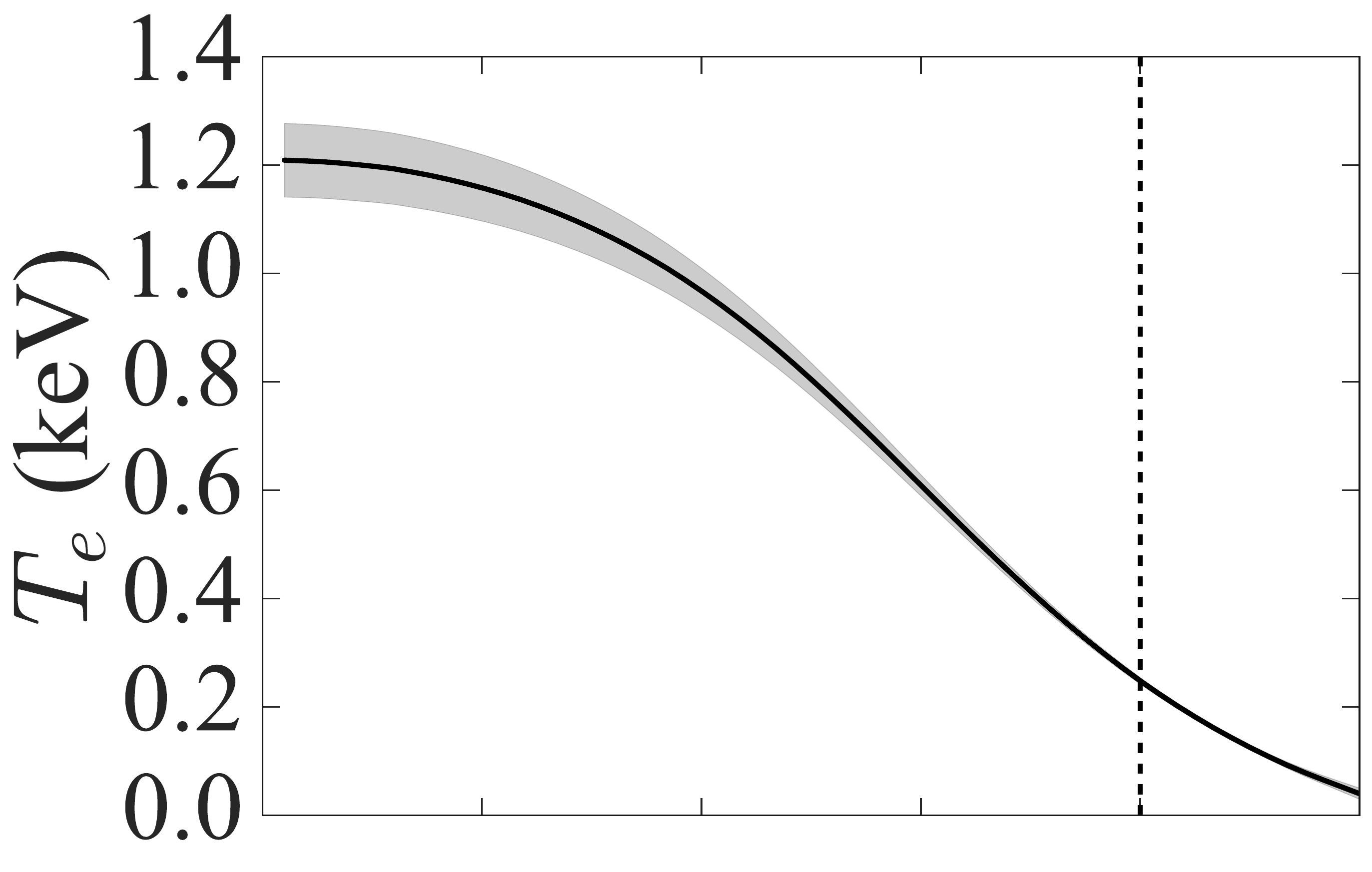}
      \caption{}
      \label{fig:te}
    \end{subfigure}
    \begin{subfigure}[t]{0.35\textwidth}
      \includegraphics[width=\textwidth]{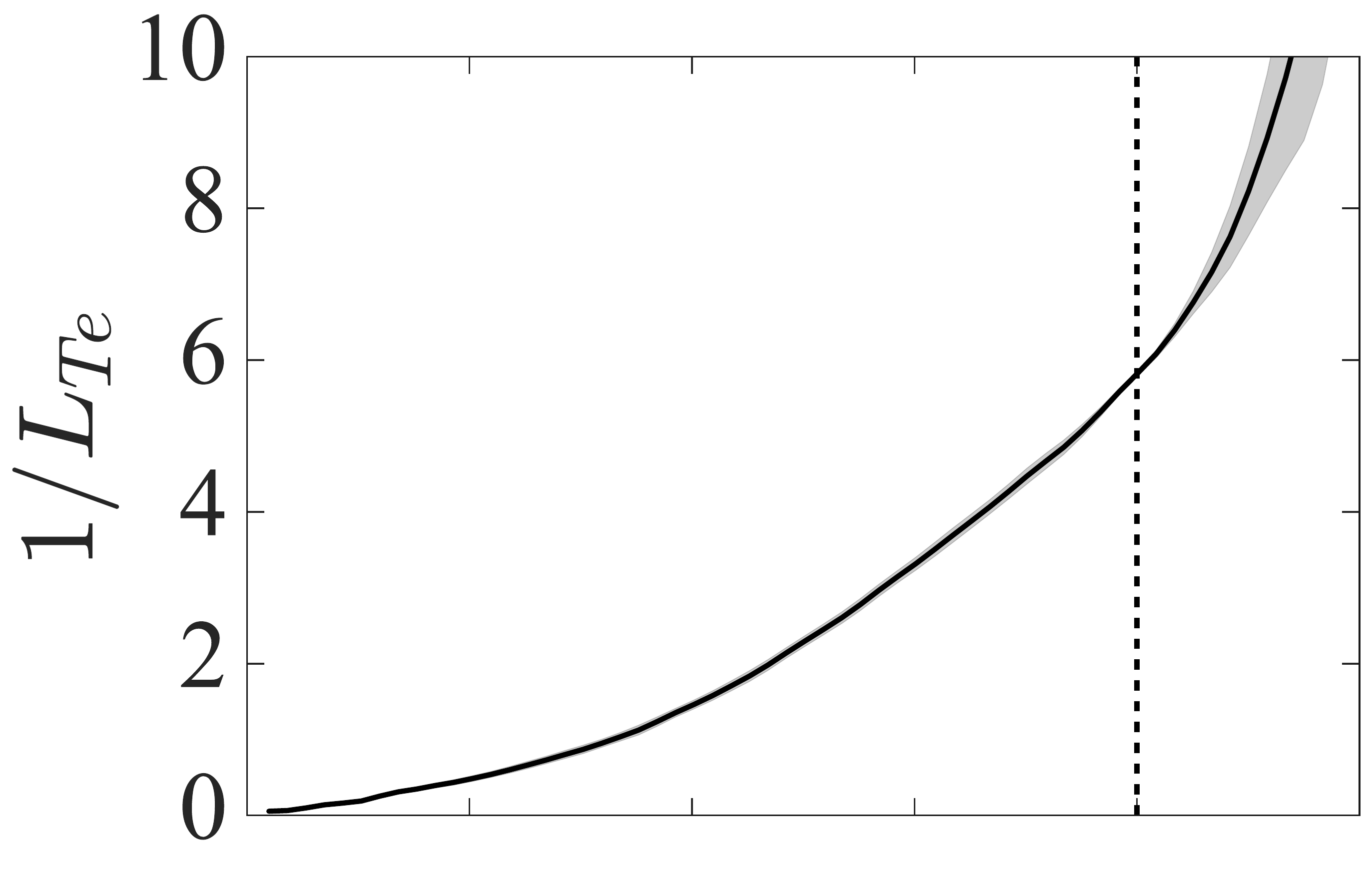}
      \caption{}
      \label{fig:te_prime}
    \end{subfigure}
    \begin{subfigure}[t]{0.35\textwidth}
      \includegraphics[width=\textwidth]{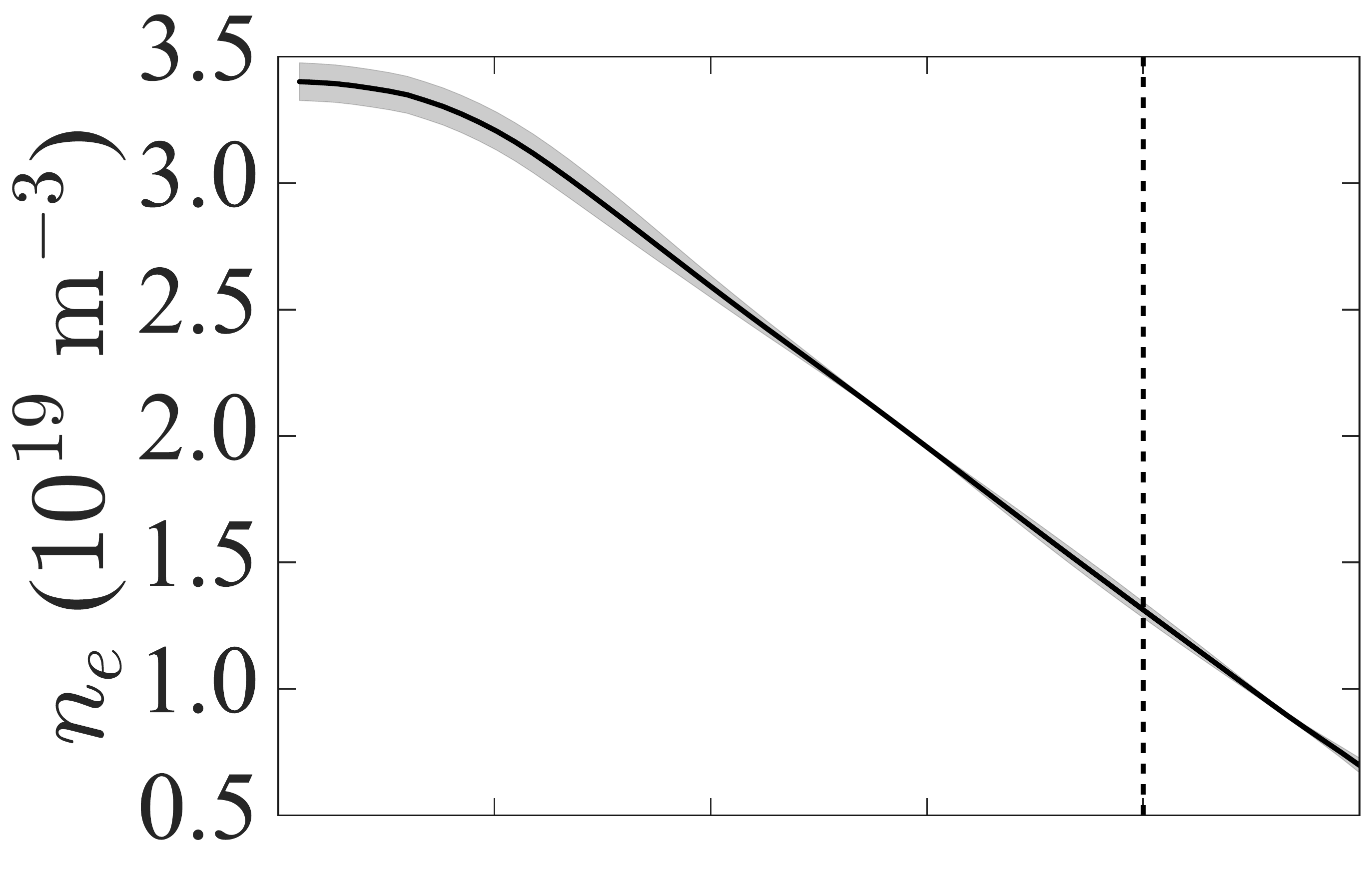}
      \caption{}
      \label{fig:ne}
    \end{subfigure}
    \begin{subfigure}[t]{0.35\textwidth}
      \includegraphics[width=\textwidth]{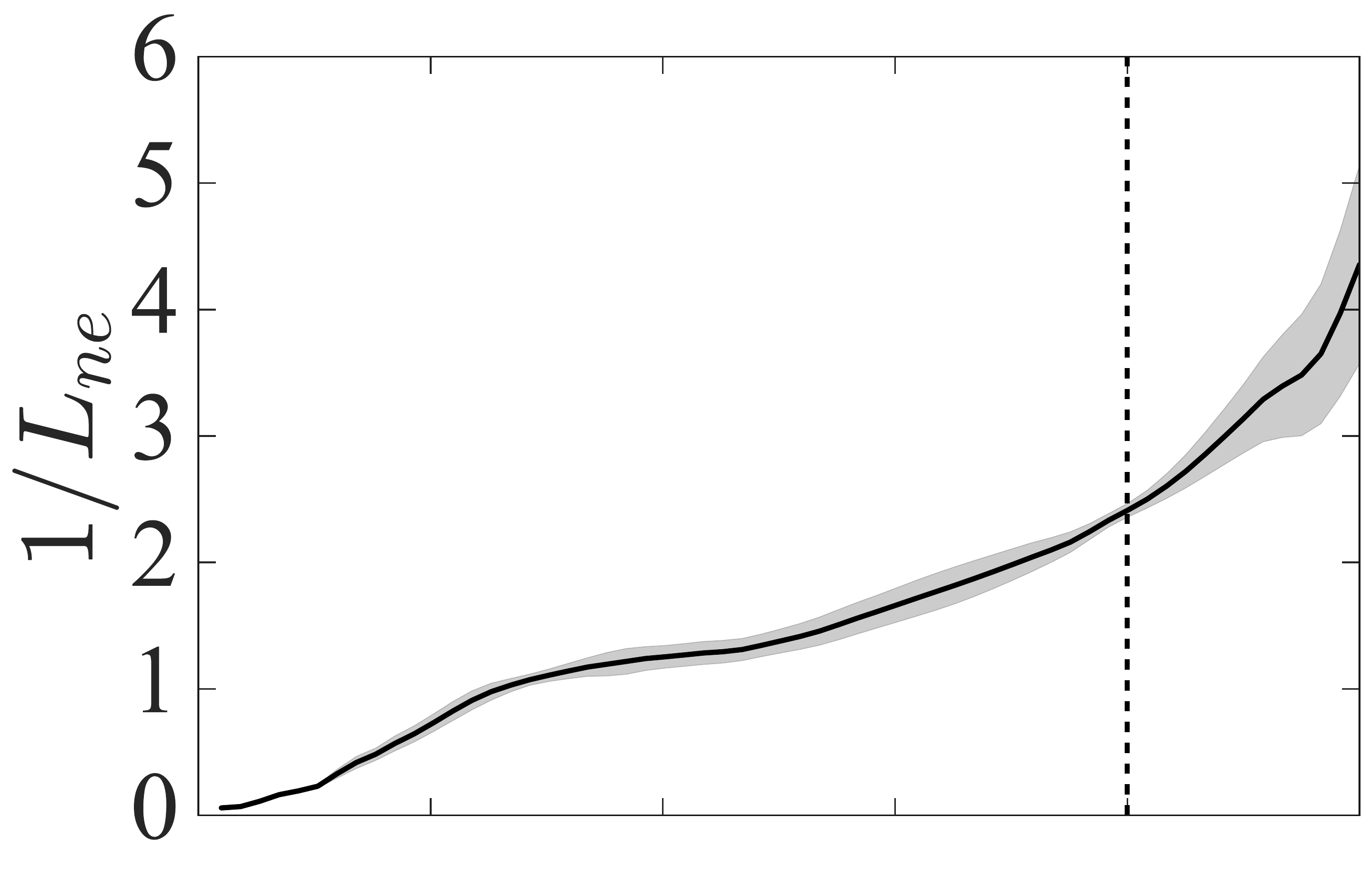}
      \caption{}
      \label{fig:ne_prime}
    \end{subfigure}
    \begin{subfigure}[t]{0.37\textwidth}
      \includegraphics[width=\textwidth]{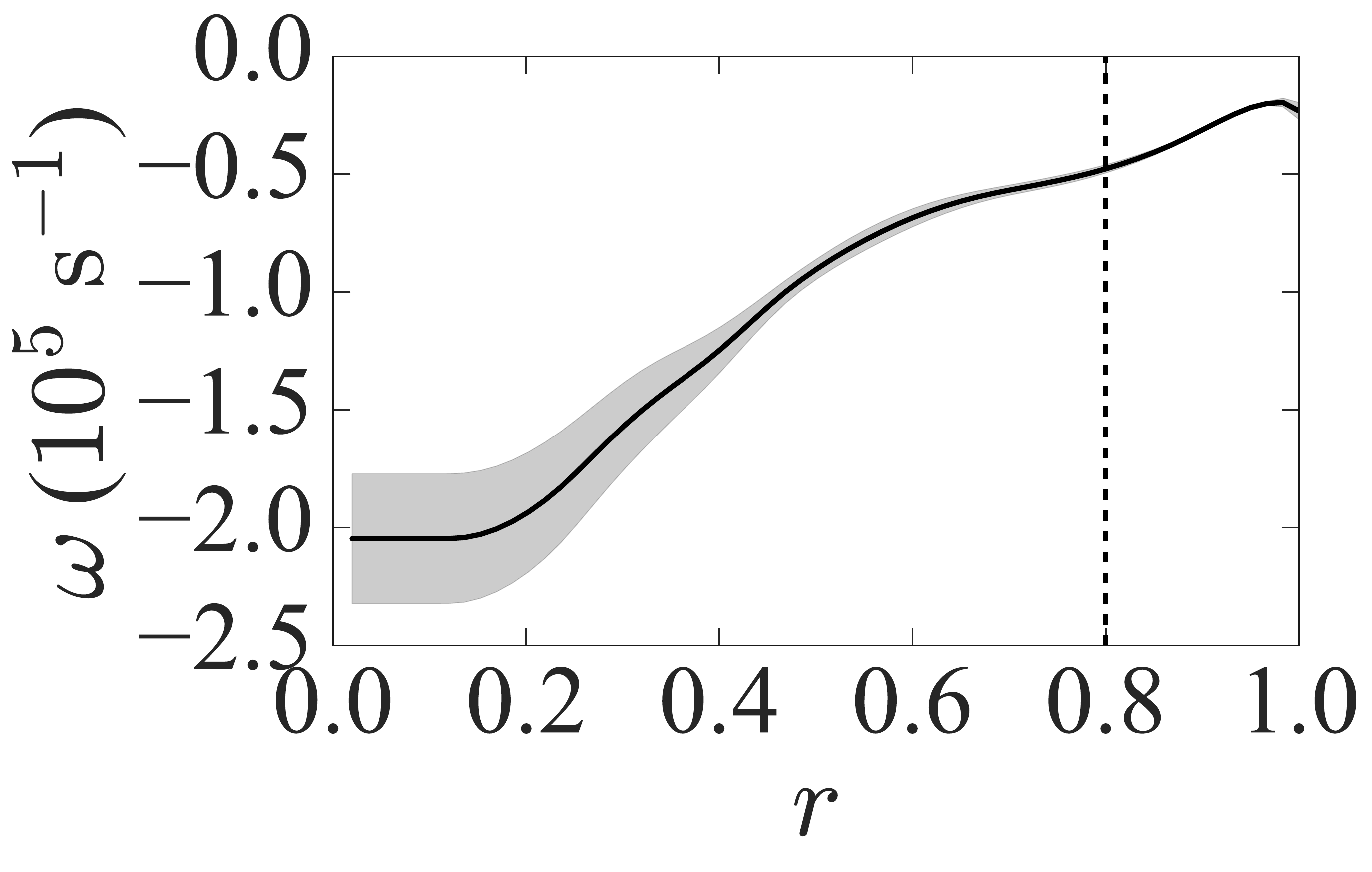}
      \caption{}
      \label{fig:omega}
    \end{subfigure}
    \begin{subfigure}[t]{0.37\textwidth}
      \includegraphics[width=\textwidth]{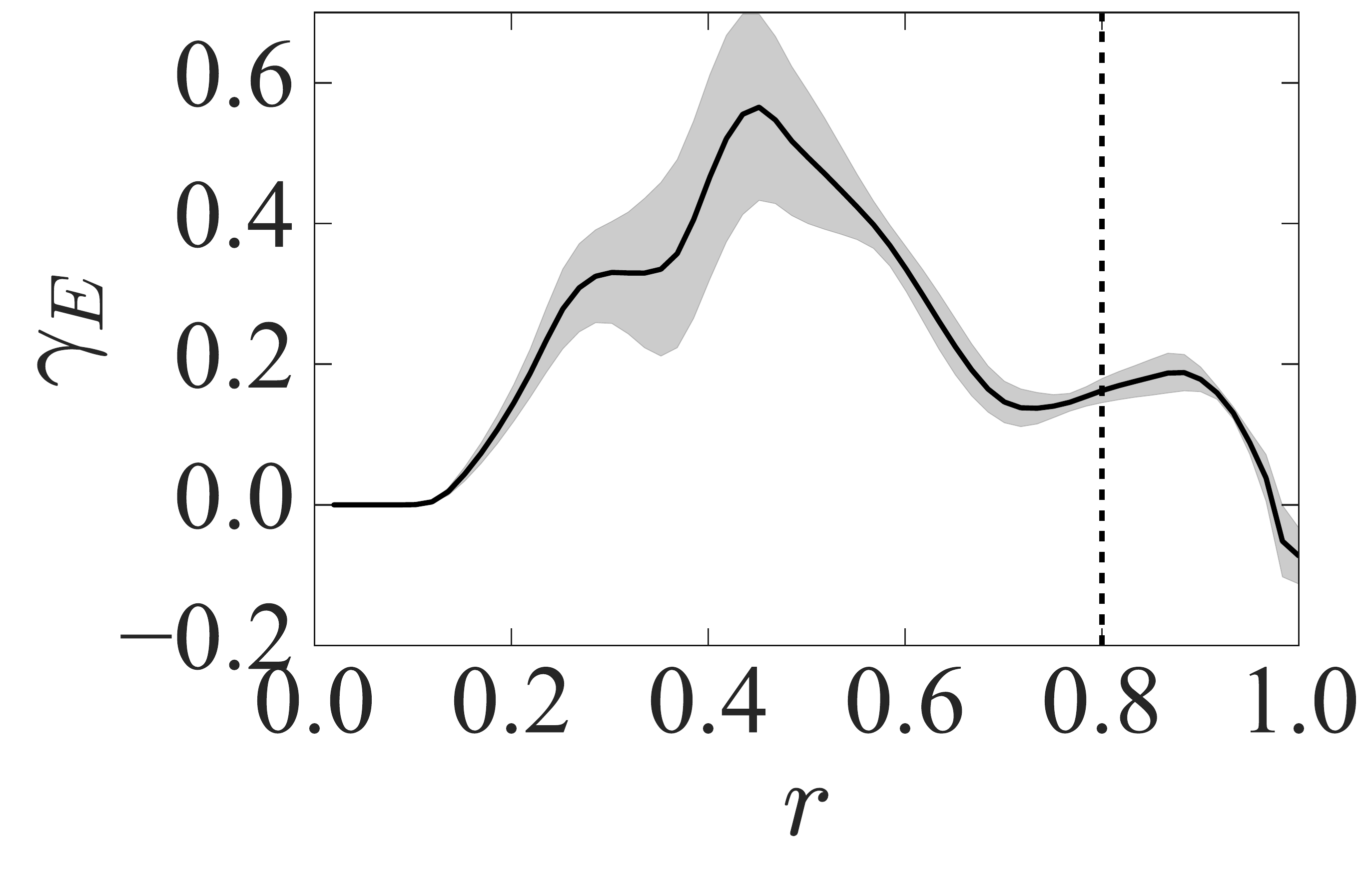}
      \caption{}
      \label{fig:g_exb}
    \end{subfigure}
    \caption[Experimental profiles]{Radial profile measurements from MAST
      discharge \#27274 (see Section~\ref{sec:mast_diagnostics}) of
      \subref*{fig:ti} the ion temperature, $T_i$,
      \subref*{fig:ti_prime} the ion temperature gradient, $1/L_{Ti}$,
      calculated using~\eqref{ti_prime},
      \subref*{fig:te} the electron temperature, $T_e$,
      \subref*{fig:te_prime} the electron temperature gradient, $1/L_{Te}$,
      calculated using~\eqref{te_prime},
      \subref*{fig:ne} the electron density, $n_e$,
      \subref*{fig:ne_prime} the electron density gradient, $1/L_{ne}$,
      calculated using~\eqref{ne_prime},
      \subref*{fig:omega} the toroidal angular frequency, $\omega$, and
      \subref*{fig:g_exb} the flow shear, $\gamma_E$, calculated
      using~\eqref{flow_shear}. The dashed line in each plot indicates $r=0.8$
      and the shaded regions indicate the standard deviation of the profiles
      over a $20$-ms time window around $t = 0.25$~s.
    }
    \label{fig:profiles}
  \end{figure}

  The left-hand column of \figref{profiles} shows the radial profiles of $T_i$,
  $T_e$, $n_e$, and $\omega$ (with the sign determined as in
  Section~\ref{sec:sign_omega}), as functions of $r$. The gradient scale
  lengths~\eqref{ti_prime}--\eqref{ne_prime} and flow shear~\eqref{flow_shear}
  are plotted as functions of $r$ in the right-hand column
  in~\figref{profiles}. The dashed lines indicate $r=0.8$ and the equilibrium
  values at this radial location are given in Table~\ref{tab:equil_params}. The
  profiles in \figref{profiles} represent a $20$-ms time average around $t =
  0.25$~s and the shaded areas indicate the standard deviations.

\begin{figure}[t]
    \centering
    \includegraphics[width=0.6\linewidth]{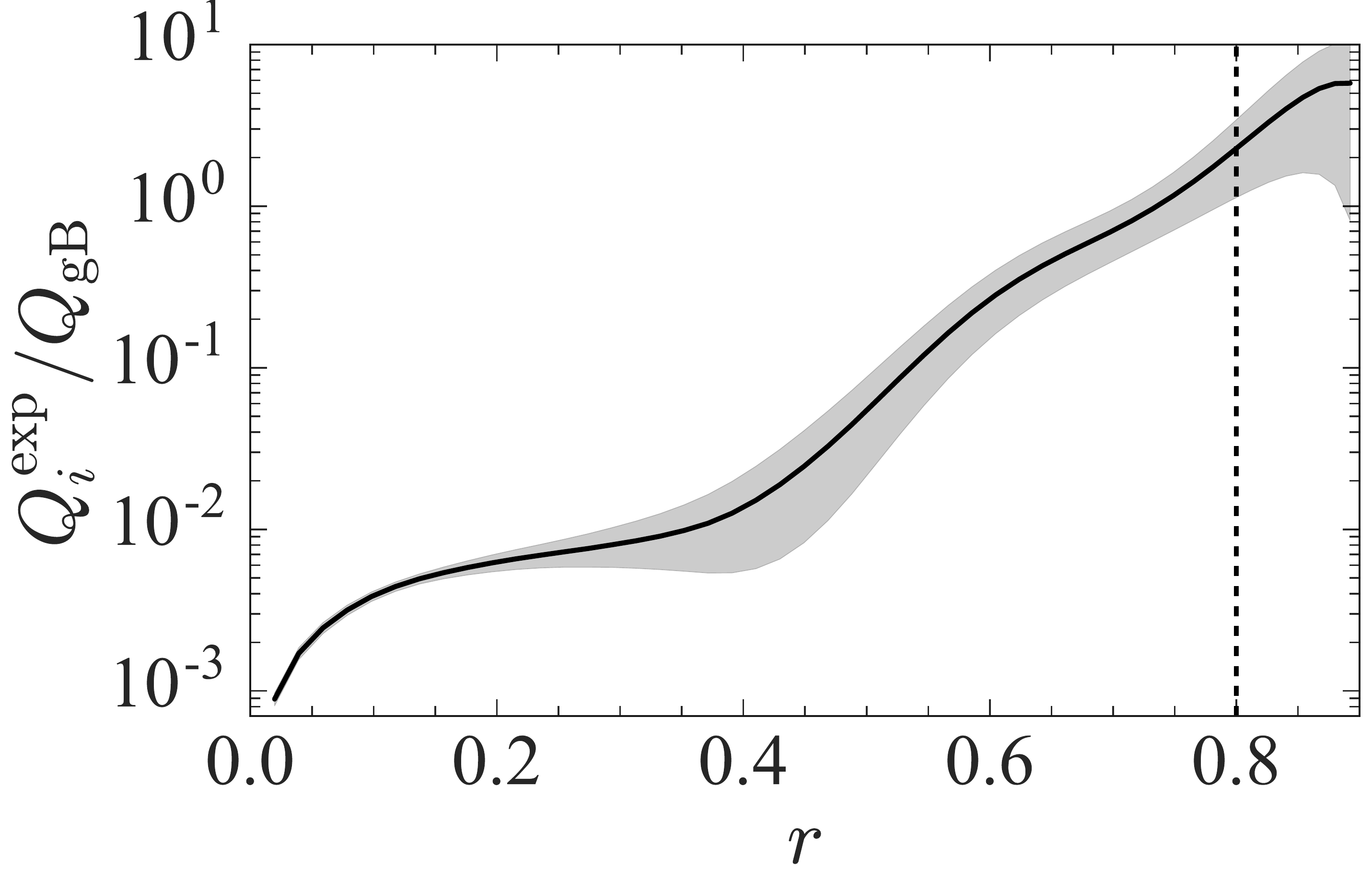}
    \caption[Experimental heat flux profile]{Experimental ion heat flux
      determined from power balance by the TRANSP analysis code as a function
      of $r$.  The dashed line indicates $r=0.8$ and the shaded region
      indicates the uncertainty estimated by TRANSP.
    }
    \label{fig:q_exp}
  \end{figure}

  The profile of the ion heat flux $Q_i^{\exp}$ was calculated by using the
  equilibrium profiles and magnetic geometry as input to a TRANSP analysis, which
  calculated $Q^{\exp}_i$ as a function of $r$ by equating it to the net
  deposited power within the flux surface labelled by $r$. The profile of
  $Q_i^{\exp}$ as a function of $r$ is shown in~\figref{q_exp}. In this work,
  we normalise the heat flux to the gyro-Bohm value defined by
  \begin{equation}
    Q_{\mathrm{gB}} = n_i T_i v_{\mathrm{th}i} \frac{\rho_i^2}{a^2}.
    \label{q_gb}
  \end{equation}
  From~\figref{q_exp}, we find that the experimental level of heat flux at $r =
  0.8$ is $Q^{\exp}_i/Q_{\mathrm{gB}} = 2 \pm 1$.

  {\renewcommand{\arraystretch}{1.5}%
  \begin{table}
    \centering
    \caption{Equilibrium values for MAST discharge \#27274 at $t = 0.25$~s.}
    \begin{tabular}{r c}
      \toprule
      Name & Value \\
      \midrule
      Electron density $n_e (= n_i)$ & $1.31 \times 10^{19}$~m$^{-3}$ \\
      Electron temperature $T_e$ & $0.24$ keV \\
      Half diameter of LCFS $a$ & $0.58$~m \\
      Ion gyroradius $\rho_i$ & $6.08 \times 10^{-3}$~m \\
      Ion temperature $T_i$ & $0.22$ keV \\
      Toroidal magnetic field $B_{\phi}(r=0)$ & $0.46$~T \\
      Toroidal angular frequency $\omega$ & $4.71 \times 10^{4}$~s$^{-1}$ \\
      \bottomrule
    \end{tabular}
    \label{tab:equil_params}
  \end{table}}

\subsection{Sign of $\omega$ and $\gamma_E$}
\label{sec:sign_omega}

  Determining the appropriate sign of $\gamma_E$ is essential when running
  numerical simulations and comparing with experimental measurements, such as
  from the BES diagnostic. Given that $r_0$ and $q_0$ are positive numbers, the
  sign of $\gamma_E$ is completely determined by the sign of $\dv*{\omega}{r}$,
  as in \eqref{flow_shear}.  The sign of $\dv*{\omega}{r}$ is determined
  by the convention used in the experiment.  For MAST, the
  directions of $\vb*{u}$ and $\vb*{B}$ are defined with respect to the plasma
  current $\vb*{I}_p$, which is in the toroidal direction at the magnetic
  axis~\cite{Field2011}:
  \begin{align}
    \sgn(\vb*{B} \cdot \vb*{I}_p) &= -1, \\
    \sgn(\vb*{u} \cdot \vb*{I}_p) &= 1,
    \label{mast_signs}
  \end{align}
  i.e, $\vb*{B}$ and $\vb*{u}$ are in opposite directions. We will be
  simulating this experimental configuration using the GS2 code and we employ the
  GS2 sign conventions, which is to define $\vb*{B}$ in the direction of
  increasing $\phi$~\cite{HighcockThesis}, and determine other signs with
  respect to increasing $\phi$:
  \begin{align}
    \sgn (\vb*{B} \cdot \nabla \phi) \equiv 1 \\
    \sgn(\omega) = \sgn (\vb*{u} \cdot \nabla \phi).
    \label{gs2_signs}
  \end{align}
  Therefore, given that $\vb*{u}$ and $\vb*{B}$ are in opposite directions,
  $\sgn(\omega) = -1$ and $\dv*{\omega}{r} > 0$, as shown in \figref{omega}. We
  conclude that for the MAST configuration we are investigating, the appropriate
  sign of the flow shear is
  \begin{equation}
    \sgn(\gamma_E) > 0.
    \label{sig_gexb}
  \end{equation}

\section{Beam emission spectroscopy}
\label{sec:bes}
  \begin{figure}[t]
    \centering
    \includegraphics[width=0.6\linewidth]{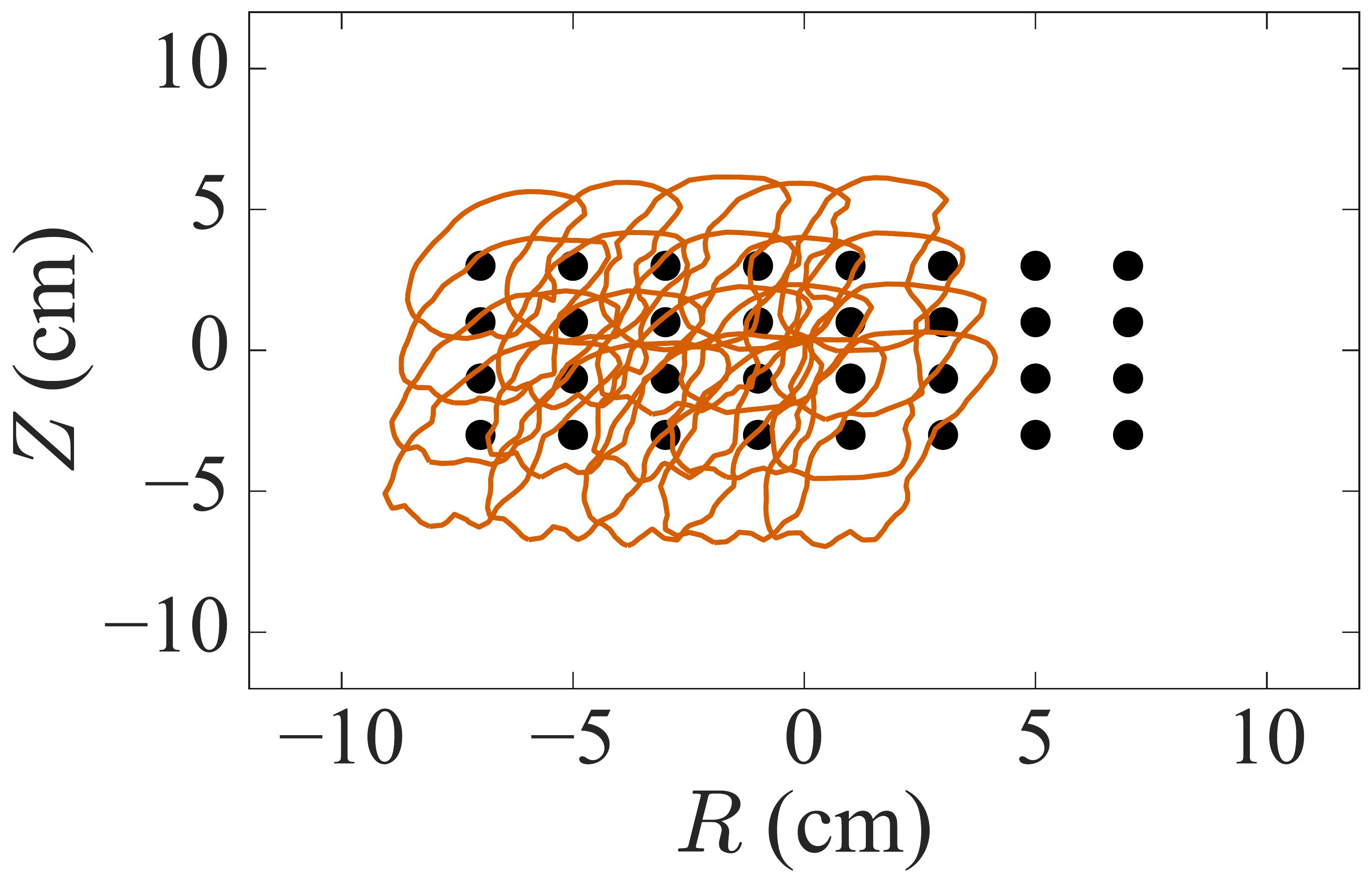}
    \caption[MAST point-spread functions]{
      Point-spread functions for MAST discharge \#27274 at $t=0.25$~s centred
      around $1.35$~m.  The points indicate the BES channels associated with
      each PSF.  Approximately half of the BES is outside the plasma volume and
      no PSFs are calculated for those channels.
    }
    \label{fig:psf}
  \end{figure}
  Turbulent eddies in tokamak plasmas are
  anisotropic due to the strong background magnetic field~\cite{Barnes2011,
  Ghim2013}. In the parallel direction, turbulent eddies have a length scale
  comparable to the system size, which in a torus is the \emph{connection
  length} $qR$, i.e., $l_\parallel \sim qR \sim 1$~m~\cite{Barnes2011}. In the
  direction perpendicular to the magnetic field, ITG-unstable turbulent
  structures have a typical length scale of the order of the ion gyroradius
  $l_\perp \sim \rho_i \sim 1$~cm.  Therefore, in the plane perpendicular to
  the magnetic field, we are interested in two-dimensional measurements of
  fluctuating quantities at approximately the scale of $\rho_i$.  Beam emission
  spectroscopy is a diagnostic technique that was developed to address this
  need. Specifically, the BES diagnostic on MAST~\cite{Field2009, Field2012}
  is designed to measure ion-scale density fluctuations in a radial-poloidal
  plane.  Density fluctuations are inferred from D$_\alpha$ emission produced
  by the NBI beam as it penetrates the plasma. The fluctuating intensity of the
  D$_\alpha$ emission $\delta I$, is proportional to the local plasma density
  at the corresponding viewing location, and the two quantities are related via
  point-spread functions (PSFs)~\cite{Ghim2012, Field2014, Fox2016},
  \begin{equation}
    \delta I_j = \int P_j(R - R_j, Z - Z_j) \delta n (R,Z) \dd R \dd Z,
    \label{psf_eqn}
  \end{equation}
  where $\delta n(R,Z)$ is the fluctuating (laboratory-frame) density field,
  $R$ and $Z$ are the radial and poloidal coordinates, and $P_j(R - R_j, Z - Z_j)$
  is the PSF for the BES channel $j$. The PSFs depend on the magnetic
  equilibrium, beam parameters, viewing location, and plasma profiles and as a
  result, have to be calculated explicitly for each measurement. The PSFs
  for MAST discharge \#27274 at $t = 0.25$~s are shown in \figref{psf}. Note
  that only part of the BES is inside the plasma volume for this discharge [see
  \figref{flux_surfaces}], hence, only approximately half the PSFs are
  calculated.  Recent work~\cite{Fox2016}, has shown that the PSFs play an
  important role in the measurement of turbulence and the precise form that
  they take determines a lower bound on the BES resolution as well as affecting
  the measurement of the turbulent structures and density fluctuation levels --
  effects that we will also consider in this work. For further details on the
  MAST BES system the reader is referred to
  Ref.~\cite{Field2009,Field2012,Ghim2012} and for a detailed study of the
  effect of PSFs on the measurement of turbulent structures to
  Ref.~\cite{Fox2016}.

\chapter{Modelling plasma turbulence}
\label{sec:gk_modelling}

\section{Introduction}

  To model the scenario described in Chapter~\ref{sec:exp_setup}, we use
  gyrokinetics. The aim of modelling plasma turbulence, using
  gyrokinetics or any other theoretical framework, is to predict the
  properties of turbulent fluctuations given a description, or measurement, of
  the equilibrium conditions inside a fusion device (e.g., temperatures,
  densities, flows, etc.). Above all, we are interested in the turbulent
  transport of particles, momentum, and heat due to turbulence, since this is
  significantly enhanced by turbulence in an experimental plasma, and can
  adversely affect potential fusion performance.

  The gyrokinetic equation is derived from the Fokker-Planck equation; however,
  a number of important approximations are employed that are specifically
  relevant to fusion plasmas in tokamaks, and, crucially, result in a reduction
  of the number of phase-space dimensions from six to five. The approximations
  made are, in short: only considering time scales longer than the
  gyrofrequency, but shorter than the time scales over which the equilibrium
  profiles vary; only considering spatial scales which are larger than the
  gyroradius, but smaller than the scale over which equilibrium profiles vary;
  and assuming that turbulent structures are elongated along the magnetic field
  lines. The formulation of \emph{local} gyrokinetics takes this approximation
  one step further by introducing the ``local approximation'': that turbulence
  at a given radial location depends only on the equilibrium quantities and
  their first derivatives at that radial location. This allows a further
  reduction of computational cost. In order for this local approximation to be
  valid, we require that $\rho_i/a \ll 1$, where we assume that other important
  length scales in the system, such as $L_{Ti}$, are of the same order as $a$.
  For the MAST discharge and radial location described in
  Chapter~\ref{sec:exp_setup}, one finds $\rho_i/a \sim 1/100$, where $\rho_i
  \approx 6 \times 10^{-3}$~m and $a\approx 0.6$~m. While this is a reasonably
  small number (which we formally assume to be zero in the local formulation of
  gyrokinetics), previous work has shown that non-local effects can reduce the
  level of turbulent transport at values similar to
  $1/100$~\cite{McMillan2010}. To test whether non-local effects change the
  level of turbulence, one could run a $\rho^*$ scan using a global gyrokinetic
  code. There is also ongoing work to extend GS2 to include finite radial
  effects, such as profile variation, which may be used to test their effect on
  MAST turbulence.

  To solve the gyrokinetic system of equations we use the local gyrokinetic GS2
  code~\cite{Kotschenreuther1995, Dorland2000, HighcockThesis}, which has been
  under active development since the 1990s, when the algorithm for solving the
  linear gyrokinetic problem was first developed.  Taking advantage of the
  local approximation as well as of the axisymmetric nature of tokamak plasmas,
  GS2 solves the gyrokinetic equation in a region known as a ``flux tube'', a
  thin radial region that follows the magnetic field where equilibrium
  quantities and their first derivatives are assumed be constant.

  This chapter is organised as follows. In Section~\ref{sec:tor_geometry}, we
  review the toroidal geometry relevant to plasmas in tokamak devices and
  define an appropriate coordinate system. In Section~\ref{sec:gk_theory}, we
  give an overview of gyrokinetics and the approximations that are required to
  derive the gyrokinetic equation. In Section~\ref{sec:gs2} we give an overview
  of the GS2 code along with parts of the implementation that are relevant to
  our study. Finally, we give the specific numerical setup for the study that
  is the main purpose of this work in Section~\ref{sec:num_setup}.

\section{Toroidal geometry}
\label{sec:tor_geometry}

\begin{figure}[t]
    \centering
    \includegraphics[width=\linewidth]{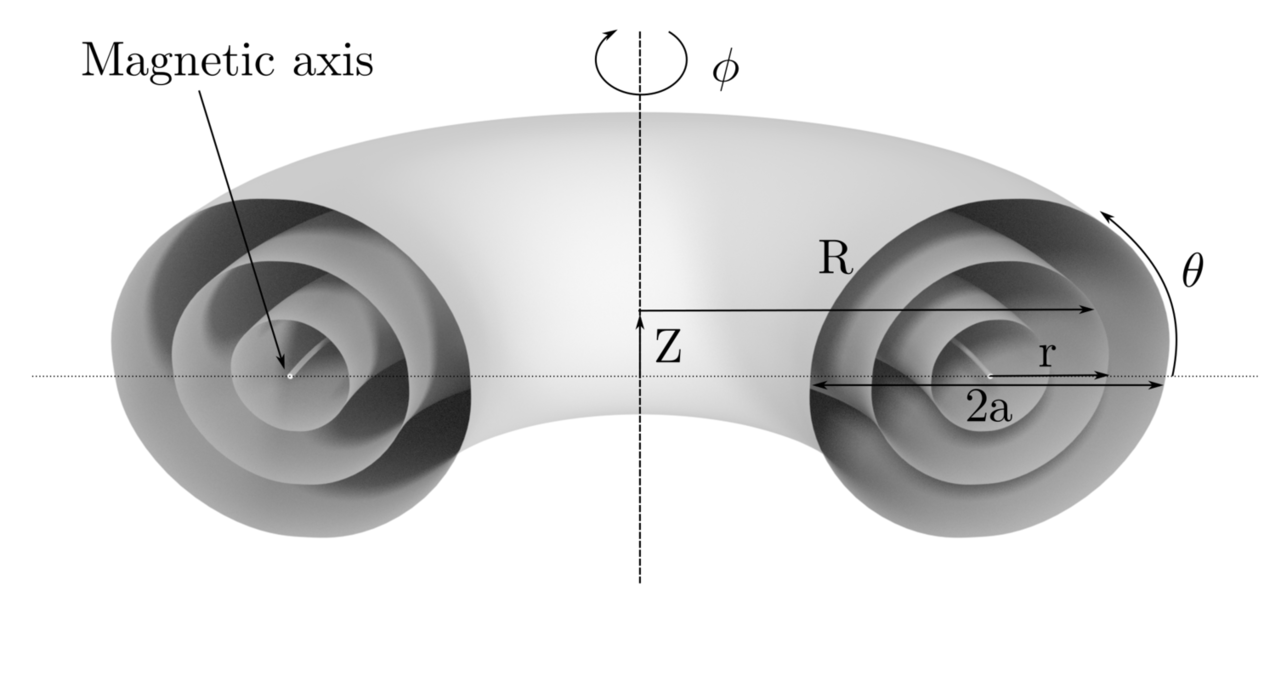}
    \caption[Toroidal geometry]{
      Illustration of circular nested flux surfaces of constant $\psi$
      highlighting the axisymmetric toroidal geometry of a tokamak. Also shown
      is the magnetic axis (which need not be at the geometric centre of any
      flux surface for a finite Shafranov shift), the major radius $R$, the
      poloidal height $Z$ above the midplane of the machine, the minor radius
      $r$, the diameter of the LCFS at the height of the magnetic axis $2a$,
      the toroidal angle $\phi$, and the poloidal angle $\theta$.
    }
    \label{fig:tor_geometry}
  \end{figure}

  In a tokamak, magnetic field lines lie on
  nested toroidal surfaces of constant $\psi$ called flux surfaces. These
  surfaces are roughly axisymmetric, and in such cases one may write the
  magnetic field as:
  \begin{equation}
    \vb*{B} = B_\phi R \nabla \phi + \nabla \psi \times \nabla \phi,
    \label{axis_B}
  \end{equation}
  where $B_\phi$ is the toroidal component of the magnetic field.
  \Figref{tor_geometry} is an illustration of the nested flux surfaces of
  constant $\psi$ in a system with circular flux surfaces, along with the
  coordinates we will use in this work: the major radius $R$, the poloidal
  height $Z$ above the midplane of the machine, the toroidal angle $\phi$, the
  minor radius $r$ (which is simply the distance from the magnetic axis in the
  case of concentric circular flux surfaces, but $r=D/2a$ in the case of more
  complicated flux surface shapes, such as MAST), the diameter of the LCFS at
  the height of the magnetic axis $2a$, and the poloidal angle $\theta$. The
  LCFS is the flux surface just inside the separatrix which separates flux
  surfaces with open and closed field lines [see \figref{flux_surfaces}].

\section{Local gyrokinetic description}
\label{sec:gk_theory}

  Gyrokinetics~\cite{Frieman1982, Sugama1998, Abel2013} describes the
  time-evolution of turbulent plasma in the toroidal geometry described in
  Section~\ref{sec:tor_geometry}. The derivation of the gyrokinetic equation
  has been extensively covered and the reader is referred to
  Ref.~\cite{Abel2013}, and references therein, for a detailed review. In this
  section, we will provide only an overview.

\subsection{The Fokker-Planck equation}
  Our starting point is the Fokker-Planck equation that describes the
  evolution of the distribution function of species $s$, $f_s$. In simplified
  terms, $f_s$ is the probability that there is a particle of species $s$ at
  a given location $\vb*{r}$ and travelling at a given speed $\vb*{v}$. The
  Fokker-Planck equation for the evolution of $f_s$ is given by
  \begin{equation}
    \pd{f_s}{t} + \vb*{v}\cdot \nabla f_s + \frac{Z_s e}{m_s}
    \qty(\vb*{E} + \frac{1}{c} \vb*{v} \times \vb*{B}) \cdot \pd{f_s}{\vb*{v}}
    = C[f_s],
    \label{fokker_planck}
  \end{equation}
  where $Z_s e$ is the charge of species $s$ as a multiple of the fundamental
  charge $e$, $m_s$ is the mass of species $s$, $c$ is the speed of light,
  $\vb*{E}$ and $\vb*{B}$ are the electric and magnetic fields, respectively,
  and $C[f_s]$ is the Landau collision operator. In theory, one could solve
  \eqref{fokker_planck} directly; however, $f_s(t, \vb*{r}, \vb*{v})$ is a
  six-dimensional function (plus time) and solving \eqref{fokker_planck}
  is impractical for the conditions of a magnetically confined fusion
  plasma. The gyrokinetic description makes several simplifying assumptions and,
  importantly, reduces the number of dimensions from six to five, resulting in
  a more tractable problem.

\subsection{The gyrokinetic orderings and assumptions}
  We start by splitting $f_s$ into an equilibrium part $F_s$, and a fluctuating
  part $\delta f_s$:
  \begin{equation}
    f_s = F_s + \delta f_s.
    \label{fs_expansion}
  \end{equation}
  We then make the following assumptions:
  \begin{itemize}
    \item perturbations of the distribution function and background electric
      and magnetic fields are small compared to their equilibrium values;
    \item the frequency of the turbulent fluctuations,
      $\omega_{\mathrm{turb}}$, is small compared to the frequencies at which the
      particles gyrate around the magnetic field $\Omega_s$, but large
      compared to the rate at which the equilibrium quantities change
      $\tau_E^{-1}$;
    \item the turbulent structures are anisotropic and, as such, vary more
      quickly across magnetic field lines compared to along the magnetic field;
      and
    \item the spatial scale of the turbulence perpendicular to the magnetic
      field is of the order of the gyroradius $\rho_s$, and is much smaller
      than the scale over which the equilibrium quantities vary, $a$.
  \end{itemize}
  We define the gyrokinetic parameter as
  \begin{equation}
    \epsilon_{\mathrm{GK}} \equiv \frac{\rho_i}{a},
    \label{gk_param}
  \end{equation}
  and impose the following order on the small parameters identified
  above~\cite{Frieman1982, Abel2013}:
  \begin{equation}
    \frac{|\delta \vb*{B}|}{|\vb*{B}|} \sim
    \frac{|\delta \vb*{E}|}{|\vb*{E}|} \sim
    \frac{\delta f_s}{f_s} \sim
    \frac{k_\parallel}{k_\perp} \sim
    \frac{\omega_{\mathrm{turb}}}{\Omega_i} \sim
    \frac{\rho_i}{a} = \epsilon_{\mathrm{GK}},
    \label{gk_orderings}
  \end{equation}
  where $k_\parallel$ and $k_\perp$ are the typical parallel and perpendicular
  wavelengths of the turbulence, respectively, and $\Omega_i = Z_i e B/m_i c$
  is the gyrofrequency of the ions.

  As this point we translate into a frame rotating with the plasma at
  velocity $\vb*{u}$.  Following from the above  assumptions, it can be shown
  that, to lowest order in $\epsilon_{\mathrm{GK}}$, $\vb*{u}$ is in the
  toroidal direction and independent of the species. It is defined such that
  \begin{equation}
    \vb*{u} = \omega(\psi) R^2 \nabla \phi.
    \label{tor_vel}
  \end{equation}
  We now convert from $(\vb*{r}, \vb*{v})$ to the following variables,
  which reflect the roughly helical motion of the particles in the plasma, and
  the conserved quantities of that motion: the guiding-centre position
  $\vb*{R}_s$, the particle energy $\vareps_s$, the magnetic moment $\mu_s$,
  the gyrophase $\xi$, and the sign of the parallel velocity $\sigma$:
  \begin{align}
    \vb*{R}_s &= \vb*{r} - \frac{\vu*{b} \times \vb*{w}}{\Omega_s} \\
    \vareps_s &= \frac{1}{2} m_s w^2 \\
    \mu_s &= \frac{m_s w_\perp^2}{2 B} \\
    \sigma &= \frac{w_\parallel}{|w_\parallel|}
    \label{gk_variables}
  \end{align}
  where $\vu*{b} = \vb*{B}/B$ is a unit vector in the direction of the magnetic
  field, $w = |\vb*{w}|$ is the velocity shifted into the rotating
  frame~\cite{Abel2013}
  \begin{equation}
    \vb*{w} = \vb*{v} - \vb*{u} = w_\parallel + w_\perp (\cos \xi \vb*{e}_2 -
      \sin \xi \vb*{e}_1),
    \label{peculiar_vel}
  \end{equation}
  $w_\parallel$ and $w_\perp$ are the parallel and perpendicular components of
  $\vb*{w}$, and $\vb*{e}_1$ and $\vb*{e}_2$ are arbitrary orthogonal unit
  vectors perpendicular to the magnetic field.

  Finally, we will formally assume that the Mach number $M$ of
  the plasma rotation is small, but that the flow shear is large enough to
  affect the plasma dynamics:
  \begin{equation}
    \frac{R\omega}{v_{{\mathrm{th}}i}} \equiv M \ll 1,\quad
    |a\nabla\ln\omega| \sim \frac{1}{M}.
    \label{flow_scaling}
  \end{equation}
  This allows us to formulate local gyrokinetics on a rotating surface,
  neglecting effects such as the Coriolis and centrifugal force, but retaining
  the effect of flow shear.

\subsection{The gyrokinetic equation}
\label{sec:gk_eqn}
  Using the gyrokinetic orderings \eqref{gk_orderings} and assuming
  that the plasma is sufficiently collisional, it can be shown that the
  background distribution function of species $s$, $F_s$, is a Maxwellian
  distribution, to lowest order,
  \begin{equation}
    F_s = F_{Ms} \equiv
    n_s {\qty(\frac{m_s}{2 \pi T_s})}^{3/2} \exp\qty(- \frac{\vareps_s}{T_s}),
    \label{maxwellian}
  \end{equation}
  where $n_s$ and $T_s$ are the density and temperature of
  species $s$, respectively. Furthermore, it may be shown that, to the first
  order in $\epsilon_{\mathrm{GK}}$, the fluctuating part of the perturbed
  distribution function $\delta f_s$ can be written
  \begin{equation}
    \delta f_s = - \frac{Z_s e \varphi}{T_s} F_{Ms} +
      h_s (t, \vb*{R}_s, \mu_s, \vareps_s, \sigma),
    \label{pert_dist}
  \end{equation}
  where $\varphi$ is the perturbed electrostatic potential and
  $h_s (t, \vb*{R}_s, \mu_s, \vareps_s, \sigma)$ is the gyrophase-independent
  distribution function of Larmor rings that will completely determine the
  plasma dynamics in the gyrokinetic formulation. As $h_s$ is independent of
  the gyrophase, we have effectively removed one of the velocity space
  dimensions (with velocity space now described only by $\vareps_s$ and
  $\mu_s$) and reduced the problem to five dimensions instead of six, and in
  doing so, significantly reduced the computational requirements.

  Applying the gyrokinetic orderings to the Fokker-Planck equation
  \eqref{fokker_planck}, we obtain the gyrokinetic equation, which describes
  the evolution of the gyrophase-independent distribution function $h_s$
  \begin{equation}
    \begin{split}
      &\left(\pdv{}{t} + \vb*{u} \cdot \nabla \right) \left(h_s
      - \frac{Z_s e \ensav{\varphi}{\vb*{R}s}}{T_s} F_s\right) +
      \left(w_\parallel \vu*{b} + \vb*{V\!}_{{\mathrm{D}}s} +
      \ensav{\vb*{V\!}_E}{\vb*{R}s}\right)
      \cdot \nabla{h_s} - \ensav{C[h_s]}{\vb*{R}s} \\
      &\quad=
      -\ensav{\vb*{V\!}_E}{\vb*{R}s} \cdot \nabla r
      \left[\dv{\ln n_s}{r} + \left(\frac{\vareps_s}{T_s} -
        \frac{3}{2}\right)\dv{\ln T_s}{r}
      + \frac{m_s w_\parallel}{T_s}\frac{R B_\phi}{B}\dv{\omega}{r}\right]F_{Ms},
      \label{gk}
    \end{split}
  \end{equation}
  where $\ensav{\ldots}{\vb*{R}s}$ is an average over the particle orbit at
  constant guiding centre position $\vb*{R}_s$,
  \begin{equation}
    \vb*{V\!}_{{\mathrm{D}}s} = \frac{c}{Z_s e B}\vu*{b}\times \left[ m_s w_\parallel^2
              \vu*{b} \cdot \nabla \vu*{b} + \mu_s \nabla B \right]
    \label{v_drift}
  \end{equation}
  is the magnetic drift velocity,
  \begin{equation}
    \vb*{V\!}_E = \frac{c}{B}\vu*{b}\times \nabla \varphi
    \label{v_exb}
  \end{equation}
  is the perturbed \exb drift velocity, and $C[h_s]$ is the linearised
  collision operator~\cite{Abel2008a,Barnes2008}.

  To close our system of equations, we use the quasineutrality condition
  \begin{equation}
    \sum_s Z_s\delta n_s = 0
    \quad\Rightarrow\quad
    \sum_s \frac{Z_s^2 e \varphi}{T_s} n_s = \sum_s Z_s \int \dd^3 \vb*{w}
      \ensav{h_s}{\vb*{r}},
    \label{quasineutrality}
  \end{equation}
  where $\ensav{\ldots}{\vb*{r}}$ indicates a gyroaverage at constant
  $\vb*{r}_s$, to calculate $\varphi$ using $h_s$.

  The right-hand side of~\eqref{gk} represents the advection by the
  gyroaveraged $\vb*{E} \times \vb*{B}$ velocity of the Maxwellian equilibrium
  distribution function, which is characterised by $n_s$, $T_s$, and $\omega$.
  The equilibrium quantities $n_s$, $T_s$, and $\omega$ are functions only of
  the poloidal magnetic flux $\psi$. However, for the purposes of this work,
  we have converted this dependence from $\psi$ to the Miller coordinate $r =
  D/2a$ introduced previously. Since $r$ is also a flux-surface label, it is
  simple to relate gradients in $\psi$ and $r$ via
  \begin{equation}
    \nabla r = \dv{r}{\psi} \nabla \psi.
    \label{grad_relation}
  \end{equation}
  The right-hand side of~\eqref{gk} contains terms proportional to $\dv*{\ln
  T_s}{r}$ and $\dv*{\omega}{r}$, which are related to the parameters $\kappa_T$
  and $\gamma_E$, defined by~\eqref{ti_prime} and~\eqref{flow_shear},
  respectively. These terms are sources of free energy in the system and
  are responsible for the ITG and PVG instabilities. The stabilising effect of
  $\gamma_E$ on $h_s$ is contained in the term proportional to $\vb*{u} \cdot
  \nabla$ and is further discussed in Section~\ref{sec:gamma_stab}. In deriving
  \eqref{gk}, we have also assumed that the fluctuations are purely
  electrostatic, i.e., no fluctuating magnetic fields (see
  Section~\ref{sec:num_setup} for further details).

\subsection{Flow shear stabilisation}
\label{sec:gamma_stab}

  \begin{quote}
    \emph{This section is based on Appendix A of Ref.~\cite{Schekochihin2012}.}
  \end{quote}

  As noted in Section~\ref{sec:gk_eqn}, flow shear enters~\eqref{gk} as a
  \emph{destabilising} term on the right-hand side, but for the values of
  $\gamma_E$ that we will be considering, this effect is small compared to the
  destabilising effect of the ITG (see~\cite{Schekochihin2012} for further
  details). However, flow shear also enters our system as a \emph{stabilising}
  term, as we will now explain using a simplified magnetic geometry.

  Consider a locally straight and uniform magnetic field that has constant
  magnitude, no curvature, and no shear. We define a local Cartesian coordinate
  system with unit vectors (note we do note use these definitions throughout
  this work, we define a related but slightly different coordinate system in
  Section~\ref{sec:gs2}):
  \begin{equation}
    \vu*{x} = \frac{\nabla \psi}{B_\theta R}, \qquad
    \vu*{y} = \frac{\vu*{z} \times \nabla \psi}{B_\theta R}, \qquad
    \vu*{z} = \vu*{b}.
    \label{slab_coords}
  \end{equation}
  We choose our local coordinate $x$ such that $x=0$ at some reference flux
  surface labelled by $\psi_0$. In the vicinity of this flux surface, we may then
  Taylor expand $\psi$ in terms of this local radial coordinate as
  $\psi(x) \approx \psi(x=0) + x \dv*{\psi}{x} = \psi_0 + x B_\theta R$. The
  toroidal angular frequency is a function of $\psi$ only and we can again
  Taylor expand in $x$ (since we assume in~\eqref{flow_scaling} that the scale
  over which $\omega$ changes is much smaller than $a$) to get $\omega \approx
  \omega_0 + x B_\theta R \dv*{\omega}{\psi}$. Now consider the $\vb*{u} \cdot
  \nabla$ term on the left-hand side of~\eqref{gk}, where $\vb*{u}$ is given
  by~\eqref{tor_vel}. Using the axisymmetric representation of the magnetic
  field in a torus \eqref{axis_B}, we can write
  \begin{equation}
    \vb*{u} = \omega R^2\nabla\phi
    \approx \left(\omega_0 R + x B_\theta R^2 \dv{\omega}{\psi} \right)
    \left(\frac{B_\phi}{B}\vu*{b} + \frac{B_\theta}{B}\vu*{y}\right),
    \label{rot_vel}
  \end{equation}
  If we now go to the frame rotating with the flux surface at the
  rate $\omega_0$ and also use the fact that, in gyrokinetics, gradients of
  fluctuating quantities parallel to $\vu*{b}$ are always small compared to
  those perpendicular to it, we find
  \begin{equation}
    \vb*{u}\cdot\nabla \approx
    x \frac{B_{\theta}^2 R^2}{B} \dv{\omega}{\psi}\vu*{y}\cdot\nabla
    = \left(\frac{q R B_{\theta}}{rB}|\nabla r|\right) x \gamma_E
      \frac{v_{{\mathrm{th}}i}}{a} \vu*{y}\cdot\nabla,
    \label{u_grad}
  \end{equation}
  with $\gamma_E$ as defined in~\eqref{flow_shear}. The prefactor enclosed in
  the parentheses is close to unity and so $\gamma_E$ is the
  normalised shear that acts on the distribution function. The presence of this
  shear will have a stabilising effect on the turbulence.

\section{Overview of GS2}
\label{sec:gs2}

  In this work, we used the local gyrokinetic code
  GS2\footnote{\url{http://gyrokinetics.sourceforge.net}}~\cite{Kotschenreuther1995,
  Dorland2000,HighcockThesis} to solve the system of equations given
  by~\eqref{gk} and~\eqref{quasineutrality} to give us the time evolution of
  $h_s(t, \vb*{R}_s, \vareps_s, \mu_s, \sigma)$ and $\varphi(t, \vb*{R}_s)$.
  With knowledge of $h_s$ and $\varphi$, one can calculate a range of physical
  characteristics of the turbulence, e.g., density-, flow-,
  temperature-fluctuation fields, particle, momentum, and heat transport, and
  so on. Of particular interest is the ion density fluctuation field,
  \begin{equation}
    \frac{\delta n_i}{n_i} = \frac{1}{n_i} \int \dd^3 \vb*{w} \ensav{h_i}{\vb*{r}},
    \label{delta_n}
  \end{equation}
  and the radially outwards, time-averaged turbulent heat flux carried by the
  ions (for reasons which have been given previously),
  \begin{equation}
    Q_i = \left\langle\frac{1}{V}\int \dd^3 \vb*{r} \int \dd^3 \vb*{w}
    \frac{m_i v^2}{2} h_i \vb*{V\!}_E \cdot \nabla r \right\rangle,
    \label{q_def}
  \end{equation}
  where $V$ is the volume enveloping a given flux surface and
  $\langle \ldots \rangle$ is a flux-surface average. $Q_i$ can be
  normalised to the gyro-Bohm heat flux given in~\eqref{q_gb}. It is
  a feature of the asymptotic ordering on which gyrokinetic theory is based
  that $Q_i/Q_{\mathrm{gB}}$ is a number of order unity~\cite{Abel2013}.

  In this section, we will review aspects of the GS2 code that are pertinent to
  our study. The geometry of the nested flux surfaces in GS2 is described by
  the Miller specification~\cite{Miller1998}, which is detailed in
  Section~\ref{sec:miller}. The Miller specification consists of nine
  parameters that control the aspects of the magnetic field lines and
  flux-surface shapes such as the safety factor, elongation, triangularity, and
  so on. In Section~\ref{sec:gs2_geometry}, we define the coordinate system
  relative to the magnetic flux surfaces used in GS2. By making the ``local
  approximation'' (Section~\ref{sec:local_approx}), GS2 is able to solve the
  gyrokinetic equation on a single flux surface in a region known as a flux
  tube, which follows a single magnetic field line described by the Miller
  parameters. In Section~\ref{sec:collisions}, we detail the calculation of the
  ion-ion and electron-ion collision frequencies from equilibrium parameters
  and show how we can account for enhanced ion-ion collisionality due impurity
  ions without treating them as additional kinetic species in our simulations.
  The implementation of flow shear and its effect on turbulence is detailed
  in Section~\ref{sec:flow_shear}. Finally, we show the form of hyperviscosity
  used in GS2 to damp plasma dynamics at large values of $k_\perp$ and explain
  how this is beneficial in our simulations.  For a detailed review of the
  algorithms and numerical implementations that are used in GS2 to solve the
  gyrokinetic equation, the reader is referred to~\cite{HighcockThesis}, and
  references therein.

  \subsection{Geometry}
  \label{sec:gs2_geometry}

  \subsubsection{The Miller flux-surface specification}
  \label{sec:miller}

  Throughout this work we have used the Miller specification~\cite{Miller1998}
  of the magnetic equilibrium. The Miller specification is a nine-parameter
  parametrisation of up-down symmetric flux surfaces suitable for the
  description of MAST flux surfaces\footnotemark.
  Table~\ref{tab:miller_params} lists the definitions of the Miller parameters.
  As explained in Section~\ref{sec:mast_diagnostics}, experimental flux
  surfaces from MAST were obtained from an MSE-constrained EFIT equilibrium, or
  more conveniently, from a TRANSP output file, where TRANSP used the EFIT
  equilibrium as input. For reference, we also list in Table~\ref{tab:miller_params} the associated variable names
  of the Miller parameters as they are listed or calculated from the TRANSP
  analysis output. The Miller parameter values
  and associated GS2 input parameters for our study are detailed, along with
  other equilibrium parameters, in Section~\ref{sec:num_setup}.
  \footnotetext{For the specification of up-down \emph{asymmetric} flux
  surfaces the reader is referred to recent work by Ball
  \emph{et. al.}~\cite{Ball2014} that extends the Miller specification.}
  {\renewcommand{\arraystretch}{1.25}%
  \begin{table}
    \centering
    \caption{The Miller parametrisation of flux surfaces along with
      their associated variable names in the TRANSP output file. The
      derivatives of geometric quantities are calculated by manually taking a
      derivative with respect to $r$ (after transforming from the TRANSP
      $\rho_{\mathrm{tor}}$ grid onto an $r$ grid).
    }
    \begin{tabular}{r c c}
      \toprule
      Name & Definition & TRANSP variable \\
      \midrule
      Elongation & $\kappa$ & \texttt{ELONG} \\

      Elongation derivative & $\kappa' = \dv*{\kappa}{r}$ &
      $\dv*{r}(\texttt{ELONG})$ \\

      Magnetic shear & $\hat{s} = r_0/q_0 \dv*{q}{r}$ &
      $r_0/q_0 \dv*{r}(\texttt{Q})$ \\

      Major radius & $R_{N} = R/a$ & $\texttt{RMAJM}/a$ \\

      Miller radial coordinate & $r_0 = {D/2a}$ & calc. using \texttt{RMAJM} \\

      Safety factor & $q_0 = \pdv*{\psi_\mathrm{tor}}{\psi_{\mathrm{pol}}}$ &
      \texttt{Q} \\

      Shafranov Shift & $1/a \dv*{R}{r}$ & $1/a\dv*{r}(\texttt{RMJMP})$ \\

      Triangularity & $\delta$ & \texttt{TRIANG} \\

      Triangularity derivative & $\delta' = \dv*{\delta}{r}$ &
      $\dv*{r}(\texttt{TRIANG})$ \\
      \bottomrule
    \end{tabular}
    \label{tab:miller_params}
  \end{table}}

  \subsubsection{GS2 Coordinate system}

  We saw in Section~\ref{sec:tor_geometry} that the magnetic field lines in a
  tokamak form well-defined, nested flux surfaces of constant magnetic field
  and, hence, constant $\psi$.  As well as this, magnetic field lines, in the
  absence of magnetic islands and other similar effects (the typical
  configuration in a tokamak), do not cross each other. Therefore, we can use
  these two observations to define a coordinate system, following
  Ref.~\cite{Beer1995a,HighcockThesis}.

  The first natural basis vector is the direction of the magnetic field,
  $\vu*{b} = \vb*{B}/B$. As stated in Section~\ref{sec:gk_theory}, equilibrium
  quantities are functions only of the poloidal flux $\psi$ because that they
  are constant on a given flux surface (no poloidal dependence) and that the
  system is axisymmetric (no toroidal dependence).  Therefore, we can use the
  gradient of $\psi$ to define the radial coordinate with basis vector:
  \begin{equation}
    \vu*{\psi} = \frac{\nabla \psi}{\abs{\nabla \psi}}.
    \label{rad_basis}
  \end{equation}
  Finally, we define a third coordinate, $\alpha$, with basis vector
  \begin{equation}
    \vu*{\alpha} = \frac{\nabla \alpha}{\abs{\nabla \alpha}},
    \label{alpha_basis}
  \end{equation}
  such that $\vb*{B} = \nabla \alpha \times \nabla \psi$ (using the Clebsch
  representation of the magnetic field~\cite{Kruskal1958}). It was shown
  in~\cite{Kruskal1958} that $\alpha$ is a function of the form
  \begin{equation}
    \alpha = \phi + q(\psi) \theta + \nu(\theta, \psi),
    \label{alpha}
  \end{equation}
  where $\nu$ is a function which depends on the geometry and is periodic in
  $\phi$ and $\theta$~\cite{Kruskal1958}.

  \subsubsection{The local approximation}
  \label{sec:local_approx}

\begin{figure}[t]
    \centering
    \includegraphics[width=0.6\linewidth]{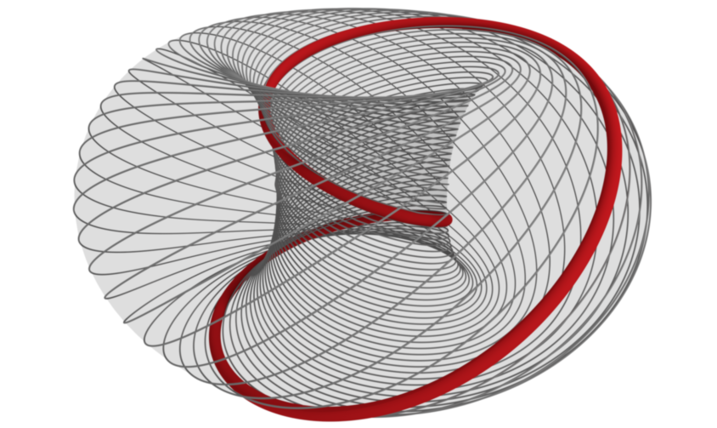}
    \caption[MAST magnetic field lines]{
      Magnetic field lines that lie on the flux
      surface at $r = 0.8$ (setting $q=2$ so that field lines are closed for
      visualisation purposes).  The field line marked in red is the centre line
      of the GS2 flux tube that we use to simulate the plasma. The GS2 flux
      tube itself is approximately rectangular at the outboard midplane but
      twists as it follows the magnetic field line due to the magnetic shear.
      The flux tube follows the field line once around the flux surface in the
      poloidal direction.
    }
    \label{fig:flux_tube}
  \end{figure}

  Using the above coordinate system we define the coordinates used in GS2 after
  employing the ``local approximation''. Due to the fast motion of particles
  along the magnetic field lines and the relatively slow drift across them,
  turbulent structures are anisotropic in the parallel and perpendicular
  directions to the field line.  Specifically, turbulent structures in a
  tokamak are elongated along field lines, with length scales of the order of
  the connection length $l_\parallel \sim q R$, and are much shorter in the
  perpendicular directions, with length scales of the order of  the ion
  gyroradius $l_\perp \sim \rho_i$. GS2 takes advantage of this anisotropy by
  solving the gyrokinetic equation in a region known as a ``flux
  tube''~\cite{Beer1995a}. A flux tube is chosen to be several turbulence
  decorrelation lengths long in both the perpendicular and parallel directions,
  i.e., long enough to avoid spurious interactions of turbulence with the edges
  of the box, but still short enough to be highly resolved.  \Figref{flux_tube} shows
  the MAST flux surface and magnetic field lines at $r = 0.8$ with one field
  line highlighted in red to represent a flux tube. The actual flux tube is
  approximately rectangular at the outboard midplane and is highly twisted
  along the field line due to the magnetic shear (this is not shown in
  \figref{flux_tube} for clarity).  Assuming axisymmetry, along with the
  anisotropy of the fluctuations, implies that we are in fact capturing the
  dynamics of the entire flux surface.  Simulating only a single flux tube in
  this way leads to dramatic savings in computational cost.

  The local approximation in gyrokinetics assumes that the gradients of
  equilibrium quantities (such as those shown in \figref{profiles}) are
  constant across the radial simulation domain. It is also assumed that,
  provided the simulation domain in the plane perpendicular to the magnetic
  field is significantly larger than the spatial scales of the turbulence, it
  is acceptable to take periodic boundary conditions in the radial and binormal
  directions. For these two directions, the two perpendicular coordinates used
  in GS2 are $x$ and $y$, which measure the distance from the magnetic field
  line located at $(\psi_0, \alpha_0)$~\cite{HighcockThesis}:
  \begin{align}
    x &= a \frac{q_0}{r_0} (\psi_N - \psi_{0N}), \label{gs2_x}\\
    y &= a \eval{\dv{\psi_N}{r}}_{r_0} (\alpha - \alpha_0), \label{gs2_y}
  \end{align}
  where $\psi_N = \psi/a^2 B_{\mathrm{ref}}$ is the normalised poloidal flux.
  In the parallel direction, the poloidal angle $\theta$ is used in GS2 (noting
  that any coordinate that is not fixed at fixed $\psi$ and $\alpha$ can be
  used as a parallel coordinate and noting in addition, some geometric
  quantities are more convenient to calculate when using $\theta$ as a parallel
  coordinate~\cite{HighcockThesis}).

  \subsubsection{Spectral coordinates}
  \label{sec:spectral_coords}

  In the absence of flow shear, the gyrokinetic equation \eqref{gk} has no
  explicit dependence on $x$ or $y$ and can be solved using spectral methods in
  these directions.  Spectral methods are computationally efficient and can be
  used to enforce the conservation properties required by the system exactly.
  More specifically, GS2 uses a pseudo-spectral algorithm with only the
  nonlinear term being calculated in $(x,y,\theta)$ coordinates.  Otherwise,
  perturbed quantities have the following spectral
  representation~\cite{Beer1995a}
  \begin{equation}
    A = \sum_{k_x,k_y} \hat{A}(t,\theta)_{k_x,k_y} e^{i(k_x x + k_y y)}
      \equiv \mathcal{F}^{-1}[\hat{A}(t,\theta)],
    \label{pert_spectral}
  \end{equation}
  where $k_x$ and $k_y$ are the perpendicular coordinates used by GS2 in
  spectral space, and $\mathcal{F}^{-1}$ is the inverse Fourier transform.

  One important caveat regarding the use of spectral coordinates and the
  conversion of GS2 perturbed quantities from spectral to real space (as we do
  in this work) is the normalisation convention used when performing the
  Fourier transform. GS2 uses the open-source
  FFTW\footnote{\url{http://www.fftw.org/}} package to transform between $(x,
  y)$ and $(k_x, k_y)$ representations. FFTW performs the following
  calculations\footnote{\url{http://www.fftw.org/doc/What-FFTW-Really-Computes.html}}:
  \begin{align}
    \text{Forward: } \hat{A}(k) &= \mathcal{F}[A(x)] =
      \sum_{j=0}^{n-1} A_j e^{-2 \pi j k i/n}, \label{fftw_calc_forward} \\
      \text{Backward: } A(x) &= \mathcal{F}^{-1}[\hat{A}(k)] =
        \sum_{j=0}^{n-1} \hat{A}_j e^{2 \pi j k i/n}, \label{fftw_calc_backward}
  \end{align}
  where $A$ is the real-space representation, $\hat{A}$ is the spectral-space
  representation, and $\mathcal{F}$ is the forward Fourier transform. We see
  that there is no implicit normalisation applied by the FFTW library, meaning
  that applying a forward (going from real to spectral space) and then a backward
  (going from spectral to real space) transform will multiply the input by $n$.
  Therefore, the following normalisation is commonly used:
  \begin{align}
    \hat{A}(k) &= \mathcal{F}[A(x)], \label{gs2_fft_forward} \\
    A(x) &= \frac{\mathcal{F}^{-1}[\hat{A}(k)]}{n}. \label{gs2_fft_backward}
  \end{align}
  In contrast, GS2 uses the following normalization:
  \begin{align}
    \hat{A}(k) &= \frac{\mathcal{F}[A(x)]}{n}, \\
    A(x) &= \mathcal{F}^{-1}[\hat{A}(k)].
    \label{gs2_fft_norm}
  \end{align}
  In other words, when converting GS2 fields from spectral space to real space,
  no normalisation is necessary and care must be taken when using FFT packages
  external to GS2 since they may be using the normalisations given in
  equations~\eqref{fftw_calc_forward} and~\eqref{fftw_calc_backward}.

  \subsubsection{GS2 variable Normalisations}
  Before detailing aspects of the GS2 algorithm, we note the normalisations
  used in GS2 and this work. The normalisations used within GS2 are chosen such
  that all quantities are of order unity. Table~\ref{tab:normalising} lists the
  normalising quantities and Table~\ref{tab:normalised} lists
  the main normalised quantities used within GS2~\cite{HighcockThesis}.
  {\renewcommand{\arraystretch}{1.25}%
  \begin{table}
    \centering
    \caption{Normalising quantities used in GS2.}
    \begin{tabular}{r c}
      \toprule
      Quantity & Definition \\
      \midrule
      $a$ & Half the diameter of the LCFS at the height of the magnetic axis \\
      $B_{\mathrm{ref}}$ & Toroidal magnetic field strength at $r=0$ \\
      $v_{\mathrm{th}i}$ & $\sqrt{2 T_i / m_i}$ \\
      $Z_i \equiv 1$ & Charge number of ion species \\
      $m_i$ & Mass of ion species \\
      $\Omega_i$ & $Z_i e B_{\mathrm{ref}} / m_i c$ \\
      $\rho_i$ & $v_{\mathrm{th}i}/\Omega_i$ \\
      \bottomrule
    \end{tabular}
    \label{tab:normalising}
  \end{table}}

  {\renewcommand{\arraystretch}{1.25}%
  \begin{table}
    \centering
    \caption{The main normalised quantities used in this
      work~\cite{HighcockThesis}.
    }
    \begin{tabular}{r c c}
      \toprule
      Name & Normalised definition \\
      \midrule
      Binormal coordinate & $y/\rho_i$ \\
      Binormal wavenumber & $k_y \rho_i$ \\
      Charges & $Z_s/Z_i$ \\
      Densities & $n_s/n_i$ \\
      Density gradients & $\kappa_{ns} = 1/L_{ns}$ \\
      Flow shear & $\gamma_E = (r_0/q_0) \dv*{\omega}{r} (a/v_{\mathrm{th}i})$ \\
      Magnetic field & $B/B_{\mathrm{ref}}$ \\
      Masses & $m_s/m_i$ \\
      Perturbed electrostatic potential & $\varphi/(\rho_i/a)(T_i/e)$ \\
      Radial coordinate & $x/\rho_i$ \\
      Radial wavenumber & $k_x \rho_i$ \\
      Temperatures & $T_s/T_i$ \\
      Temperature gradients & $\kappa_{Ts} = 1/L_{Ts}$ \\
      Time & $t/(a/v_{\mathrm{th}i})$ \\
      \bottomrule
    \end{tabular}
    \label{tab:normalised}
  \end{table}}

  \subsection{Collisions}
  \label{sec:collisions}

  The fundamental effect of turbulence is to transfer energy from large spatial
  scales at which energy is injected to small scales where energy is
  dissipated, which leads to heating. As well as the transfer of energy due to
  turbulence, there are several mechanisms that lead to phase-space mixing,
  which produce small-scale structure and large gradients in velocity space
  (see~\cite{Schekochihin2008} and references therein). It is these large
  gradients in velocity space that eventually bring collisions into effect
  regardless of how small the collisionality is. Therefore, in any plasma
  turbulence simulation some form of dissipation must be included to smooth out
  the small-scale structure that develops in velocity space. While dissipation
  due to collisions is the primary physical dissipation mechanism in kinetic
  plasmas, artificial dissipation is also possible, and useful, in numerical
  simulations (see Section~\ref{sec:hyperviscosity}).

  Recent work~\cite{Abel2008a, Barnes2008} has led to the implementation of a
  linearised Fokker-Planck collision operator in GS2 that satisfies the
  following important properties,
  \begin{inparaenum}[(i)]
    \item smooths out small-scale structure in velocity space;
    \item obeys Boltzmann's H-theorem (the condition that collisional processes
      are irreversible and cannot decrease entropy); and
    \item conserves particles, momentum, and energy.
  \end{inparaenum}
  This collision operator includes the effect of both pitch-angle scattering
  and energy diffusion because small-scale structure can be generated in
  both $v_\perp$ and $v_\parallel$ by phase mixing. The level of collisional
  dissipation in GS2 is set by the collision frequencies calculated as follows.

  In GS2, velocity space is represented by the particle energy $\vareps_s$ and
  the pitch-angle variable $\lambda'_s = \mu_s/\vareps_s$. The associated
  input parameters which control the grid sizes are \texttt{negrid} and
  \texttt{ngauss}. These parameters are only related to the real grid sizes used
  by GS2, because the exact magnetic geometry also plays a role through the
  calculation of bounce points of trapped particles (see
  Ref.~\cite{HighcockThesis} for further details).  The input parameters that
  control the strength of the collisional dissipation in GS2 are the collision
  frequencies for each species.  The electron-ion collisionality is calculated
  via~\cite{Hammett2003, Abel2008a}
  \begin{equation}
    \texttt{vnewk\_2} = \nu_{ei} \frac{a}{v_{\mathrm{th}i}} =
    \frac{4 \pi n_e e^4 \ln \Lambda}{(2T_e)^{3/2} m_e^{1/2}} \frac{a}{v_{\mathrm{th}i}},
    \label{elec_coll}
  \end{equation}
  where~\cite{Huba2016}
  \begin{equation}
    \ln \Lambda = 24 - \ln (10^4 \sqrt{\frac{n_e^{1/2}}{10}} T_e^{-1}),
    \label{loglam}
  \end{equation}
  is the Coulomb logarithm where $n_e$ is in units of $10^{19}$~m$^{-3}$ and
  $T_e$ is in keV, and \texttt{vnewk\_2} is the GS2 parameter denoting the
  electron collision frequency. We can derive a convenient form
  of~\eqref{elec_coll} by converting to cgs units and eliminating physical
  constants~\cite{Hammett2003}:
  \begin{equation}
    \texttt{vnewk\_2} \approx 2.7913 \times 10^{-3}
    \frac{n_e \ln \Lambda a A_i^{1/2}}{T_{e}^{3/2} T_i^{1/2}},
    \label{elec_coll_simple}
  \end{equation}
  where $A_i$ is the atomic mass of the ion species in units of the proton mass
  $m_p$ and $T_i$ is in units of keV.

  In this work, we have simulated only a single ion species. However, the
  experiment contains several different ion impurities, such as C$^{+6}$ carbon
  impurity ions, and beam ions, that may affect the ion equilibrium profiles
  and ion-ion collision frequencies. Unfortunately, including additional
  gyrokinetic ion species in our simulations is prohibitively expensive for the
  extensive parameter scan performed in this work. Instead, it is possible to
  improve the realism of our simulations by creating an aggregate ion species,
  instead of simulating a pure deuterium plasma. We achieve this by calculating
  an effective ion charge,
  \begin{equation}
    Z_{\mathrm{eff}} = \frac{\sum_j n_j Z_j^2}{|\sum_j n_j Z_j|},
    \label{zeff}
  \end{equation}
  where the summation is over all ion species present in the experiment, and
  $n_j$ and $Z_j$ are the density and charge of ion species $j$, respectively.
  This parameter is denoted \texttt{zeff} in GS2 and the value, determined from
  the experiment, is given in Section~\ref{sec:num_setup}. This leads to the
  following enhancement of the ion-ion collision frequency~\cite{Hammett2003}
  \begin{equation}
    \texttt{vnewk\_1} = \texttt{vnewk\_2} \times  Z_i^2 Z_\mathrm{eff}
    \qty(\frac{m_e}{m_i})^{1/2} \qty(\frac{T_e}{T_i})^{3/2},
    \label{ion_coll}
  \end{equation}
  where \texttt{vnewk\_1} is the GS2 parameter denoting the ion-ion collision
  frequency.  The calculated values for the above collision frequencies that
  were inputs to our simulations are listed in Table~\ref{tab:sim_params} in
  Section~\ref{sec:num_setup}.

  \subsection{Real-space effect of flow shear}
  \label{sec:flow_shear}

\begin{figure}[t]
    \centering
    \includegraphics[width=0.7\linewidth]{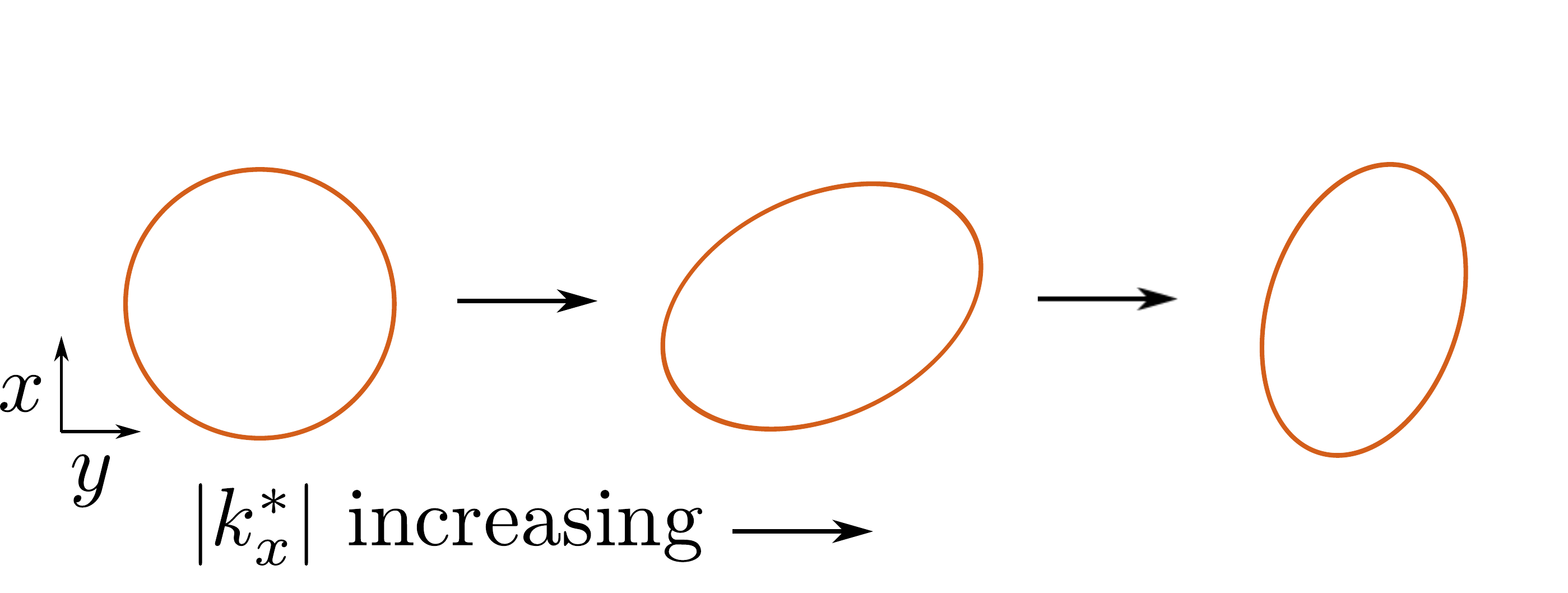}
    \caption[Physical effect of $\gamma_E$]{
      Illustration of the effect of flow shear of turbulent structures.  As
      $k_x^*$ increases in time there is increased radial structure and
      displacement in the $y$ direction.
    }
    \label{fig:flow_shear_effect}
  \end{figure}

  Flow shear is implemented in GS2 by allowing $k_x$ to vary with
  time~\cite{Hammett2006}:
  \begin{equation}
    k_x^*(t) = k_x - \gamma_E k_y t.
    \label{kx_time}
  \end{equation}
  In simplified terms, GS2 shifts the fluctuation fields along the $k_x$
  dimension as a function of time (see~\cite{HighcockThesis} for a complete
  review of the GS2 flow shear algorithm). This leads to finer radial structure
  and a displacement of fluctuations in the $y$ direction, as
  illustrated in \figref{flow_shear_effect}. However, complications arise in
  this implementation as a result of the fixed $k_x$ grid in GS2,
  which causes jumps in the displacement of fluctuations in the $y$ direction
  at the radial extremes of the box as we will now explain.

  When $k_x^*$ changes by $\delta k_x = \gamma_E k_y \Delta t$, where $\Delta
  t$ is a GS2 time step, the value of the GS2 fluctuation fields at $k_x$ would
  ideally be shifted to $k_x \pm \delta k_x$. However, the $k_x$ grid
  is fixed in GS2 (with a grid separation of $\Delta k_x$) and so the
  fluctuation fields must be shifted by at least $\Delta k_x$. This issue is
  resolved in GS2 by keeping track of the difference between the exact shift in
  $k_x$ and the grid spacing $\Delta k_x$: when the exact shift is less than
  $\Delta k_x/2$, no shifting takes place but the value is recorded and added
  to the size of the shift at the next time step. This process is repeated
  until the shift is greater than or equal to $\Delta k_x/2$, at which point
  all fluctuation fields are shifted by $\Delta k_x$.

  The distribution function calculated by GS2 is of the form
  \begin{equation}
    h \sim \exp[i (k_x^* x + k_y y)].
    \label{gs2_h}
  \end{equation}
  Substituting for $k_x^*$ using~\eqref{kx_time}, we get $h \sim \exp[i (k_x x
  + k_y y - \gamma_E k_y x t)]$ and we can identify the wave frequency
  $\omega_h = \gamma_E k_y x$ to calculate the group velocity
  \begin{equation}
    \vb*{v}_g = \pdv*{\omega_h}{\vb*{k}} = -\gamma_E x \vu*{y}.
    \label{v_group}
  \end{equation}
  Writing $\vb*{v}_g = \Delta y / \Delta t$, we find the displacement of
  fluctuations in the $y$ direction, for an ideal $k_x$ shift of $\delta k_x =
  \gamma_E k_y \Delta t$,
  \begin{equation}
   \Delta y =  - \frac{ \delta k_x x}{k_y}.
   \label{y_disp}
  \end{equation}
  However, $\delta k_x$ is forced to match the fixed $k_x$
  grid with a spacing $\Delta k_x = 2 \pi / L_x$, where $L_x$ is the size of
  the box in the $x$ direction. Using $k_y = 2 \pi / \lambda_y$,
  where $\lambda_y$ is the wavelength of a given $k_y$ mode, we can finally
  write the displacement due to the flow shear as,
  \begin{equation}
   \Delta y = \lambda_y \frac{x}{L_x}.
   \label{y_disp_final}
  \end{equation}
  This means that at the edges of the radial domain, where $x = \pm L_x/2$, the
  displacement in the $y$ direction for every shift in $k_x$ due to the flow
  shear is $\Delta y = \pm \lambda_y/2$. The rate of shifting is dependent on
  $k_y$ according to~\eqref{kx_time} and so the largest modes (smallest
  $k_y$s) will be acted on more infrequently than smaller modes
  (larger $k_y$s). However, the largest modes are then shifted by half the
  size of their wavelength according to~\eqref{y_disp_final}. This causes
  visual separation (or multiplication) of structures at the edges of the GS2
  domain in real space in a way that may affect our correlation analyses
  performed in Chapter~\ref{sec:struc_of_turb}.

  We emphasise that the separation of turbulent structures we have described
  above is only present in the real-space representation of the GS2
  distribution function. Given that GS2 performs calculations (apart from the
  calculation of nonlinear interactions) in Fourier space, this does not
  present a problem to the overall calculation. We note that the implementation
  of flow shear in GS2 is correct in the limit of infinitely small $\Delta k_x$
  and so it is sufficient to check convergence with $\Delta k_x$ to be
  confident of our results. Ideally, some form of interpolation could be used
  to smooth out these shifts in $k_x$ and a future program of work is planned
  to implement this in GS2.

  \subsection{Implementation of hyperviscosity}
  \label{sec:hyperviscosity}

  In addition to the dissipation caused by collisions (see
  Section~\ref{sec:collisions}), it is possible to dissipate energy
  artificially at moderately small spatial scales, rather than having to
  resolve the entire spatial cascade of energies. However, this has to be done
  in such a way so as not to affect the turbulent transport that we are
  trying to predict by running simulations.  The benefit of such artificial
  dissipation is that it allows us to damp the dynamics at small scales where we do
  not expect the contribution to the transport to be large, but would require
  significant computational resources to resolve (see
  Appendix~\ref{App:hyperviscosity}).  Hyperviscosity is one such technique for
  artificially damping turbulent dynamics at small scales (large wavenumbers).
  Whereas collisional dissipation acts on large gradients in velocity space,
  hyperviscosity directly damps large wavenumbers.

  The GS2 implementation is based on a 2D Smagorinsky-like hyperviscosity
  subgrid model~\cite{Belli2006}.  It is a fourth-order damping model applied
  to the non-adiabatic part of the distribution function at every time step,
  with the result that a perturbed quantity like the
  electrostatic potential $\varphi$ is multiplied at each time step by
  \begin{equation}
    \exp \left [ -D_{\mathrm{hv}} S \Delta t
                   {\left (\frac{k_\perp}{k_{\perp, \max}}\right)}^4 \right],
    \label{hypervisc}
  \end{equation}
  where $D_{\mathrm{hv}}$ is a constant coefficient controlling the
  strength of the hyperviscosity (denoted by \texttt{d\_hypervisc} in
  GS2), $k_\perp^2 = k_x^2 + k_y^2$, $k_{\perp,\max}$ is the largest
  perpendicular wavenumber in the simulation, and $S$ is the $x$-$y$ averaged
  shearing rate, defined in terms of the perturbed \exb drift velocity
  $\vb*{V}_E = (c/B) \vu*{b} \cross \nabla \ensav{\varphi}{\vb*{R}_s}$
  as~\cite{Belli2006}
  \begin{equation}
    \begin{split}
      S^2(\theta) & = \left< {\left(\dv{V_{Ex}}{x} \right)}^2 +
                             {\left(\dv{V_{Ey}}{y} \right)}^2 +
                              \frac{1}{2}{\left(\dv{V_{Ex}}{x} +
                              \dv{V_{Ey}}{y} \right)}^2 \right>_{x,y} \\
                  & = \sum_{k_x} \sum_{k_y} k_\perp^4 \frac{c}{B} |\varphi|^2,
    \end{split}
    \label{hyp_shear_rate}
  \end{equation}
  where $\left<\cdots \right>_{x,y}$ indicates an average over $x$-$y$ space.
  We see that the damping rate in \eqref{hypervisc} is a function of $k_\perp$
  and thus damps large wavenumbers most strongly.

  Equation~\eqref{hyp_shear_rate} shows that the damping due to hyperviscosity
  depends on the amplitude of $\varphi$. This is beneficial when focusing on
  nonlinear simulations since it reduces the importance of choosing the right
  value of $D_{\mathrm{hv}}$, i.e., the damping rate will change dynamically with
  the amplitude of the plasma dynamics.  However, it complicates the study of
  the linear dynamics, where the amplitude of $\varphi$ grows exponentially in
  time -- with the implication that hyperviscous damping would have an
  ever-increasing effect. Whereas in a saturated nonlinear simulation, the
  damping due to hyperviscosity would be roughly constant (since $\varphi$ is
  roughly constant). For this reason, there are two methods for using
  hyperviscosity in GS2, controlled by the input flag \texttt{const\_amp}:
  \begin{itemize}
    \item \texttt{const\_amp = True}: The shearing rate \eqref{hyp_shear_rate}
      $S = 1$ and the level of damping will
      only depend on the value of $D_{\mathrm{hv}}$ and the wavenumber.
    \item \texttt{const\_amp = False}: The level of damping will depend on the
      fluctuation amplitude of $\varphi$ via~\eqref{hyp_shear_rate}.
  \end{itemize}

  In this work, we are interested in both the linear and nonlinear behaviour,
  and so our simulations were all run with \texttt{const\_amp = False}. This
  allows us to study linear growth rates and be sure they are relevant to our
  nonlinear simulations. When using hyperviscosity, it is important to study
  its effect on linear growth rates and turbulent transport. We investigate
  this in Appendix~\ref{App:hyperviscosity} and show that by damping electron
  spatial scales we are able to keep simulations resolutions modest while not
  significantly affecting the turbulent transport. This was further tested by
  sensitivity scan for nonlinear simulations: assessing that the precise value
  of $D_{\mathrm{hv}}$ did not affect any measured quantities.

\section{Numerical set-up}
\label{sec:num_setup}

\begin{figure}[t]
    \centering
    \includegraphics[width=0.6\linewidth]{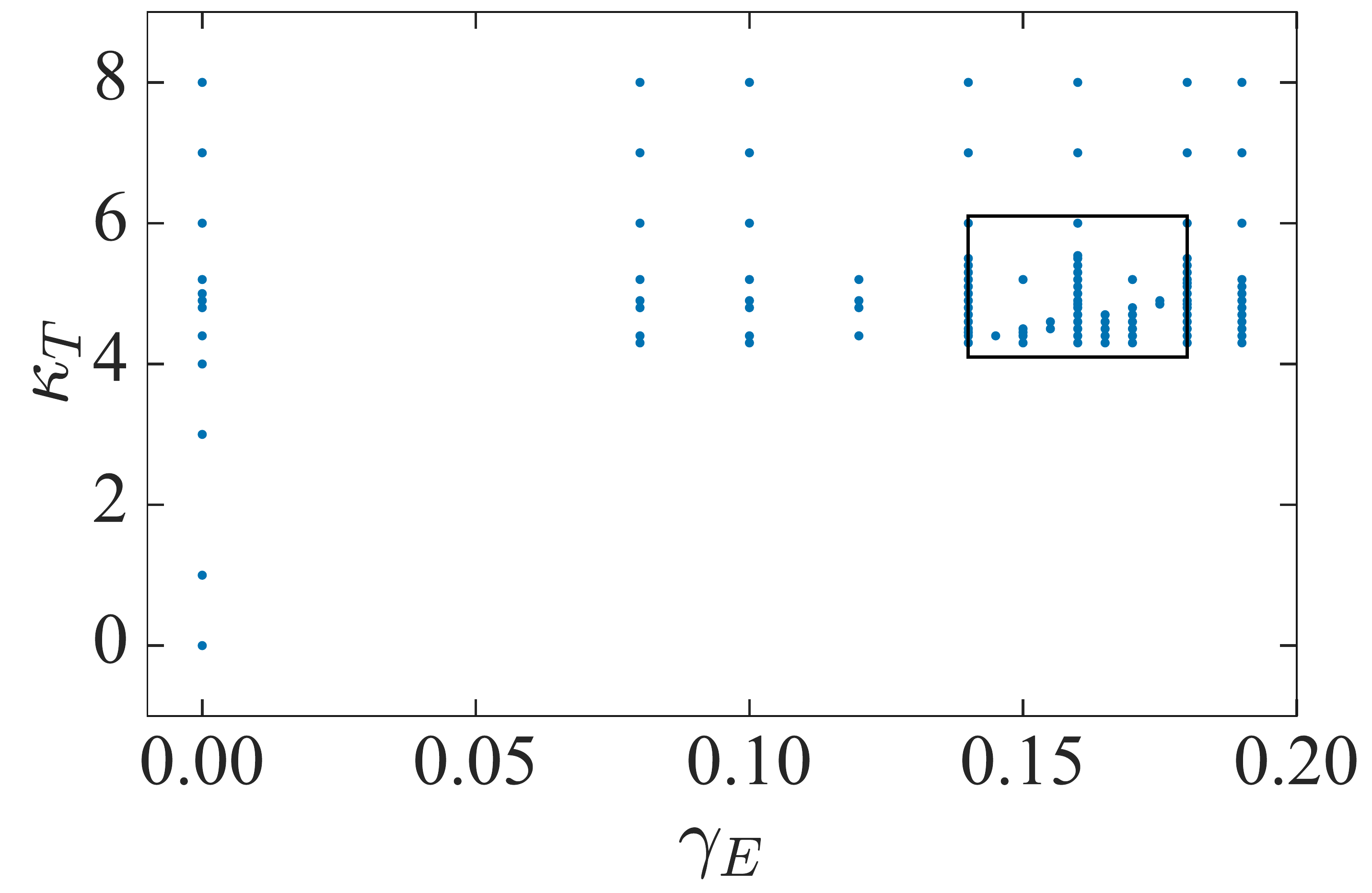}
    \caption[$(\kappa_T, \gamma_E)$ values in parameter scan]{
      Equilibrium values of $\kappa_T$ and $\gamma_E$ for the parameter scan in
      this study. The highlighted region indicates the region of experimental
      uncertainty. Simulations in the region $0 < \gamma_E < 0.08$ were not
      reliable due to resolution constraints (see main text).
    }
    \label{fig:scan_scatter}
  \end{figure}

  The MAST equilibrium parameters used in our simulations were extracted
  from the MAST diagnostics and EFIT equilibrium, as explained in
  Section~\ref{sec:exp_profiles}. In practise, these diagnostic measurements
  and equilibria are cleaned and serve as input to a TRANSP analysis to
  calculate the transport coefficients.  As a result, the output from a TRANSP
  analysis contains all the information necessary to run GS2 simulations. To
  extract these parameters, an open-source, and freely available
  package\footnote{\url{https://github.com/ferdinandvanwyk/transp_to_gs2}} was
  developed that reads a TRANSP output file and calculates all the required GS2
  parameters.  The equilibrium parameters at $r=0.8$ and $t=0.25$~s, for the
  MAST discharge \#27274 we will be investigating in this work, are listed in
  Table~\ref{tab:sim_params}.  The two nominal experimental values for the
  parameters we vary in this study were $\kappa_T = 5.1 \pm 1$ and $\gamma_E =
  0.16 \pm 0.02$; however, we also scanned outside the region of experimental
  uncertainty in order to map out the turbulence threshold more fully. Overall,
  our parameter scan consisted of 76 simulations over the regions $\kappa_T \in
  [4.3,8.0]$ and $\gamma_E \in [0, 0.19]$.  \Figref{scan_scatter} shows the
  parameter values for the full parameter scan in this study, where the
  highlighted region indicates parameters that lie within the experimental
  uncertainty. Due to resolution constraints, we were not able to simulate
  between $0 < \gamma_E \lesssim 0.08$ (as explained in
  Appendix~\ref{App:res_flow_shear}).

  {\renewcommand{\arraystretch}{1.25}%
  \begin{table}[t]
    \centering
    \caption{GS2 equilibrium parameters calculated from diagnostic measurements
      and from the EFIT equilibrium of the MAST discharge \#27274 and
      appropriately normalised. The nominal experimental values for $\kappa_T$
      and $\gamma_E$ are $\kappa_T = 5.1 \pm 1$ and $\gamma_E = 0.16 \pm 0.02$.
      The reference magnetic field is the toroidal magnetic field strength at
      the magnetic axis, i.e., $B_{\mathrm{ref}} = B_{\phi}(r=0)$. See
      Appendix~\ref{App:gs2_input_file} for an example GS2 input file with
      these parameters.
    }
    \begin{tabular}{r c c}
      \toprule

      Quantity & GS2 variable & Value \\
      \midrule

      $\beta = {8\pi n_i T_i}/{B_\mathrm{ref}^2}$ & \texttt{beta} &
      0.0047 \\

      $\beta' = \pdv*{\beta}{r}$ & \texttt{beta\_prime\_input}
      & -0.12 \\

      Eff. ion charge $Z_{\mathrm{eff}}
      = {\sum_i n_i Z_i^2/|\sum_i n_i Z_i|}$ & \texttt{zeff} & 1.59 \\

      Elec.-ion collisionality $\nu_{ei}$ & \texttt{vnewk\_2} & 0.59 \\

      Elec. density $n_{eN} = n_e/n_i$ &
      \texttt{dens\_2} & 1.00 \\

      Elec. density grad. $1/L_{ne} = - \dv*{\ln n_e}{r}$ &
      \texttt{fprim\_2} & 2.64 \\

      Elec. mass $m_{eN} = m_e/m_i$ &
      \texttt{mass\_2} & $1 / (2 \times 1836)$ \\

      Elec. temp. $T_{eN} = T_e/T_i$ &
      \texttt{temp\_2} & 1.09 \\

      Elec. temp. grad. $1/L_{Te} = - \dv*{\ln T_e}{r}$ &
      \texttt{tprim\_2} &  5.77\\

      Elongation $\kappa$ & \texttt{akappa} & 1.46 \\

      Elongation derivative $\kappa' = \dv*{\kappa}{r}$ &
      \texttt{akappri} & 0.45 \\

      Flow shear $\gamma_E = (r_0/q_0) \dv*{\omega}{r} (a/v_{{\mathrm{th}}i})$ &
      \texttt{g\_exb} & [0, 0.19] \\

      Ion collisionality $\nu_i$ & \texttt{vnewk\_1} & 0.02 \\

      Ion density $n_{iN} = n_i/n_i$ & \texttt{dens\_1} & 1.00 \\

      Ion density grad. $1/L_{ni} = - \dv*{\ln n_i}{r}$ &
      \texttt{fprim\_1} & 2.64 \\

      Ion mass $m_{iN} = m_i/m_i$ & \texttt{mass\_1} & 1.00 \\

      Ion temp. $T_{iN} = T_i/T_i$ & \texttt{temp\_1} & 1.00 \\

      Ion temp. grad. $\kappa_T \equiv 1/L_{Ti} = - \dv*{\ln T_i}{r}$ &
      \texttt{tprim\_1} & [4.3, 8.0] \\

      Magnetic shear $\hat{s} = r_0/q_0\dv*{q}{r}$ & \texttt{s\_hat\_input} & 4.00\\

      Magnetic field reference point $R_\mathrm{geo}$ & \texttt{r\_geo} & 1.64 \\

      Major radius $R_{N} = R/a$ & \texttt{rmaj} & 1.49 \\

      Miller radial coordinate $r_0 = {D/2a}$ & \texttt{rhoc} & 0.80 \\

      Safety factor $q_0 = \pdv*{\psi_\mathrm{tor}}{\psi_{\mathrm{pol}}}$ &
      \texttt{qinp} & 2.31\\

      Shafranov Shift $1/a \dv*{R}{r}$ & \texttt{shift} & -0.31 \\

      Triangularity $\delta$ & \texttt{tri} & 0.21 \\

      Triangularity derivative $\delta' = \dv*{\delta}{r}$ &
      \texttt{tripri} & 0.46 \\

      \bottomrule
    \end{tabular}
    \label{tab:sim_params}
  \end{table}}

  Previous investigations~\cite{Roach2009,Field2011} of similar MAST
  discharges have found that electrons play an important role in driving
  turbulence in MAST, even at ion scales. Our study confirmed these findings:
  in Appendix~\ref{App:linear_sims} we present a series of linear simulations
  with $\gamma_E=0$ while varying $\kappa_T$. We show that the maximum linear
  growth rates at ion scales for simulations with a kinetic electron species is
  $\sim$~2--3 times larger than linear simulations with adiabatic electrons.
  Initial simulations with adiabatic electrons confirmed that sustained
  turbulence required $\kappa_T$ significantly higher than even the upper
  estimate based on the experimental uncertainties.  Accordingly, we have
  included electrons in our simulations as a kinetic species. Given that our
  simulations contained only two kinetic species (deuterium ions and
  electrons), it follows from the quasineutrality condition that they must have
  the same density and density gradient, i.e., $n_i$ = $n_e$ and $L_{ni} =
  L_{ne}$.

  Previous work investigating electromagnetic effects
  in MAST plasmas~\cite{Applegate2004,Roach2005a}, found that electromagnetic
  effects were only significant at $r\sim0.5$, where $\beta\geq0.1$.
  In the outer-core region we consider in this work, where $\beta \sim 0.005$,
  these effects are not significant and we are thus able to assume the plasma
  is electrostatic.

  We determined the appropriate grid sizes for our nonlinear
  simulations using the results from the linear simulations without flow shear
  presented in Appendix~\ref{App:hyperviscosity} and~\ref{App:linear_sims}.
  Without hyperviscosity, we found strong linear growth at both ion and
  electron scales without a clear separation -- suggesting expensive
  multiscale simulations are required.  However, we are only interested in ion
  scales (given that the BES diagnostic measures turbulent dynamics at this
  scale), while still including the effect of kinetic electrons. Therefore, we
  have made use of hyperviscosity and show in Appendix~\ref{App:hyperviscosity}
  that we can truncate our nonlinear simulations at $k_y \rho_i \gtrsim 2$,
  where we have chosen $k_y \rho_i \sim 3$, and verified that changes in
  this cut-off scale or the number of $k_y$ modes (where we were only able to
  test with $\sim20\%$ more $k_y$ modes due to cost constraints) do not
  significantly affect the turbulence.

  In the $x$ direction, we have chosen our grid based on the grid spacing
  $\Delta k_x$ such that we could resolve reasonably small values of $\gamma_E$
  (as explained above and in Section~\ref{sec:flow_shear}). Again, we have
  verified that changes in $k_{x,\max}$ or the number of $k_x$ modes (where we
  increased the number by $50\%$) do not significantly affect the turbulence. In
  both the $x$ and $y$ directions, we chose the truncation scale to be somewhat
  higher than necessary to ensure a sufficient ``inertial range'' between the
  injection and dissipation scales and such that favourable parallelisation was
  achieved when decomposing our grids over supercomputing nodes.

  In the parallel direction we chose the smallest grid that adequately resolved
  the eigenfunction and ensured that it reached very small values at the edges
  of the parallel domain. The cost of GS2 simulations is a strong function of
  the parallel resolution and so minimising parallel resolution was key to
  being able to run such a large numerical study.

  In velocity space, we again chose grid sizes as small as possible in order
  to minimise computational cost. We tested this by ensuring that the
  velocity-space integrals had small errors when velocity-space grid points
  were added or taken away.

  Table~\ref{tab:resolution_params} lists the GS2 resolution input parameters
  used for our nonlinear simulations. We note that the pseudo-spectral method
  employed by GS2 requires additional Fourier modes to prevent
  aliasing~\cite{Orszag1970}. As a result, the number of physical grid points
  were $85 \times 32 \times 20$ in the radial, binormal, and parallel
  directions (while the number of grid points in the code was $128 \times 96
  \times 20$), and $27 \times 16$ pitch-angle and energy-grid points,
  respectively. We chose the box sizes in $x$ and $y$ to be $L_x \approx 200
  \rho_i$ and $L_y \approx 62\rho_i$, respectively, while $\theta \in [-\pi,
  \pi]$. We note that while $L_x$ is comparable to the size of MAST, the
  turbulence predicted by GS2 can only be compared to experimental MAST
  turbulence at $r=0.8$.  All of the GS2 parameters summarised in this section
  can be found in the example GS2 input file in
  Appendix~\ref{App:gs2_input_file}.
  {\renewcommand{\arraystretch}{1.5}%
  \begin{table}
    \centering
    \caption{Resolution parameters used in our nonlinear simulations. See
      Appendix~\ref{App:gs2_input_file} for and example GS2 input file with
      these parameters.
    }
    \begin{tabular}{r c c}
      \toprule
      Name & GS2 variable & Value \\
      \midrule
      No. of $k_x$ modes & \texttt{nx} & 128 \\
      No. of $k_y$ modes & \texttt{ny} & 96 \\
      $\theta$ grid points & \texttt{ntheta} & 20 \\
      $\vareps_s$ grid points & \texttt{negrid} & 16 \\
      $\lambda'_s$ grid points & \texttt{ngauss} & 8 \\
      $x$ box size parameter & \texttt{x0} & 10 \\
      $y$ box size parameter & \texttt{y0} & 10 \\
      No. of $2\pi$ parallel segments & \texttt{nperiod} & 1 \\
      Hyperviscosity coefficient & \texttt{d\_hypervisc} & 9 \\
      \bottomrule
    \end{tabular}
    \label{tab:resolution_params}
  \end{table}}

\chapter{Nonlinear simulations}
\label{sec:nl}

\section{Introduction}
\label{sec:part_2_intro}
  In this chapter, we present the results of a parameter scan in $\kappa_T$ and
  $\gamma_E$. We focus on the prediction of the ion heat flux $Q_i$ and make
  comparisons with experimental estimates of the ion heat flux $Q_i^{\exp}$
  calculated from TRANSP results. In a fusion reactor we would like to
  maximise the core temperature (and hence the temperature gradient between the
  edge and the core) at a given heat flux. In local simulations, the heat
  flux is a useful measure of the level of turbulence and we would, therefore,
  like to explore how the heat flux changes with the equilibrium parameters
  that we vary and whether our simulations are in agreement with experimental
  measurements. This will allow us to gain confidence in our models and
  eventually make predictions for the optimal parameters to maximise the fusion
  power for a given reactor.  We exclusively vary $\kappa_T$ and $\gamma_E$,
  while keeping all other equilibrium quantities constant. In other words, we
  do not self-consistently recalculate other equilibrium quantities that would
  be needed to support the values of $\kappa_T$ and $\gamma_E$ that we use.
  However, this allows us to isolate the effect of these two parameters on MAST
  turbulence. We demonstrate in Section~\ref{sec:heat_flux} that GS2 is able to
  match the experimental heat flux at equilibrium values within the
  experimental uncertainty and that the experiment lies close to the turbulence
  threshold.

  We showed in Sections~\ref{sec:gk_eqn} and~\ref{sec:gamma_stab}, that the ITG
  is a source of free energy, which drives instabilities, while flow shear has
  a stabilising effect on turbulence. In the absence of a background flow shear,
  numerical studies have suggested that ITG-unstable plasma reaches a
  statistically steady-state in the following way~\cite{Waltz1994, Dimits2000,
  Rogers2000}.  Linear modes are unstable due to the ITG instability and grow
  exponentially in time. Once the modes have sufficient amplitude, they interact
  nonlinearly to give rise to a turbulent state. The nonlinear interactions
  spontaneously generate ``zonal flows'' (poloidally symmetric flows with
  finite radial wavenumber). The zonal flows give rise to an \exb
  shear and have a suppressing effect on turbulence. When the nonlinear
  interaction is sufficiently suppressed, linear growth due to the ITG
  instability returns and the process repeats.

  In the presence of a background
  flow shear, the situation may become more complicated. It has been
  shown, in simple geometries, that the turbulence can become
  subcritical~\cite{Newton2010,Schekochihin2012,Landreman2015},
  i.e., large initial perturbations are required to ignite turbulence, as
  opposed to only requiring infinitesimal perturbations in conventional
  supercritical turbulence. In
  Section~\ref{sec:subcritical}, we show that the turbulence for the MAST
  configuration we are investigating is subcritical. We study the linear
  dynamics and estimate the conditions necessary to ignite turbulence,
  namely the transient-amplification factor and time. Studying the real-space
  structure of turbulence (Section~\ref{sec:struc_analysis}), we show that
  coherent, long-lived structures dominate the saturated state close to the
  turbulence threshold. Furthermore, the fluctuations in the system have a
  clear minimum amplitude needed to sustain turbulence. We present a novel
  structure counting analysis and show that the number of turbulent structures
  increases rapidly as one moves away from the turbulence threshold into more
  strongly driven regimes. Finally, we show that far from the turbulence
  threshold, the turbulence is similar to turbulence in the absence of flow
  shear, characterised by many interacting eddies.  This suggests that the
  observed nonlinear state dominated by coherent structures is an intermediate
  state between completely suppressed turbulence and the zonal-flow regulated
  scenarios observed in conventional ITG-unstable plasmas.  We estimate the
  $\vb*{E} \times \vb*{B}$ shear due to the zonal flows
  (Section~\ref{sec:zf_shear}) and show that it is small compared to the
  background flow shear close to the turbulence threshold, but becomes
  comparable and eventually dominates over the flow shear far from the
  threshold, again resembling a system in the absence of flow shear.

\section{Heat flux}
\label{sec:heat_flux}

  We performed a parameter scan in $\kappa_T$ and $\gamma_E$ around their
  respective experimental values to investigate the resulting changes in
  turbulent transport.  The
  experimental values and associated uncertainties were $\kappa_T = 5.1 \pm 1$
  and $\gamma_E = 0.16 \pm  0.02$. However, we also performed simulations
  outside the experimental uncertainty ranges to aid our understanding of how
  the nature of the turbulence changes with $\kappa_T$ and $\gamma_E$ and, in
  particular, how it is different near to versus far from the
  (nonlinear) stability threshold. Our entire study covered $\kappa_T \in [4.3,
  8.0]$ and $\gamma_E \in [0, 0.19]$ and consisted of 76 simulations. All
  simulations were run until they reached a statistical steady state, i.e.,
  until the running time average became independent of time. Averages were
  taken over a time period of approximately $200$--$400~(a/v_{\mathrm{th}i})$
  (which corresponds to $\sim 800$--$1600~\mu$s) and in many cases longer.

\begin{figure}[t]
    \centering
    \includegraphics[width=0.75\linewidth]{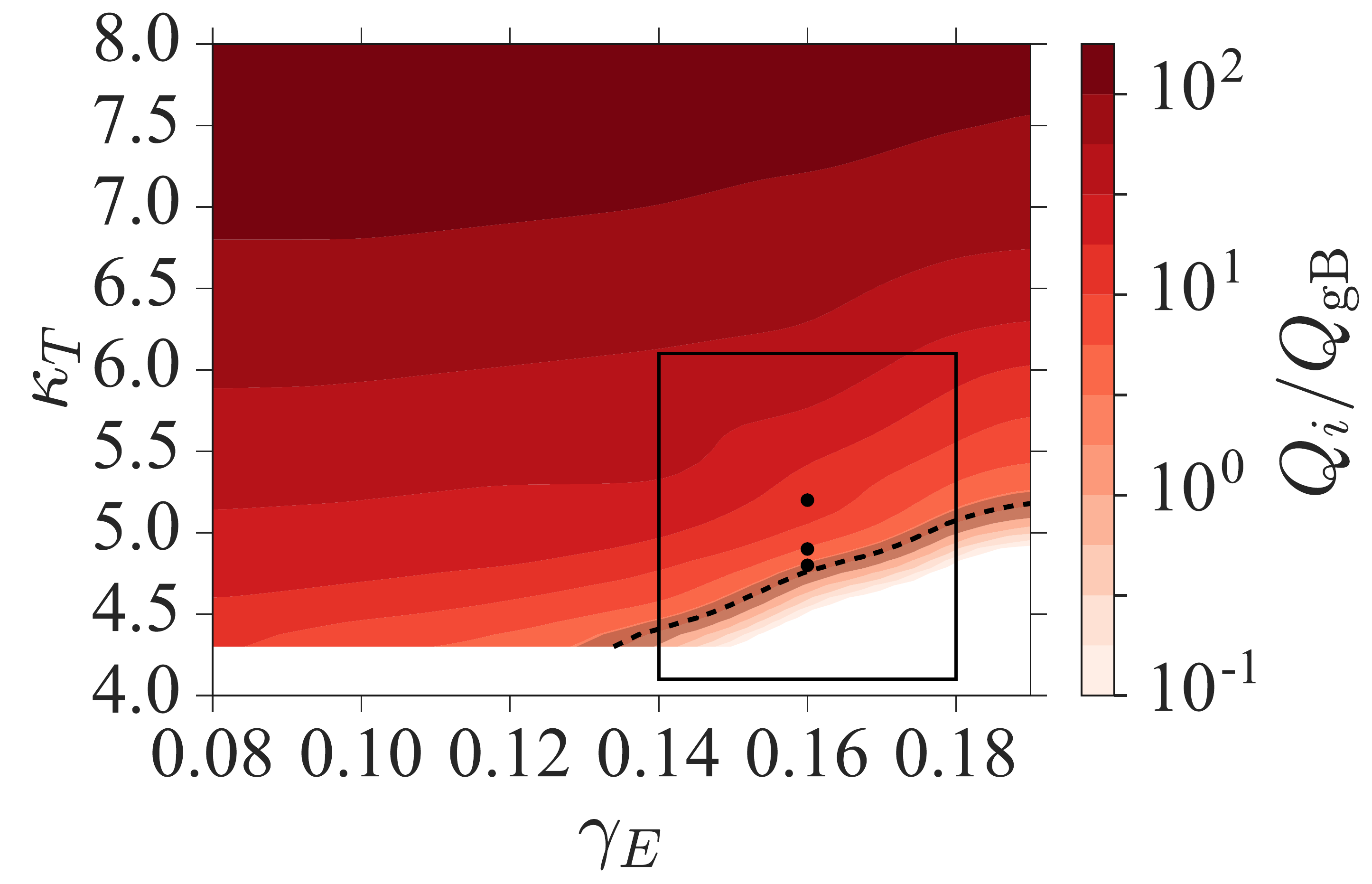}
    \caption[Contour plot of $Q_i/Q_{\mathrm{gB}}$ versus $\kappa_T$ and $\gamma_E$]{
      $Q_{i}/Q_{\mathrm{gB}}$ as a function of $\kappa_T$ and $\gamma_E$ for
      all simulations with $\gamma_E>0$. The rectangular region indicates the range
      in $\kappa_T$ and $\gamma_E$ consistent with the experiment within
      measurement uncertainties. The dashed line indicates the value of
      $Q_{i}^{\exp}/Q_{\mathrm{gB}}$ and the shaded area the experimental
      uncertainty. The experiment is clearly near the turbulence threshold
      defined by $(\kappa_T, \gamma_E)$. The points indicate the parameter
      values for which the density-fluctuation fields are shown
      in~\figref{density_fluctuations}.
    }
    \label{fig:contour_heatmap}
  \end{figure}

\begin{figure}[t]
    \centering
    \includegraphics[width=0.6\linewidth]{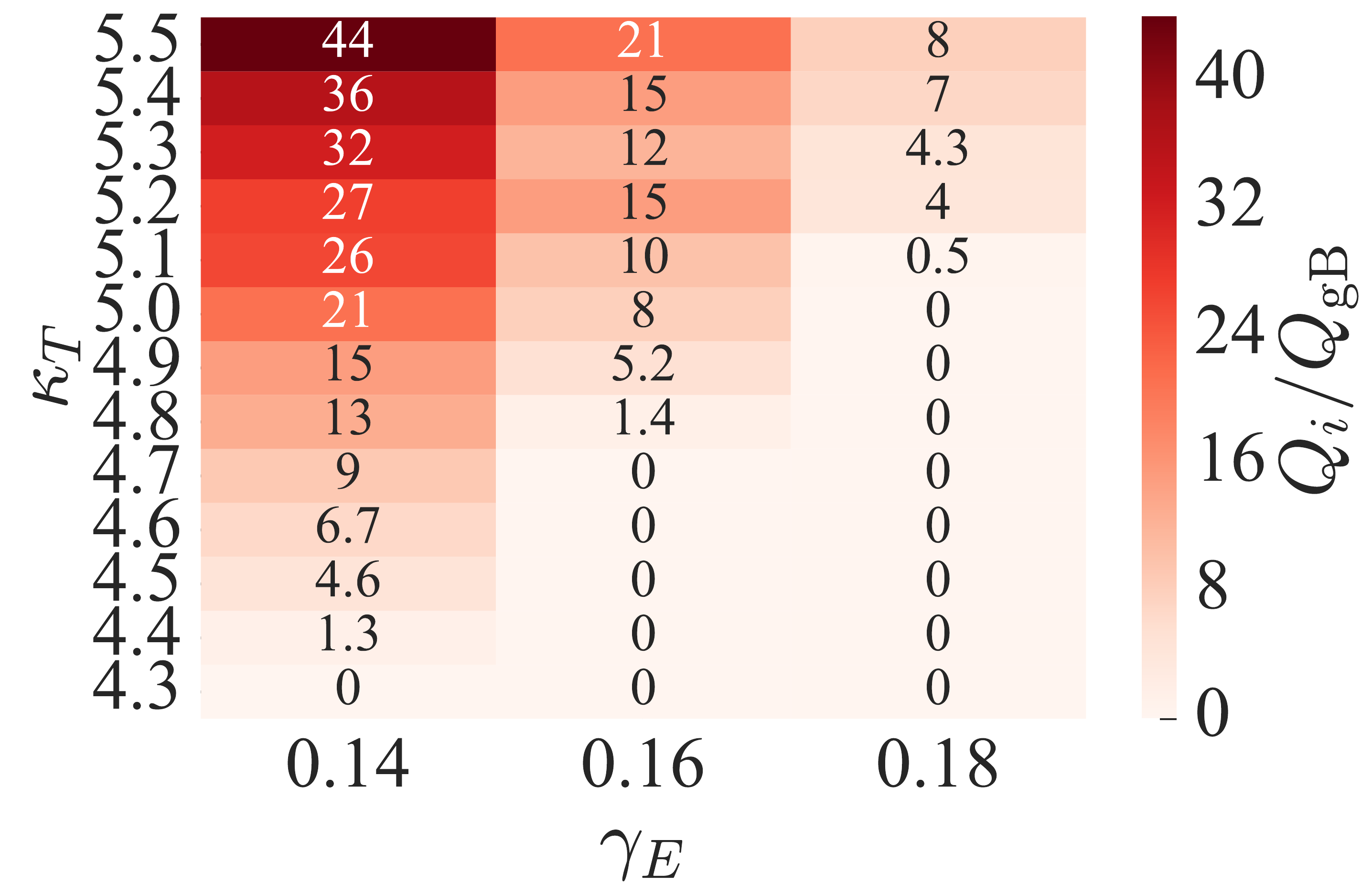}
    \caption[Values of $Q_i/Q_{\mathrm{gB}}$ versus $\kappa_T$ and $\gamma_E$]{
      Values of the ion heat flux $Q_i/Q_{\mathrm{gB}}$ as a function of
      $\kappa_T$ and $\gamma_E$ for part of the region of experimental
      uncertainty around the turbulence threshold. It is clear that the system
      is subject to ``stiff transport'' as shown by the dramatic increase in
      heat flux for small changes in our equilibrium gradient stability
      parameters.
    }
    \label{fig:value_heatmap}
  \end{figure}

  \Figsref{contour_heatmap}{value_heatmap} show the
  anomalous ion heat flux versus $\kappa_T$ and $\gamma_E$ found in our
  simulations. \Figref{contour_heatmap} shows the full parameter scan with the
  rectangular region indicating the extent of the experimental errors in each
  equilibrium parameter. The dashed line indicates the value of experimental
  heat flux, $Q_i^{\exp}/Q_{\mathrm{gB}}$, and the shaded region the
  experimental uncertainty.  \Figref{contour_heatmap} demonstrates two of the
  important conclusions of this work:
  \begin{inparaenum}[(i)]
    \item GS2 is able to match the experimental heat flux within the
      experimental uncertainties of $\kappa_T$ and $\gamma_E$, and
    \item the experiment regime is located close to the turbulence threshold
      (defined as the separating line between the regions of parameter space
      with $Q_i = 0$ and $Q_i > 0$).
  \end{inparaenum}
  \Figref{value_heatmap} shows part of the region of experimental uncertainty
  around the turbulence threshold giving the specific values of
  $Q_i/Q_{\mathrm{gB}}$ in each simulation. It demonstrates that transport is
  ``stiff'', i.e.,\ that relatively small changes in the equilibrium parameters
  lead to large changes in $Q_i/Q_{\mathrm{gB}}$ as one moves away from the
  turbulence threshold.  From~\figref{value_heatmap}, we can identify several
  simulations that represent the marginally unstable cases in our parameter
  scan: $(\kappa_T, \gamma_E) = (4.4, 0.14), (4.8, 0.16), (5.1, 0.18)$. We will
  consider these parameter values when studying the conditions necessary to
  reach a saturated turbulent state in Section~\ref{sec:subcritical}.

\begin{figure}[t]
    \centering
    \begin{subfigure}{0.49\linewidth}
      \includegraphics[width=\linewidth]{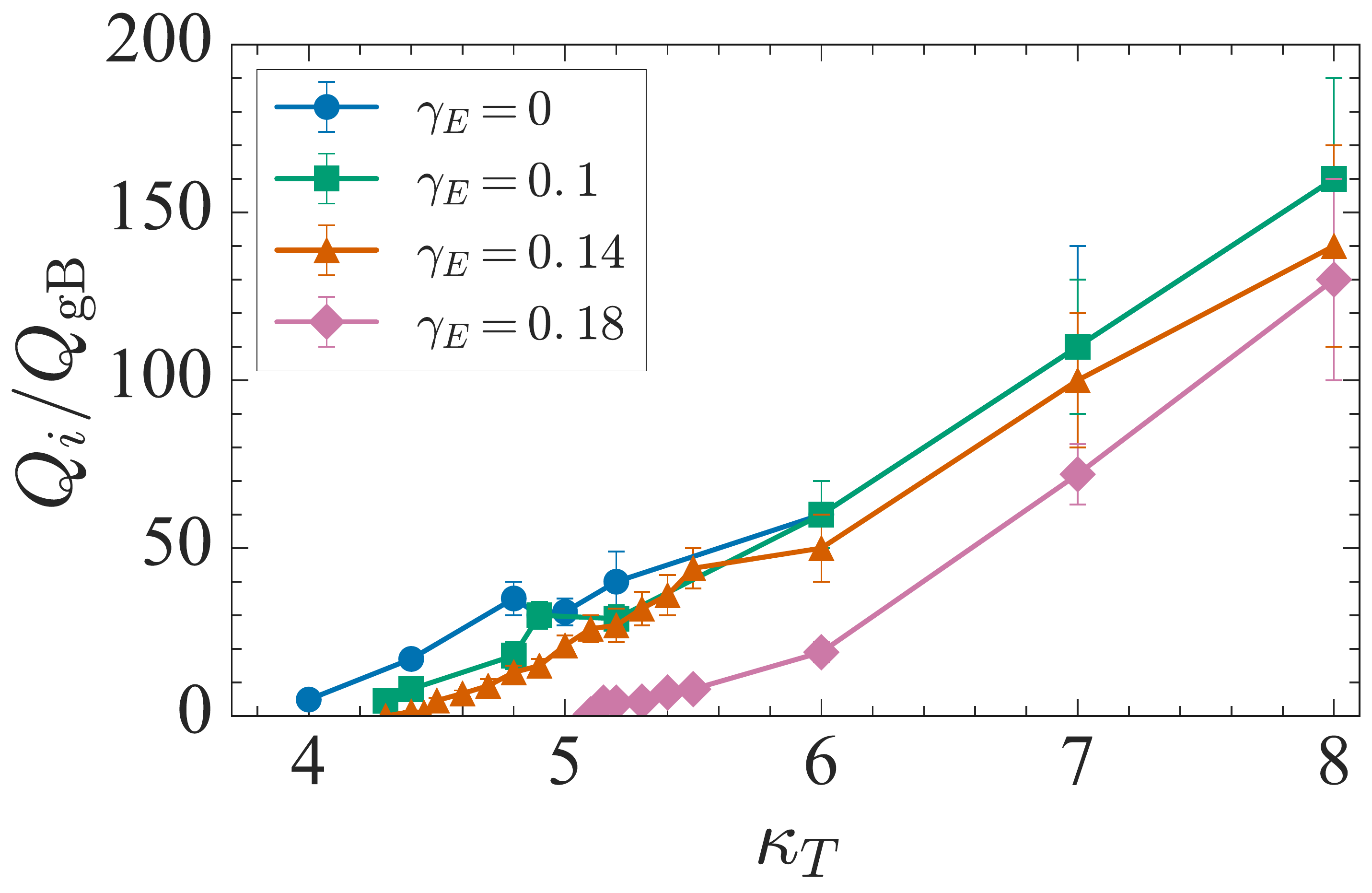}
      \caption{}
      \label{fig:q_vs_tprim}
    \end{subfigure}
    \hfill
    \begin{subfigure}{0.49\linewidth}
      \includegraphics[width=\linewidth]{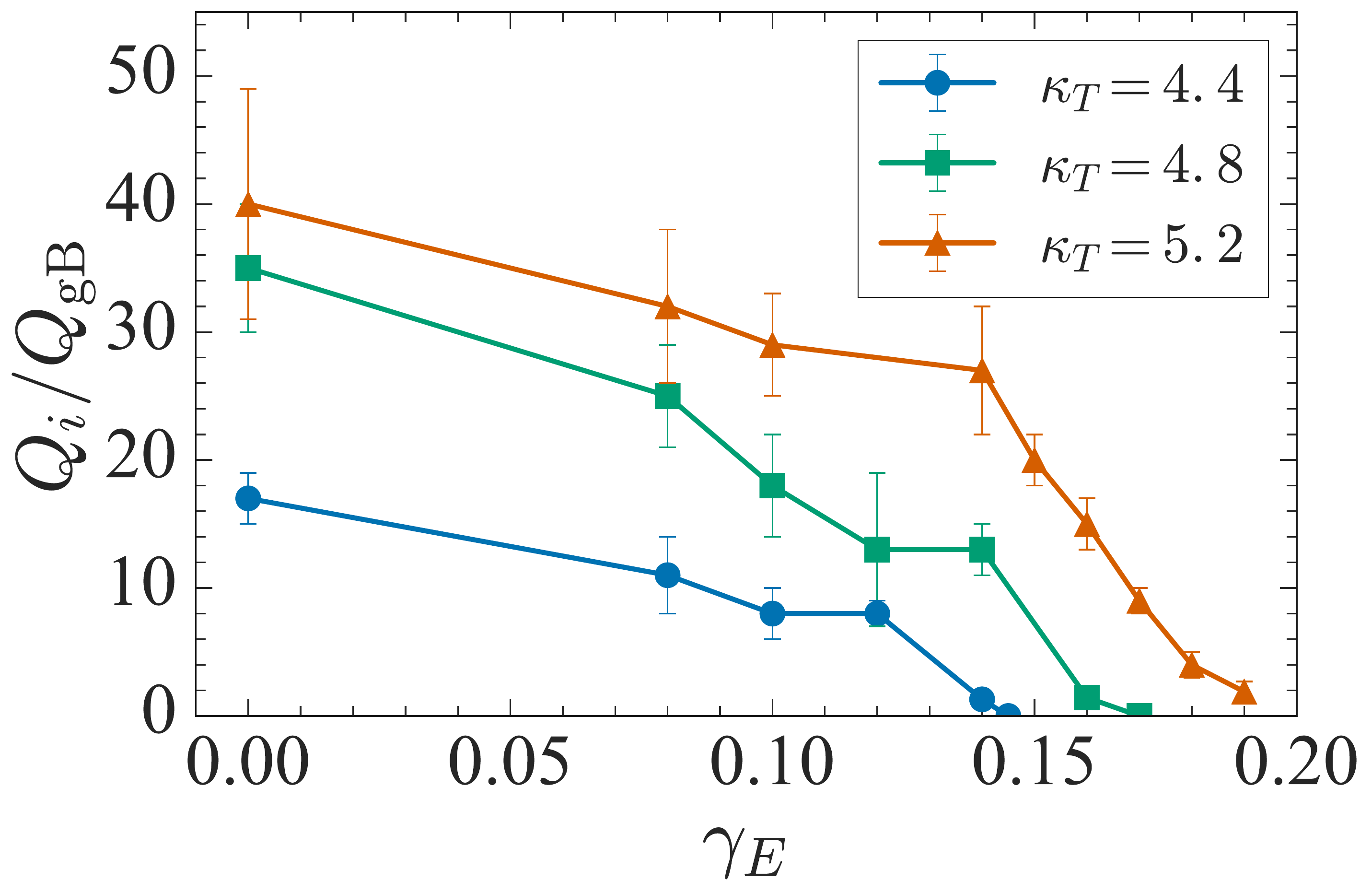}
      \caption{}
      \label{fig:q_vs_gexb}
    \end{subfigure}
    \caption[$Q_i/Q_{\mathrm{gB}}$ line plots versus $\kappa_T$ and $\gamma_E$]
      {\subref*{fig:q_vs_tprim} $Q_i/Q_{\mathrm{gB}}$ as a function of
      $\kappa_T$ for several values of $\gamma_E$ (including $\gamma_E=0$).
      \subref*{fig:q_vs_gexb} $Q_i/Q_{\mathrm{gB}}$ as a function of $\gamma_E$
      for several values of $\kappa_T$.
    }
    \label{fig:q_line_plots}
  \end{figure}

  The plots in \figref{q_line_plots} give another view of the data
  in~\figref{contour_heatmap} and also demonstrate the stiffness of the
  transport. \Figref{q_vs_tprim} shows the values of
  $Q_i/Q_{\mathrm{gB}}$ for several values of $\gamma_E$ (including
  $\gamma_E=0$) as a function of $\kappa_T$, whereas \figref{q_vs_gexb} shows
  $Q_i/Q_{\mathrm{gB}}$ as a function of $\gamma_E$ for several values of
  $\kappa_T$.  We see that an $O(1)$ change in $\kappa_T$ gives rise to an
  $O(10)$ change in $Q_i/Q_{\mathrm{gB}}$, and even more dramatically for
  changes in $\gamma_E$, which requires only an $O(0.1)$ change to cause
  $O(10)$ changes in the turbulent heat flux.  The important conclusion from
  \figref{q_vs_tprim} is that the presence of flow shear does not significantly
  affect the transport stiffness, i.e., the rate of increase of
  $Q_i/Q_{\mathrm{gB}}$ with respect to $\kappa_T$, but only changes the
  threshold value of $\kappa_T$ above which turbulence is present. This
  increase in critical ITG without a change in the stiffness of
  $Q_i/Q_{\mathrm{gB}}$ with respect to $\kappa_T$ has been observed in
  numerical simulations of simplified ITG-unstable plasmas in the presence of
  flow shear~\cite{Highcock2010, Barnes2011a}. It is also in agreement with
  experimental~\cite{Mantica2009,Mantica2011} and numerical~\cite{Citrin2014}
  findings in the outer core of the JET experiment, which also showed that ion
  heat transport stiffness is not affected by an increase in $\gamma_E$, but
  may increase the critical ITG threshold.

\begin{figure}[t]
    \centering
    \includegraphics[width=0.6\linewidth]{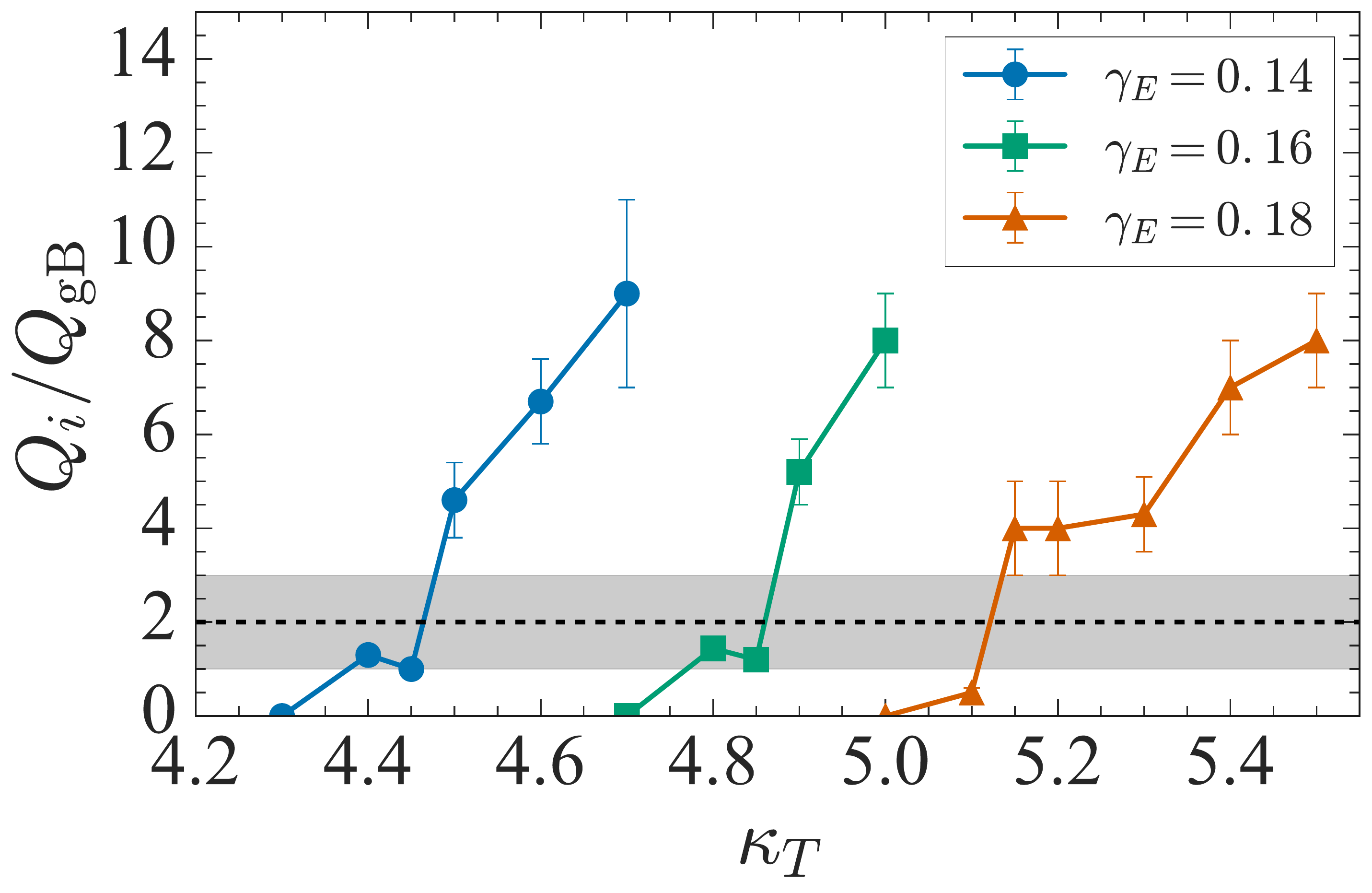}
    \caption[$Q_i/Q_{\mathrm{gB}}$ line plots within experimental uncertainty]
      {$Q_i/Q_{\mathrm{gB}}$ as a function of $\kappa_T$ strictly within
      experimental uncertainty of $\kappa_T$ and $\gamma_E$, and close to the
      turbulence threshold.  The shaded region indicates the experimental heat
      flux $Q_i^{\exp}/Q_{\mathrm{gB}} = 2 \pm 1$, determined
      from~\figref{q_exp}.
    }
    \label{fig:q_vs_tprim_marginal}
  \end{figure}

  \Figref{q_vs_tprim_marginal} shows $Q_i/Q_{\mathrm{gB}}$ as a function of
  $\kappa_T$ strictly within the region of measurement uncertainty of $\kappa_T$ and
  $\gamma_E$, close to the turbulence threshold. The dashed line and shaded
  region indicate $Q_i^{\exp}/Q_{\mathrm{gB}}$ and its associated uncertainty.
  We see that there is a range of $\kappa_T$ and $\gamma_E$ values where we
  might expect $Q_i/Q_{\mathrm{gB}}$ to match $Q_i^{\exp}/Q_{\mathrm{gB}}$, and
  we have a number of individual simulations that match the value of
  $Q_i^{\exp}/Q_{\mathrm{gB}}$. A list of these is given in
  Table~\ref{tab:exp_match_sims} . We will investigate these simulations
  further when we make more detailed comparisons with the experiment.
  \begin{table}
    \centering
    \caption{Parameter values for simulations that match the experimental
      heat flux, $Q_i^{\exp}/Q_{\mathrm{gB}} = 2 \pm 1$.
    }
    \begin{tabular}{c c c}
      \toprule
      $\kappa_T$ & $\gamma_E$ & $Q_i/Q_{\mathrm{gB}}$ \\
      \midrule
      4.4 & 0.14 & $1.3 \pm 0.1$ \\
      4.45 & 0.14 & $1.0 \pm 0.1$ \\
      4.8 & 0.16 & $1.44 \pm 0.05$ \\
      4.85 & 0.16 & $1.2 \pm 0.1$ \\
      5.15 & 0.18 & $4 \pm 1$ \\
      5.2 & 0.18 & $4 \pm 1$ \\
      \bottomrule
    \end{tabular}
    \label{tab:exp_match_sims}
  \end{table}

\section{Subcritical turbulence}
\label{sec:subcritical}
  We have found that in all our simulations with $\gamma_E>0$, a finite initial
  perturbation was required in order to ignite turbulence and reach a saturated
  turbulent state. In subcritical systems~\cite{Trefethen1993,Schekochihin2012,
  Highcock2012,Landreman2015}, linear modes are formally stable, but may
  be transiently amplified by a given factor over a given time. If the
  transient amplification is sufficient for nonlinear interactions to become
  significant before the modes decay, then a turbulent state may persist,
  provided the fluctuation amplitudes do not fall below the critical values
  (by way of random fluctuations that characterise the turbulent state) that
  prevent them being transiently amplified once again to amplitudes where
  nonlinear interactions are dominant.

  In our simulations, the amplitude of the initial condition required was found
  to depend on how far the system was from the turbulence threshold, i.e.,
  simulations far from the turbulence threshold required a smaller initial
  perturbation because they were shown to amplify transiently growing modes by a
  larger factor (see below). This suggests that the turbulence threshold
  identified in Section~\ref{sec:heat_flux} in terms of $\kappa_T$ and
  $\gamma_E$ is also a function of the amplitude of the initial condition.
  However, in this work, we have assumed that the fluctuations in the
  experiment (e.g., due to large-scale MHD modes or more virulent turbulence on
  neighbouring flux surfaces) can generate arbitrarily large perturbations as
  an initial condition to our system. For this reason, we have used the largest
  initial perturbation allowed by the numerical algorithm used in GS2 in this
  work, i.e., as large as possible without forcing the system to evolve the
  distribution function with time steps so small that the simulations would
  require prohibitively long simulation times. The
  nonlinear simulations presented in Section~\ref{sec:heat_flux} were run with
  such large initial conditions. Thus, for the regions where we have
  indicated $Q_i = 0$, we could not ignite turbulence using even the largest
  initial condition allowed by the GS2 algorithm. We will demonstrate the
  subcritical nature of the turbulence in this section by investigating the
  effect of changing the amplitude of the initial perturbation in both linear
  and nonlinear simulations.

  \subsection{Minimum initial perturbation amplitude}

  GS2 initialises the distribution function (both wavenumbers and velocity
  space) with random complex numbers between $-0.5$ and $0.5$, and scales these
  numbers via the input parameter \texttt{phiinit}. We start by considering the
  nonlinear time evolution of $Q_i/Q_{\mathrm{gB}}$ at the nominal equilibrium
  parameters $(\kappa_T, \gamma_E) = (5.1, 0.16)$ varying the value of
  \texttt{phiinit}, shown in \Figref{phiinit}. These equilibrium parameter
  values represent a simulation far from the turbulence threshold
  (see~\figref{contour_heatmap}) and yet, for a range of initial amplitudes, we
  see that the system decays rapidly. This is a clear demonstration that the
  turbulence is subcritical. We see that there is a certain minimum value of
  \texttt{phiinit} between $0.2$ and $0.3$, starting from which it is possible
  for the system to reach a saturated state, rather than decay. Importantly,
  for simulations that do reach a saturated state, the level of saturation does
  not depend on the amplitude of the initial perturbation. However, a large
  initial perturbation is not sufficient to guarantee that a subcritical system
  continues in a statistically steady state indefinitely, as we explain in the
  next section.

  \subsection{Finite lifetime of turbulence}
  \begin{figure}[t]
    \centering
    \begin{subfigure}[t]{0.49\textwidth}
      \includegraphics[width=\textwidth]{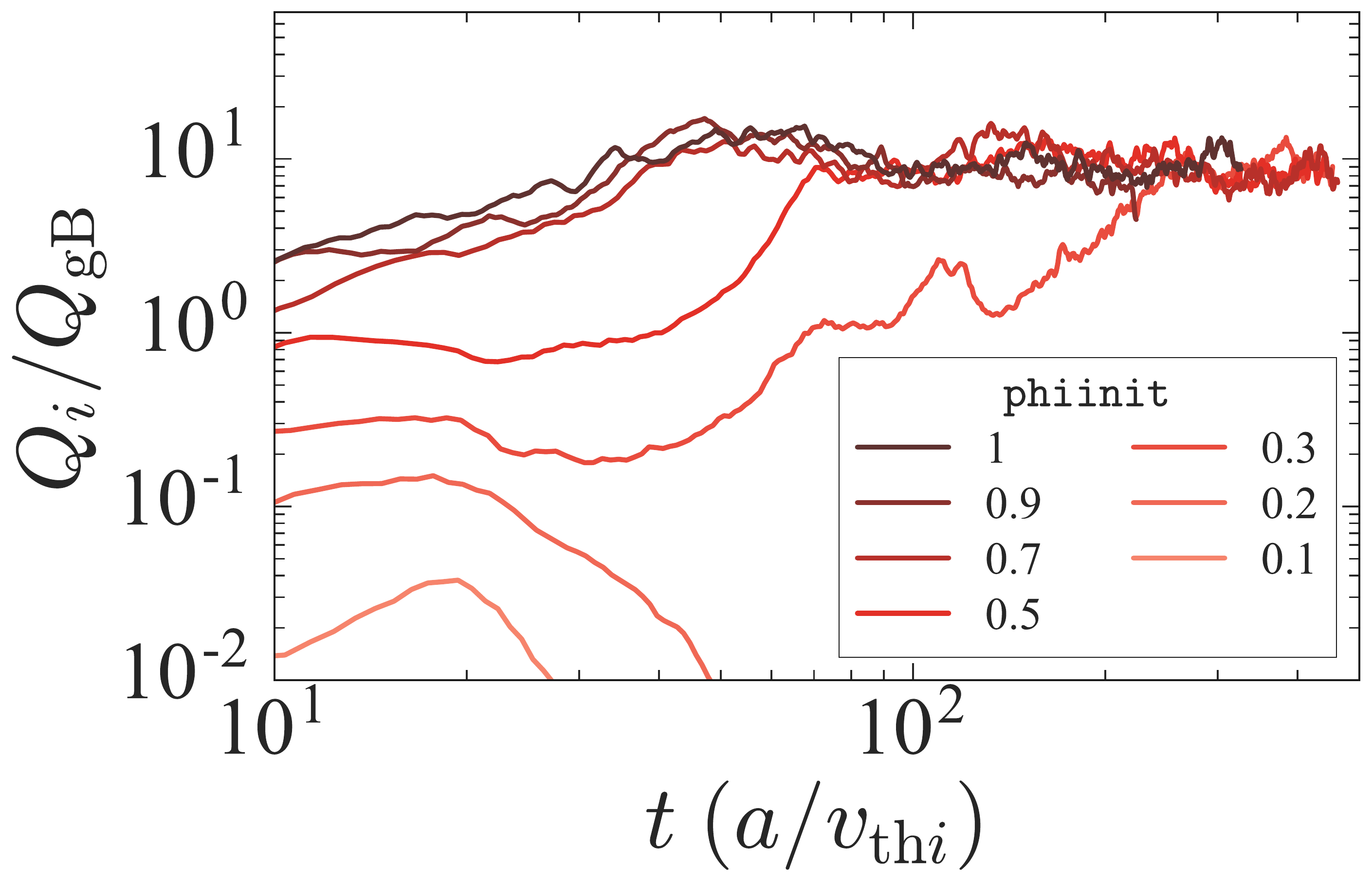}
      \caption{}
      \label{fig:phiinit}
    \end{subfigure}
    \begin{subfigure}[t]{0.49\textwidth}
      \includegraphics[width=\linewidth]{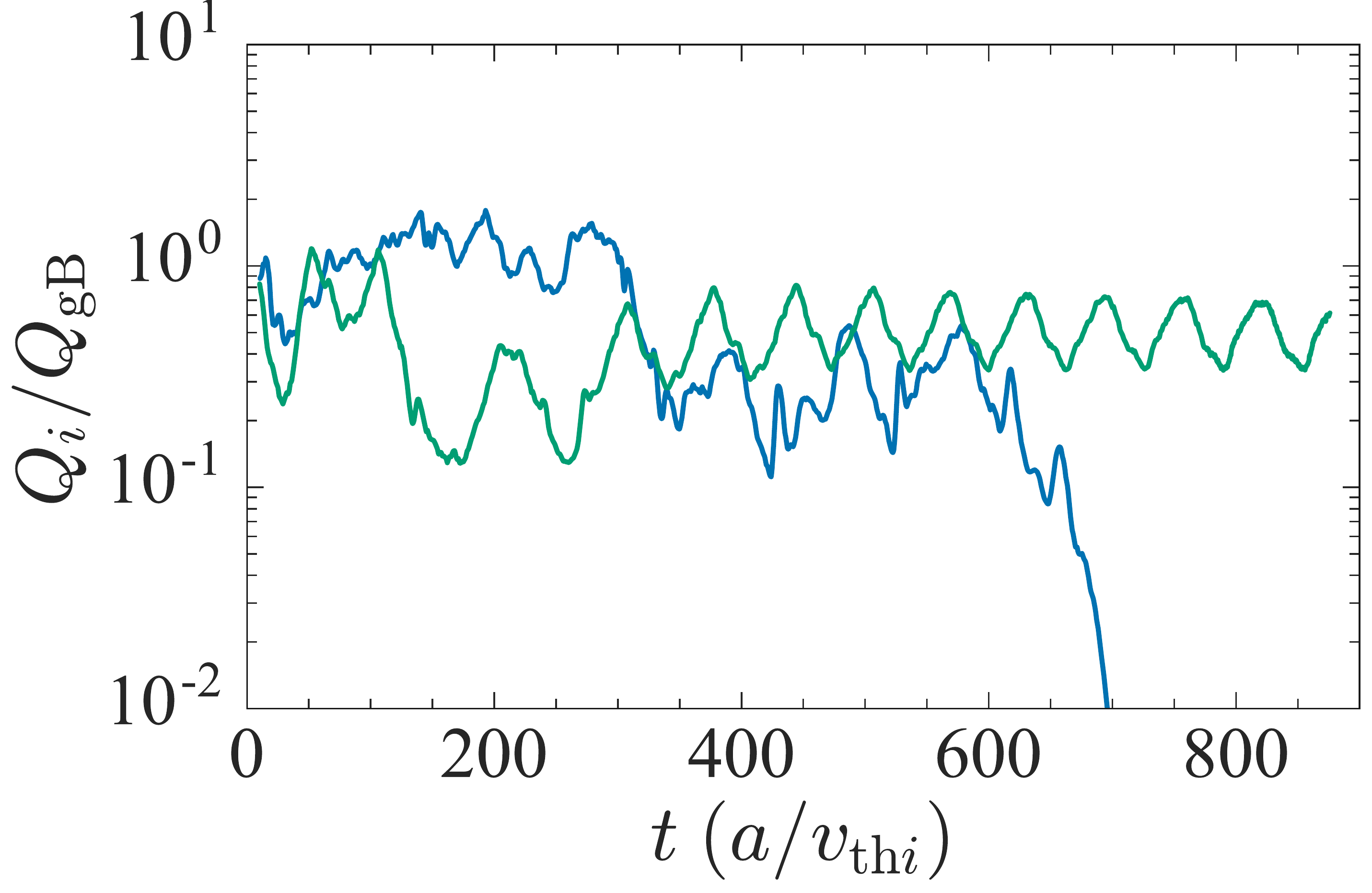}
      \caption{}
      \label{fig:subcrit_demo}
    \end{subfigure}
    \caption[Demonstration of subcritical turbulence]{
      \subref*{fig:phiinit} The ion heat flux $Q_i/Q_{\mathrm{gB}}$ as a
      function of time for different initial-condition amplitudes for
      $(\kappa_T, \gamma_E) = (5.1, 0.16)$, keeping all other parameters the
      same.  \subref*{fig:subcrit_demo} $Q_i/Q_{\mathrm{gB}}$ as a function of
      time for identical simulations at $(\kappa_T, \gamma_E ) = (5.1, 0.18)$.
      The difference between the blue and green time series is random noise
      with which GS2 initialises a simulation (having again excluded the noisy
      initial time evolution). Beyond $t=300$~$(a/v_{\mathrm{th}i})$, the
      simulations seem to converge to a similar average value before one is
      abruptly quenched due to the amplitudes falling below the critical values
      required to sustain a saturated state.
    }
  \end{figure}
  In simulations with equilibrium parameters close to the
  turbulence threshold, we found that turbulence could be quenched at a
  seemingly unpredictable time. For example, \figref{subcrit_demo} shows the
  time trace of $Q_i/Q_{\mathrm{gB}}$ for two identical simulations at the
  parameter values $(\kappa_T, \gamma_E) = (5.1, 0.18)$, close to the
  turbulence threshold. Our simulations were initialised with random noise
  in each Fourier mode (with $\mathtt{phiinit}=1$) and the only difference
  between the two simulations is the realisation of this random noise. We see
  the simulations saturate at a similar level beyond
  $t=300$~$(a/v_{\mathrm{th}i})$ before one of them abruptly decays. This is
  another indication that the system is subcritical: the decaying simulation
  has fallen below the critical amplitude to sustain turbulence. Practically,
  we decided that a simulation reached a saturated state if the heat flux
  evolved at a roughly constant value for at least
  $200$~$(a/v_{\mathrm{th}i})$.

  The finite life time of turbulence in subcritical systems is well established
  in some hydrodynamic systems, such as fluid flow in a pipe~\cite{Faisst2004}.
  By running a large number of identical pipe-flow
  experiments~\cite{Peixinho2006, Hof2006, Avila2011} and numerical
  simulations~\cite{Faisst2004, Hof2006, Avila2010, Avila2011}, it was shown
  that the ``lifetime'' of subcritical turbulence (the characteristic time it
  takes before turbulence decays to laminar flow) is a function of the Reynolds
  number. The Reynolds number in pipe flows characterises the tendency of the
  system to be turbulent and is used to quantify the ``distance from the
  turbulence threshold''.  In particular, it was shown that the larger the
  value of the Reynolds number (i.e., the further the system is from the
  turbulence threshold), the longer the turbulence is likely to persist. More
  recently, this same phenomenon of finite turbulence lifetime has been
  observed in MHD simulations of astrophysical Keplerian shear flow
  systems~\cite{Rempel2010}, where the magnetic Reynolds number characterises
  the distance from threshold and the turbulence persists longer for larger
  values.

  Given the above findings, we would also expect the turbulence to persist
  longer for larger values of $Q_i/Q_{\mathrm{gB}}$ in the subcritical
  turbulence we consider here. However, the pipe flow and astrophysical
  studies referred to above relied on running many experiments in order to
  build up sufficient statistics to determine the dependence of the turbulence
  lifetimes on the system parameters. Currently, we are neither able to run
  enough simulations nor run them for a sufficient amount of time to determine
  the turbulence lifetimes for our system, given the high resolutions demanded
  by nonlinear gyrokinetic simulations of plasmas in the core of tokamaks.
  However, this may be possible in future, given advances in computing and
  numerics or through the use of reduced models (upon being shown to be valid
  for this MAST regime).

  \subsection{Transient growth of perturbations}
  A system can reach a saturated turbulent state despite being stable to
  infinitesimal perturbations due to transient growth of perturbations.
  This transient growth is sufficient to sustain turbulence provided
  perturbations reach an amplitude sufficient for nonlinear interaction. The
  question we would like to answer now is how much transient growth is
  sufficient for the system to reach a turbulent state. We have already seen
  which values of $\kappa_T$ and $\gamma_E$ lead to a turbulent state (see
  \figref{contour_heatmap}) and we now investigate transient growth of
  perturbations via linear GS2 simulations.

\begin{figure}[t]
    \centering
    \begin{subfigure}[t]{0.49\textwidth}
      \includegraphics[width=\textwidth]{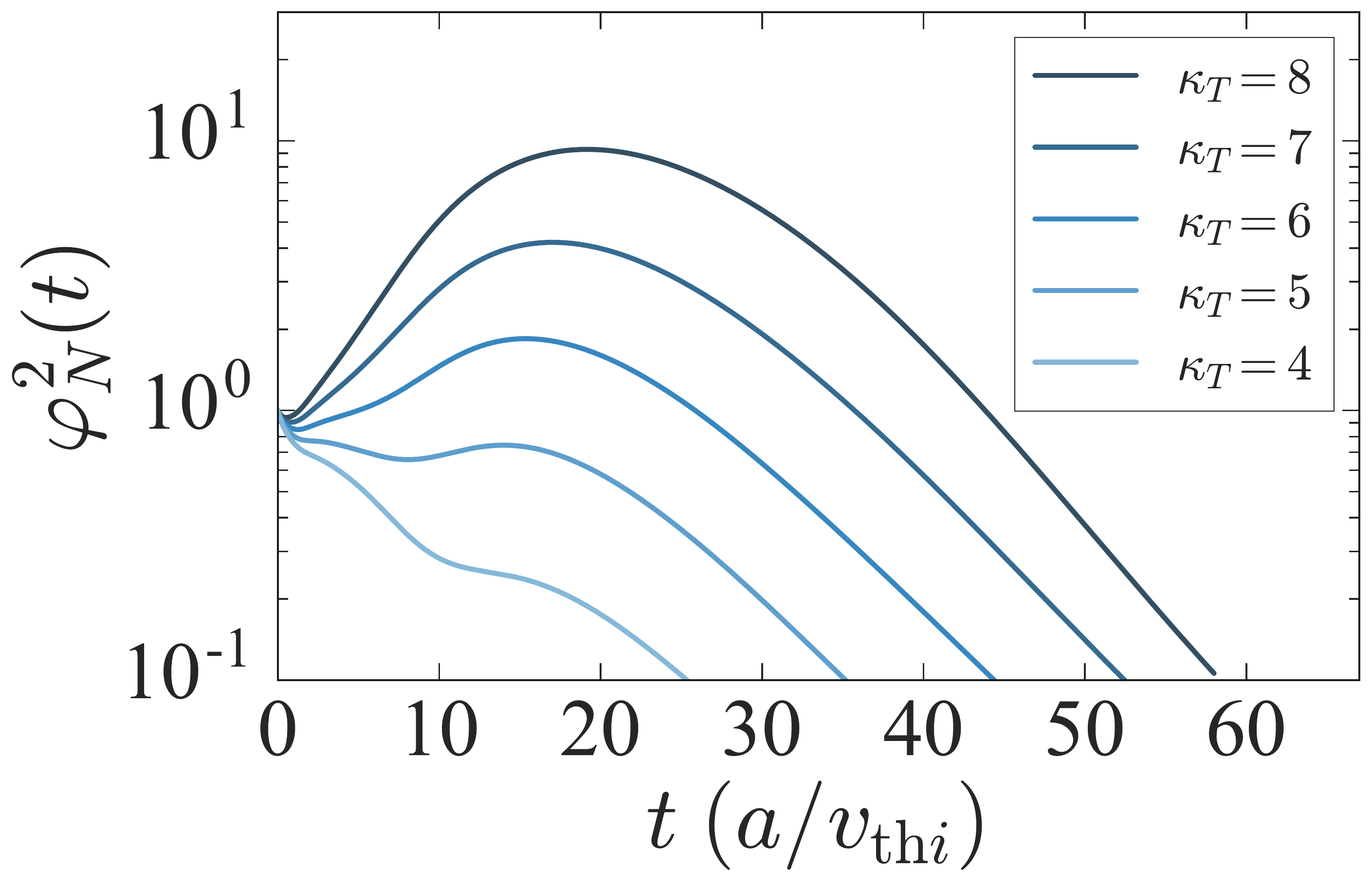}
      \caption{}
      \label{fig:transient}
    \end{subfigure}
    \hfill
    \begin{subfigure}[t]{0.49\textwidth}
      \includegraphics[width=\linewidth]{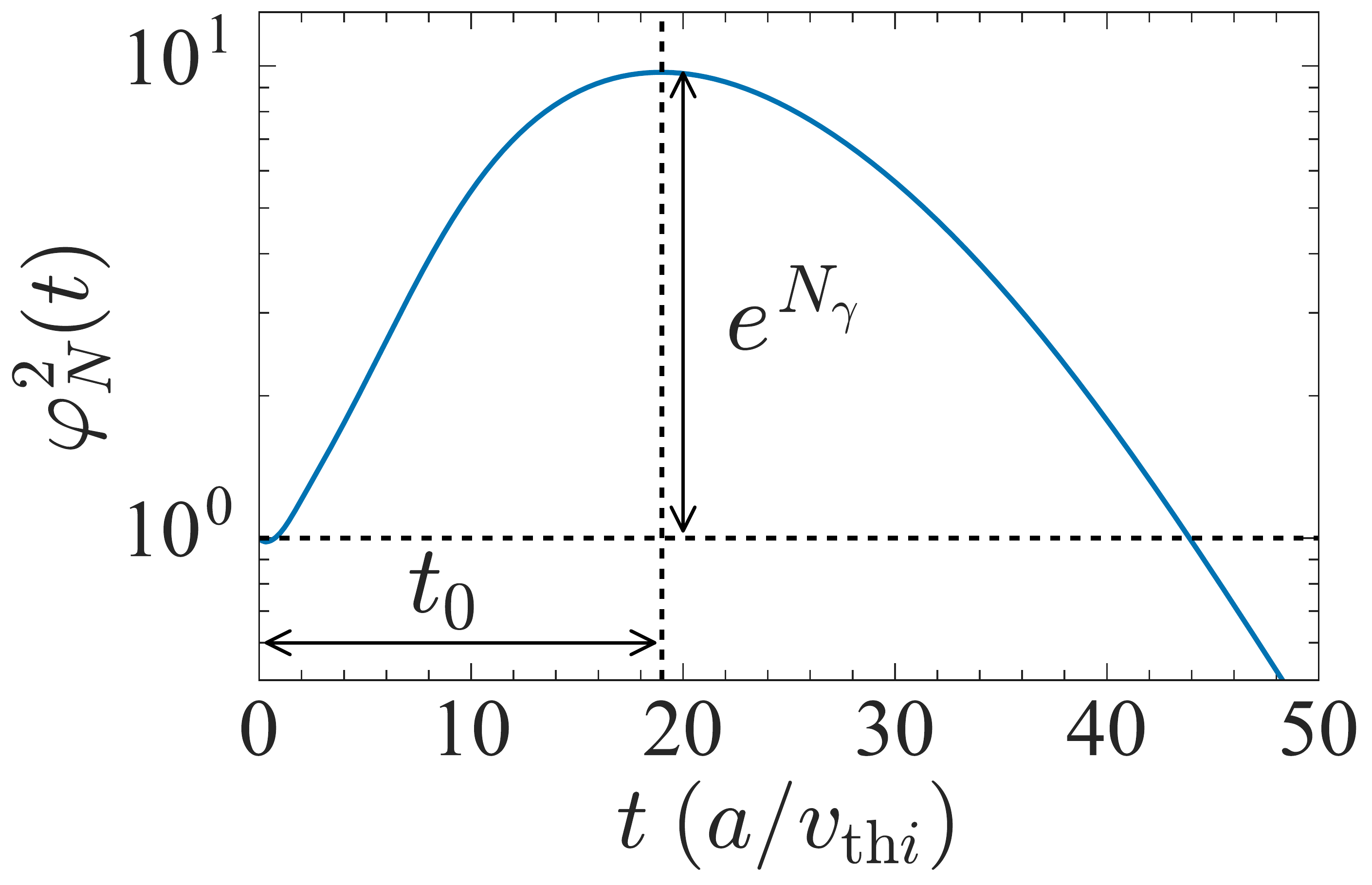}
      \caption{}
      \label{fig:phi_transient}
    \end{subfigure}
    \caption[Transient growth of electrostatic potential]{
      \subref*{fig:transient}
      Transient growth of initial perturbations of the electrostatic potential
      $\varphi^2_N(t)$ (normalised to the time at which flow shear is switched
      on ) at $\gamma_E = 0.16$, for a range of $\kappa_T$ values.
      These time evolutions were obtained from purely linear simulations for a
      binormal wavenumber $k_y \rho_i = 0.2$, approximately the wavenumber
      that gives the largest transient growth (see~\figref{N_16}), and averaged
      over $k_x$. As $\kappa_T$ is increased, the strength of the transient
      growth is also increased.
      \subref*{fig:phi_transient} $\varphi^2_N(t)$ as a function of time for a
      strongly growing mode at $(\kappa_T, \gamma_E, k_y \rho_i) = (5.1,
      0.16, 0.2)$ to further illustrate transient amplification.  The total
      amplification is given by $e^{N_\gamma}$ and the time taken to reach
      maximal amplification is $t_0$.
    }
  \end{figure}

  We performed an extensive series of linear simulations and calculated the
  time-evolution of the electrostatic potential as a function of $k_y \rho_i$,
  $\kappa_T$, and $\gamma_E$. \figref{transient} shows the time evolution of
  $\varphi$ (at $k_y \rho_i = 0.2$ and $\gamma_E=0.16$) for a range of
  $\kappa_T$, normalised to the value at the time when the flow shear is
  switched on, i.e., $\varphi_N^2(t) = \varphi^2(t)/\varphi^2(0)$, where $t=0$
  defines the time at which $\gamma_E$ is changed from $0$ to $0.16$. We have
  averaged $\varphi$ over $k_x$. \Figref{transient} illustrates the phenomenon
  of transient growth in a subcritical system and we see that, as $\kappa_T$ is
  increased, the system shows stronger transient growth.  At $\gamma_E = 0.16$,
  we saw in \figref{contour_heatmap} that turbulence could be sustained at
  $\kappa_T \approx 4.8$. \figref{transient} shows only a marginal amount of
  transient growth for $\gamma_E = 0.16$

  We investigate the linear dynamics in the absence of flow shear in
  Appendix~\ref{App:hyperviscosity} and~\ref{App:linear_sims}.

  \subsection{Characterising transient growth}
  For linear simulations such as those shown in~\figref{transient}, it is
  problematic to define a ``linear growth rate'', as we do for linear
  simulations with $\gamma_E = 0$, where $\varphi(t)$ grows exponentially.
  Methods for determining an
  ``effective'' linear growth rate have been outlined in Refs.~\cite{Roach2009}
  and~\cite{Schekochihin2012}. Here, we follow Ref.~\cite{Schekochihin2012} and
  use the ``transient-amplification factor'' as a measure of the
  vigour of the transient growth.  For a total amplification factor,
  $e^{N_\gamma}$, the amplification exponent $N_\gamma$ is defined by
  \begin{equation}
    N_\gamma = \int_0^{t_0} \dd t \gamma(t) = \frac{1}{2} \ln
    \frac{\varphi^2(t_0)}{\varphi^2(0)},
    \label{amp_exponent}
  \end{equation}
  where $t_0$ is the time taken to reach the maximum amplification, and
  $\gamma(t)$ is the time-dependent growth rate. We note that both the
  transient-amplification factor and time are functions of $k_y$:
  $N_\gamma = N_\gamma(k_y)$ and $t_0 = t_0(k_y)$, however, we will write these
  as $N_\gamma$ and $t_0$ for convenience.  The concept of transient growth is
  more clearly illustrated in \Figref{phi_transient}, which shows a typical
  linear simulation with strong amplification at $(\kappa_T, \gamma_E, k_y
  \rho_i) = (5.1, 0.16, 0.2)$. The total amplification $e^{N_\gamma}$ and the
  time taken to reach maximal amplification $t_0$, are also indicated in
  \figref{phi_transient}.

  It was shown in Ref.~\cite{Schekochihin2012} that the parameters $N_\gamma$
  and $t_0$ determine whether turbulence can be sustained in the following way.
  Perturbations grow transiently because they are swept from values of $k_x(t)$
  that are unstable to values that are stable, where $k_x(t)$
  evolves according to \eqref{kx_time}. If nonlinear interactions scatter
  energy back into the unstable modes before perturbations decay to values too
  small to be acted upon by the nonlinearity, they can be transiently amplified
  once again, and so on. In this way, a nonlinear saturated state can be
  sustained.  The typical timescale for nonlinear interactions is the nonlinear
  decorrelation time $\tau_\mathrm{NL} \sim 1/k_\perp V_E$, where
  $k_\perp$ is the typical perpendicular wavenumber, and $V_E \sim
  k_\perp (c \varphi/B)$ from \eqref{v_exb}. To sustain turbulence, transient
  growth should last at least as long as one nonlinear decorrelation time:
  \begin{align}
    \begin{split}
      t_0 &\gtrsim \tau_{\mathrm{NL}}.
      \label{schek_t0}
    \end{split}
  \end{align}
  At the same time, the rate of amplification should be comparable to the
  nonlinear decorrelation rate for a sustained turbulent state:
  \begin{equation}
    \frac{N_\gamma}{t_0} \sim \frac{1}{\tau_{\mathrm{NL}}}.
    \label{schek_gamma_eff}
  \end{equation}
  Combining \eqref{schek_t0} and \eqref{schek_gamma_eff}, we see that a
  sustained turbulent state requires
  \begin{equation}
    N_\gamma \gtrsim 1.
    \label{schek_N}
  \end{equation}
  We will now investigate the values of $N_\gamma$ and $t_0$ for
  experimentally-relevant equilibrium parameters and return to the comparison
  of $t_0$ with $\tau_{\mathrm{NL}}$ in Section~\ref{sec:corr_gs2} after
  estimating $\tau_{\mathrm{NL}}$ using the results from our correlation
  analysis.

\begin{figure}[t]
    \centering
    \begin{subfigure}[t]{0.49\textwidth}
      \includegraphics[width=\textwidth]{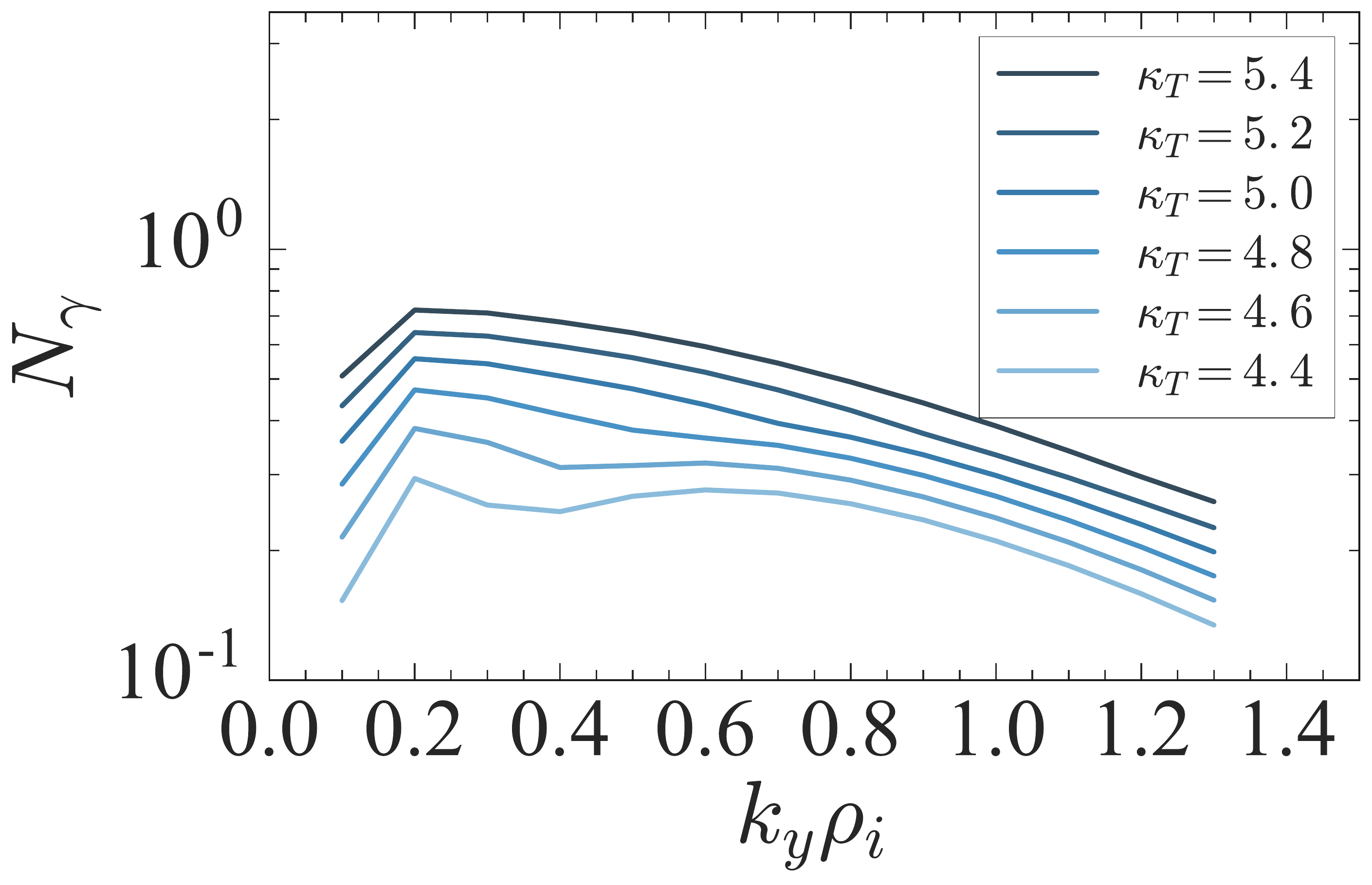}
      \caption{}
      \label{fig:N_16}
    \end{subfigure}
    \hfill
    \begin{subfigure}[t]{0.49\textwidth}
      \includegraphics[width=\textwidth]{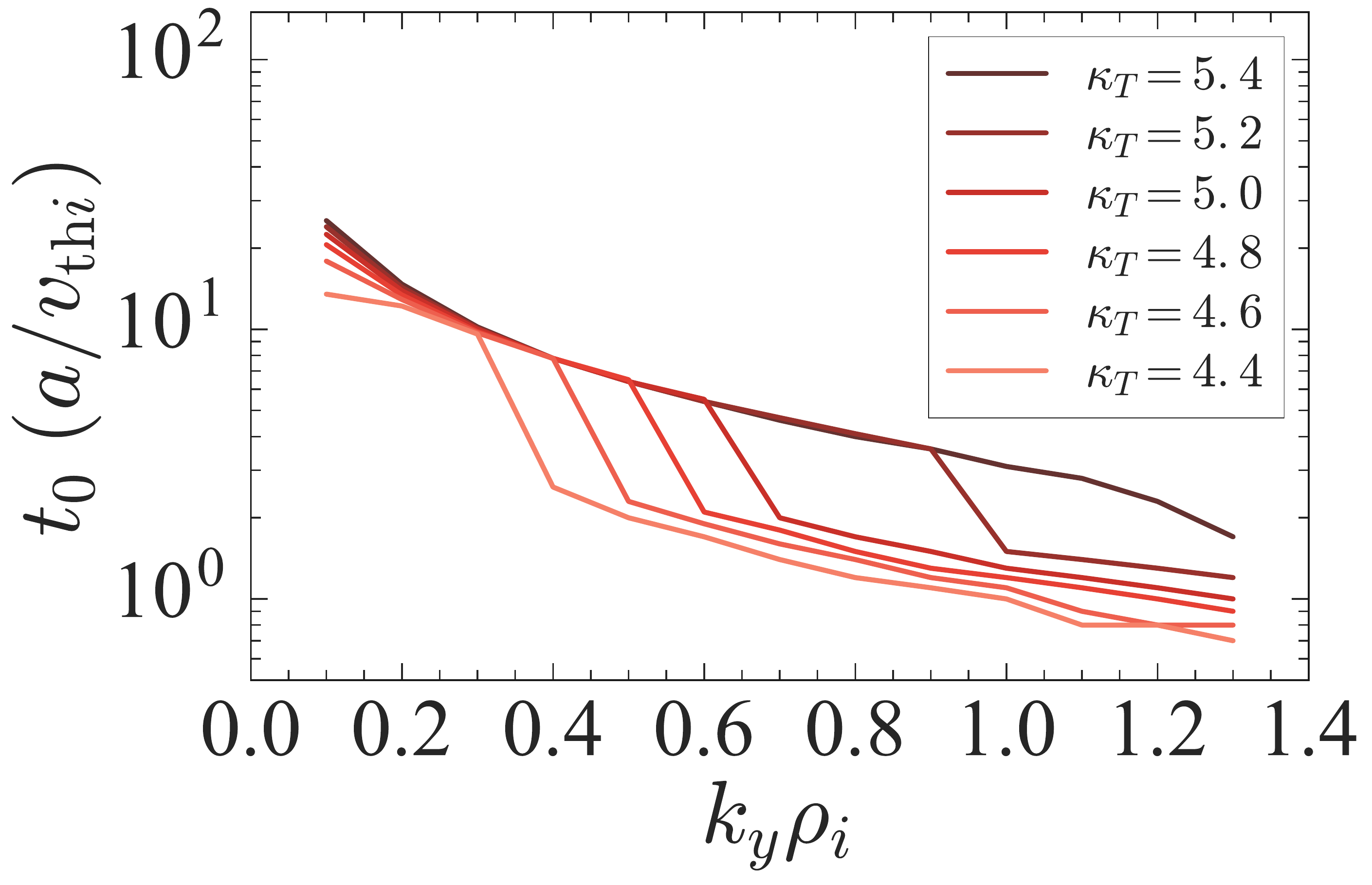}
      \caption{}
      \label{fig:t0_16}
    \end{subfigure}
    \caption[$N_\gamma$ and $t_0$ versus $k_y \rho_i$]{
      \subref*{fig:N_16} The transient-amplification factor $N_\gamma$
      \eqref{amp_exponent} for a range of values of $\kappa_T$ at $\gamma_E =
      0.16$. $N_\gamma$ increases with increasing $\kappa_T$ and increases
      smoothly as the nonlinear threshold is passed.
      \subref*{fig:t0_16} Time taken to reach maximum amplification $t_0$
      for a range of values of $\kappa_T$, also at $\gamma_E = 0.16$.
      Increasing $\kappa_T$ leads to transient amplification lasting for a
      longer time.
    }
    \label{fig:N_and_t0_16}
  \end{figure}

  Considering figures~\ref{fig:transient} and \subref{fig:phi_transient}, we
  want to estimate the critical values of $N_\gamma$ and $t_0$ above which
  turbulence is triggered and a saturated state can be established in our
  system. We note that reaching a saturated state would still require a
  sufficiently large initial perturbation, as we showed in \figref{phiinit}.
  \Figref{N_and_t0_16} shows $N_\gamma$ and $t_0$ as functions of $k_y \rho_i$
  for a range of different $\kappa_T$ values at $\gamma_E = 0.16$. The linear
  simulations are only shown up to $k_y \rho_i = 1.3$, because hyperviscosity
  effectively suppresses transient growth beyond this value (this is discussed
  in more detail in appendix~\ref{App:hyperviscosity}). As a point of
  reference, \figref{value_heatmap} previously showed that for $\gamma_E =
  0.16$, the transition to turbulence occurs at $\kappa_T = 4.8$. For the
  linear simulations in \figref{N_and_t0_16}, we see a relatively smooth
  increase in $N_\gamma$ and $t_0$ as $\kappa_T$ is increased across this
  nonlinear threshold. We see larger transient amplification and modes with
  smaller $k_y \rho_i$ experiencing amplification over a longer time period as
  $\kappa_T$ is increased. The fact that neither~\figref{N_16}
  nor~\figref{t0_16} show significant changes as the nonlinear turbulence
  threshold is passed suggests that nonlinear simulations are essential for
  predicting whether the system will exhibit turbulence for this experimental
  configuration.

  \subsection{Conditions for the onset of subcritical turbulence}
\begin{figure}[t]
    \centering
    \begin{subfigure}[t]{0.49\textwidth}
      \includegraphics[width=\textwidth]{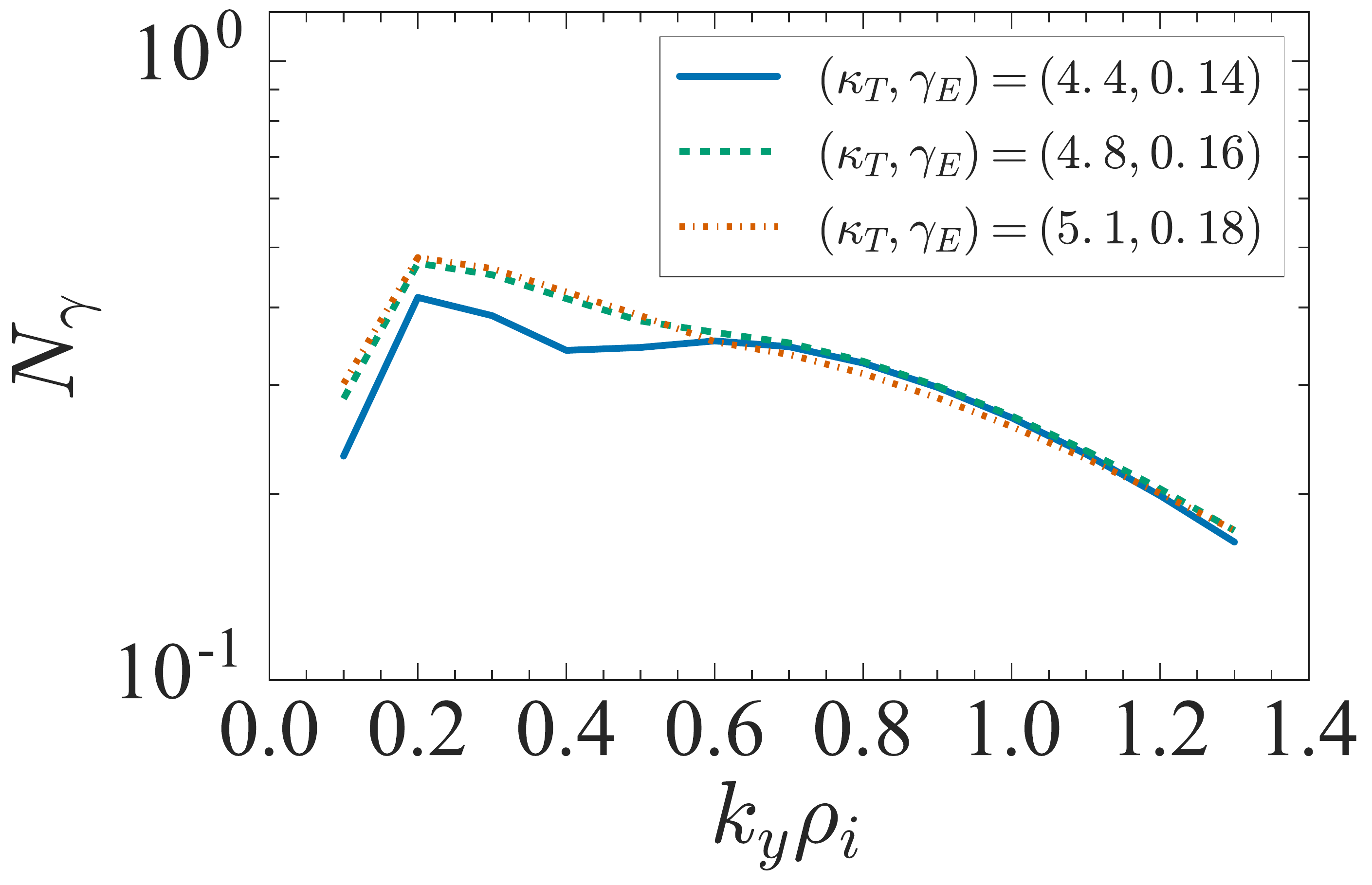}
      \caption{}
      \label{fig:N_marginal}
    \end{subfigure}
    \hfill
    \begin{subfigure}[t]{0.49\textwidth}
      \includegraphics[width=\textwidth]{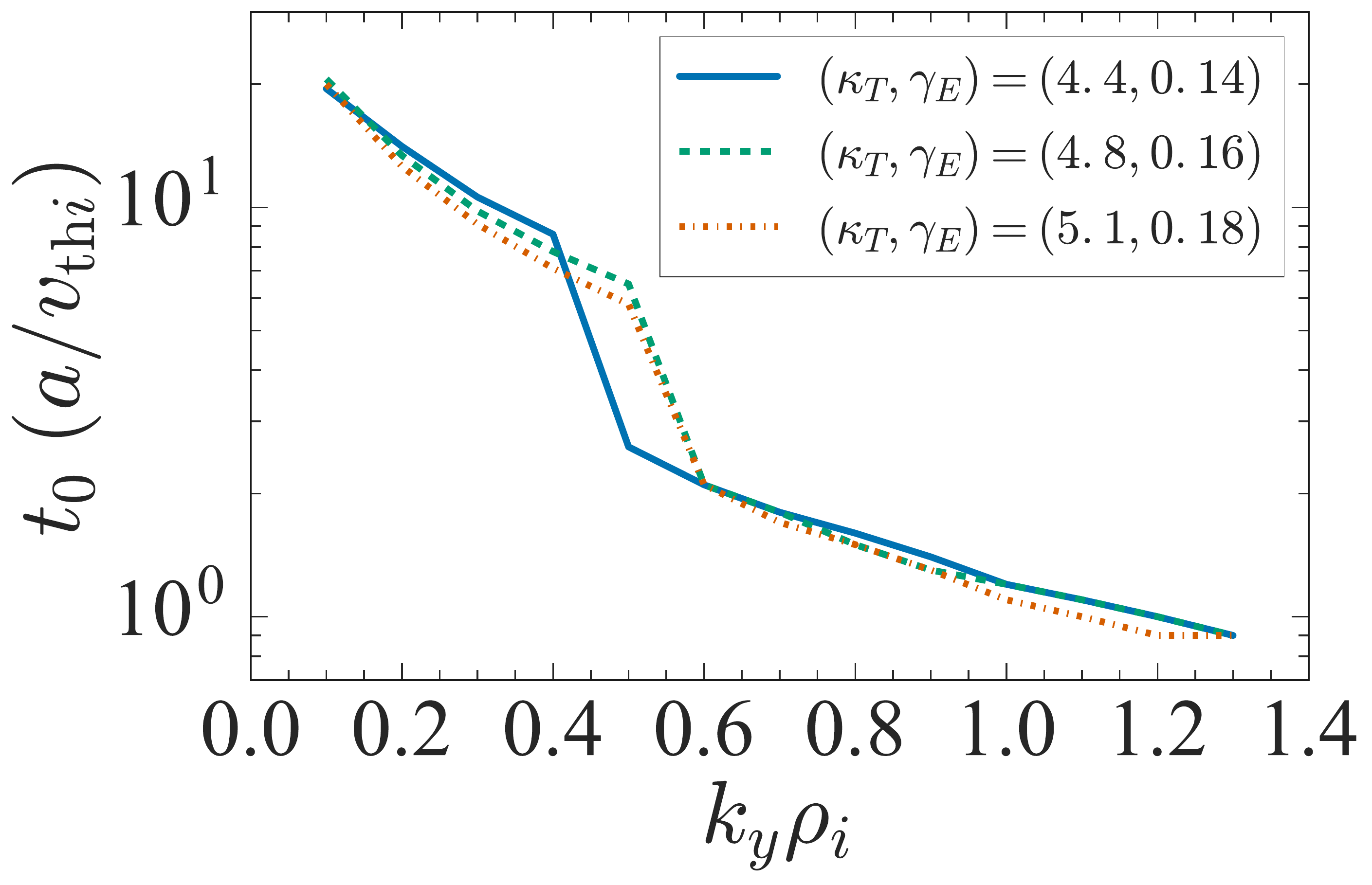}
      \caption{}
      \label{fig:t0_marginal}
    \end{subfigure}
    \caption[$N_\gamma$ and $t_0$ versus $k_y \rho_i$ for marginally unstable
             simulations]{
      \subref*{fig:N_marginal} Transient-amplification factor $N_\gamma$ and
      \subref*{fig:t0_marginal} transient-amplification time $t_0$
      for the three marginal simulations identified in
      Section~\ref{sec:heat_flux}. The values of $N_\gamma$ and $t_0$ that
      correspond to the marginally unstable equilibria are approximately the
      same, suggesting that these are the critical values required in order to
      reach a saturated turbulent state.
    }
    \label{fig:N_and_t0_marginal}
  \end{figure}

  For supercritical turbulence, the onset of turbulence is typically
  characterised by a critical value of the linear growth rate. Similarly, for
  subcritical systems, we may reasonably expect that critical values of
  $N_\gamma$ and/or $t_0$ exist that lead to a saturated turbulent state. To
  investigate the conditions for the onset of turbulence we consider $N_\gamma$
  and $t_0$ for the marginally unstable simulations identified in
  Section~\ref{sec:heat_flux}. Figures~\ref{fig:N_marginal} and
  \subref{fig:t0_marginal} show $N_\gamma$ and $t_0$ as functions of
  $k_y \rho_i$ for $(\kappa_T, \gamma_E) = (4.4, 0.14), (4.8, 0.16), (5.1,
  0.18)$. We see that both $N_\gamma$ and $t_0$ are roughly the same for our
  marginally unstable simulations, suggesting that the values shown in
  Figures~\ref{fig:N_marginal} and \subref{fig:t0_marginal} are the critical
  values necessary for the onset of turbulence. Assuming that low $k_y$ modes
  are the dominant scales in the system, it is reasonable to estimate from
  \figsref{t0_16}{t0_marginal} that the onset of turbulence requires
  $t_0 \gtrsim 10$~$(a/v_{\mathrm{th}i})$.

\begin{figure}[t]
    \centering
    \includegraphics[width=0.6\linewidth]{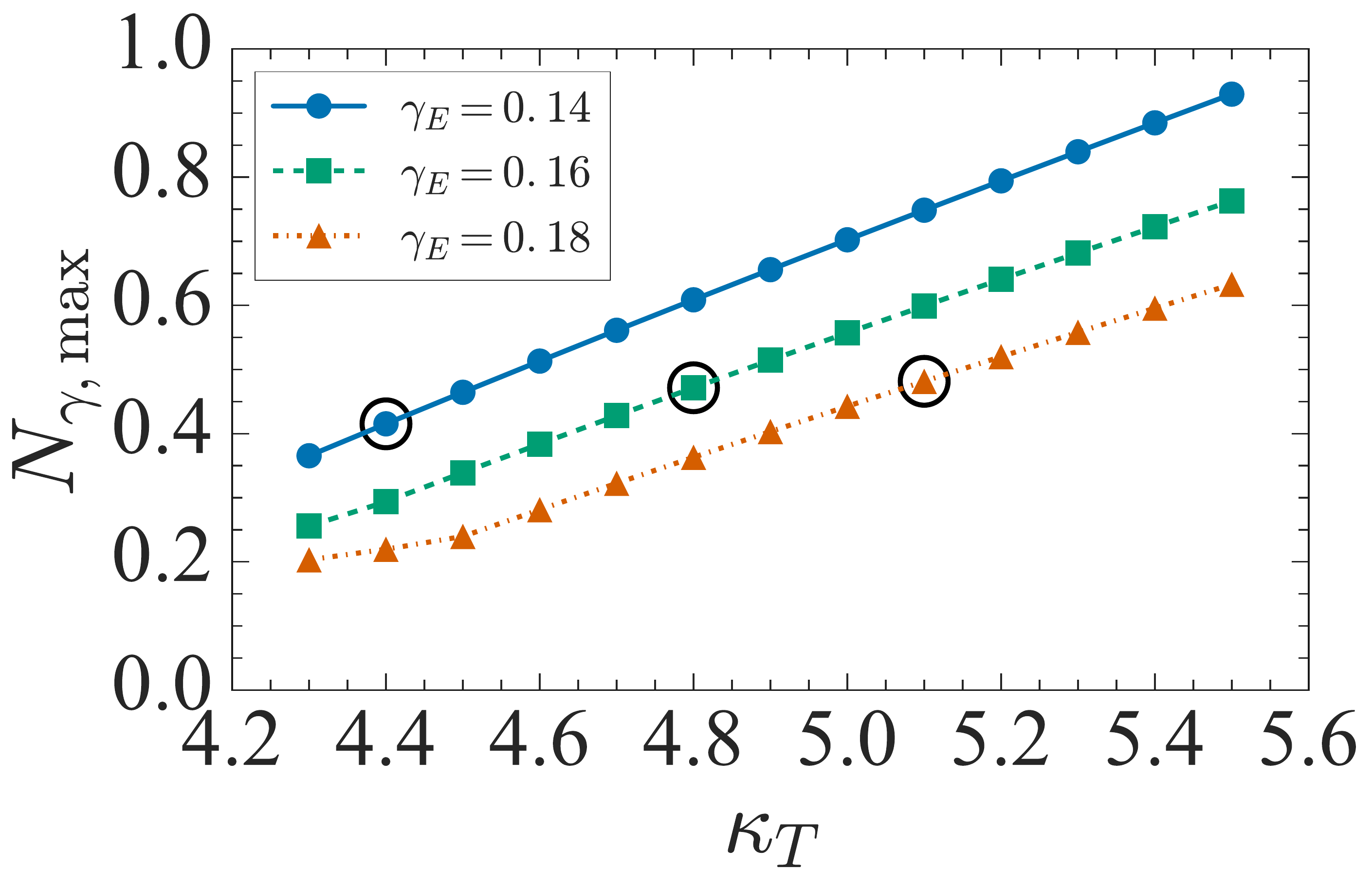}
    \caption[Maximum transient-amplification factor]{
      Maximum transient-amplification factor $N_{\gamma,\max}$ versus
      $\kappa_T$ for three values of $\gamma_E$ within the range of
      experimental uncertainty. The simulations circled in black represent the
      critical values of $\kappa_T$ above which turbulence can be sustained,
      suggesting the onset to turbulence occurs at $N_{\gamma,\max} \sim 0.4$.
    }
    \label{fig:max_trans_amp}
  \end{figure}

  To determine a critical condition for $N_\gamma$, we consider the value at
  the peak of the $N_\gamma$ spectrum, $k_y \rho_i \sim 0.2$, shown in
  \figref{N_16}.  \Figref{max_trans_amp} shows the maximum value of the
  transient-amplification factor $N_{\gamma,\max}$, as a function of
  $\kappa_T$. The marked simulations are for the critical values of $\kappa_T$
  above which turbulence can be sustained, given a sufficiently large initial
  perturbation amplitude.  \Figref{max_trans_amp} shows that $N_{\gamma,\max}$
  is linear in $\kappa_T$ for each $\gamma_E$, with higher values of $\gamma_E$
  resulting in lower values of $N_{\gamma,\max}$.  The other important feature
  is that the values of $N_{\gamma,\max}$ at the critical values of $\kappa_T$
  are similar, giving an approximate critical condition: $N_{\gamma,\max} \sim
  0.4$. We can conclude that, for the system we are investigating, the
  conditions for the onset of turbulence (given a sufficiently large initial
  perturbation) are:
  \begin{align}
    \begin{split}
      N_{\gamma,\max} &\gtrsim 0.4,\\
      t_0 &\gtrsim 10~(a/v_{\mathrm{th}i}).
      \label{linear_turb_conds}
    \end{split}
  \end{align}
  The value of $N_{\gamma,\max}$ in \eqref{linear_turb_conds} is comparable to
  that found in previous work~\cite{Schekochihin2012,Highcock2012}. We will
  return to the comparison of $t_0$ with $\tau_{\mathrm{NL}}$ after estimating
  $\tau_{\mathrm{NL}}$ in Section~\ref{sec:struc_of_turb}, where we confirm
  that $t_0 \gtrsim \tau_{\mathrm{NL}}$ and, therefore, that a sustained
  turbulent state requires an amplification time comparable to the nonlinear
  decorrelation time.

  We can summarise the linear behaviour described above as follows. Flow shear
  sweeps perturbations in time from regions of $k_x$ space where modes are
  unstable to where they are damped. This sweeping through unstable regions
  leads to the transient growth of the perturbations.  The turbulent state is
  sustained through transient amplification of sufficient strength and
  duration. We showed that the changes in $N_\gamma$ and $t_0$ are relatively
  smooth as the turbulence threshold is surpassed (determined from our
  simulations in Section~\ref{sec:heat_flux}), suggesting nonlinear simulations
  are essential in predicting the transition to turbulence.  Therefore, we will
  now investigate our nonlinear simulations further to determine the nature of
  this transition to turbulence.

\section{Structure of turbulence close to and far from the threshold}
\label{sec:struc_analysis}

  \begin{quote}
    \emph{Much of this section is based on Ref.~\cite{VanWyk2016}}.
  \end{quote}

  Having established the subcritical nature of the system, we want to
  investigate the consequences for the structure of turbulence. We
  will argue that a subcritical system such as ours supports the
  formation of coherent structures close to the turbulence threshold, that
  the heat flux is proportional to the product of number of structures and
  their maximum amplitude, and that the properties of the turbulence are
  characterised by the ``distance from threshold'' (as opposed to the specific
  values of the stability parameters $\kappa_T$ and $\gamma_E$), as measured,
  for example, by the turbulent ion heat flux.

  \subsection{Coherent structures in the near-marginal state}
  \label{sec:coherent_strucs}
  \begin{figure}[t]
    \centering
    \begin{subfigure}{0.49\linewidth}
      \includegraphics[width=\linewidth]{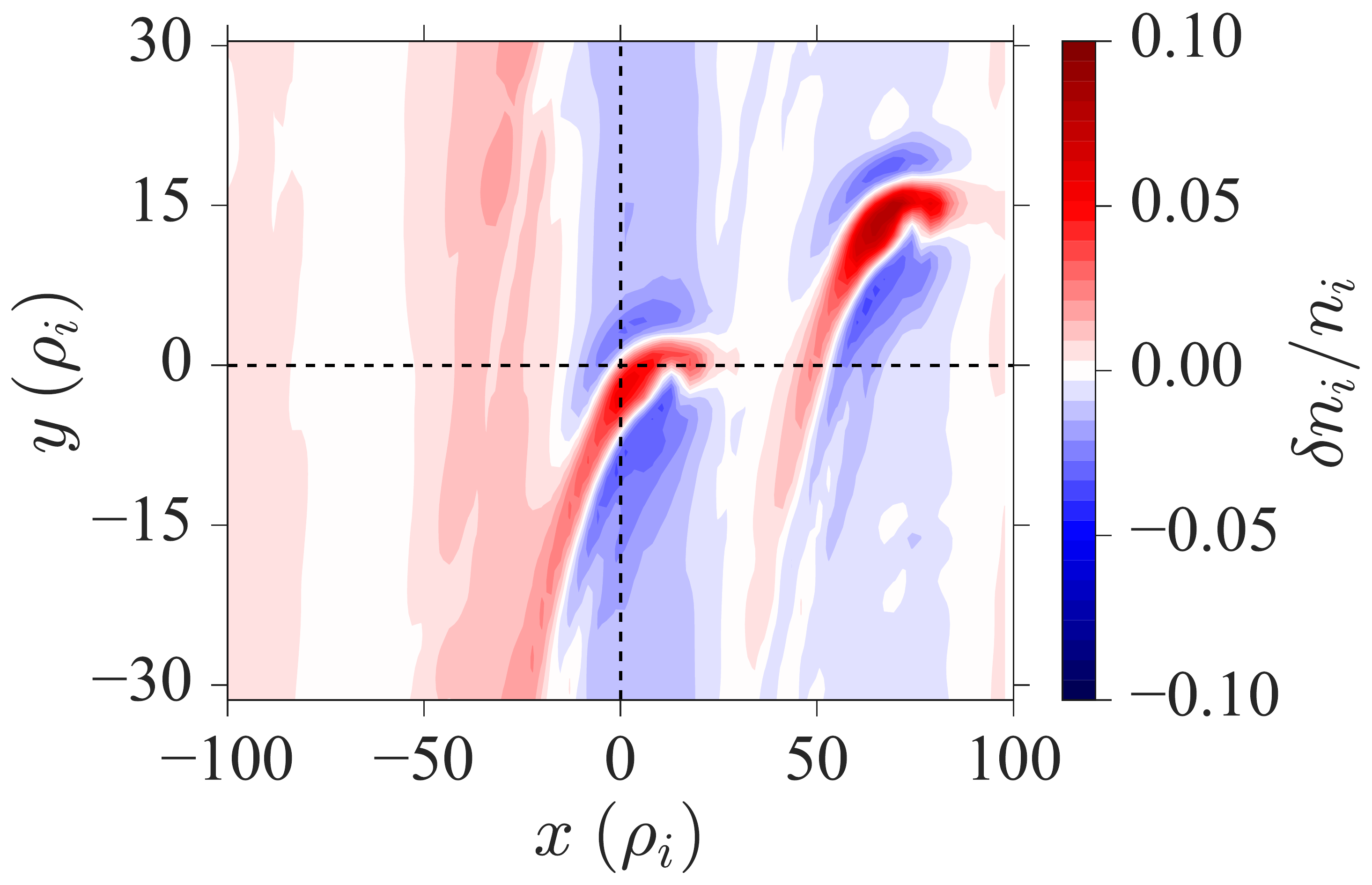}
      \caption{}
      \label{fig:marginal}
    \end{subfigure}
    \begin{subfigure}{0.49\linewidth}
      \includegraphics[width=\linewidth]{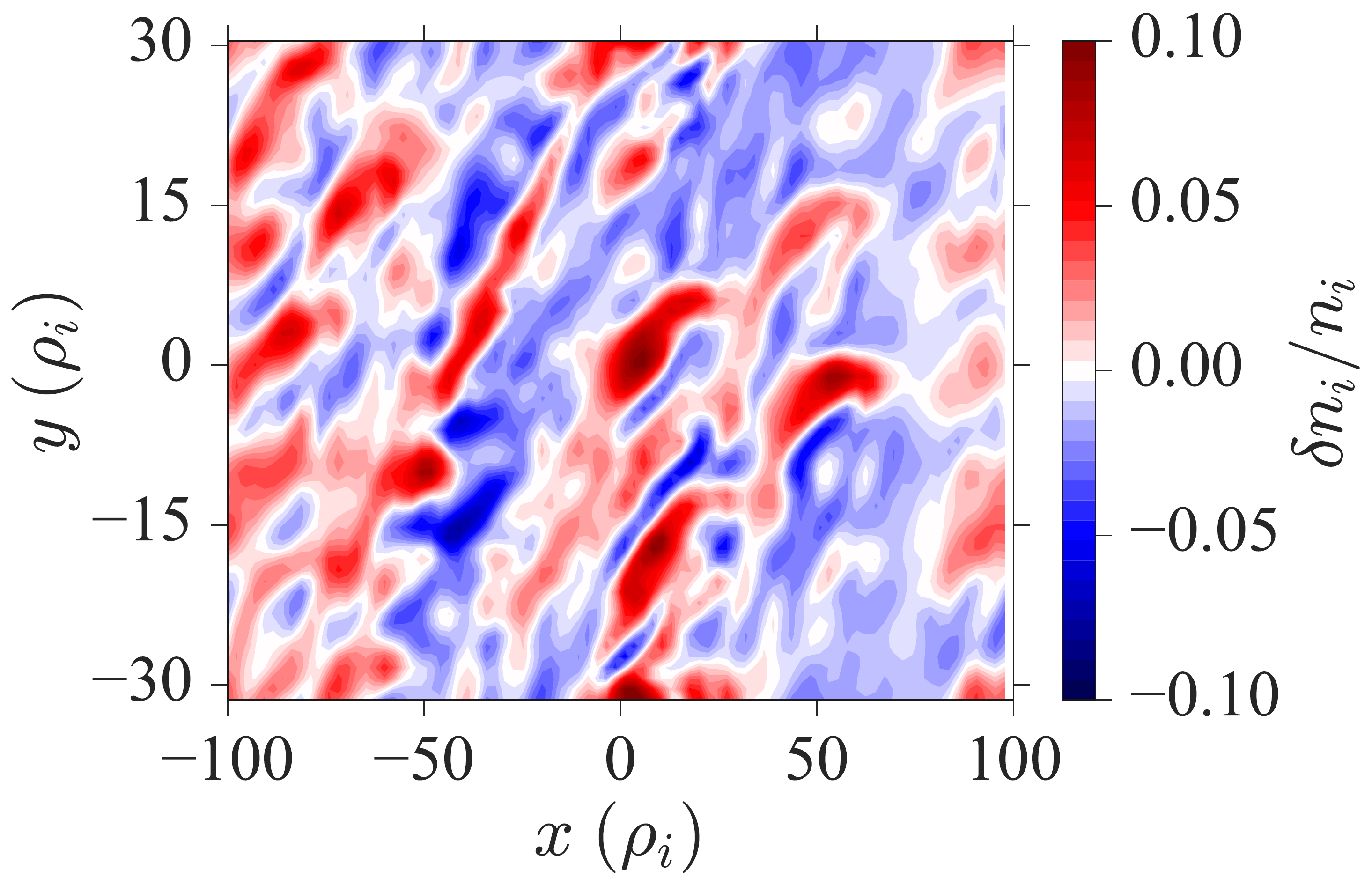}
      \caption{}
      \label{fig:intermediate}
    \end{subfigure}
    \begin{subfigure}{0.49\linewidth}
      \includegraphics[width=\linewidth]{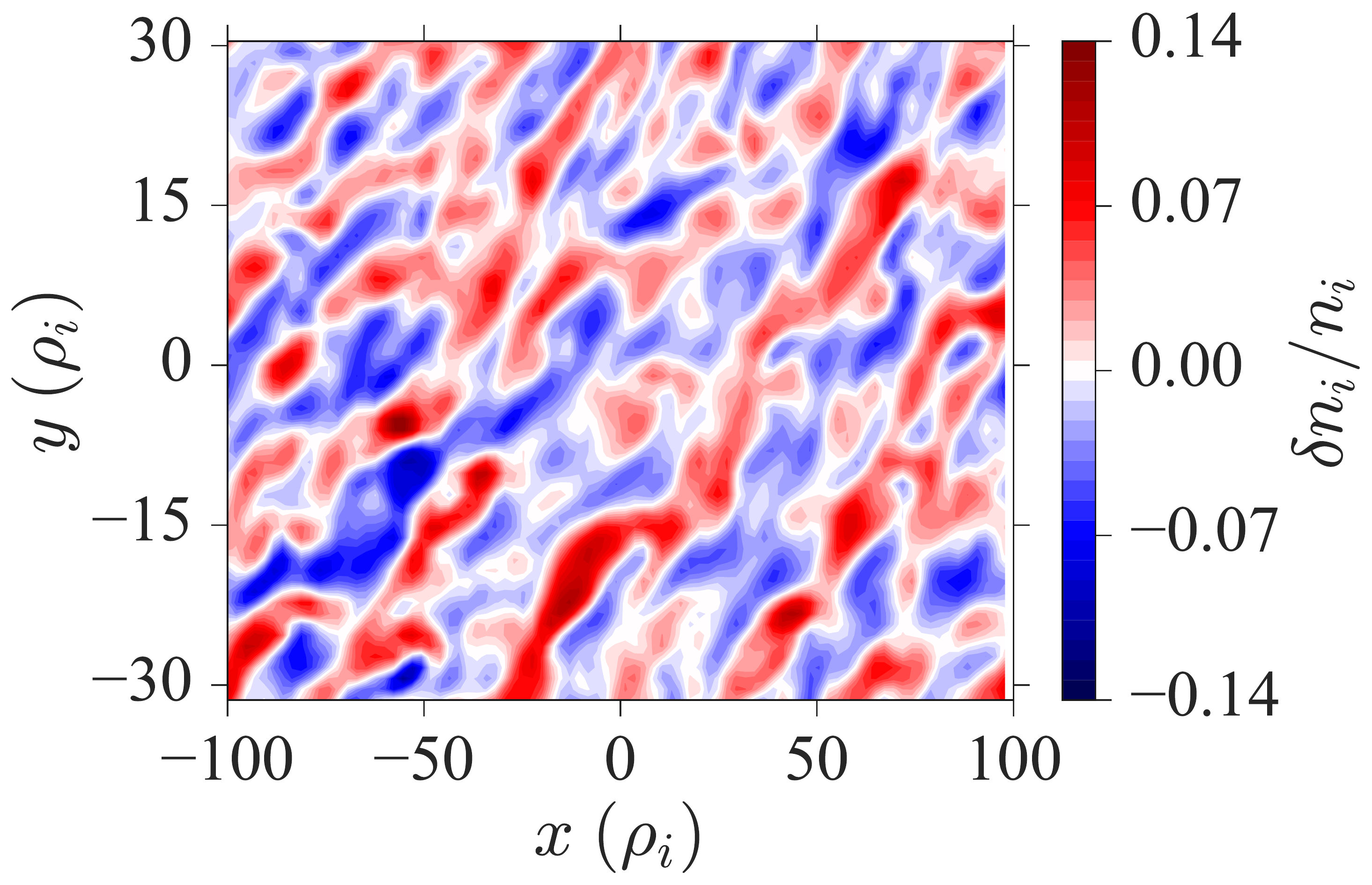}
      \caption{}
      \label{fig:strongly_driven}
    \end{subfigure}
    \begin{subfigure}{0.49\linewidth}
      \includegraphics[width=\linewidth]{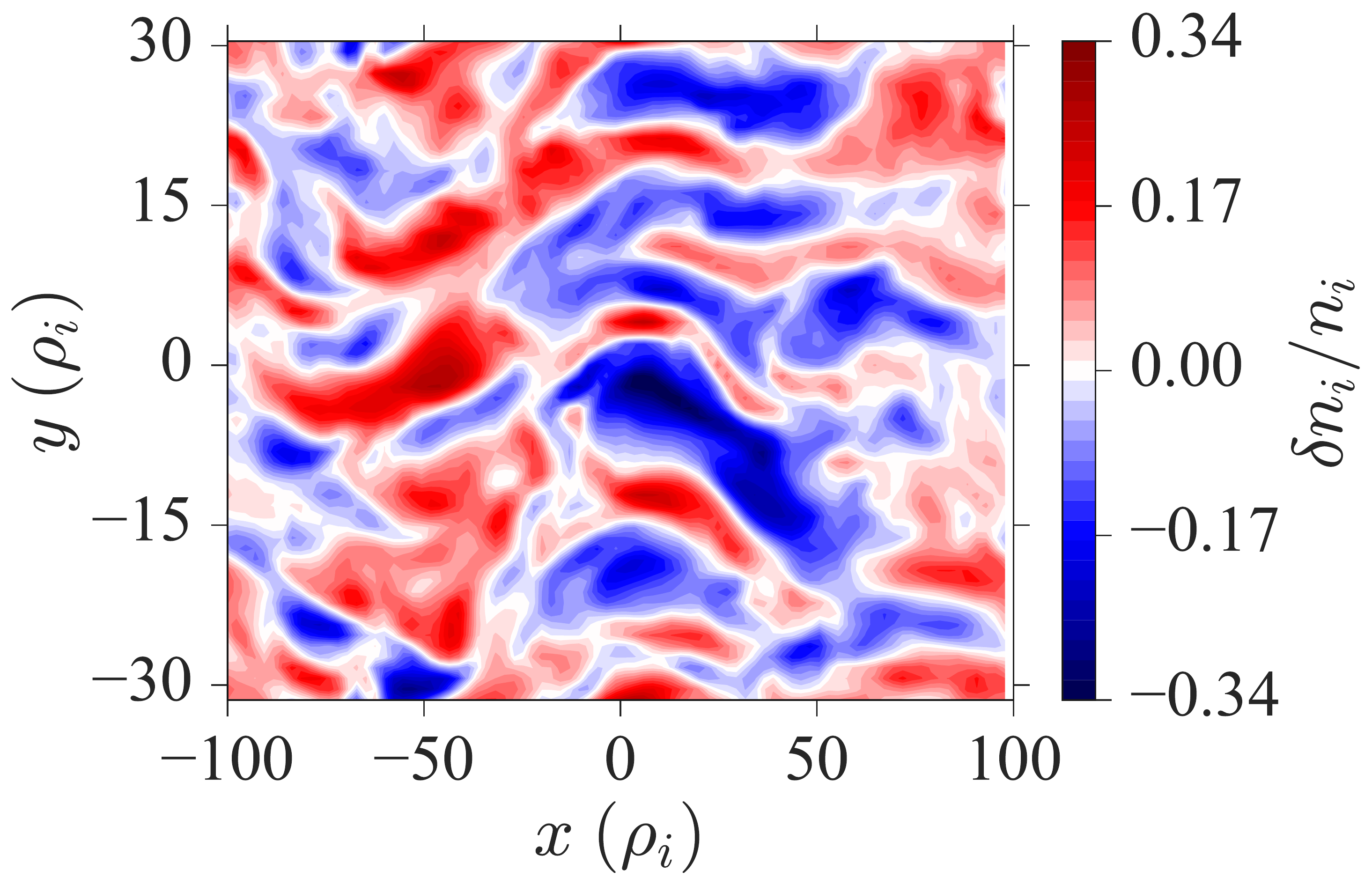}
      \caption{}
      \label{fig:no_shear}
    \end{subfigure}
    \caption[Real-space density-fluctuation fields on $(x,y)$ plane]{
      Density-fluctuation field $\delta n_i/n_i$ at the outboard midplane of
      MAST as a function of the local GS2 coordinates $x$ and $y$, for four
      combinations of stability parameters.
      \subref*{fig:marginal} Near-threshold turbulence, $(\kappa_T, \gamma_E) =
      (4.8, 0.16)$. The dashed lines indicate the planes of constant $x$ and
      $y$ used to demonstrate the parallel structure in
      \figref{parallel_density}.
      \subref*{fig:intermediate} Turbulence intermediate between the
      near-threshold and strongly driven cases, $(\kappa_T, \gamma_E) =
      (4.9,0.16)$.
      \subref*{fig:strongly_driven} Strongly driven turbulence, $(\kappa_T,
      \gamma_E) = (5.2,0.16)$.
      \subref*{fig:no_shear} Turbulence without flow shear, $(\kappa_T,
      \gamma_E) = (5.2,0)$, showing strong zonal flows.
	}
    \label{fig:density_fluctuations}
  \end{figure}
  \Figref{density_fluctuations} shows the density-fluctuation field $\delta n_i
  / n_i$ at the outboard midplane of MAST as functions of the local GS2
  coordinates $x$ and $y$ (see Appendix~\ref{App:real_space_transform} for how
  these are related to real-space $(R,Z)$ coordinates). The simulations shown
  in Figures~\subref{fig:marginal}--\subref{fig:strongly_driven} are marked by
  points in~\figref{contour_heatmap} and importantly they are all well within
  the region of experimental uncertainty. We choose four combinations of the
  stability parameters $(\kappa_T, \gamma_E)$ as the system is taken away from
  the turbulence threshold: $(4.8, 0.16)$, which is close to the turbulence
  threshold [\figref{marginal}], $(4.9, 0.16)$, an intermediate case between
  the marginal and strongly driven turbulence [\figref{intermediate}], $(5.2,
  0.16)$, a strongly driven case further from the threshold
  [\figref{strongly_driven}], and $(5.2, 0)$, a case without flow shear
  [\figref{no_shear}], representative of the basic ITG turbulence that has been
  thoroughly studied in the past~\cite{Waltz1988,Dimits1996,Rogers2000}.

\begin{figure}[t]
    \centering
    \begin{subfigure}{0.49\linewidth}
      \includegraphics[width=\linewidth]{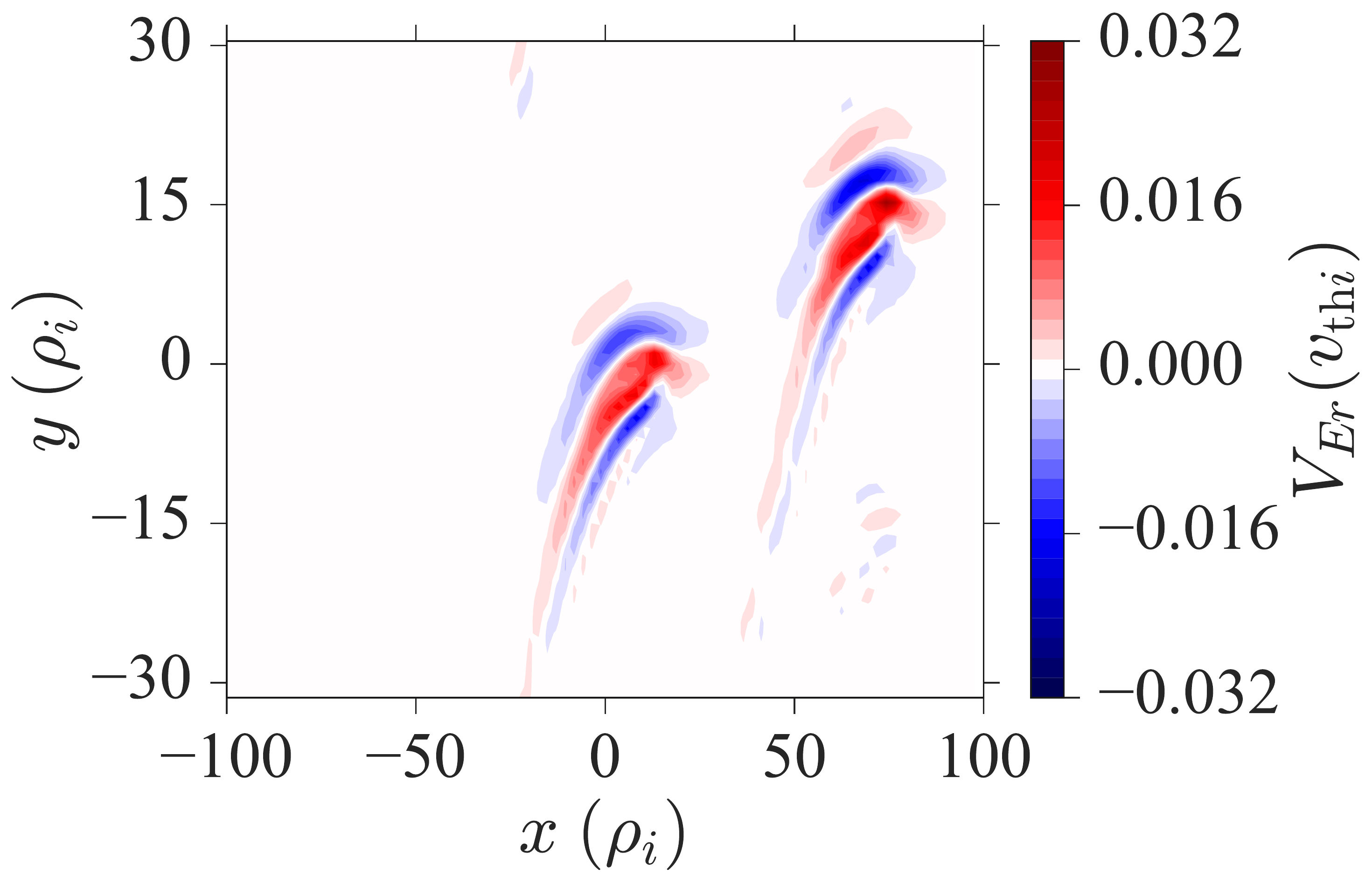}
      \caption{}
      \label{fig:vel_marginal}
    \end{subfigure}
    \begin{subfigure}{0.49\linewidth}
      \includegraphics[width=\linewidth]{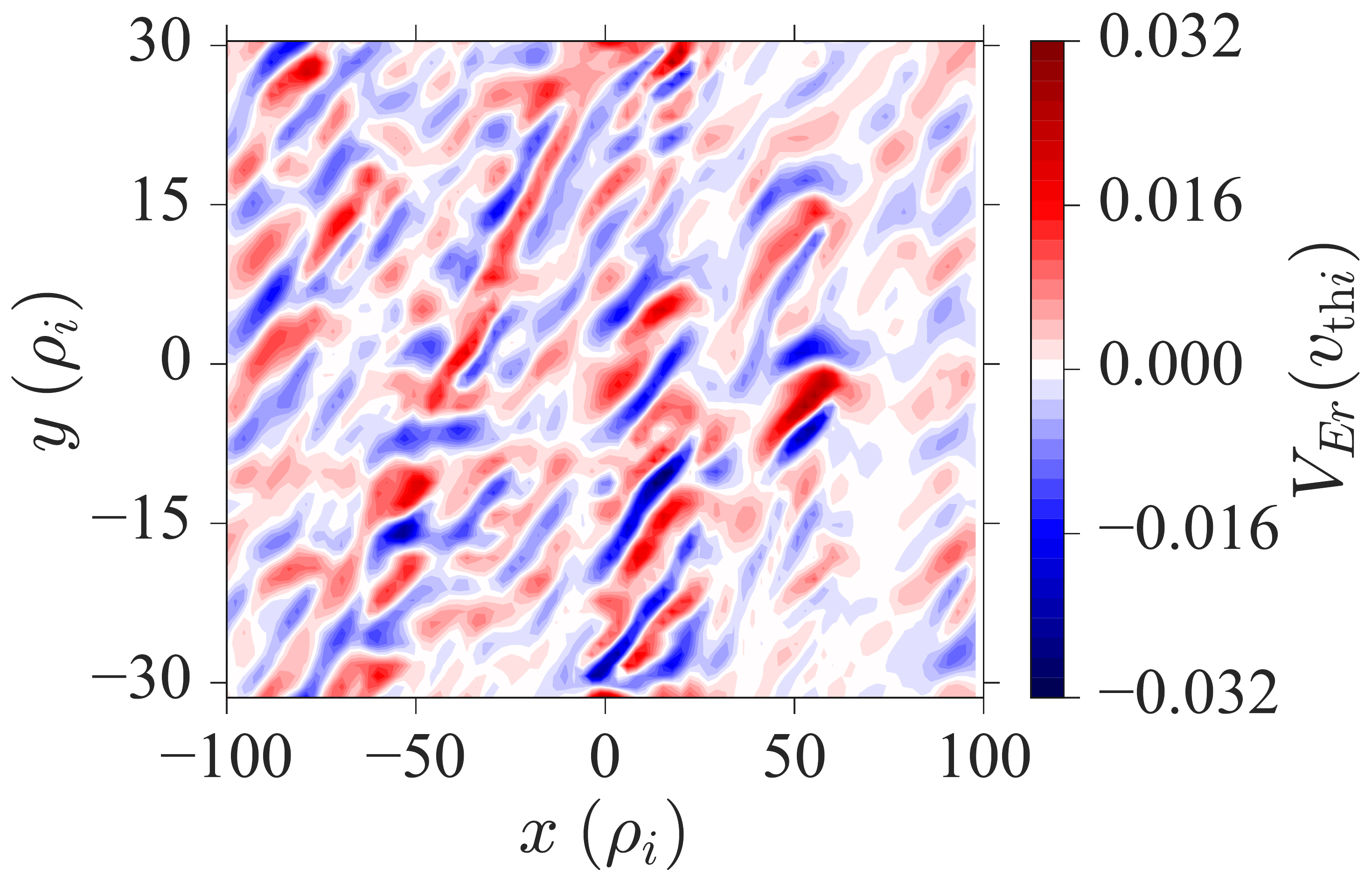}
      \caption{}
      \label{fig:vel_intermediate}
    \end{subfigure}
    \begin{subfigure}{0.49\linewidth}
      \includegraphics[width=\linewidth]{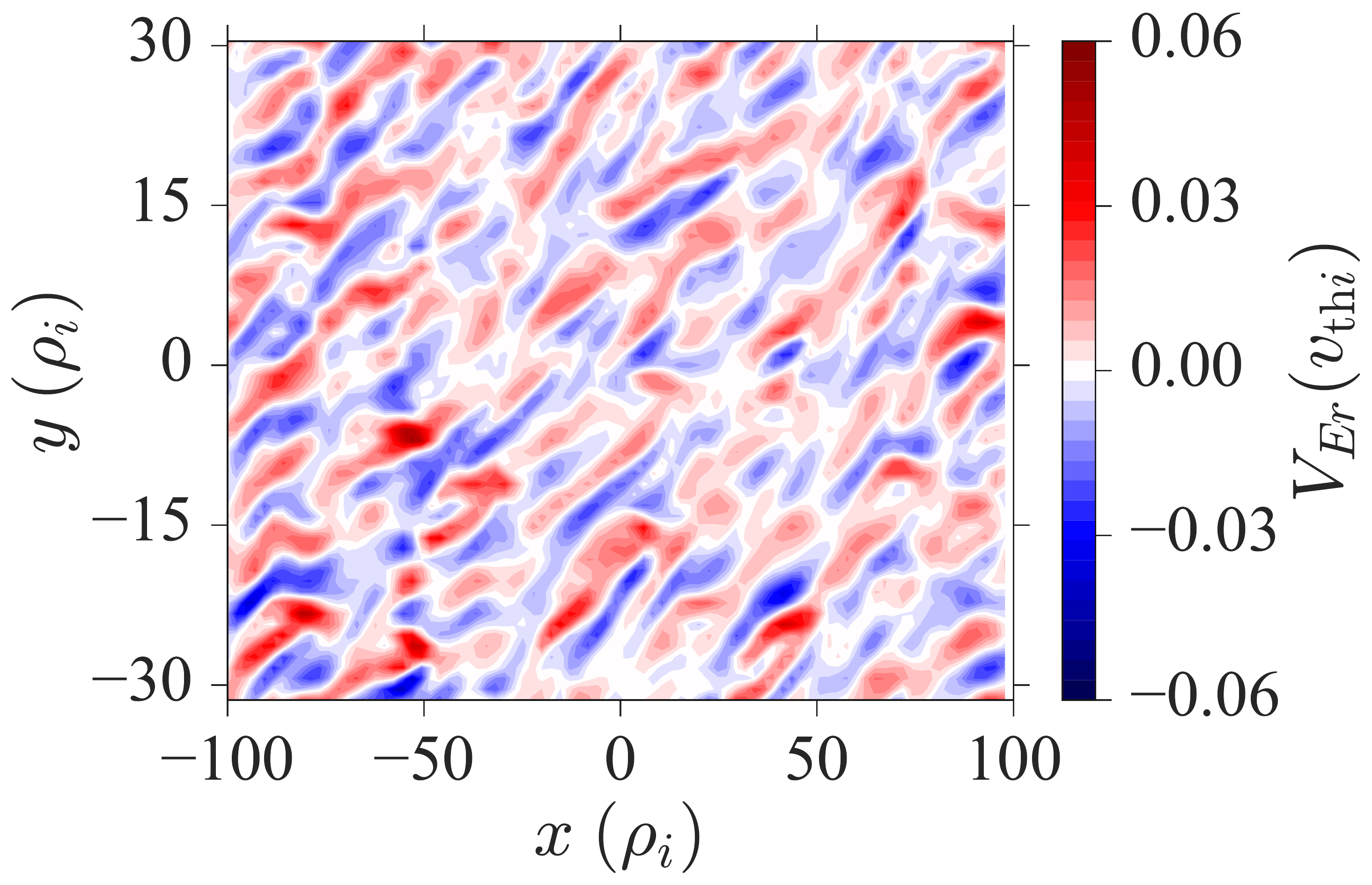}
      \caption{}
      \label{fig:vel_strongly_driven}
    \end{subfigure}
    \begin{subfigure}{0.49\linewidth}
      \includegraphics[width=\linewidth]{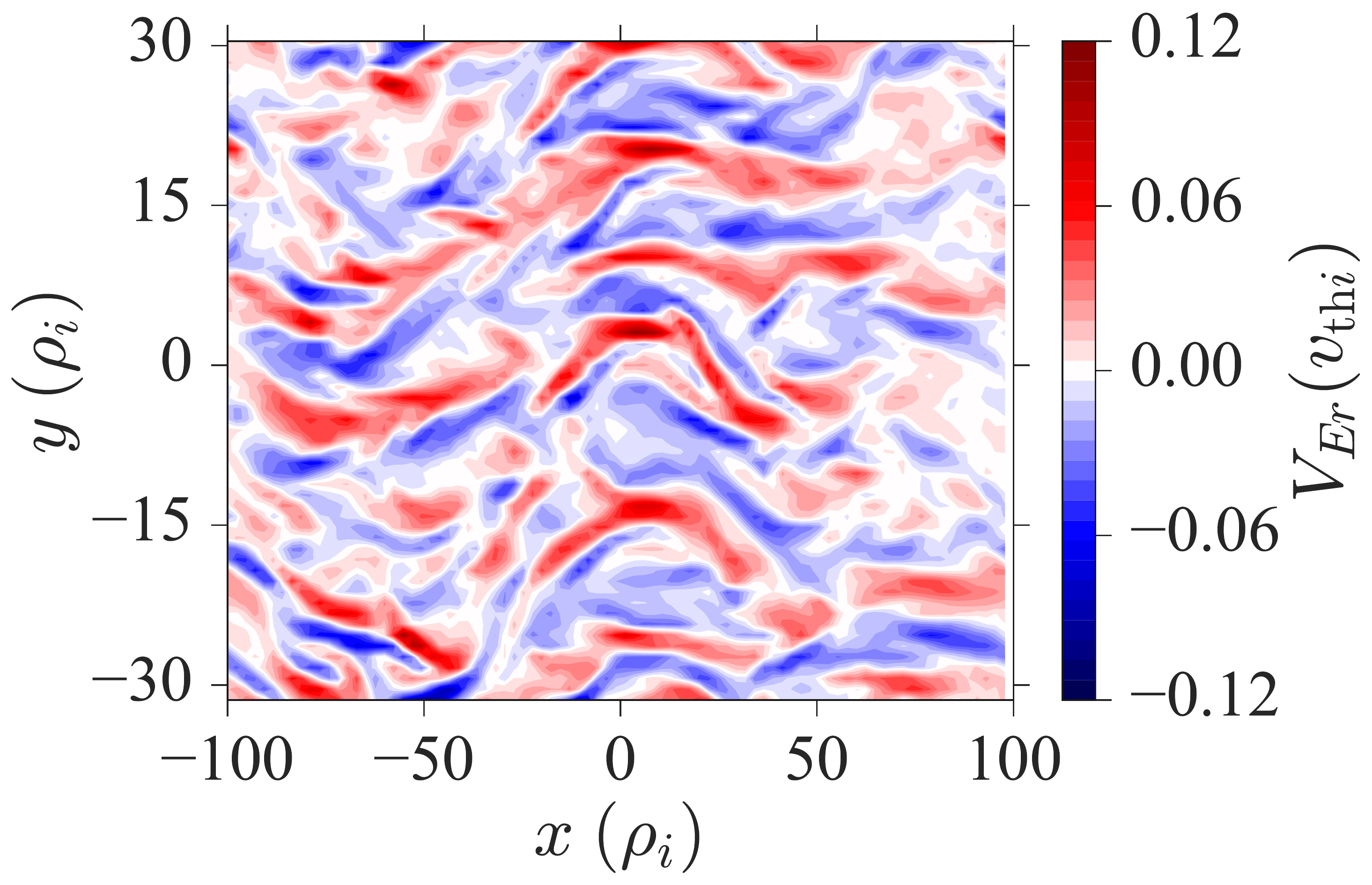}
      \caption{}
      \label{fig:vel_no_shear}
    \end{subfigure}
    \caption[Real-space radial \exb velocity on $(x,y)$ plane]{
      Radial \exb velocity $V_{Er}$ at the outboard midplane of MAST as a
      function of the local GS2 coordinates $x$ and $y$ for the same
      equilibrium parameters as in \figref{density_fluctuations}.
	}
    \label{fig:vel_fluctuations}
  \end{figure}

  We can describe the change in the nature of the density-fluctuation field as
  follows. The near-threshold state [\figref{marginal}] is dominated by intense
  (compared to the background fluctuations), coherent, and long-lived
  structures. As $\kappa_T$ is slightly increased (in this case by only 0.1),
  these structures become more numerous [\figref{intermediate}], but have
  roughly the same maximum amplitude: ${(\delta n_i/n_i)}_{\max} \sim 0.08$.
  The strongly driven state [\figref{strongly_driven}] exhibits a more
  conventional chaotic turbulent state characterised by many interacting eddies
  with larger amplitudes. The coherent structures in the marginal case are
  unlike the strongly interacting eddies that characterise the strongly driven
  turbulent state and more likely constitute nonlinear travelling wave
  (soliton-like) solution to the gyrokinetic equation. We note that these
  simulations are representative of the regions close to and far from the
  turbulence threshold, i.e.,\  in simulations near the threshold, we always
  find sparse but well-defined coherent structures that survive against a
  backdrop of weaker fluctuations. An important exception are simulations with
  $\gamma_E=0$, where we do not observe such coherent structures. As the system
  is taken away from the threshold by increasing $\kappa_T$, or decreasing
  $\gamma_E$, the structures become more numerous, while maintaining roughly
  the same amplitude, until they fill the entire domain, interact with each
  other, and break up.  For parameter values far from the threshold, we observe
  no discernible coherent structures, but rather strongly time-dependent
  fluctuations with amplitudes that increase with $\kappa_T$. For completeness,
  \figsref{vel_fluctuations}{tperp_fluctuations} show the perturbed radial \exb
  velocity $V_{Er}$ and the perpendicular temperature-fluctuation $\delta
  T_{\perp i}/T_{\perp i}$ fields.  We have calculated $V_{Er}$ velocity by
  taking the radial component of~\eqref{v_exb}, given by (see equation (3.42)
  in Ref.~\cite{HighcockThesis})
  \begin{equation}
    V_{Er} = \frac{c}{a B_{\mathrm{ref}}} \frac{1}{|\nabla \psi|}
    \qty|\pd{\psi}{r}|_{r_0} \pd{\varphi}{y}.
    \label{v_er}
  \end{equation}
  We see that the coherent structures have both high $V_{Er}$ and
  $\delta T_{\perp i}/T_{\perp i}$.
  \begin{figure}[t]
    \centering
    \begin{subfigure}{0.49\linewidth}
      \includegraphics[width=\linewidth]{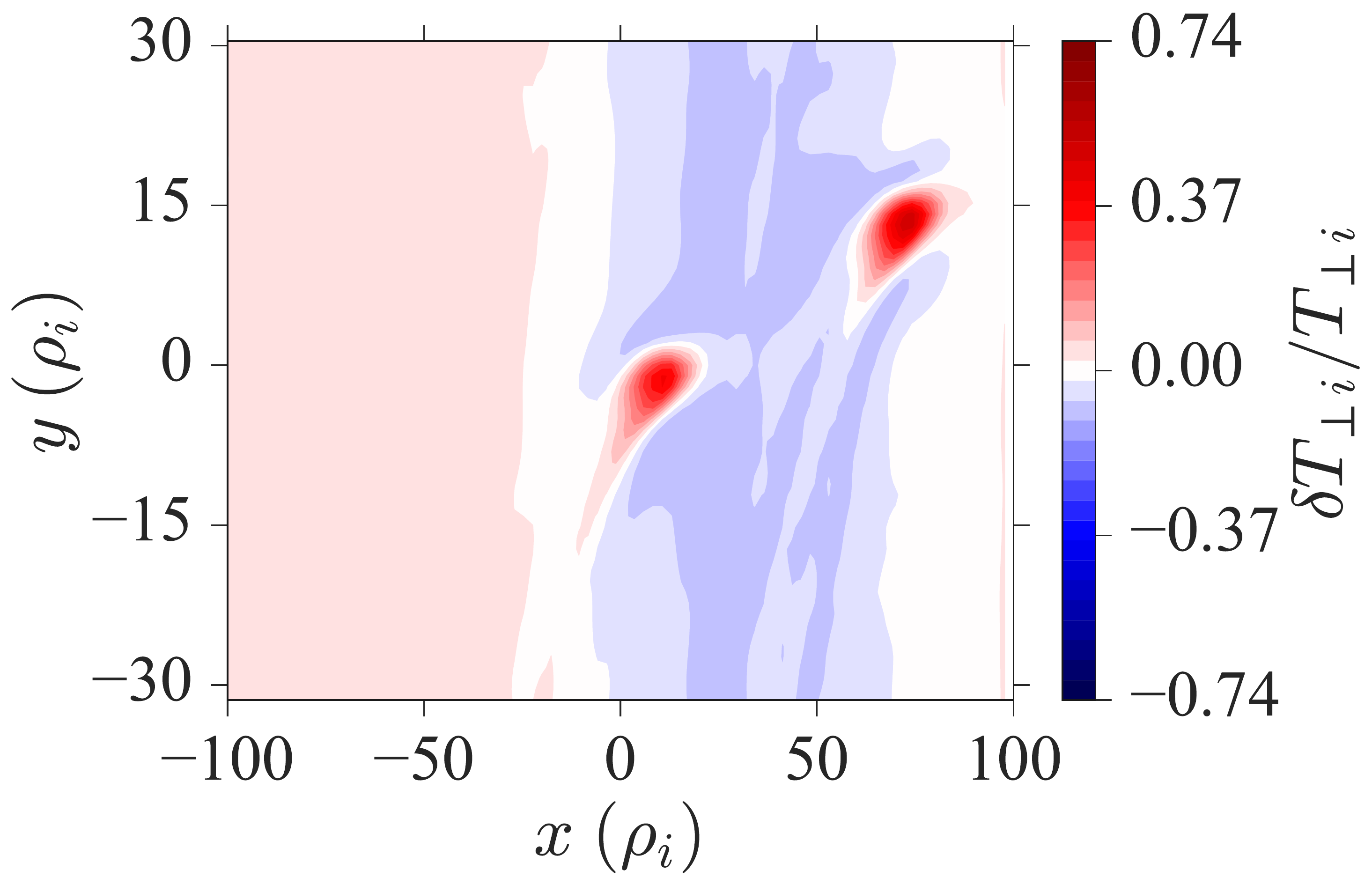}
      \caption{}
      \label{fig:tperp_marginal}
    \end{subfigure}
    \begin{subfigure}{0.49\linewidth}
      \includegraphics[width=\linewidth]{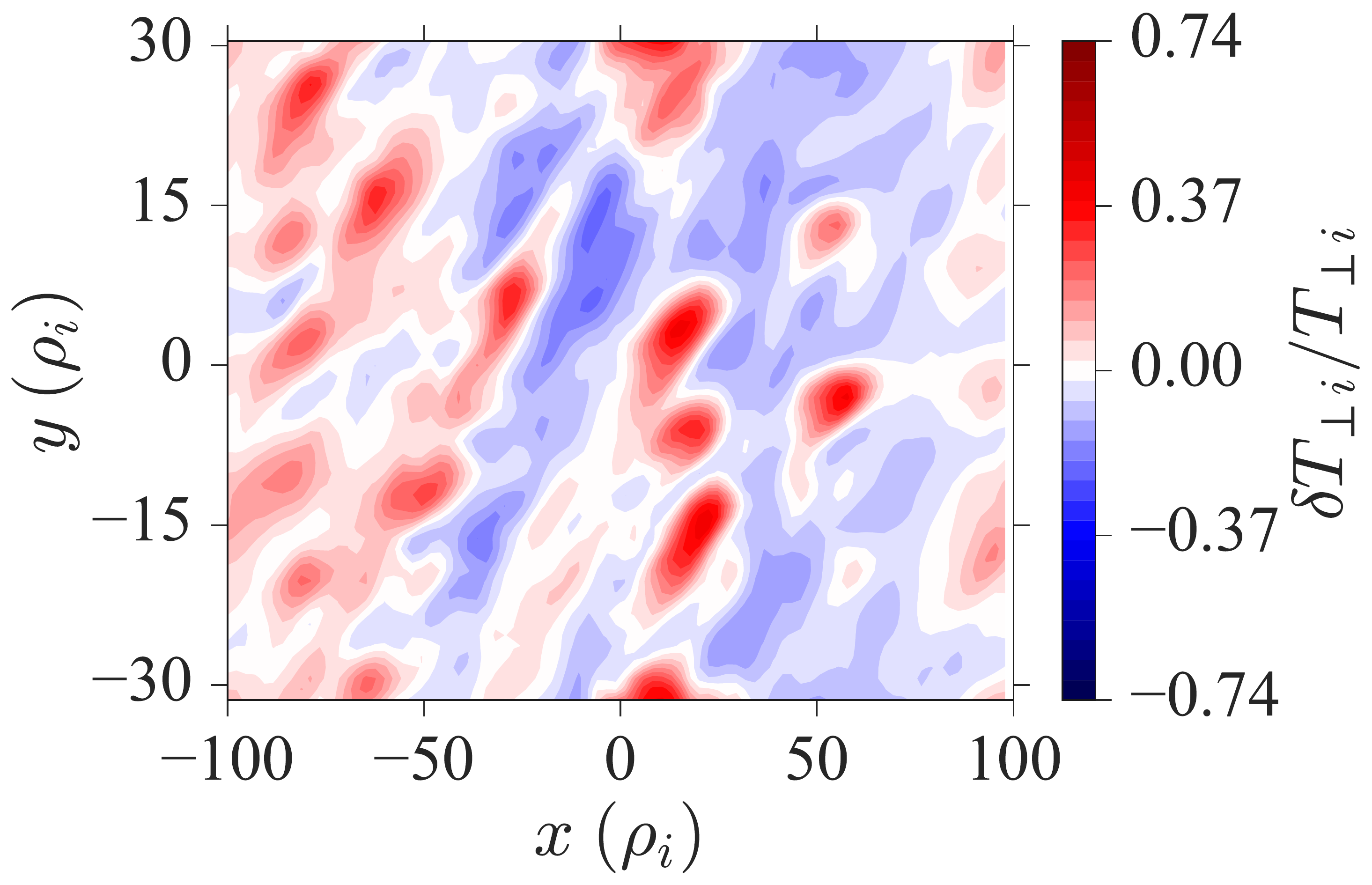}
      \caption{}
      \label{fig:tperp_intermediate}
    \end{subfigure}
    \begin{subfigure}{0.49\linewidth}
      \includegraphics[width=\linewidth]{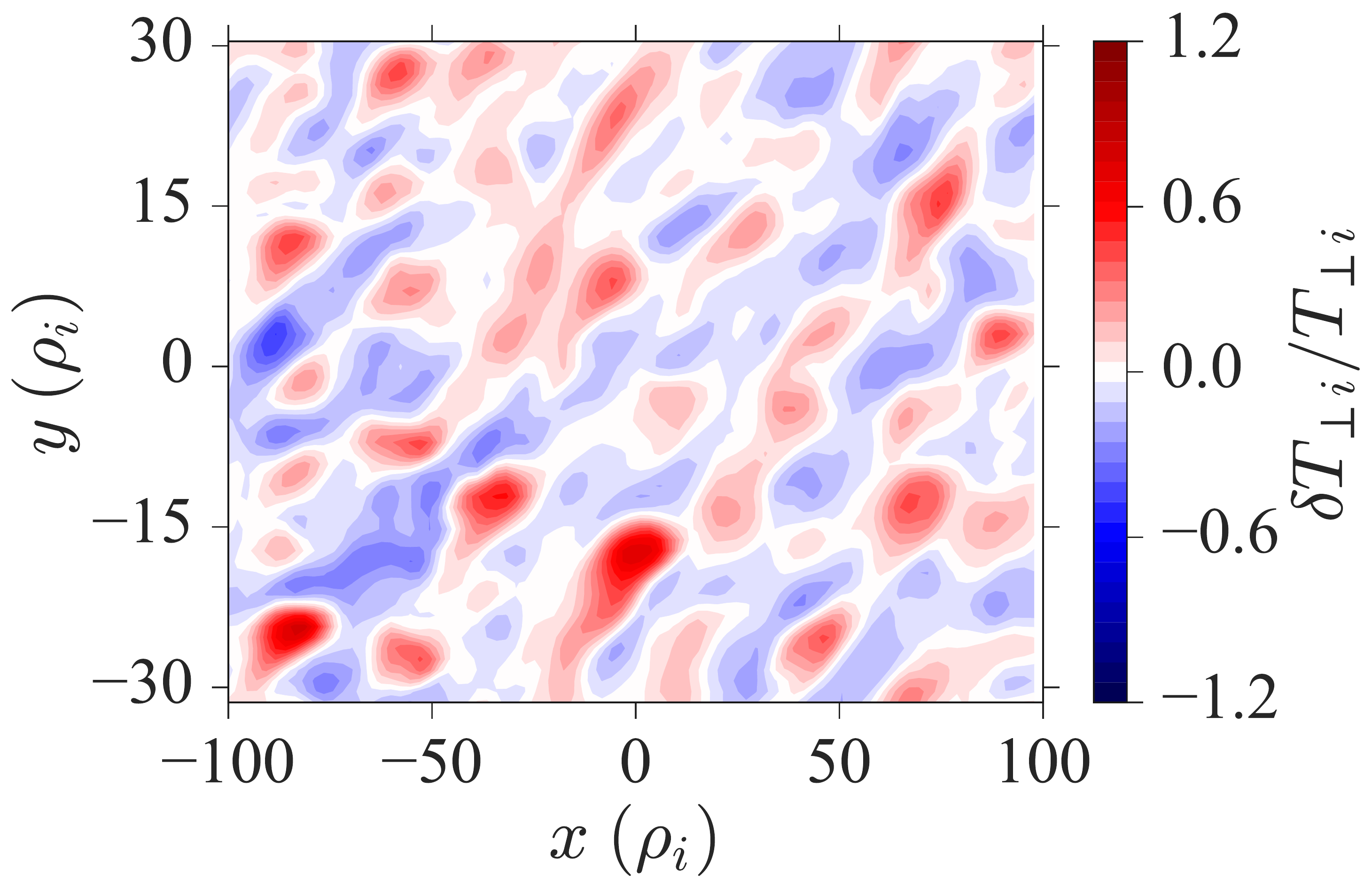}
      \caption{}
      \label{fig:tperp_strongly_driven}
    \end{subfigure}
    \begin{subfigure}{0.49\linewidth}
      \includegraphics[width=\linewidth]{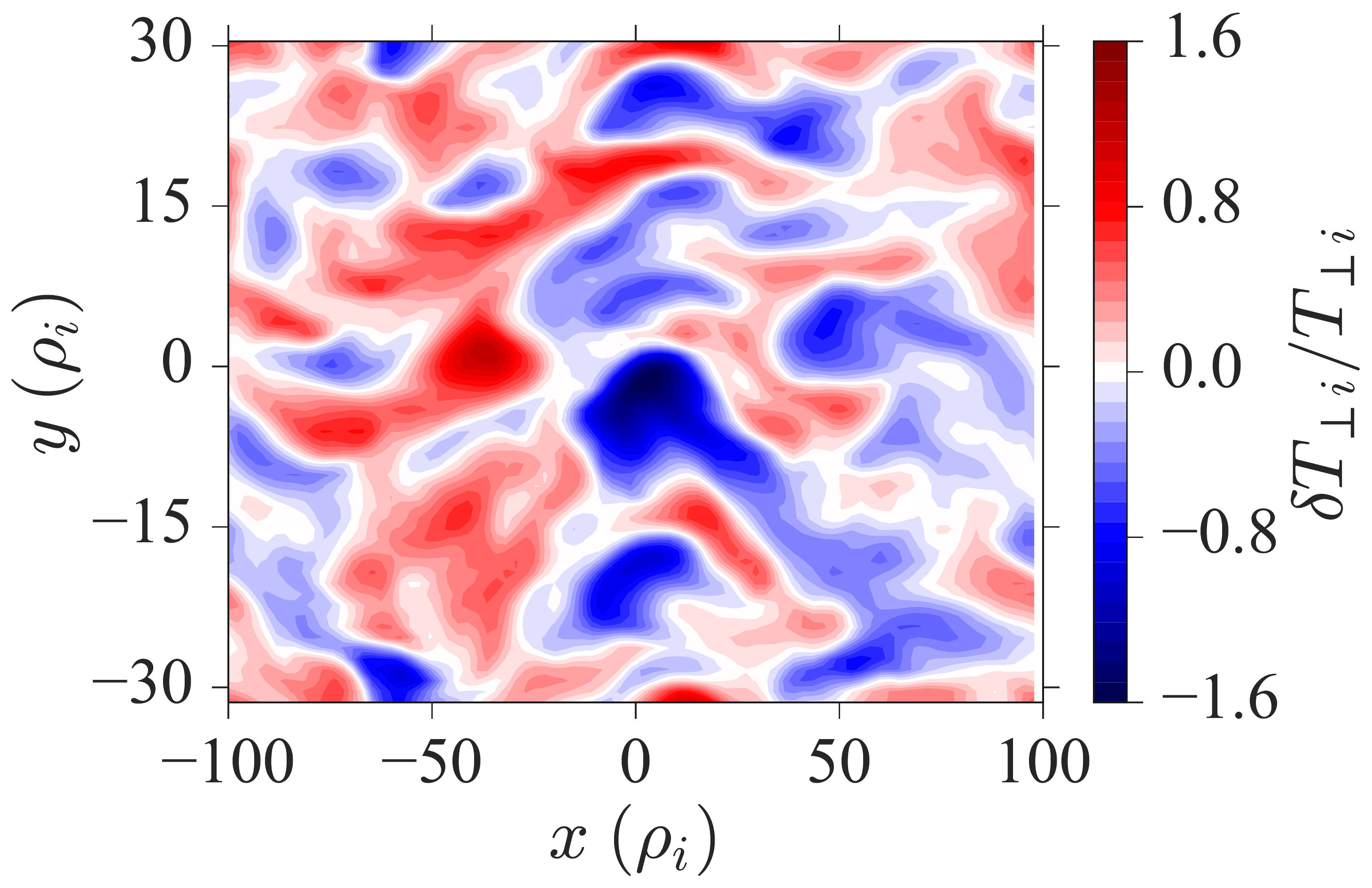}
      \caption{}
      \label{fig:tperp_no_shear}
    \end{subfigure}
    \caption[Real-space perpendicular temperature-fluctuation fields
             $T_{\perp i}$ on $(x,y)$ plane]{
     Perpendicular-temperature fluctuation field $\delta T_{\perp i}/T_{\perp
     i}$ outboard midplane of MAST as a function of the local GS2 coordinates
     $x$ and $y$ for the same equilibrium parameters as in
     \figref{density_fluctuations}.
    }
    \label{fig:tperp_fluctuations}
  \end{figure}

\begin{figure}[t]
    \centering
    \begin{subfigure}{0.49\linewidth}
      \includegraphics[width=\linewidth]{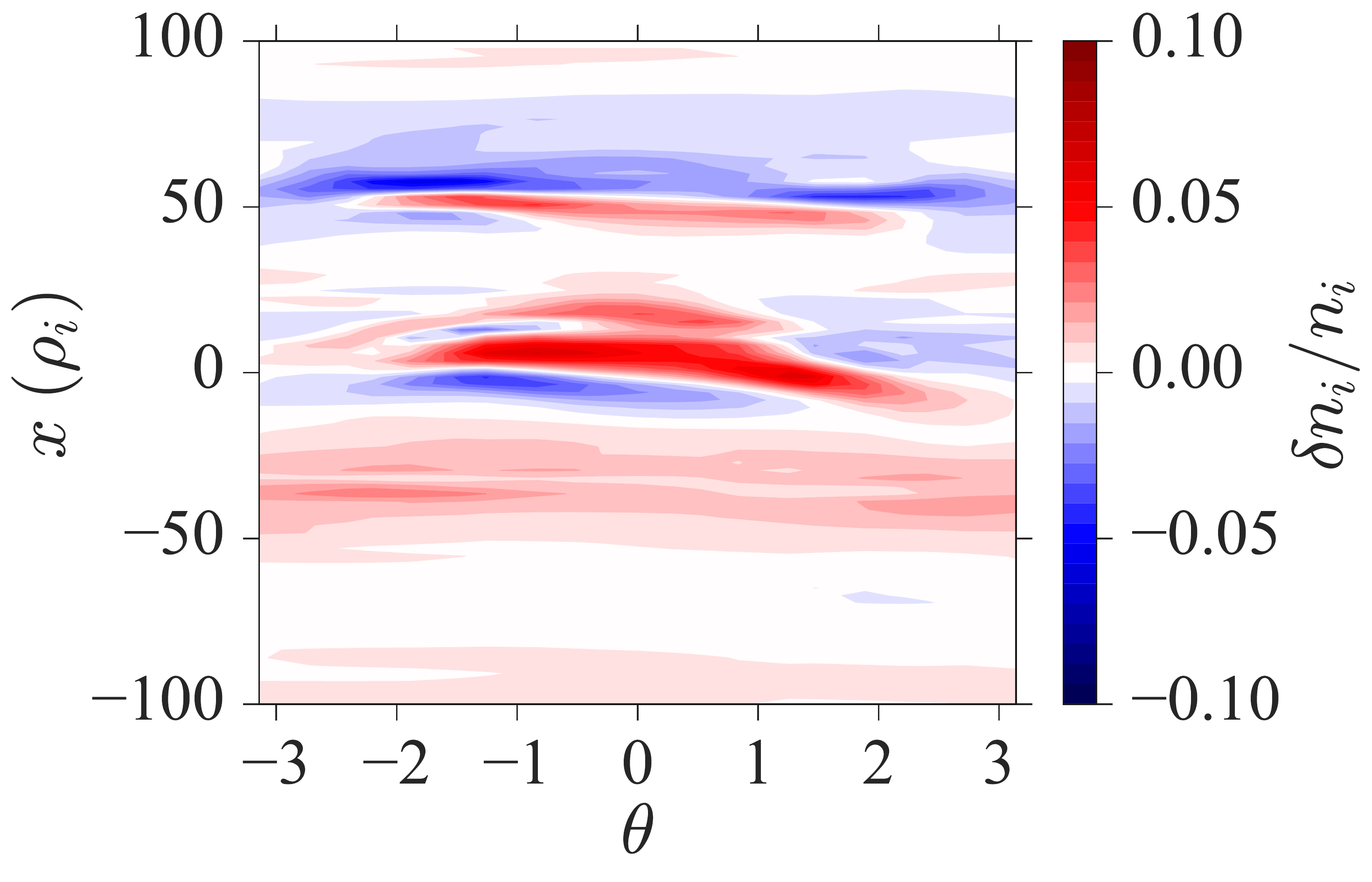}
      \caption{}
      \label{fig:marginal_xz}
    \end{subfigure}
    \hfill
    \begin{subfigure}{0.49\linewidth}
      \includegraphics[width=\linewidth]{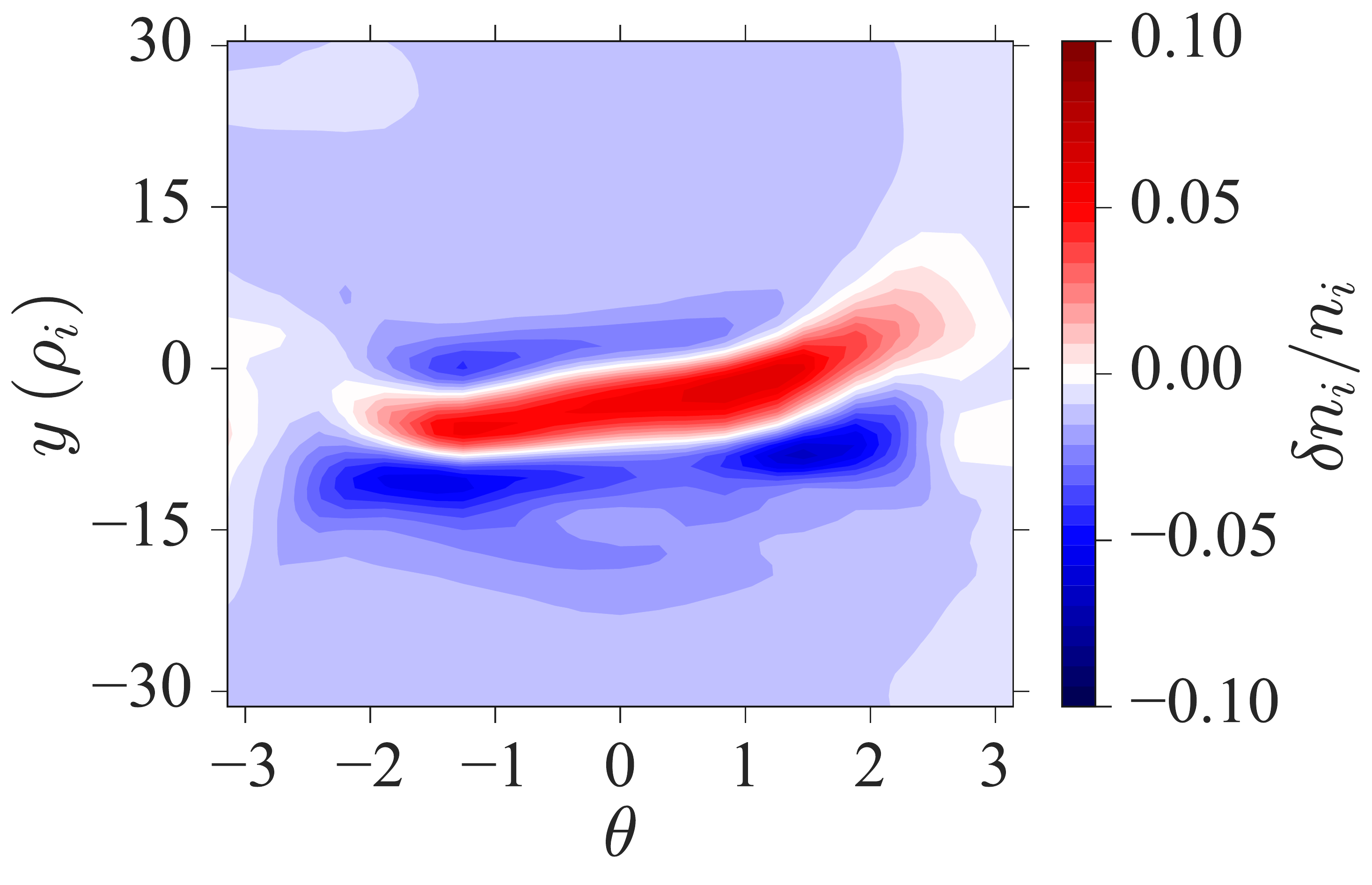}
      \caption{}
      \label{fig:marginal_yz}
    \end{subfigure}
    \caption[Real-space density-fluctuation fields in parallel direction]{
      \subref*{fig:marginal_xz} Density-fluctuation field $\delta n_i / n_i$ in
      the $x$-$z$ plane at $y=0$.
      \subref*{fig:marginal_yz} Density-fluctuation field $\delta n_i / n_i$ in
      a $y$-$z$ plane at $x=0$. Both plots are shown for the same simulation
      and at the same time as in \figref{marginal}; the corresponding planes
      are indicated by the dashed lines in \figref{marginal}. The parallel
      direction in GS2 is quantified by the poloidal angle $\theta$ (see
      Section~\ref{sec:gs2_geometry}).
    }
    \label{fig:parallel_density}
  \end{figure}

  We now consider the marginal cases, and the dynamics of the coherent
  structures, more carefully, starting with their parallel structure.
  \Figref{parallel_density} shows two views of the coherent structures in
  \figref{marginal} in the parallel direction (which in GS2 is quantified by
  the poloidal angle $\theta$; see Appendix~\ref{App:real_space_transform}) at
  constant $y$ [\figref{marginal_xz}] and at constant $x$
  [\figref{marginal_yz}]. It is clear that the coherent structures are
  elongated in the parallel direction and have an amplitude much larger than
  the ``background'' fluctuations.

\begin{figure}[t]
    \centering
    \begin{subfigure}{0.9\linewidth}
      \includegraphics[width=\linewidth]{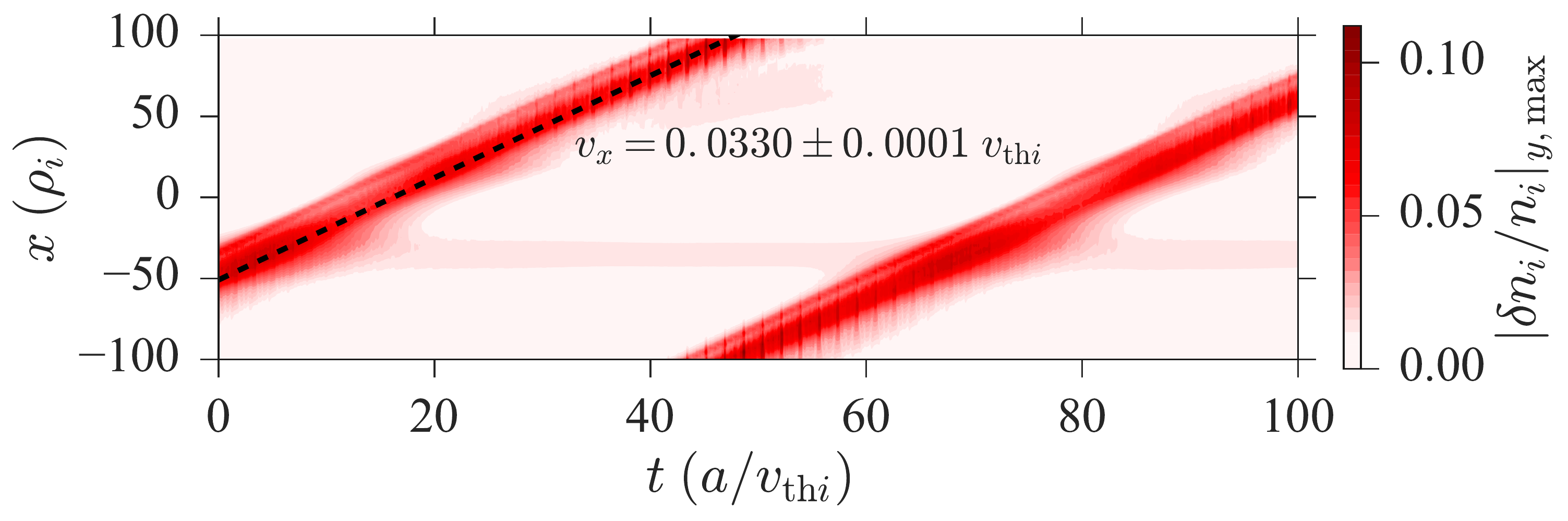}
      \caption{}
      \label{fig:marginal_xt}
    \end{subfigure}
    \begin{subfigure}{0.9\linewidth}
      \includegraphics[width=\linewidth]{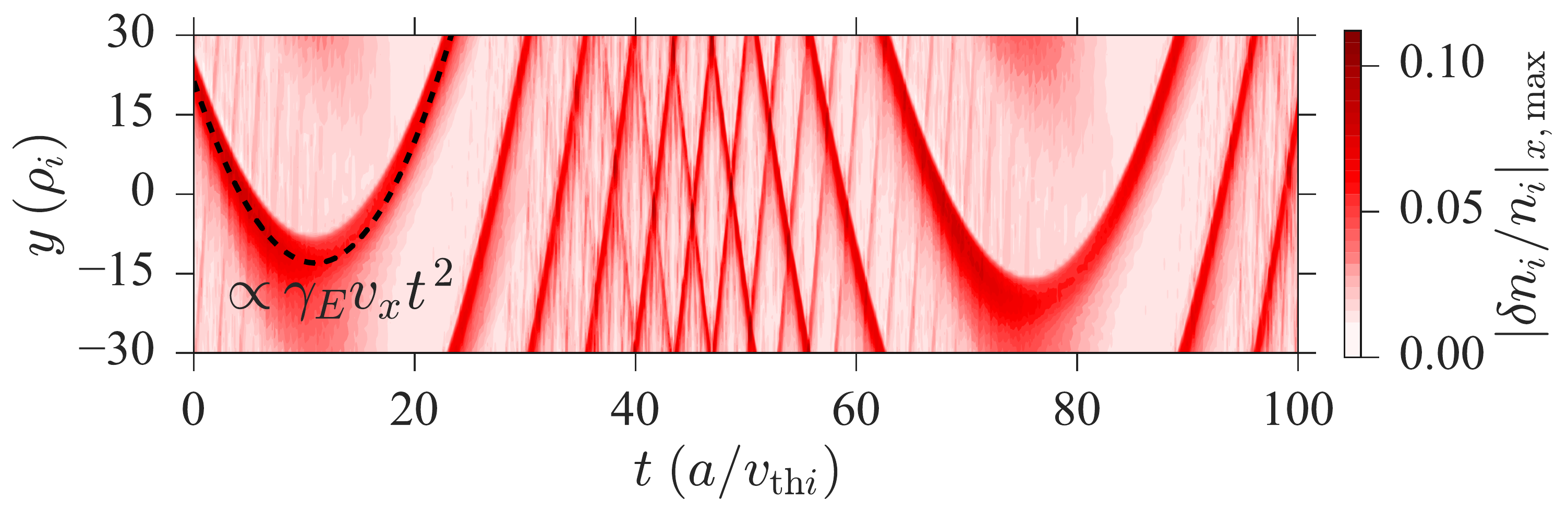}
      \caption{}
      \label{fig:marginal_yt}
    \end{subfigure}
    \caption[Radial and poloidal advection of coherent structure]{
      Density-fluctuation field $\delta n_i / n_i$ as a function of
      \subref*{fig:marginal_xt} $x$ and $t$ (taking the maximum in
      the $y$ direction) and \subref*{fig:marginal_yt} $y$ and $t$ (taking the
      maximum in the $x$ direction) for a marginally unstable case with
      $(\kappa_T, \gamma_E) = (5.1, 0.18)$, which contains only one coherent
      structure. The structure is advected both radially and poloidally. We
      note that the GS2 domain is periodic in $x$ and $y$ and so this is the
      same structure throughout the entire time period shown. The dashed line
      in \subref*{fig:marginal_xt} indicates $x=v_x t$, and in
      \subref*{fig:marginal_yt} indicates $y \propto \gamma_E v_x t^2$ showing
      that the poloidal advection is due to the flow imposed by the flow shear.
    }
  \end{figure}

  In time, the coherent structures are advected by the flow imposed by the flow
  shear in the poloidal direction, but also drift in the radial direction.
  Figures~\ref{fig:marginal_xt} and~\subref{fig:marginal_yt} show $\delta n_i /
  n_i$ for a marginal nonlinear simulation at $(\kappa_T, \gamma_E) = (5.1,
  0.18)$, which has only one coherent structure, as a function of $(t,x)$ and
  $(t,y)$ (taking the maximum value of $\delta n_i/n_i$ in the other
  direction), respectively.  \Figref{marginal_xt} shows the radial motion of
  the structure across the domain, which the structures crosses in a time of
  roughly $50~(a/v_{\mathrm{th}i})$. The radial motion of the structures in
  \figref{marginal_xt} has a constant velocity and fitting the trajectory with
  a straight line (the dashed line) gives a radial velocity of $v_x = 0.0330
  \pm 0.0001$~$v_{\mathrm{th}i}$.  \figref{marginal_yt} shows the poloidal
  advection of the structure with a much shorter poloidal crossing time of
  roughly $5~(a/v_{\mathrm{th}i})$. The poloidal motion of the structure is
  entirely due to the advection caused by the flow shear as we will now
  explain. As we saw in \figref{marginal_xt}, $v_x$ is constant and the radial
  position is given by $x(t) = v_x t$. The poloidal advection due to the flow
  shear is given by $v_y(t) = \gamma_E x(t)$ and so the direction of the flow shear
  reverses at $x=0$. Combining the expressions for $x(t)$ and $v_y(t)$ and
  integrating, we find that $y(t) \propto \gamma_E v_x t^2$, and, as shown by
  the dashed line in \figref{marginal_yt}, this explains the poloidal motion of
  the structure, which indeed reverses direction at $x=0$.  The long-lived
  nature of coherent structures close to the turbulence threshold is
  illustrated by \figref{marginal_xt} given that the GS2 domain is periodic in
  $x$ and $y$, and so the structure exists for $t > 100~(a/v_{\mathrm{th}i})$.

  \subsection{$Q_i/Q_{\mathrm{gB}}$ as an order parameter}
  The results in Section~\ref{sec:coherent_strucs} suggested that the nature of
  the turbulence is set by how far the system is from the turbulence threshold.
  Specifically, that the near threshold state is dominated by coherent
  structures that seem to increase in number and amplitude as the system is
  taken further from the threshold. This suggests that the important metric
  that should be used to quantify the state of the system is the ``distance
  from threshold'' and not the specific values of $\kappa_T$ and
  $\gamma_E$ (although both can be used to control the distance from
  threshold). $Q_i/Q_{\mathrm{gB}}$ is a strong function of $\kappa_T$ and
  $\gamma_E$, with the dependence that we showed in \figref{contour_heatmap},
  and so we can use $Q_i/Q_{\mathrm{gB}}$ as a control parameter to measure the
  distance from the turbulence threshold. In Sections~\ref{sec:max_amp}
  and~\ref{sec:struc_count}, we will quantify the changes in the
  amplitude and number of structures for our parameter scan and show that the
  distance from threshold is the relevant order parameter.

  \subsection{Maximum amplitude}
  \label{sec:max_amp}
  \begin{figure}[t]
    \centering
    \includegraphics[width=0.6\linewidth]{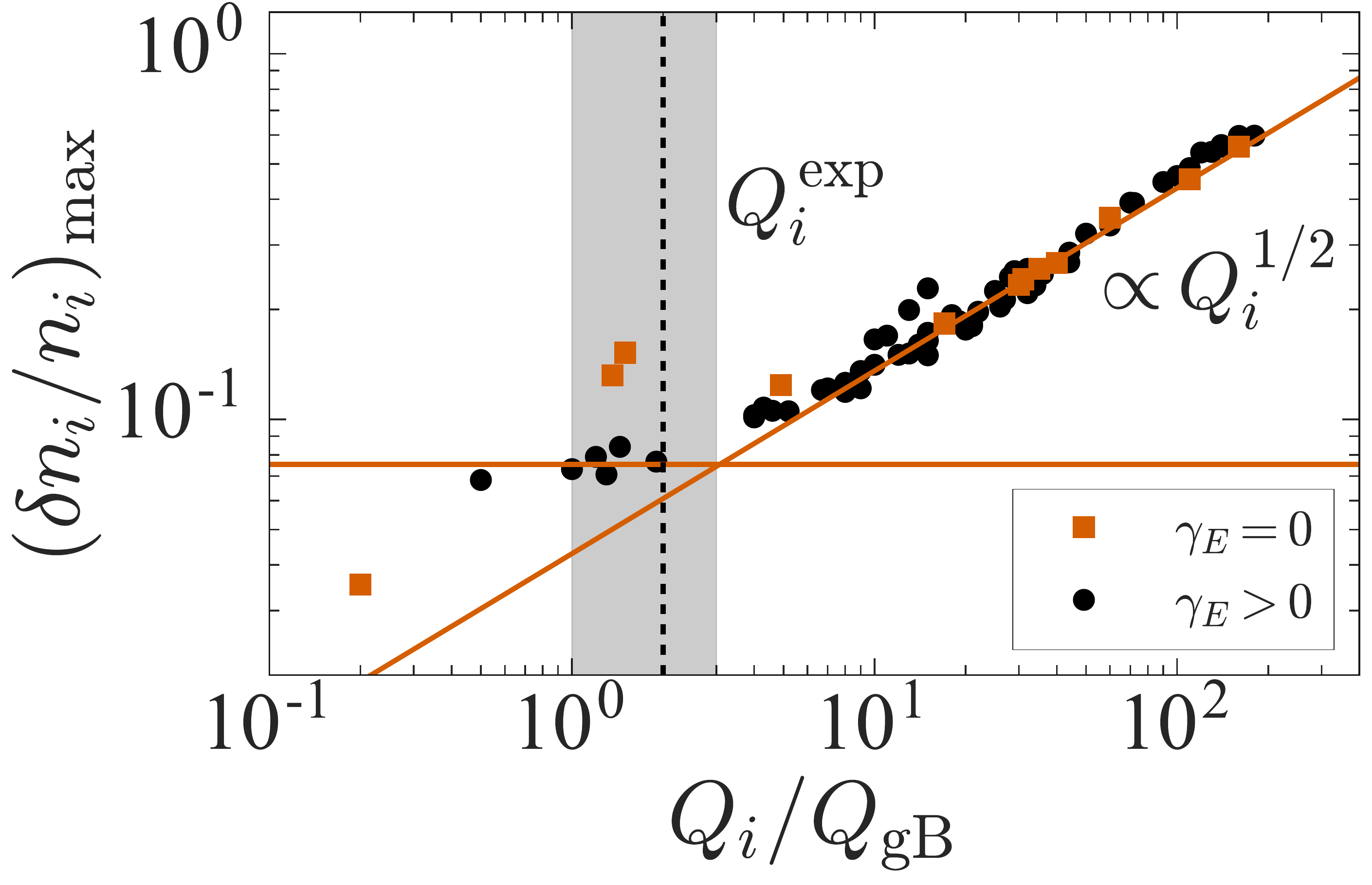}
    \caption[Scaling of maximum fluctuation amplitude with $Q_i/Q_{\mathrm{gB}}$]{
      Maximum amplitude of the density fluctuations versus
      $Q_i/Q_{\mathrm{gB}}$. The naive scaling~\eqref{q_scaling}, $Q_i^{1/2}
      \propto \delta n_i / n_i$, is shown for reference and holds far from
      threshold, whereas for small values of $Q_i/Q_{\mathrm{gB}}$ (around and
      below the experimental value $Q_i^{\exp}$), the amplitude becomes
      independent of $Q_i/Q_{\mathrm{gB}}$.
    }
    \label{fig:amplitude}
  \end{figure}
  Considering the density-fluctuation fields shown in
  \figref{density_fluctuations}, we see that a key property that changes as the
  system is taken away from the threshold is the amplitude of the eddies. We
  would like to know how the amplitude changes with the distance from
  threshold, which we quantify using $Q_i/Q_{\mathrm{gB}}$. For marginal cases,
  such as \figref{marginal}, the dominant features are structures with high
  densities compared to the background fluctuations. In order to measure the
  changes in the amplitude of these structures we want to measure the maximum
  amplitude, as opposed to an $(x,y)$-averaged quantity, which would be small
  because of the relatively small volume taken up by the coherent structures.
  Therefore, we consider the maximum amplitude (taken over $x$ and $y$),
  ${(\delta n_i/n_i)}_{\max}$, of density perturbations averaged over time in a
  given simulation. \Figref{amplitude} shows the relationship between ${(\delta
  n_i/n_i)}_{\max}$ and $Q_i/Q_{\mathrm{gB}}$ for all the simulations in our
  parameter scan.  The striking feature of~\figref{amplitude} is that ${(\delta
  n_i/n_i)}_{\max}$ hits a finite ``floor'' as $Q_i/Q_{\mathrm{gB}}$ approaches
  and goes below its experimental value. This coincides with the appearance of
  the long-lived structures shown in~\figref{marginal}. For $\gamma_E=0$
  simulations with values of $Q_i/Q_{\mathrm{gB}}$ below $Q_i^{\exp}$, we do
  not see a clear trend, and importantly do not see the flattening we see for
  $\gamma_E>0$ simulations, suggesting that the turbulence is fundamentally
  different close to the turbulence threshold (as was also suggested by the
  absence of coherent structures).

  Far from the turbulence threshold, we can estimate the expected behaviour of
  $\delta n_i / n_i$ via a naive estimate of the dependence of
  $Q_i/Q_{\mathrm{gB}}$ on $\delta n_i / n_i$ using~\eqref{q_def}:
  \begin{equation}
    \frac{Q_i}{Q_{\mathrm{gB}}}\sim \frac{a^2}{\rho_i^2}\frac{\delta
    n_i}{n_i}\frac{V_{Er}}{v_{{\mathrm{th}}i}} \sim k_y\rho_i \frac{T_e}{T_i}
    {\left(\frac{a}{\rho_i}\frac{\delta n_i}{n_i}\right)}^2,
    \label{q_scaling}
  \end{equation}
  where $(a/\rho_i) \delta n_i/n_i$ is an order-unity quantity in gyrokinetic
  theory~\cite{Abel2013}. In deriving \eqref{q_scaling}, we have used
  \eqref{v_er} and assumed that fluctuations of $\varphi$ are related (by order
  of magnitude) to the electron (and, therefore, ion) density via the Boltzmann
  response $e\varphi/T_e \sim \delta n_e/n_e$. The scaling
  $\delta n_i/n_i \propto Q_i^{1/2}$ (obtained from \eqref{q_scaling} given
  that the prefactor is order unity) is indicated by the red line in
  \figref{amplitude}, and shows that this describes the scaling far from
  threshold well. We also see that $\gamma_E=0$ and $\gamma_E>0$ simulations
  are similar far from the threshold.

  The above observations are entirely non-trivial. In the case of supercritical
  turbulence, we typically observe smaller fluctuation amplitudes all the way
  to the turbulence threshold -- there is no minimum amplitude required to
  sustain turbulence.  In contrast, \figref{amplitude} shows that for the
  subcritical we are investigating, the maximum fluctuation amplitude remains
  constant, for low heat fluxes, while the heat flux decreases because there is
  a critical value required in order to sustain a saturated nonlinear state.
  The system reconciles the requirement of finite amplitude structures while
  allowing the heat flux to decrease via a reduction of the volume taken up by
  structures. This nonlinear state has not been previously observed in fusion
  plasmas. We further study the changes in the state of the system by
  performing a structure-counting analysis in the next section, explicitly
  showing the reduction in the volume taken up by the structures.

  \subsection{Structure counting}
  \label{sec:struc_count}
  We demonstrate the change in volume taken up by finite-amplitude structures
  by measuring the typical number of these structures in our simulations
  as a function of the distance from threshold. While two-dimensional
  structures are easily discerned by the human eye (e.g., in the near-marginal
  case shown in \figref{marginal}, there are two), counting them systematically
  is a non-trivial problem often encountered in computer vision and pattern
  recognition applications. Detection of coherent structures has been
  considered before in the context of experimental measurements of
  turbulence~\cite{Muller2005,Cheng2013}; a review of various techniques is
  given in~\cite{Love2007}.
  \begin{figure}[t]
    \centering
    \begin{subfigure}{0.49\linewidth}
      \includegraphics[width=\linewidth]{density_intermediate}
      \caption{}
      \label{fig:intermediate2}
    \end{subfigure}
    \begin{subfigure}{0.49\linewidth}
      \includegraphics[width=\linewidth]{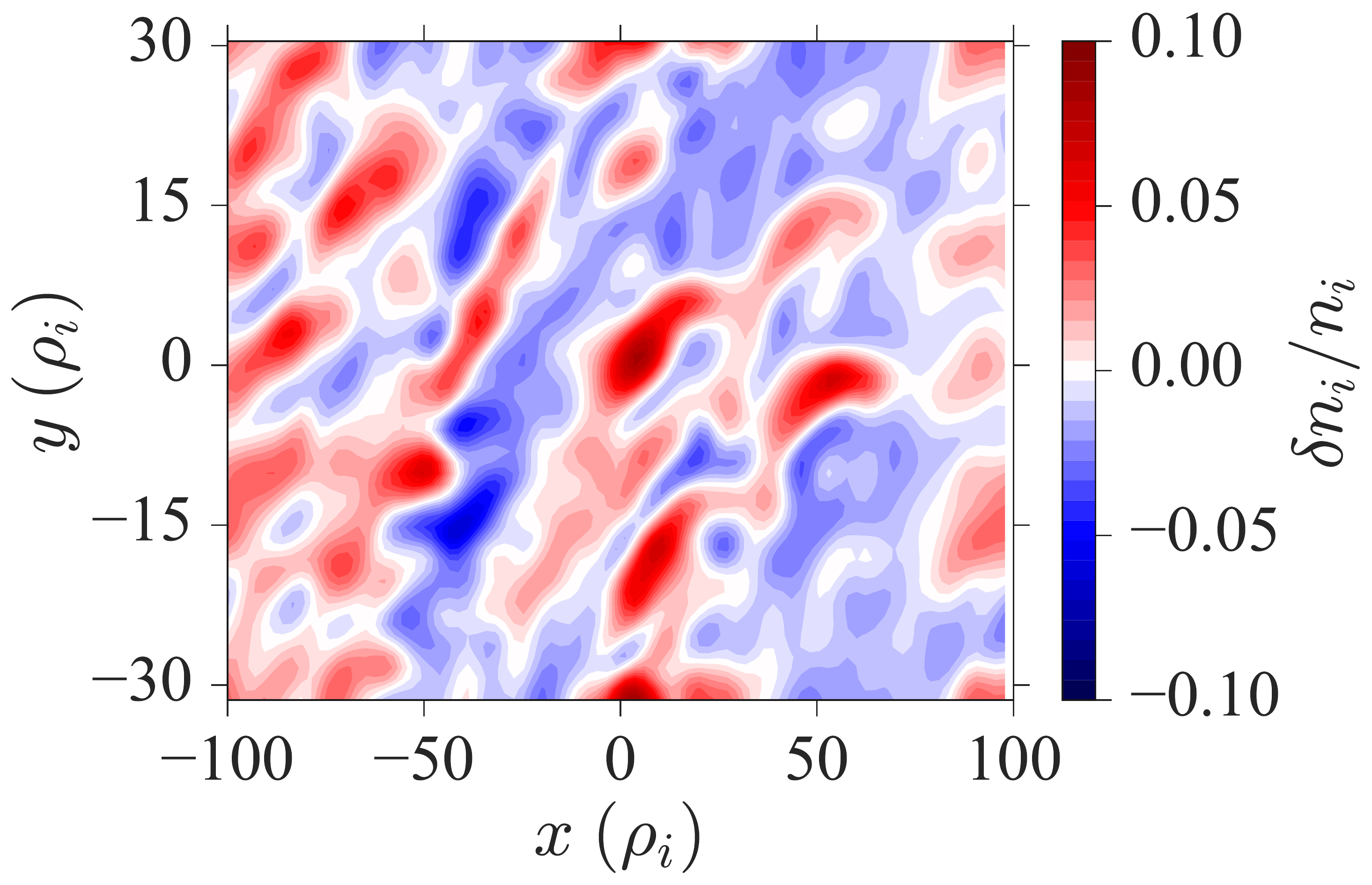}
      \caption{}
      \label{fig:post_filter}
    \end{subfigure}
    \begin{subfigure}{0.49\linewidth}
      \includegraphics[width=\linewidth]{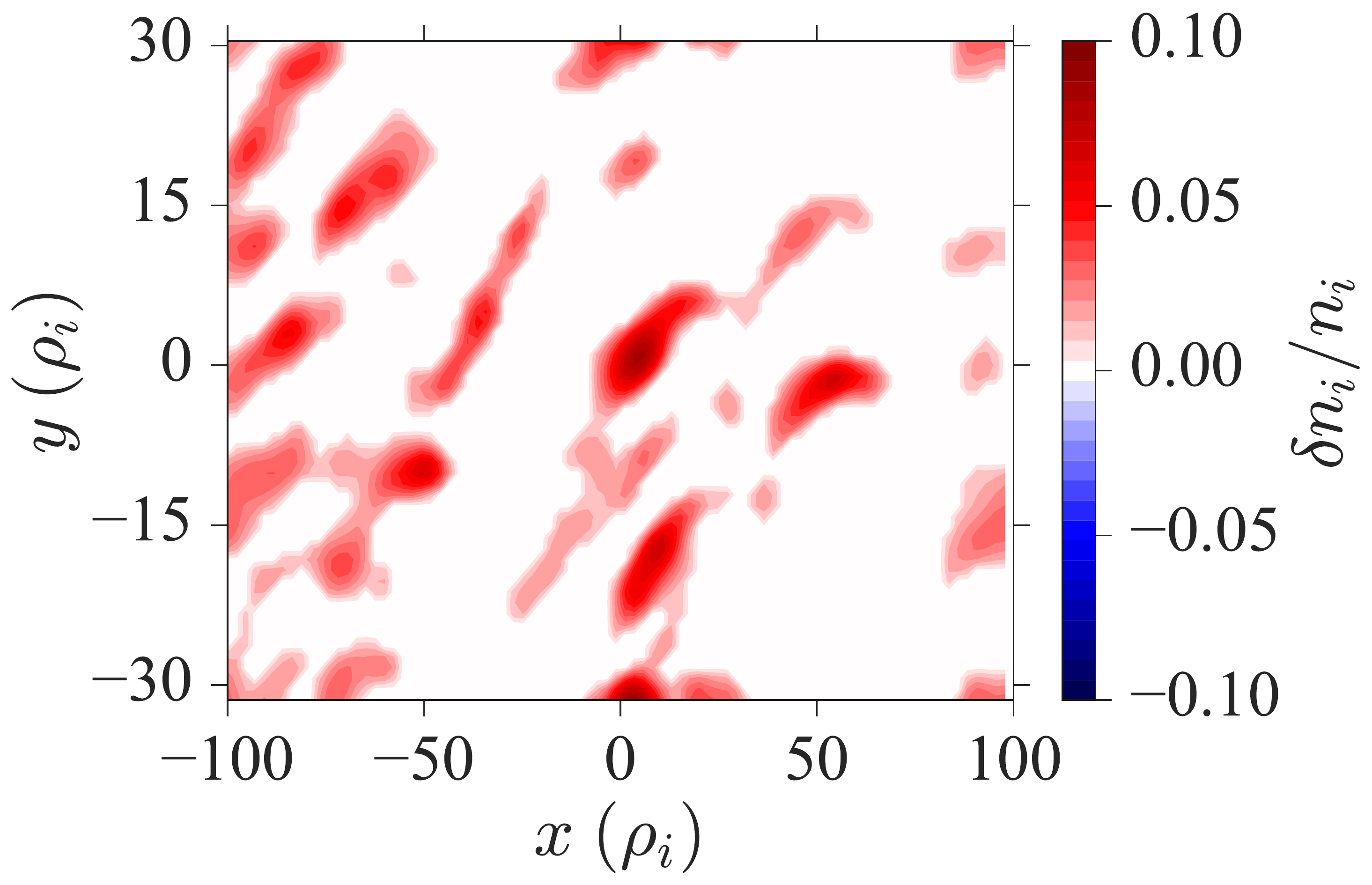}
      \caption{}
      \label{fig:post_thresh}
    \end{subfigure}
    \begin{subfigure}{0.49\linewidth}
      \includegraphics[width=\linewidth]{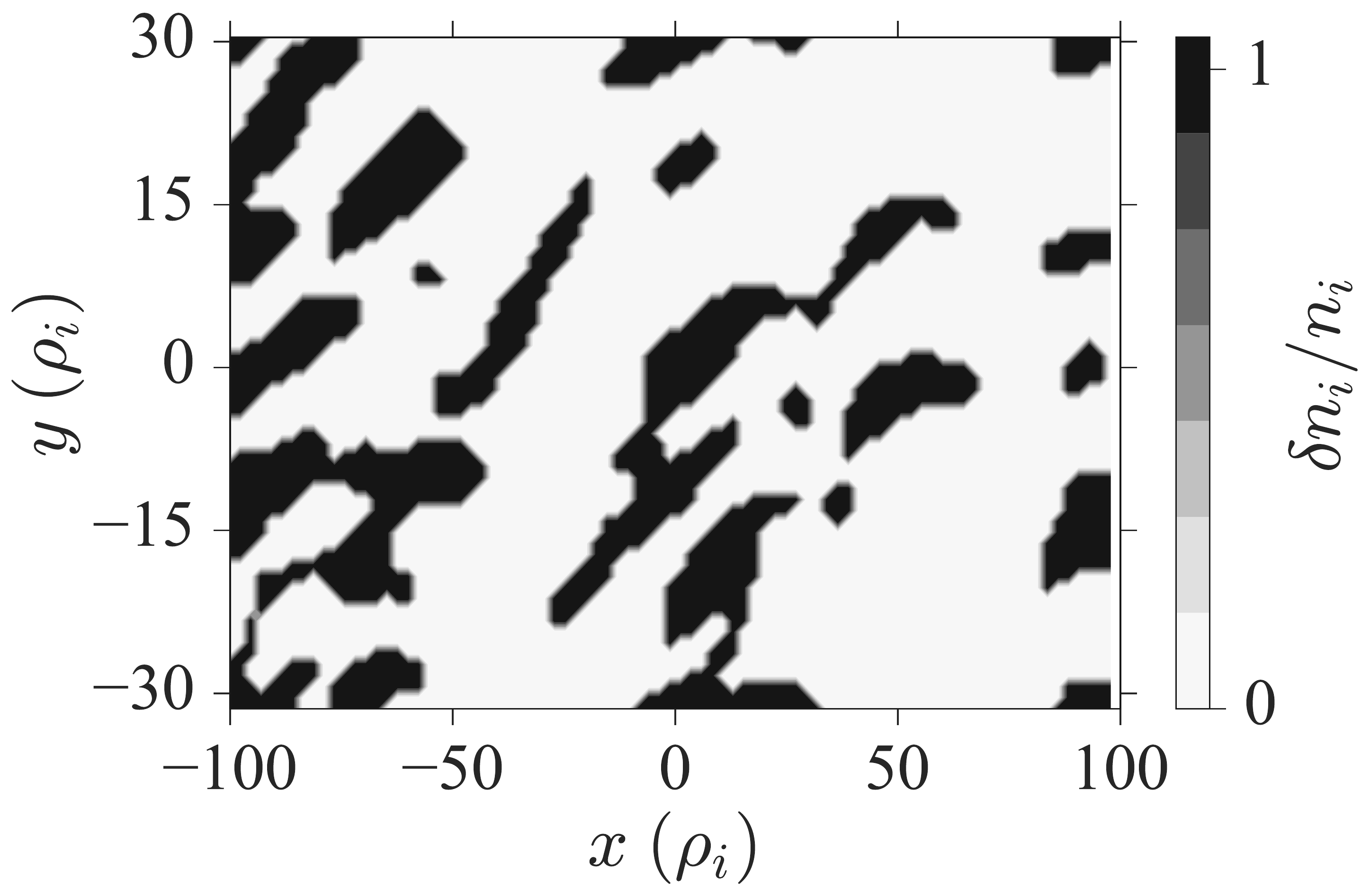}
      \caption{}
      \label{fig:struc_intermediate}
    \end{subfigure}
    \caption[Structure counting procedure]{
      Stages of the structure counting procedure:
      \subref*{fig:intermediate2} the original density-fluctuation field
      [as in \figref{intermediate}];
      \subref*{fig:post_filter} after the application of a Gaussian filter to
      smooth the structures;
      \subref*{fig:post_thresh} after the application of a 75\% threshold
      function;
      \subref*{fig:struc_intermediate} after setting $\delta n_i/n_i>0$
      values to 1 for simplicity. The image-labelling algorithm is then
      applied to \subref*{fig:struc_intermediate} and returns $19$
      structures for this case.
	}
    \label{fig:struc_count_procedure}
  \end{figure}

  Structure counting can be reduced to an image-labelling, or ``segmentation'',
  problem in the following way. We applied a Gaussian image filter (with a
  standard deviation on the order of the grid scale) as a pre-processing step
  and also removed structures below 10\% of the mean structure size as a
  post-processing step. These filtering steps are justified because we are
  interested in detecting intense, relatively large-scale structures, and
  simply applying a threshold function can lead to single points above the
  threshold scattered around the edges of structures that we are actually
  interested in counting. We then set values below a certain percentile (here
  75\% of the maximum amplitude) to 0 and above it to 1. The level of the
  threshold function is somewhat arbitrary and the number of structures will
  depend on this level, but the trend as a function of our equilibrium
  parameters did not change as we increased or decreased the level of the
  threshold function.  Choosing too low a level often leads to many structures
  being counted as only one, whereas too high a level led to only a handful of
  the most intense structures being counted. While this could be acceptable
  close to marginality, where we are interested in high-intensity structures
  compared to low-intensity background fluctuations, this would significantly
  underpredict the number of structures far from the threshold. We chose 75\%
  as a reasonable compromise. After applying a threshold function, one is left
  with an array of 1's representing our structures against a background of 0's.
  To count these structures, we employed a general-purpose image processing
  package \emph{scikit-image}~\cite{scikit-image}, which implements an
  efficient labelling algorithm~\cite{Fiorio1996}, then used by us to label
  connected regions. The structure-counting procedure is shown in
  \figref{struc_count_procedure} where the image-labelling algorithm labelled
  $19$ structures.

  \begin{figure}[t]
    \centering
    \includegraphics[width=0.6\linewidth]{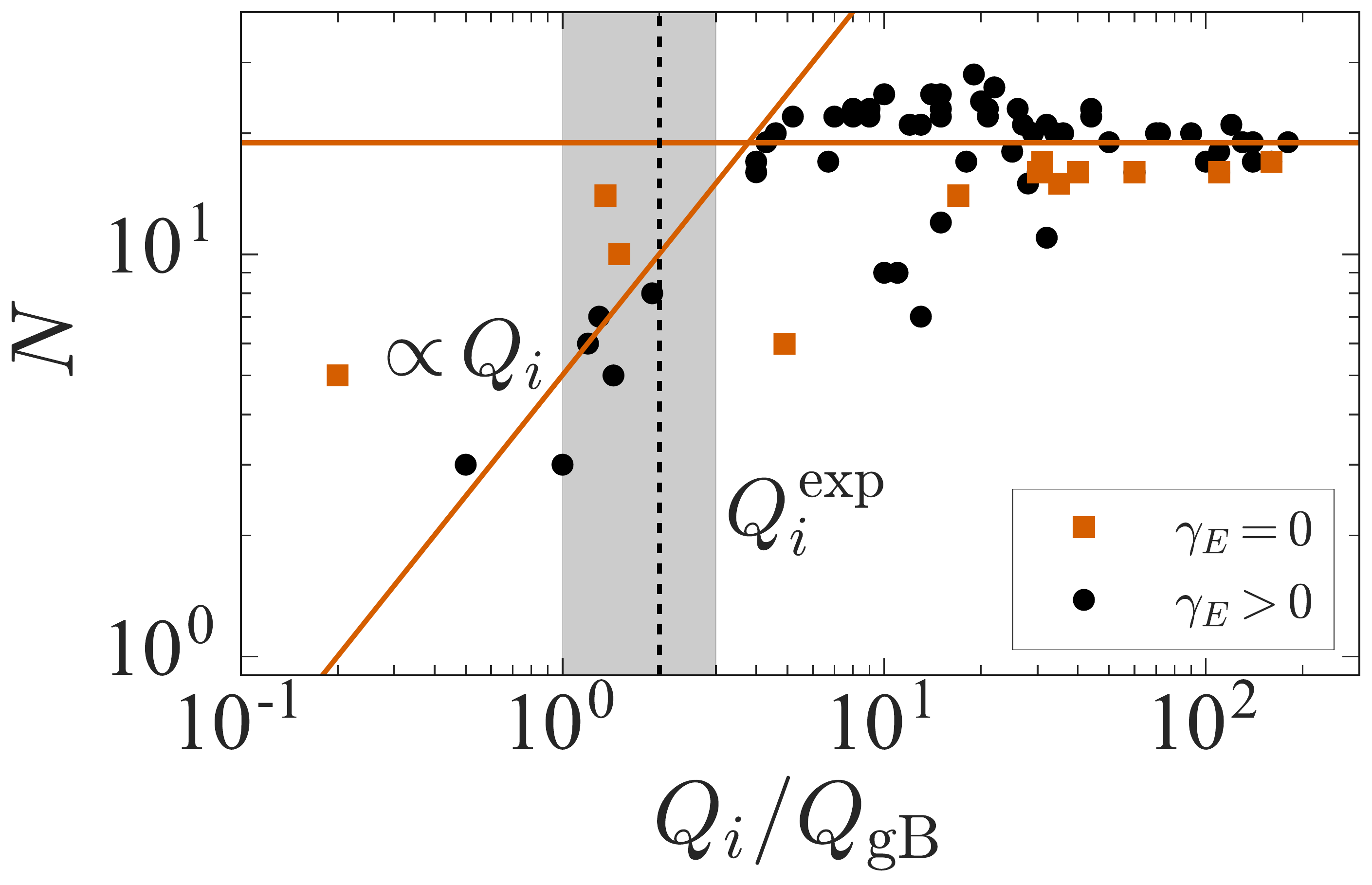}
    \caption[Scaling of number of structures with $Q_i/Q_{\mathrm{gB}}$]{
      Number of structures (defined as having an amplitude above 75\% of the
      maximum) versus $Q_i/Q_{\mathrm{gB}}$.  It grows up to and slightly
      beyond the experimental value $Q_i^{\exp}$. Eventually the volume is
      filled with structures and their number tends to a constant.  The scaling
      $Q_i \propto N$ is shown for reference.
    }
    \label{fig:nblobs}
  \end{figure}
\begin{figure}[t]
    \centering
    \includegraphics[width=0.6\linewidth]{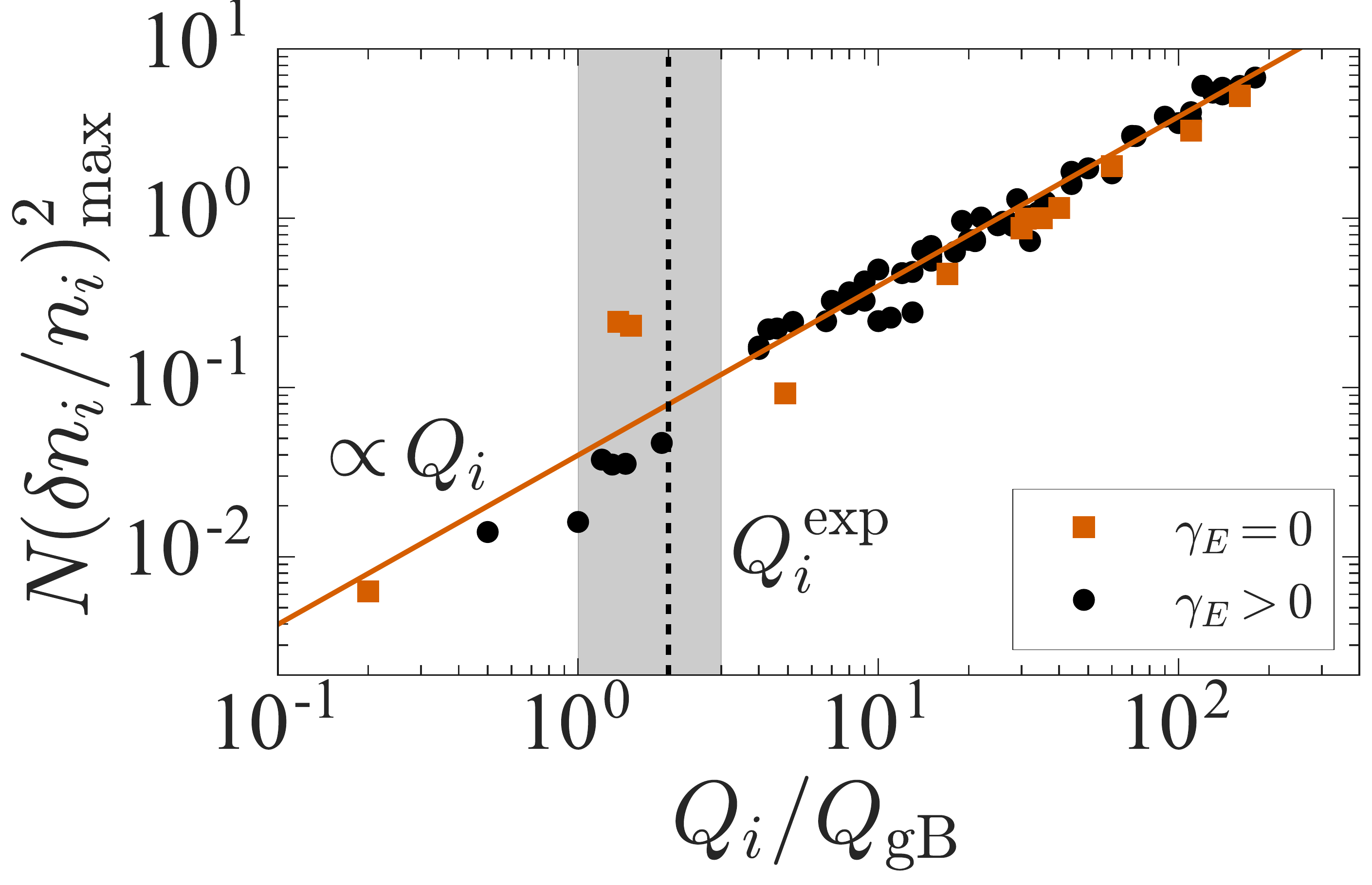}
    \caption[$N{(\delta n_i/n_i)}_{\max}^2$ scaling with $Q_i/Q_{\mathrm{gB}}$]{
      Confirmation of the scaling~\eqref{n_amp_scaling}, where the
      red line indicates a line $\propto Q_i$. We note that simulations near
      marginality are relatively difficult to saturate leading to the low
      number of simulations around $Q_i^{\exp}$. However, the trend is still
      clear even for those simulations.
    }
    \label{fig:n_amp_squared}
  \end{figure}

  \Figref{nblobs} shows the results of the above analysis: the number of
  structures $N$ with amplitudes above the 75$^{\mathrm{th}}$ percentile versus the
  ion heat flux $Q_i/Q_{\mathrm{gB}}$.  As in \figref{amplitude}, there are two
  distinct regimes: $N$ grows with $Q_i/Q_{\mathrm{gB}}$ until the structures
  have filled the simulation domain (which happens just beyond the experimental
  value of the flux), whereupon $N$ tends to a constant. Again, we see that the
  $\gamma_E=0$ and the $\gamma_E>0$ simulations are similar far from the
  threshold. Taking Figures~\ref{fig:amplitude} and~\ref{fig:nblobs} in
  combination, we have, roughly,
  \begin{equation}
    \frac{Q_i}{Q_{\mathrm{gB}}} \sim N {\qty(\frac{\delta n_i}{n_i})}_{\max}^2,
    \label{n_amp_scaling}
  \end{equation}
  i.e., near the threshold, the turbulent heat flux increases because coherent
  structures become more numerous (but not more intense), whereas far from the
  threshold, it does so because the fluctuation amplitude increases (at a
  roughly constant number of structures). This relationship is confirmed
  by~\figref{n_amp_squared}, which shows $N{(\delta n_i/n_i)}_{\max}^2$ as a
  function of $Q_i/Q_{\mathrm{gB}}$, and we see that these quantities are,
  indeed, proportional to each other.

  Thus, we have identified two types of nonlinear states depending on the
  distance from threshold: one dominated by coherent structures close to
  the threshold, and one characterised by many interacting eddies far from the
  threshold. We clearly showed that, far from the turbulence threshold,
  cases with $\gamma_E=0$ (conventional ITG-driven turbulence) have similar
  properties to $\gamma_E>0$ cases. In the next section we investigate the role
  of zonal flows in regulating turbulence and come to the same conclusions as
  above: the presence of flow shear is important close to the threshold, but
  turbulence is similar for $\gamma_E=0$ and $\gamma_E>0$ cases far from the
  threshold.

  \subsection{Shear due to zonal flows}
  \label{sec:zf_shear}

  The dominant saturation mechanism for ITG-driven turbulence is thought to be
  the stabilisation caused by zonal modes~\cite{Waltz1994, Lin1998, Dimits2000,
  Rogers2000, Diamond2005}. Zonal modes are fluctuations in the system with
  $k_y = 0$ and $k_x > 0$, i.e.,\ they have finite radial
  extent, but are poloidally symmetric. They are generated by nonlinear
  interactions in the system and contain sheared flows that can regulate
  turbulence. Previous work~\cite{Dimits2000} on the transition to turbulence
  showed that near the turbulence threshold (approached by varying the
  equilibrium parameter $\kappa_T$), turbulence is regulated by strong zonal
  flows, which can cause an upshift in the critical $\kappa_T$ required for a
  saturated turbulent state. However, in the system under investigation, the
  marginal cases seem to be dominated by the background flow shear
  [see~\figref{marginal_yt}], which also has a suppressing effect on the
  turbulence. Thus, in this section, we investigate the role played by zonal
  flows in the turbulence regimes identified in
  Sections~\ref{sec:coherent_strucs}--\ref{sec:struc_count} and show that zonal
  flows do not play an important role in the near-marginal cases but become
  more important far from the threshold, where their effect is comparable to,
  and eventually dominate over that of the background flow shear.

  In the MAST plasma we are investigating, there are two sources of shear that
  may regulate turbulence: shear due to strong toroidal rotation as a result of
  the injection of neutral particles by the NBI heating system, and shear due
  to zonal flows which are generated by nonlinear interactions. We have already
  seen that shear due to the toroidal rotation is controlled by the equilibrium
  parameter $\gamma_E$, which we vary in this study. The shear due to the zonal
  flows $V'_{\mathrm{ZF}}$ is calculated from \eqref{v_er} by considering only
  the poloidally symmetric component, and is given by
  \begin{equation}
    V'_{\mathrm{ZF}} = \frac{c}{a B_{\mathrm{ref}}} \frac{q_0}{r_0}
    \frac{1}{|\nabla \alpha|} \pdv[2]{\varphi_{\mathrm{ZF}}}{x},
    \label{v_zf_prime}
  \end{equation}
  where $V'_{\mathrm{ZF}}$ is a function only of $t$ and $x$, and
  $\varphi_{\mathrm{ZF}}$ is the poloidally symmetric component of $\varphi$.
  To determine whether the zonal shear will dominate over $\gamma_E$ we
  calculate the RMS value of the zonal shear, $\gamma_{\mathrm{ZF}}$:
  \begin{equation}
    \gamma_{\mathrm{ZF}} = \left< V^{\prime 2}_{\mathrm{ZF}}\right>^{1/2}_{t,x},
    \label{g_zf}
  \end{equation}
  where $\ensav{\cdots}{t,x}$ indicates an average over $t$ and $x$. We can
  now compare $\gamma_{\mathrm{ZF}}$ with $\gamma_E$ to determine the relative
  importance of each as a function of our equilibrium parameters.

  \begin{figure}[t]
    \centering
      \begin{subfigure}{0.51\linewidth}
        \includegraphics[width=\linewidth]{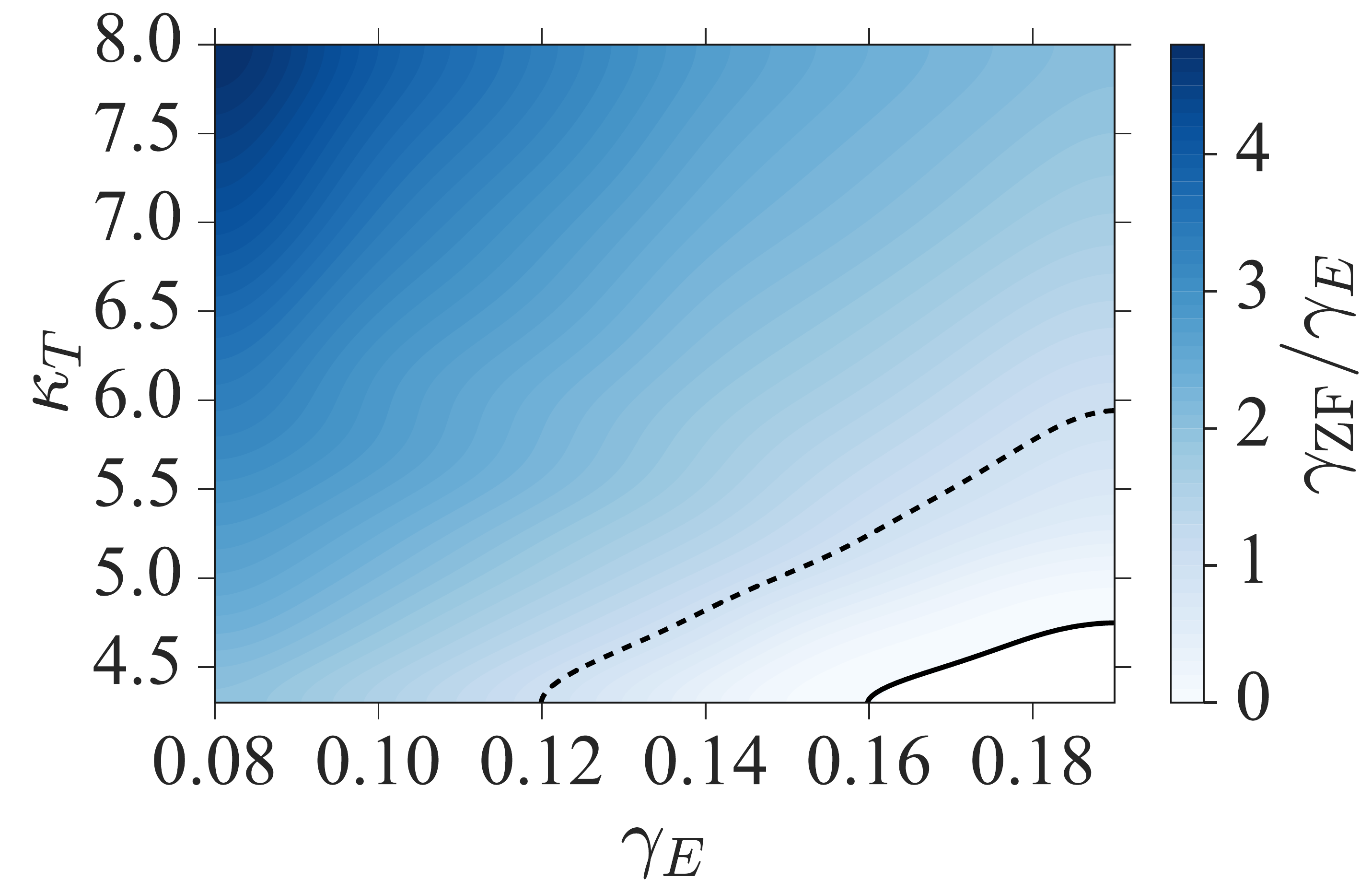}
        \caption{}
        \label{fig:zf_shear_contour}
      \end{subfigure}
      \hfill
      \begin{subfigure}{0.47\linewidth}
        \includegraphics[width=\linewidth]{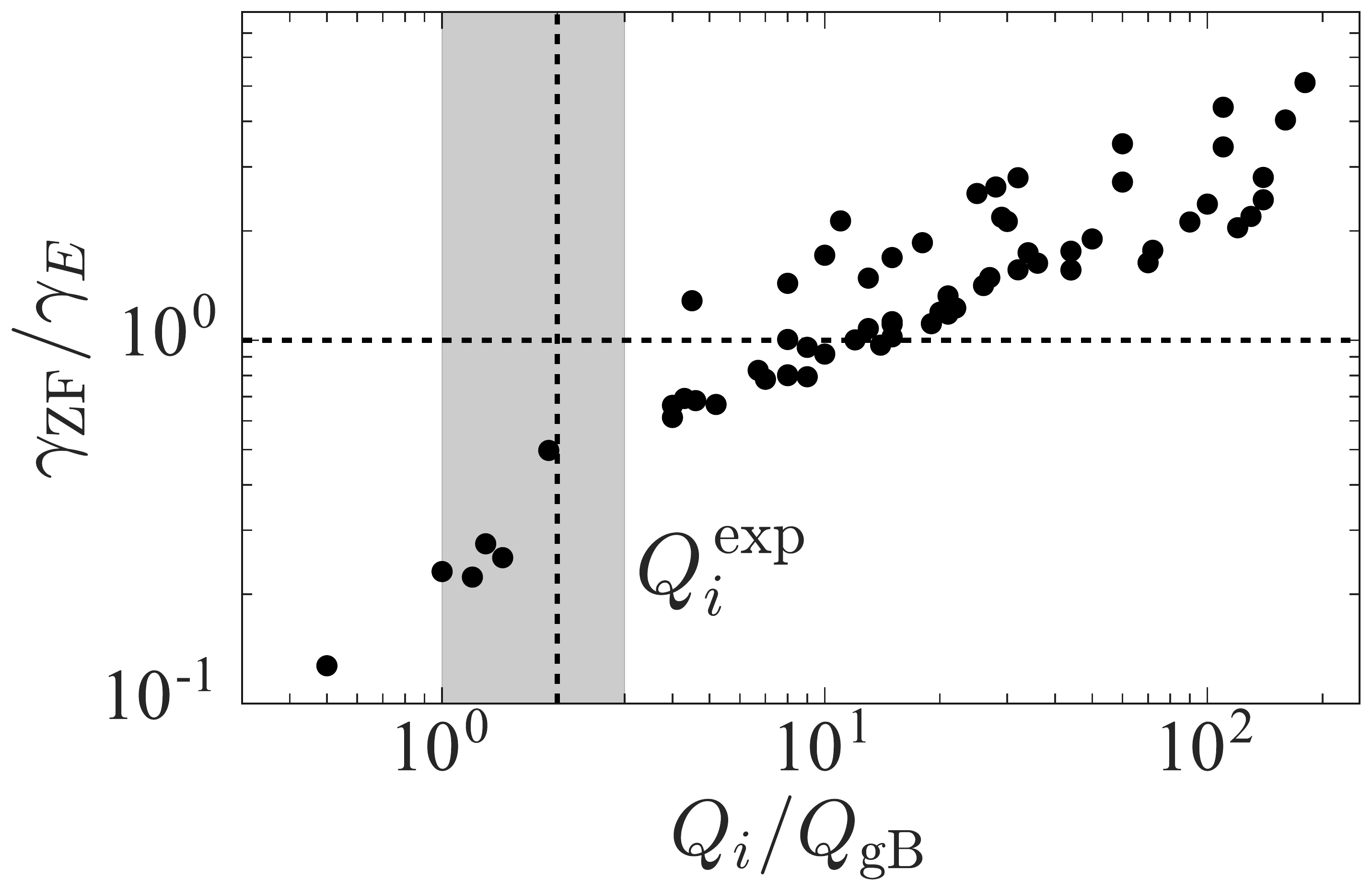}
        \caption{}
        \label{fig:zf_shear_q_scatter}
      \end{subfigure}
      \caption[Ratio of zonal shear to background flow shear]{
        \subref*{fig:zf_shear_contour} The ratio of zonal shear to background
        flow shear $\gamma_{\mathrm{ZF}}/\gamma_E$ over the same range of
        $\kappa_T$ and $\gamma_E$ as shown in \figref{contour_heatmap}. The
        effects of zonal shear and flow shear are comparable when
        $\gamma_{\mathrm{ZF}}/\gamma_E \sim 1$. The white region in the lower
        right-hand corner indicates the region where there is no turbulence,
        i.e., $Q_i=0$ (see \figref{contour_heatmap}), and the dashed black line
        indicates $\gamma_{\mathrm{ZF}}/\gamma_E = 1$.
        \subref*{fig:zf_shear_q_scatter} $\gamma_{\mathrm{ZF}}/\gamma_E$ as a
        function of $Q_i/Q_{\mathrm{gB}}$.  The vertical dashed line indicates
        the value of the experimental heat flux and the horizontal dashed line
        indicates $\gamma_{\mathrm{ZF}}/\gamma_E = 1$.
      }
    \label{fig:zf_shear}
  \end{figure}
  \Figref{zf_shear_contour} shows the ratio of the zonal shear to the flow
  shear, $\gamma_{\mathrm{ZF}}/\gamma_E$, as a function of $\kappa_T$ and
  $\gamma_E$ over the same parameter range as shown in
  \figref{contour_heatmap}. The effects of $\gamma_{\mathrm{ZF}}$ and $\gamma_E$
  are comparable where $\gamma_{\mathrm{ZF}}/\gamma_E \sim 1$, which is
  indicated by the dashed line. We see
  that the regime in which $\gamma_{\mathrm{ZF}}$ and $\gamma_E$ become comparable
  occurs some distance away from the turbulence threshold. Therefore, close to
  the threshold (small $\gamma_{\mathrm{ZF}}/\gamma_E$), we expect the shear due
  to the background flow do dominate, while far from the threshold (large
  $\gamma_{\mathrm{ZF}}/\gamma_E$), we expect the shear due to the zonal flows
  to dominate.

\begin{figure}[t]
     \centering
     \includegraphics[width=0.6\linewidth]{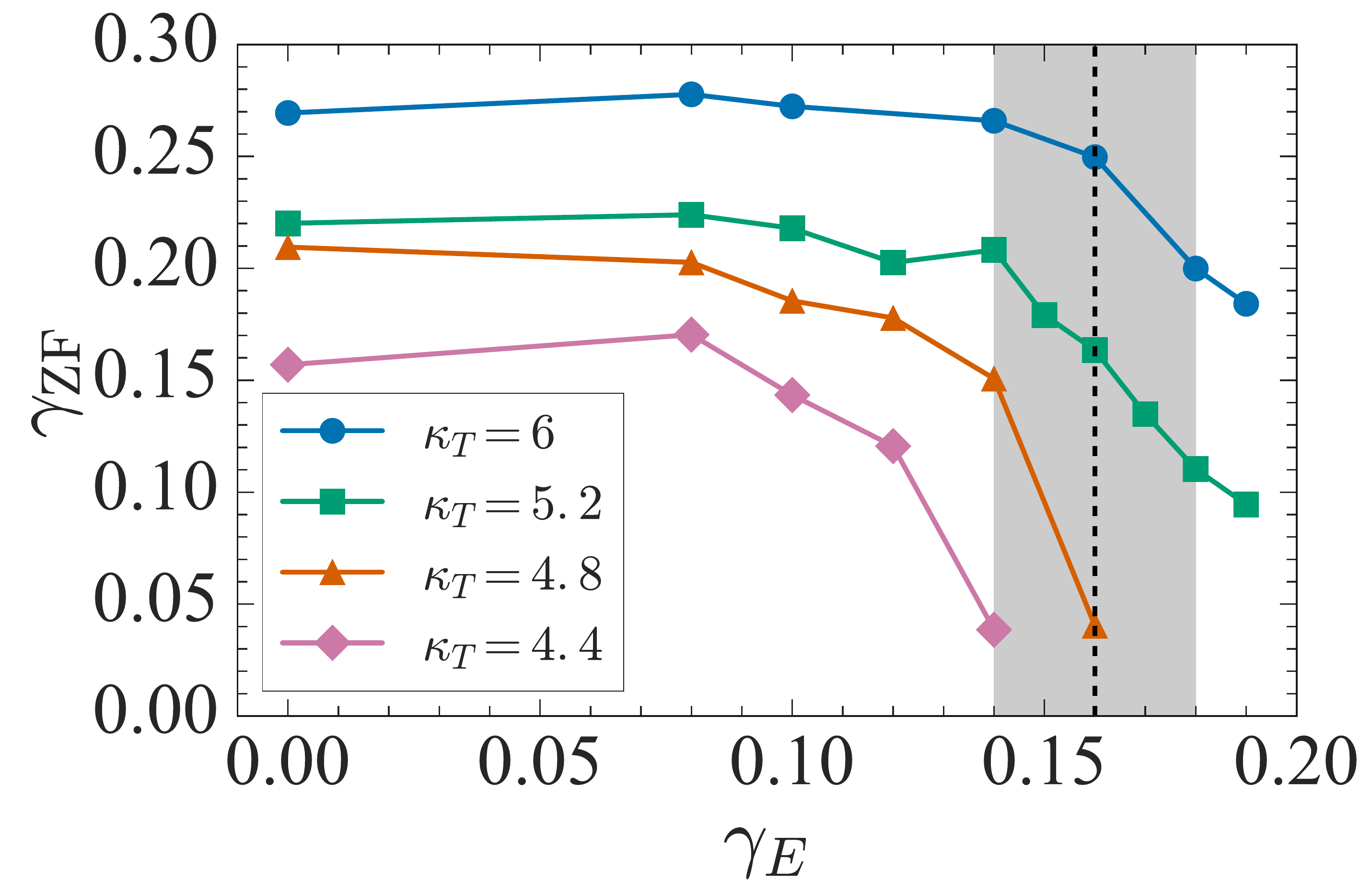}
     \caption[Zonal shear versus background flow shear]{
       Zonal shear $\gamma_{\mathrm{ZF}}$ as a function of background flow
       shear $\gamma_E$ showing that zonal flow regulation of turbulence is
       comparable between low $\gamma_E$ (high $Q_i/Q_{\mathrm{gB}}$) cases
       and $\gamma_E=0$ cases.
     }
     \label{fig:zf_shear_lines}
   \end{figure}

  Similar to our findings in Section~\ref{sec:struc_count},
  \figref{zf_shear_contour} suggests that the change in
  $\gamma_{\mathrm{ZF}}/\gamma_E$ is effectively a function of the
  distance from the turbulence threshold because (after comparing to
  \figref{contour_heatmap}) we see that regions of similar heat flux have
  similar values of $\gamma_{\mathrm{ZF}}/\gamma_E$.
  \Figref{zf_shear_q_scatter} shows this dependence explicitly:
  $\gamma_{\mathrm{ZF}}/\gamma_E$ as a function of $Q_i/Q_{\mathrm{gB}}$.
  The vertical dashed line indicates $Q_i^{\exp}/Q_{\mathrm{gB}}$ and we see
  that $\gamma_{\mathrm{ZF}}/\gamma_E$ is small around this value. This
  suggests that zonal shear plays a weaker role than $\gamma_E$ in
  regulating experimentally relevant turbulence for this MAST configuration.
  Therefore, near-threshold and far-from-threshold turbulence are
  distinguished by the fact that $\gamma_E$ is important close to the
  threshold, whereas the $\gamma_{\mathrm{ZF}}$ dominates far from the
  turbulence threshold. Far from the threshold the turbulence is likely similar
  to conventional ITG-driven turbulence in the absence of background flow
  shear. This is demonstrated in \figref{zf_shear_lines} which shows
  $\gamma_{\mathrm{ZF}}$ as a function of $\gamma_E$. We see that for low
  $\gamma_E$ and/or high $\kappa_T$ (i.e., cases far from the threshold),
  $\gamma_{\mathrm{ZF}}$ is comparable to cases where $\gamma_E=0$ and so zonal
  flows are the likely mechanism for regulating turbulence in these
  simulations.

\section{Summary}

  In this chapter we performed a parameter scan in $\kappa_T$ and $\gamma_E$
  and showed that the experimental ion heat flux is consistent with equilibrium
  parameters $(\kappa_T,\gamma_E)$ close to the turbulence threshold. We
  demonstrated that in the presence of a background flow shear, the system
  is subcritical: above a certain critical value of $\kappa_T$, and below a
  critical value of $\gamma_E$, a large initial perturbation is required to
  ignite turbulence. We studied the real-space structure of turbulence and
  found novel features of the transition to a turbulent state in an
  experimentally relevant fusion plasma when the system is
  subcritical. For equilibrium parameters near the threshold, the density and
  temperature fluctuations (and hence heat flux) are concentrated in
  long-lived, intense coherent structures. We demonstrated that flow shear (as
  opposed to zonal shear) is important at these experimentally relevant
  parameters. As the equilibrium parameters $(\kappa_T, \gamma_E)$ depart
  slightly from their critical values into the more strongly driven regime, the
  number of these structures increases rapidly while their amplitude stays
  roughly constant (in contrast to the conventional supercritical turbulence,
  where the amplitude increases with $\kappa_T$ because arbitrarily
  low-amplitude turbulence can be supported). Increasing $\kappa_T$ or
  decreasing $\gamma_E$ further leads to the structures filling the simulation
  domain and any further increase in the heat flux is caused by an increase in
  fluctuation amplitude. The latter regime is similar to the conventional
  plasma turbulence where zonal flows are the dominant mechanism for regulating
  turbulence.

\chapter{Correlation analysis and comparison with experimental results}

\label{sec:struc_of_turb}

\section{Introduction}
In Chapter~\ref{sec:nl}, we discussed the results of our nonlinear simulations
in terms of the observed transport and identified the conditions needed to
sustain a turbulent state. In this chapter, we would like to make more
quantitative comparisons with direct experimental measurements of the turbulent
fluctuations. We are interested in doing such comparisons with experimental
measurements in order to gain confidence in the predictions made by our
simulations. Only once the numerical predictions have been extensively checked
against existing experimental data in a range of different devices, can we
attempt to make predictions of turbulence in future devices. This study is
focused on MAST, but forms an important part of the wider effort of validating
numerical models against experimental data.  More broadly, we are
interested in understanding the nature of turbulence itself and how it behaves
in tokamaks as equilibrium quantities are varied, such as the flow shear and
ITG as we do in this study. Ultimately, we want to find equilibrium
configurations that maximise the fusion power and, by necessity, minimise the
turbulence. However, in order to do this, we need to understand the key drivers
of turbulence and how the turbulence responds to changes in equilibrium
parameters. It has only recently become possible to extend the study of
turbulence from the transport of particles, momentum and heat, to the
physical structure by measuring, for example, the density fluctuations. Beam
emission spectrometry is one such technique and it is with measurements from
this diagnostic that we compare our simulation predictions in this work.

The BES diagnostic on MAST infers density fluctuations on a poloidal
$(R,Z)$-plane from D$_\alpha$ emission by excited neutral particles injected by
the NBI heating system.  Correlation-analysis techniques were
developed~\cite{Ghim2013} to measure the radial correlation length, $l_R$, the
poloidal correlation length, $l_Z$, and the correlation time, $\tau_c$, of
these measured density fluctuations. The results of such a correlation analysis
for the MAST discharges that we consider in this work were reported
in Ref.~\cite{Field2014}. Also reported in Ref.~\cite{Field2014} were the first
comparisons of BES measurements with global, nonlinear particle-in-cell
simulations using the NEMORB code~\cite{Jolliet2007}, which found the
following. The simulations explicitly showed that kinetic electrons, flow
shear, and collisions between plasma particles played an important role in
predicting the turbulence found in MAST -- effects that we have included. In
the outer-core region, which we consider in this work, global simulations with
the physics effects listed above did not predict a turbulent state, possibly
due to the boundary conditions, forcing fluctuations to be zero at the plasma
boundary. However, at inner radii there was some agreement between
simulations and experiment with respect to the heat flux, density fluctuation
levels, and perpendicular correlation lengths. The correlation time, on the
other hand, was found to be on average two orders of magnitude larger in the
simulations compared to the experimental measurements over the whole radius.
The inability of global gyrokinetic simulations to predict turbulence in a
region where the BES diagnostic clearly finds the plasma to be turbulent as
well as the significant overprediction of the correlation time may suggest that
the resolution requirements for simulations of MAST plasmas are higher than
those currently allowed by global simulations.

In this work, we have used local gyrokinetic
simulations because they offer two desirable features compared to
global gyrokinetic simulations: they only attempt to simulate plasma turbulence
at a single radius and as a result allow increased resolution for resolving the
turbulence, and they avoid the complications of having to speculate on the
boundary conditions in the inner core and at the plasma edge. It is the goal of
this study to evaluate the merits of local gyrokinetic simulations in
predicting the turbulence in MAST, both in terms of averaged quantities such as
transport and in quantitative comparisons of the statistics of turbulent
fluctuations.

In this chapter, we will make such quantitative comparisons between the
fluctuations predicted by our simulations and those measured by the BES
diagnostic. Before being able to make comparisons between our simulations and
experimental measurements we converted our density fluctuation data from
flux-tube geometry to a poloidal plane, further explained in
Appendix~\ref{App:real_space_transform}. We review the correlation-analysis
techniques (Section~\ref{sec:corr_overview}) and experimental results
(Section~\ref{sec:corr_exp}) in Ref.~\cite{Field2014} and then present two types of
correlation analysis of our nonlinear simulations.  The first will be of GS2
density fluctuations with a ``synthetic BES diagnostic'' applied to simulate
what would be measured by a real BES diagnostic (Section~\ref{sec:corr_synth}).
We will consider the results from nonlinear simulations with values of
$(\kappa_T, \gamma_E)$ within the experimental-uncertainty range and compare
them with the experimental results.  The second analysis will be of the raw GS2
density fluctuations as a function of $Q_i/Q_{\mathrm{gB}}$, done for our
entire parameter scan (Section~\ref{sec:corr_gs2}), emphasising the extent to
which it is the distance from the threshold rather than individual values of
$\kappa_T$ or $\gamma_E$ that determine the statistical characteristics of the
density fluctuations.

\section{Correlation analysis}
  \label{sec:corr_overview}
  We start by giving an overview of the correlation-analysis techniques used
  in Refs.~\cite{Ghim2013,Field2014}. We will also present an alternative measurement
  of the poloidal correlation length $l_Z$, taking advantage of the increased
  resolution available in the poloidal direction from our simulations. While
  there is no experimental estimate of the parallel correlation length
  $l_\parallel$ available from the BES data, we are able to use the
  three-dimensional data available from GS2 to extend the correlation analysis
  to the parallel direction.

  The two-point spatio-temporal correlation function is, by definition,
  \begin{multline}
    C(\Delta R, \Delta Z, \Delta \lambda, \Delta t) = \\
    \frac{\left< \delta n_i/n_i(R, Z, \lambda, t) \delta n_i/n_i(R+\Delta R, Z+\Delta  Z, \lambda+\Delta \lambda, t+\Delta t)\right>}
    {{\qty[\left< \delta n_i^2/n_i^2(R, Z, \lambda, t) \right> \left<\delta n_i^2/n_i^2(R+\Delta R, Z+\Delta  Z, \lambda+\Delta \lambda, t+\Delta t)\right>]}^{1/2}},
    \label{corr_fn}
  \end{multline}
  where $\delta n_i/n_i$ is the density-fluctuation field calculated by GS2
  (which has a mean of zero) and $\Delta R$, $\Delta Z$, $\Delta
  \lambda$ are the radial, poloidal, and parallel separations, respectively
  between the two reference points, $\Delta t$ is the time lag, and
  $\left<\ldots\right>$ is an ensemble average, that is, an average over all
  possible pairs of points that have the appropriate separation and time lag.
  Note that the ensemble averages in the plane perpendicular to the magnetic
  field are calculated at $\theta=0$, i.e., they are not averaged over
  $\theta$. Note also that we divide our data in the time domain into window
  of $\sim 100$--$400$~$\mu$s, and the calculated separate ensemble averages
  in each time window. This allows us to estimate the variance of the
  correlation parameters we calculate.

  However, instead of calculating the full correlation function
  \eqref{corr_fn}, we will estimate individual correlation lengths and times
  (which we will define below) by performing a one-dimensional correlation
  analyses separately in each direction.  All of the representative correlation
  functions that are plotted in the sections that follow will be for the
  equilibrium parameters $(\kappa_T, \gamma_E) = (5.1, 0.16)$ over a real-space
  domain of $20\times20$~cm$^2$ (see Appendix~\ref{App:real_space_transform}).

  \subsection{Radial correlation length}
  \label{sec:radial_corr}
\begin{figure}[t]
    \centering
    \includegraphics[width=0.6\linewidth]{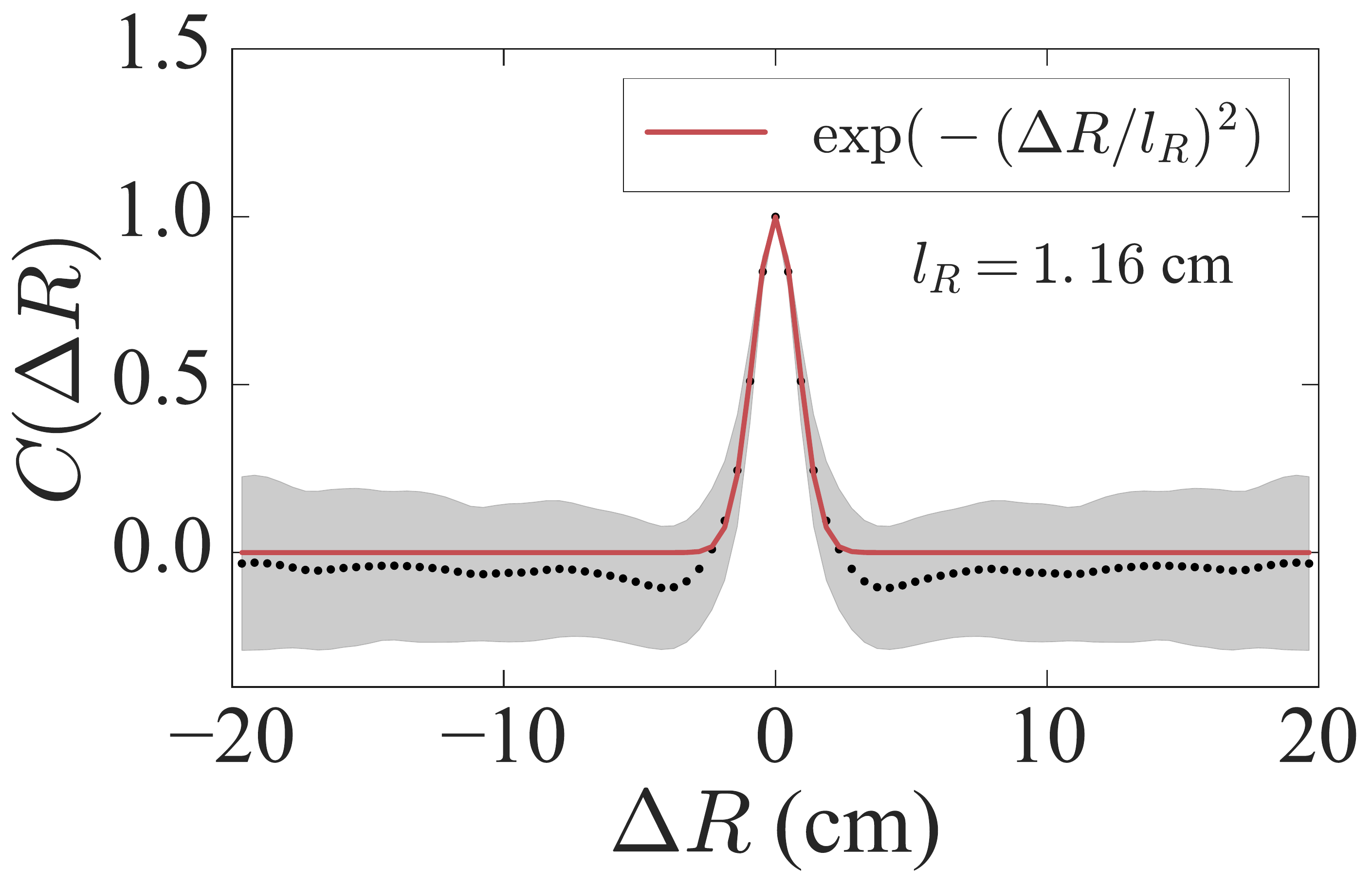}
    \caption[Radial correlation function]{
      A representative radial correlation function fitted with the
      function~\eqref{radial_fit} (red line). The points show the correlation
      function $C(\Delta R)$ averaged over $t$ and $Z$ and the shaded region
      shows the associated standard deviation.
    }
    \label{fig:radial_fit}
  \end{figure}

  The radial correlation length $l_R$ is estimated by fitting the correlation
  function $C(\Delta R, \Delta Z = 0, \lambda(\theta=0), \Delta t = 0)$ with a
  Gaussian function:
  \begin{equation}
    f_R(\Delta R) = \exp \qty[- {\qty(\frac{\Delta R}{l_R})}^2].
    \label{radial_fit}
  \end{equation}
  Following experimental observations in, this fitting function is adopted on
  the assumption that fluctuations have no wave-like structure in the radial
  direction~\cite{Ghim2013,Field2014}. Unlike in the fitting functions used for
  experimental data, no parameters are necessary here to account for global
  offsets, usually due to large-scale, global MHD modes, which do not appear in
  our simulations, where the mean density fluctuation over the whole domain is
  zero. A representative example of the fitting procedure for the radial
  correlation function is shown in~\figref{radial_fit}.
  The points show the measured correlation function and the red line the fit
  \eqref{radial_fit}. We took an average over $t$ and $Z$ and assumed that
  radial correlations do not change with $t$ and $Z$ (i.e., that the system is
  statistically homogeneous in time and in the poloidal direction). The shaded
  region indicates the standard deviation calculated over the integrals of $t$
  and $Z$ used in this averaging. We expect that $C(\Delta R) \to 0$ as $\Delta
  R$ increases (and similarly for subsequent correlation functions in the other
  directions) because the fluctuations have a mean of zero over the
  computational domain.

  \subsection{Poloidal correlation length}
  \label{sec:poloidal_corr}
\begin{figure}[t]
    \centering
    \begin{subfigure}{0.49\linewidth}
      \includegraphics[width=\linewidth]{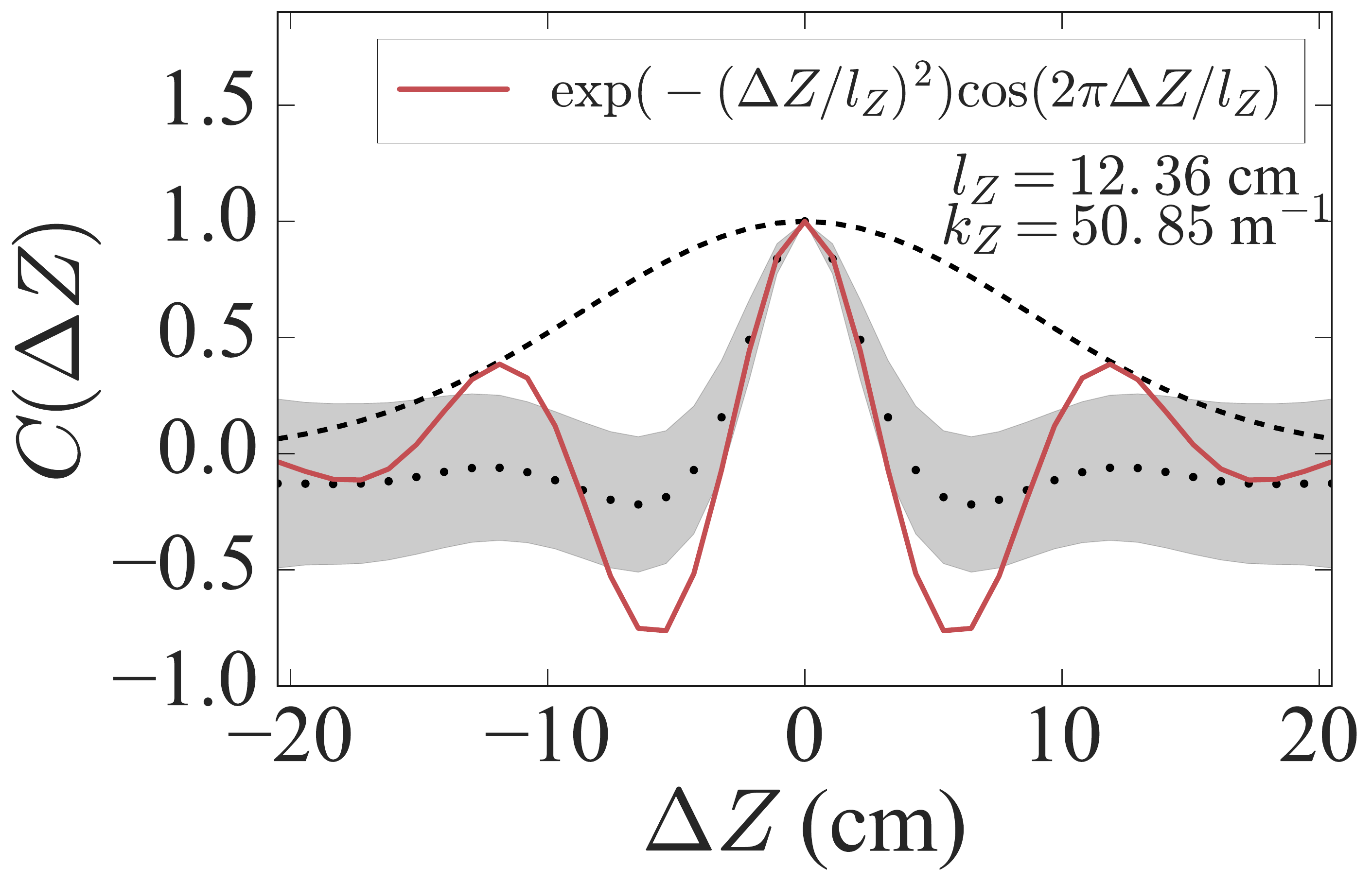}
      \caption{}
      \label{fig:poloidal_fixed_fit}
    \end{subfigure}
    \hfill
    \begin{subfigure}{0.49\linewidth}
      \includegraphics[width=\linewidth]{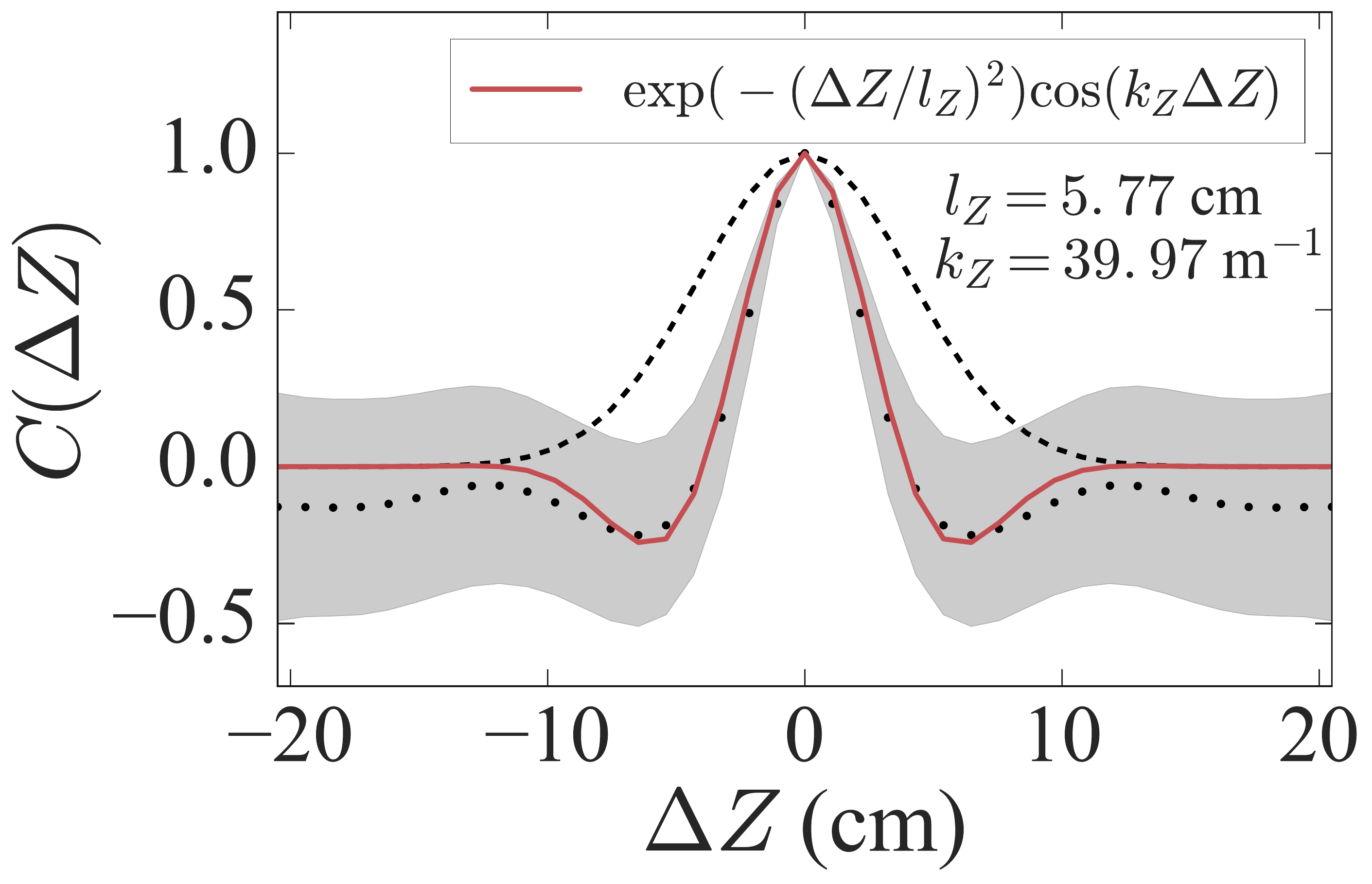}
      \caption{}
      \label{fig:poloidal_free_fit}
    \end{subfigure}
    \caption[Poloidal correlation function]{
      Representative poloidal correlation function fitted with the
      function~\eqref{poloidal_fit} (red line) keeping the poloidal wavenumber
      $k_Z$ \subref*{fig:poloidal_fixed_fit} fixed to $k_Z = 2 \pi / l_Z$,
      \subref*{fig:poloidal_free_fit} as a free fitting parameter.  The points
      in each plot show the correlation function $C(\Delta Z)$ averaged over
      $t$ and $R$ and the shaded regions show the associated standard
      deviation. The dashed lines indicate the Gaussian envelope $\exp(-(\Delta
      Z/l_Z))$
    }
    \label{fig:poloidal_fit}
  \end{figure}

  The poloidal correlation length is calculated by assuming wave-like
  fluctuations in the poloidal direction and fitting
  $C(\Delta R = 0, \Delta Z, \lambda(\theta=0), \Delta t = 0)$ with an
  oscillating Gaussian function of the form
  \begin{equation}
    f_Z(\Delta Z) = \cos \qty(2 \pi k_Z \Delta Z)
                    \exp \qty[-{\qty(\frac{\Delta Z}{l_Z})}^2],
    \label{poloidal_fit}
  \end{equation}
  where $k_Z$ is the poloidal wavenumber. References~\cite{Ghim2013,Field2014}
  found that with only four poloidal channels, the BES diagnostic could not fix
  $l_Z$ and $k_Z$ separately in a meaningful way. As a result, when fitting
  experimental data, the wavenumber is fixed to the value $k_Z = 2 \pi / l_Z$.
  In our GS2 simulations, we can have many more points in the poloidal
  direction, allowing us to compare fits with $k_Z$ both as a free fitting
  parameter and fixed in the way described above. \Figref{poloidal_fit} shows a
  representative poloidal correlation function from our simulations along with
  a fitted function~\eqref{poloidal_fit}, both with fixed $k_Z = 2 \pi / l_Z$
  [\figref{poloidal_fixed_fit}] and free $k_Z$ [\figref{poloidal_free_fit}].
  The red lines in each plot indicate the fit \eqref{poloidal_fit} and the
  dashed lines indicate the Gaussian envelope $\exp(-(\Delta Z/l_Z))$.
  We have taken an average over the variables $t$ and $R$. We see that the fit
  with $k_Z$ as a free parameter approximates the correlation function better
  and predicts a shorter $l_Z$. For consistency with previous work, we will
  show the correlation results for both cases in Section~\ref{sec:corr_gs2}.

  \subsection{Correlation time}
  \label{sec:time_corr}
  In the presence of toroidal rotation, turbulent structures are advected in
  the poloidal direction with an apparent velocity $v_Z$ given
  by~\cite{Ghim2012}
  \begin{equation}
    v_{Z} = R \omega_0 \tan \vartheta,
    \label{v_z}
  \end{equation}
  where $\vartheta$ is the magnetic-field pitch-angle (see
  Appendix~\ref{App:real_space_transform}). Following Ref.~\cite{Ghim2012}, we
  can use this to calculate the correlation time $\tau_c$ by tracking turbulent
  structures as they move poloidally and measuring their temporal
  decorrelation. This method assumes that the temporal decorrelation dominates
  over any effects due to the finite parallel correlation length, as we will
  now explain. While turbulent structures are elongated along the field lines,
  they rotate rapidly in the toroidal direction. Measurements taken at a single
  point (or a poloidal plane) will measure the correlation time as a
  combination of two effects:
  \begin{inparaenum}[(i)]
    \item true decorrelation of turbulent structures in time, and
    \item structures of finite parallel length moving past the measurement
      point.
  \end{inparaenum}
  Both of these two effects will appear as structures decorrelating in time
  but are indistinguishable in stationary measurements of turbulence. In order
  for the true decorrelation of structures (the quantity we are interested in)
  to dominate over the movement of structures past the detector
  we require that~\cite{Ghim2013}
  \begin{equation}
    \tau_c \ll l_\parallel \cos \vartheta / R \omega_0.
    \label{time_assumption}
  \end{equation}
  In Section~\ref{sec:pol_par_corr}, we will confirm that this condition is
  indeed satisfied.

\begin{figure}[t]
    \centering
    \includegraphics[width=0.6\linewidth]{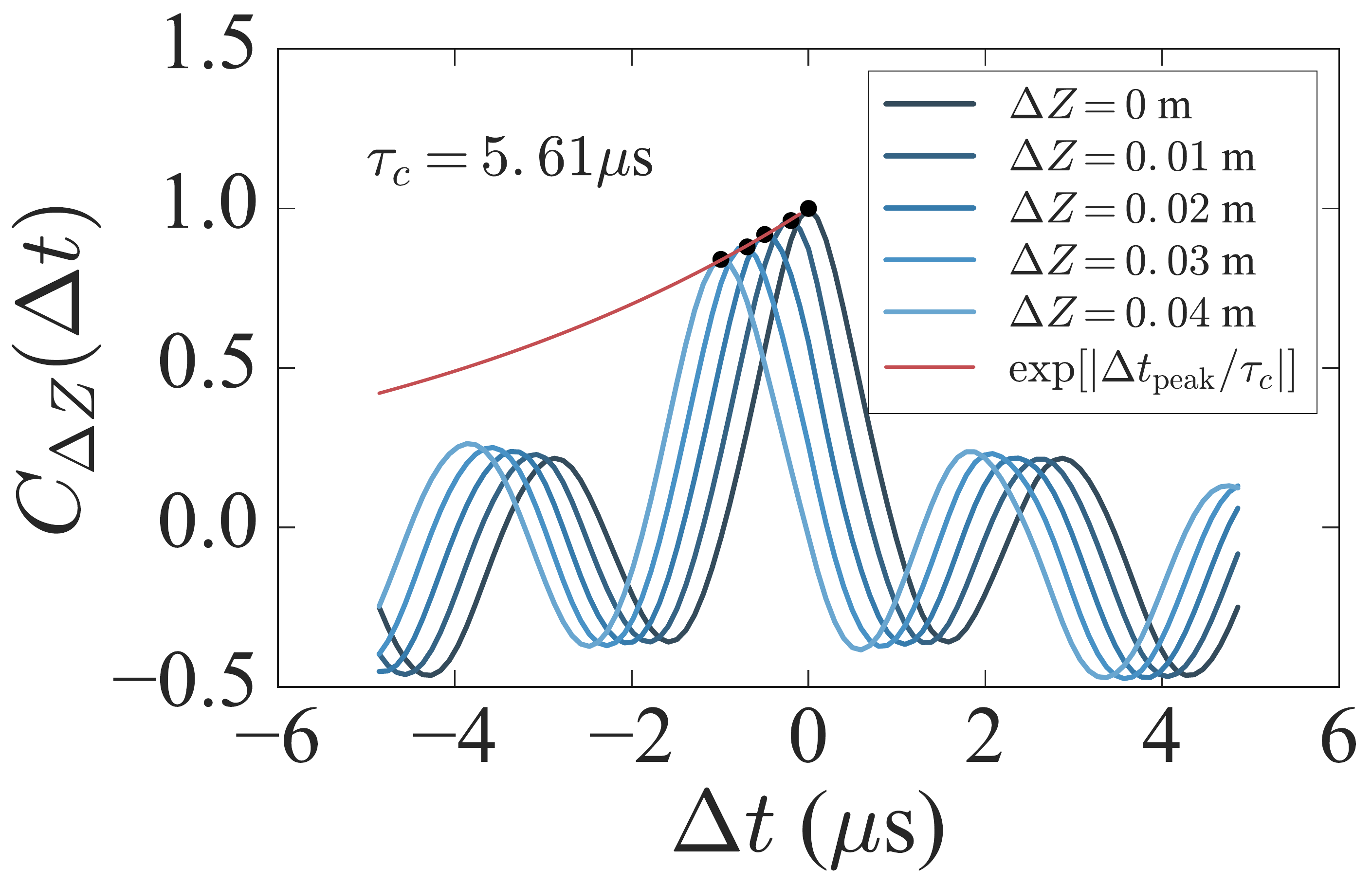}
    \caption[Time correlation function]{
      Time correlation functions $C_{\Delta Z} (\Delta t)$ for several poloidal
      separations $\Delta Z$. The points indicate the maximum value of
      $C(\Delta t)$ for a given $\Delta Z$, and the red line indicates the
      function~\eqref{time_fit} fitted to those points.
    }
    \label{fig:time_fit}
  \end{figure}

  The correlation time $\tau_c$ is calculated using the so-called
  ``cross-correlation time delay'' technique~\cite{Durst1992, Ghim2012,
  Fox2016}. Following this method, we calculate the correlation function
  $C_{\Delta Z}(\Delta t) = C(\Delta R = 0, \Delta Z, \lambda(\theta=0), \Delta
  t)$ for several poloidal separations $\Delta Z$, including $\Delta Z = 0$, as
  shown in \figref{time_fit}. As the structures are advected poloidally, they
  decorrelate and the peak of the correlation function at a given $\Delta Z$,
  i.e., the value of $C_{\Delta Z}(\Delta t)$, decreases for increasing $\Delta
  Z$.  The correlation time $\tau_c$ is then defined as the characteristic
  exponential decay time of the peaks of the correlation functions. Namely, we
  fit $C_{\Delta Z}(\Delta t = \Delta t_{\mathrm{peak}})$ with the function
  \begin{equation}
    f_\tau(\Delta Z) =
      \exp \qty[- \qty|\frac{\Delta t_{\mathrm{peak}}(\Delta Z)}{\tau_c}|],
    \label{time_fit}
  \end{equation}
  as shown for a representative correlation function in \figref{time_fit},
  where the blue lines show correlation functions $C_{\Delta Z}(\Delta t)$ for
  different poloidal separations and the red line shows the fit
  \eqref{time_fit}.

  \subsection{Parallel correlation length}
  \label{sec:par_corr}
\begin{figure}[t]
    \centering
    \includegraphics[width=0.6\linewidth]{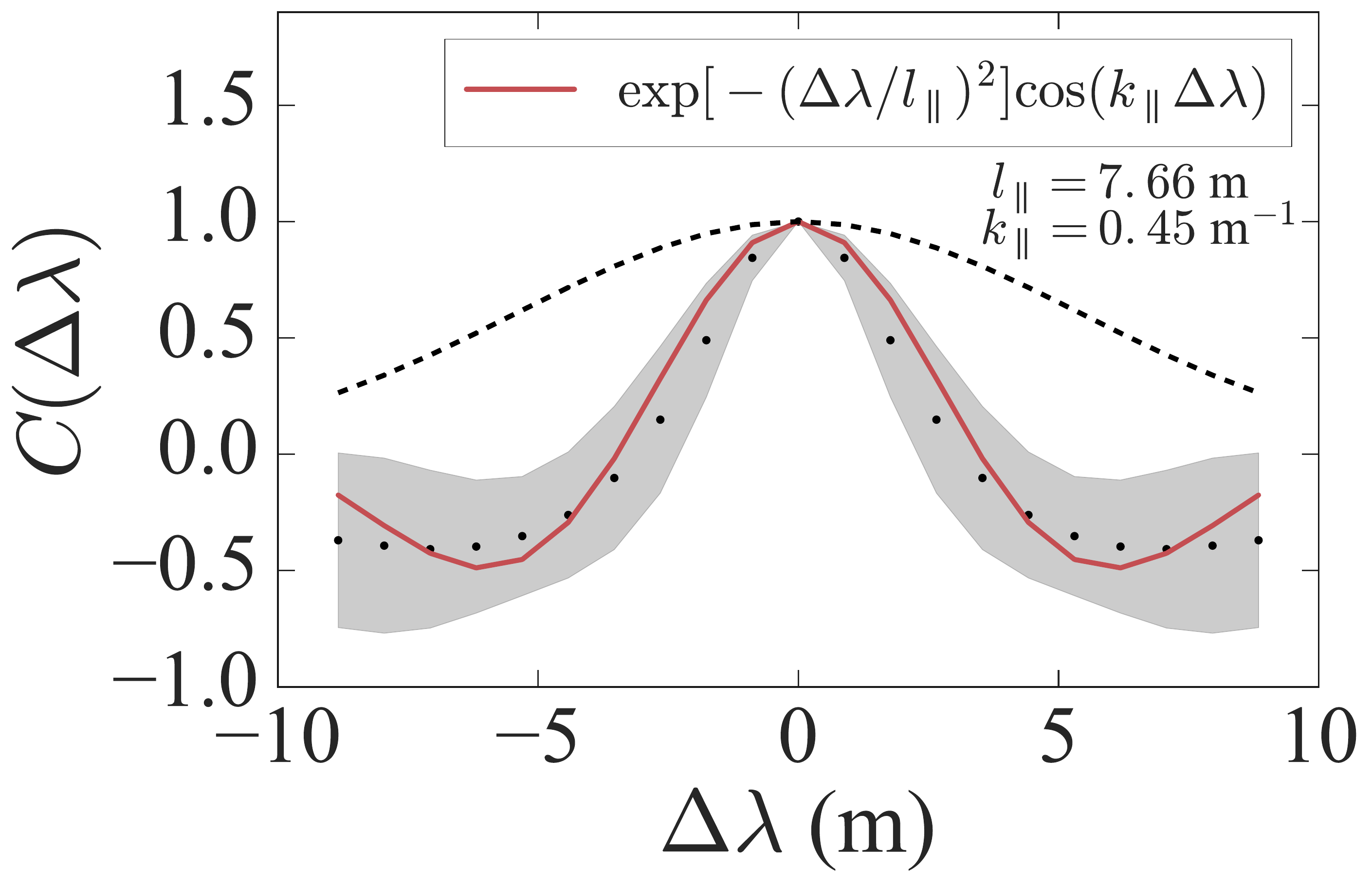}
    \caption[Parallel correlation function]{
      Representative parallel correlation function fitted with the oscillating
      Gaussian function~\eqref{parallel_fit} (red line). The points show the
      correlation function $C(\Delta \lambda)$ averaged over $(t,R,Z)$ and the
      shaded region shows the associated standard deviation. The dashed line
      shows the Gaussian envelope $\exp(-(\Delta \lambda/k_\parallel))$.
    }
    \label{fig:parallel_fit}
  \end{figure}

  Since GS2 simulations supply the full 3D density-fluctuation field (unlike
  BES measurements), we are able to study the parallel structure of the
  turbulence. To do this, we convert the fluctuation field from the GS2
  parallel coordinate $\theta$ to a real-space coordinate $\lambda(\theta)$
  along the field line, as discussed in Appendix~\ref{App:real_space_transform}.
  We then calculate the correlation function $C(\Delta R=0, \Delta Z=0, \Delta
  \lambda, \Delta t = 0)$ and take an average over $(R, Z, t)$.
  We fit the correlation function with an oscillating Gaussian function of the
  form
  \begin{equation}
    f_\parallel(\Delta \lambda) = \cos \qty(2 \pi k_\parallel \Delta \lambda)
                  \exp \qty[- {\qty(\frac{\Delta \lambda}{l_\parallel})}^2],
    \label{parallel_fit}
  \end{equation}
  where $k_\parallel$ is the parallel wavenumber. A representative example of
  the fitting procedure for the radial correlation function is shown
  in~\figref{parallel_fit}, where the red line indicates the fit
  \eqref{parallel_fit} and the dashed line shows the Gaussian envelope
  $\exp(-(\Delta \lambda/k_\parallel))$.

  \subsection{Density-fluctuation amplitude}
  \label{sec:rms_density}
  The final simulation prediction we can compare with the experimental results
  in Ref.~\cite{Field2014}, is the RMS density fluctuation at the outboard midplane
  averaged over the $(t,R,Z)$:
  \begin{equation}
    \qty(\frac{\delta n_i}{n_i})_{\mathrm{rms}} =
    \left< \frac{\delta n_i^2(t,R,Z)}{n_i^2} \right>_{t,R,Z}^{1/2}.
    \label{dn_rms}
  \end{equation}

\section{Experimental BES results}
  \label{sec:corr_exp}
  \begin{figure}[t]
    \centering
    \begin{subfigure}{0.49\linewidth}
      \includegraphics[width=\linewidth]{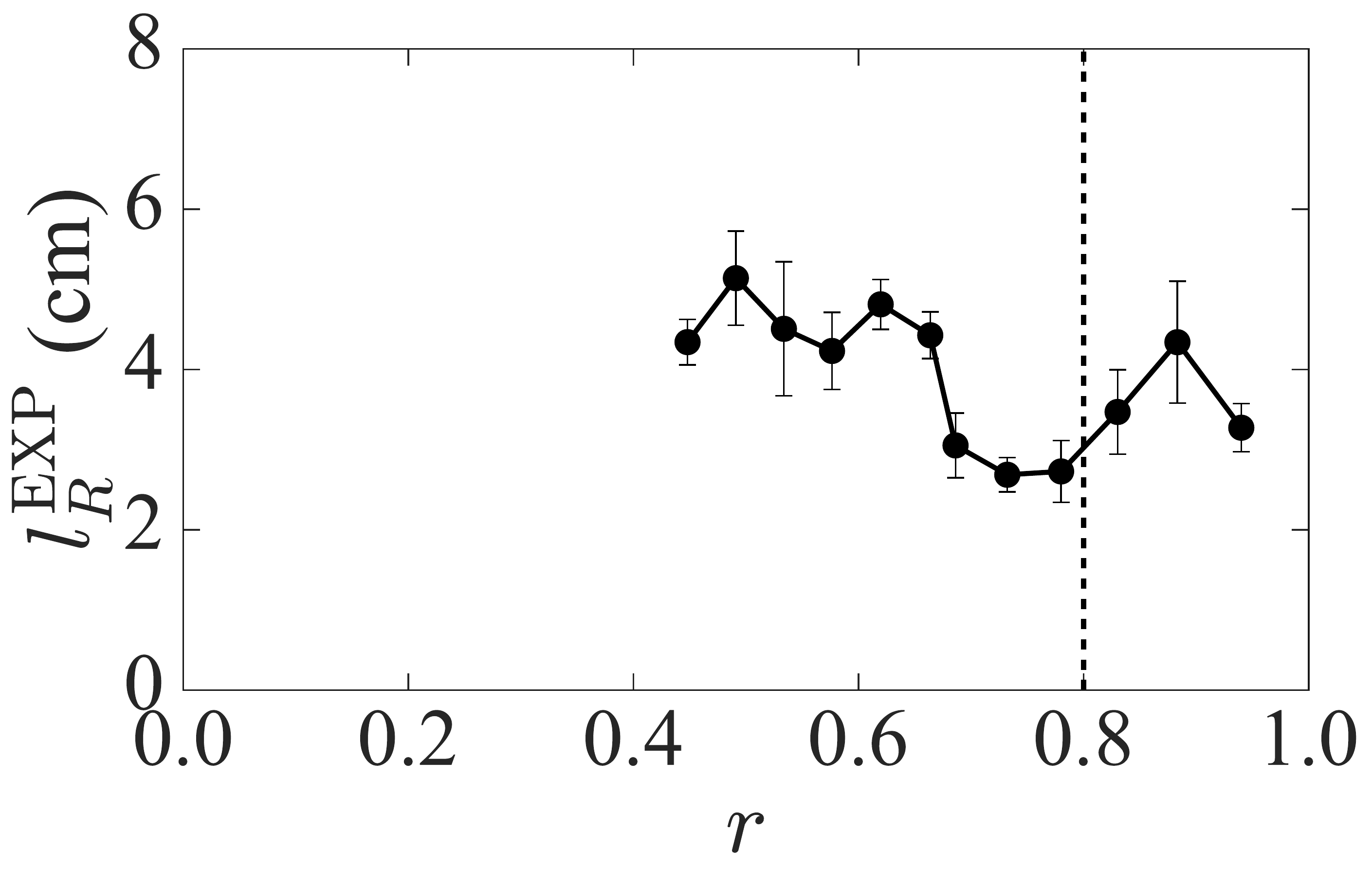}
      \caption{}
      \label{fig:lr_exp}
    \end{subfigure}
    \hfill
    \begin{subfigure}{0.49\linewidth}
      \includegraphics[width=\linewidth]{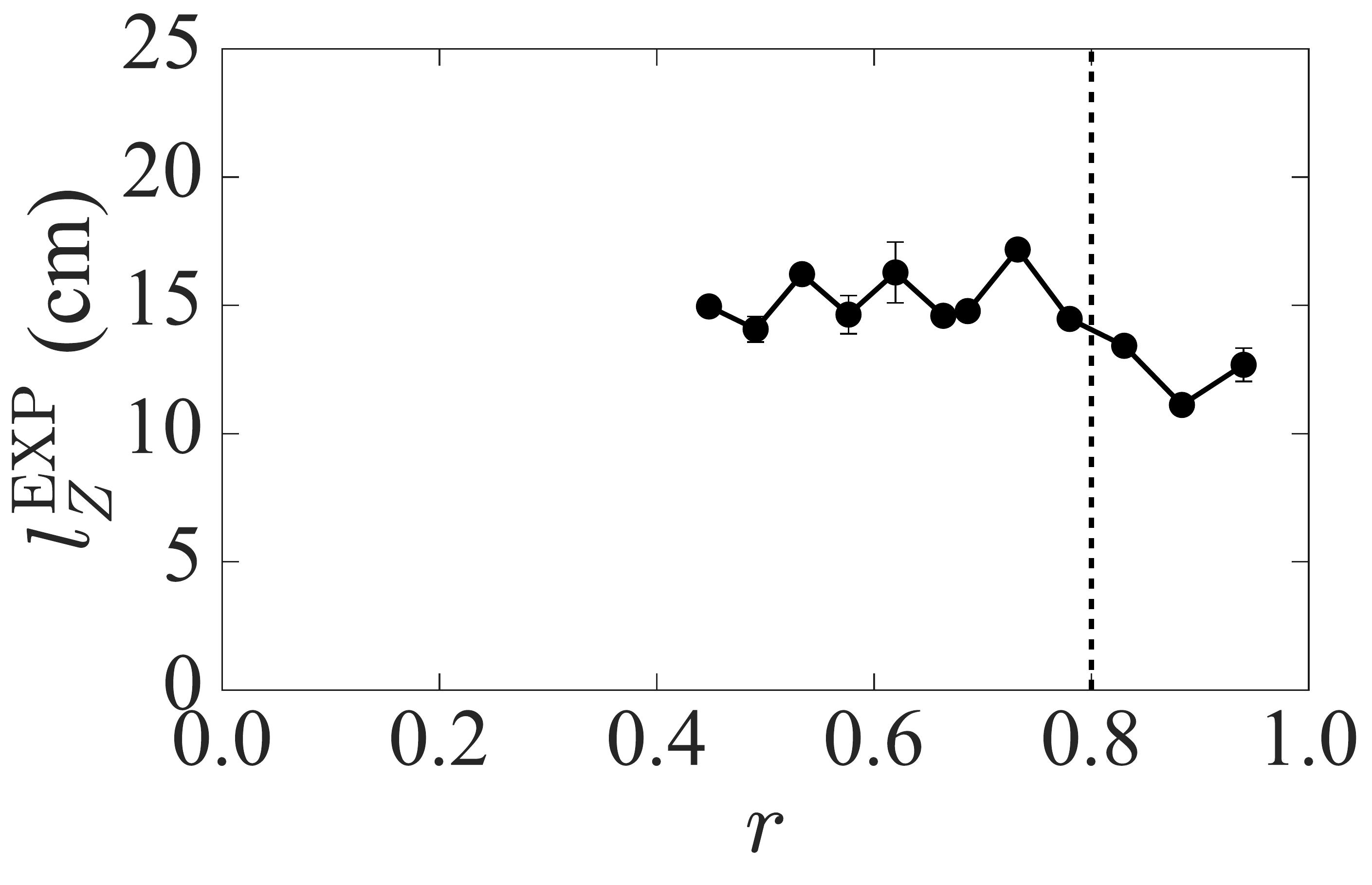}
      \caption{}
      \label{fig:lz_exp}
    \end{subfigure}
    \begin{subfigure}{0.49\linewidth}
      \includegraphics[width=\linewidth]{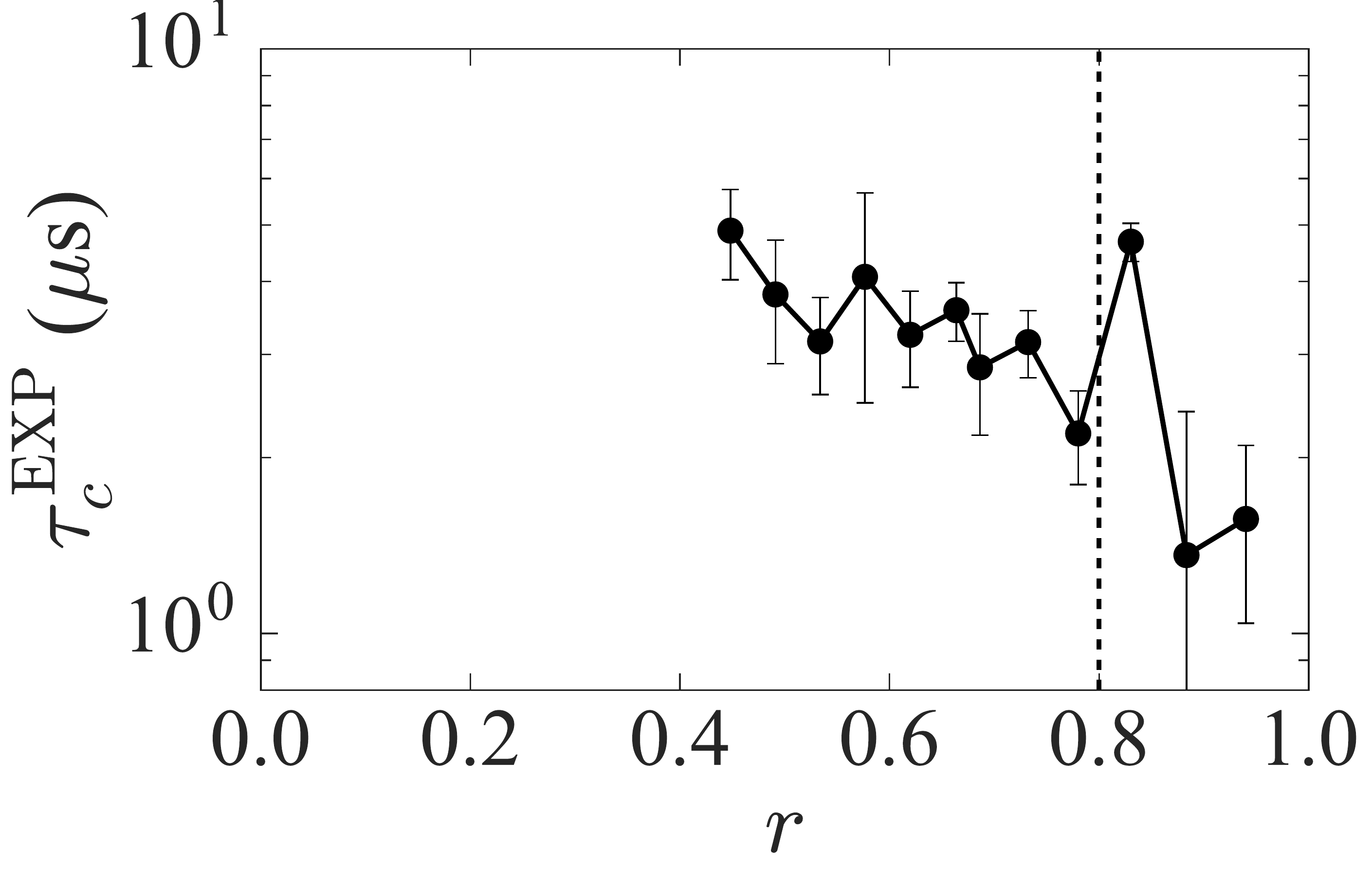}
      \caption{}
      \label{fig:tau_exp}
    \end{subfigure}
    \hfill
    \begin{subfigure}{0.49\linewidth}
      \includegraphics[width=\linewidth]{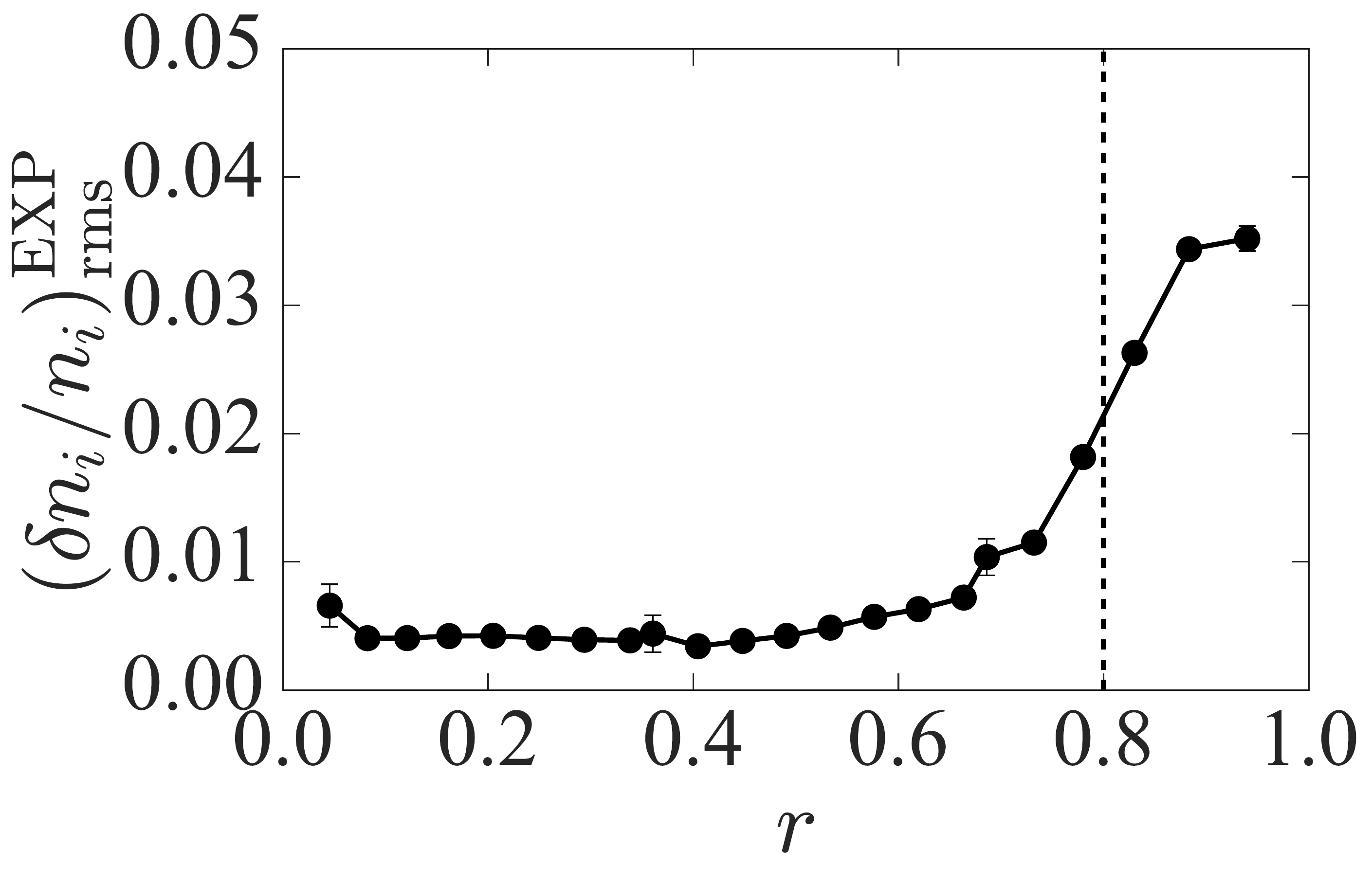}
      \caption{}
      \label{fig:n_exp}
    \end{subfigure}
    \caption[Experimental correlation results]{
      Results of the correlation analysis of BES data from MAST discharges
      \#27272, \#27268, and \#27274 combined to give correlation properties of
      the turbulence as functions of $r=D/2a$. These results are the same as
      those previously presented in~\cite{Field2014}. The values of the
      correlation parameters were not available at $r \lesssim 0.4$, because
      turbulence was suppressed in this region.  The vertical dashed line
      indicates the radius corresponding to the local equilibrium
      configurations for which we performed our simulations.
    }
    \label{fig:exp_corr_results}
  \end{figure}
  Before applying the correlation analysis to our simulations, we review the
  experimental results from MAST discharge \#27274, with which we will be
  comparing, first presented in~\cite{Field2014}. As discussed in
  Section~\ref{sec:exp_profiles}, MAST discharge \#27274 forms part of a set
  of three discharges, which measured correlation properties over the whole
  outer radius.  \Figref{exp_corr_results} shows the experimental results
  obtained for the radial correlation length $l_R^{\mathrm{EXP}}$, the poloidal
  correlation length $l_Z^{\mathrm{EXP}}$, the correlation time
  $\tau_c^{\mathrm{EXP}}$, and the RMS density fluctuations ${\qty(\delta n_i /
  n_i)}^{\mathrm{EXP}}_{\mathrm{rms}}$ as functions of $r = D/2a$. The vertical
  dashed line in each plot indicates the radius at which we performed our
  simulations and the corresponding values with which we will compare. From
  these results, we find the following (after interpolating between the
  experimental data points):
  \begin{align}
    \begin{split}
    l_R^{\mathrm{EXP}} &= 3 \pm 0.4~\mathrm{cm}, \\
    l_Z^{\mathrm{EXP}} &= 14.06 \pm 0.09~\mathrm{cm}, \\
    \tau_c^{\mathrm{EXP}} &= 3.2 \pm 0.4~\mu\mathrm{s}, \\
    {\qty(\frac{\delta n_i}{n_i})}^{\mathrm{EXP}}_{\mathrm{rms}} &= 0.0214 \pm 0.0006.
    \label{exp_results}
    \end{split}
  \end{align}
  We will be comparing the correlation parameters calculated from our
  simulations in the following sections to those in \eqref{exp_results}.

\section{Correlation analysis with synthetic diagnostic}
  \label{sec:corr_synth}
  In order to compare our simulations with the BES measurements, a number of
  data transformations were necessary. We mapped our density fluctuations
  ``measured'' in the outboard midplane (at $\theta = 0$) from GS2 $(x, y)$
  coordinates onto a poloidal $(R,Z)$-plane as explained in
  Appendix~\ref{App:real_space_transform}.  We also transformed from the
  rotating plasma frame, the frame in which our simulations were performed, to
  the laboratory frame, as also  explained in
  Appendix~\ref{App:real_space_transform}.  We then applied a synthetic
  diagnostic to our density fluctuations, including the point-spread functions
  (described in Section~\ref{sec:bes}) to model instrumentation effects and
  atomic physics, to add artificial noise similar to that found in the
  experiment, and to map the density-fluctuation field onto an $8 \times 4$
  grid similar to the arrangement of BES channels. An important feature of
  the analysis of experimental data is the presence of a filter to remove
  high-energy radiation present in the experiment. We have included this filter
  for consistency in the analysis of synthetic data produced applying the
  synthetic diagnostic to our simulation data. The results without this filter
  are presented and discussed in Appendix~\ref{App:no_spike}.

  \Figref{synth_corr_results} shows the radial correlation length
  $l_R^{\,\mathrm{SYNTH}}$, poloidal correlation length
  $l_Z^{\,\mathrm{SYNTH}}$, correlation time $\tau_c^{\,\mathrm{SYNTH}}$, and
  RMS density fluctuation $\qty(\delta n_i /
  n_i)^{\,\mathrm{SYNTH}}_{\mathrm{rms}}$ calculated from our simulations with
  the synthetic diagnostic applied using the correlation analysis described in
  Section~\ref{sec:corr_overview}.  These values should agree with the
  experimentally measured correlation parameters in \eqref{exp_results}
  because the equilibrium parameters $\kappa_T$ and $\gamma_E$ at which the
  results shown in \figref{synth_corr_results} were obtained are strictly
  within the experimental-uncertainty range of these parameters.  The dashed
  lines and shaded areas in \figref{synth_corr_results} indicate the
  experimental values and associated errors given in \eqref{exp_results} . The
  circled points indicate the simulations that matched the experimental level
  of heat flux (listed in Table~\ref{tab:exp_match_sims}).
  \begin{figure}[t]
    \centering
    \begin{subfigure}{0.49\linewidth}
      \includegraphics[width=\linewidth]{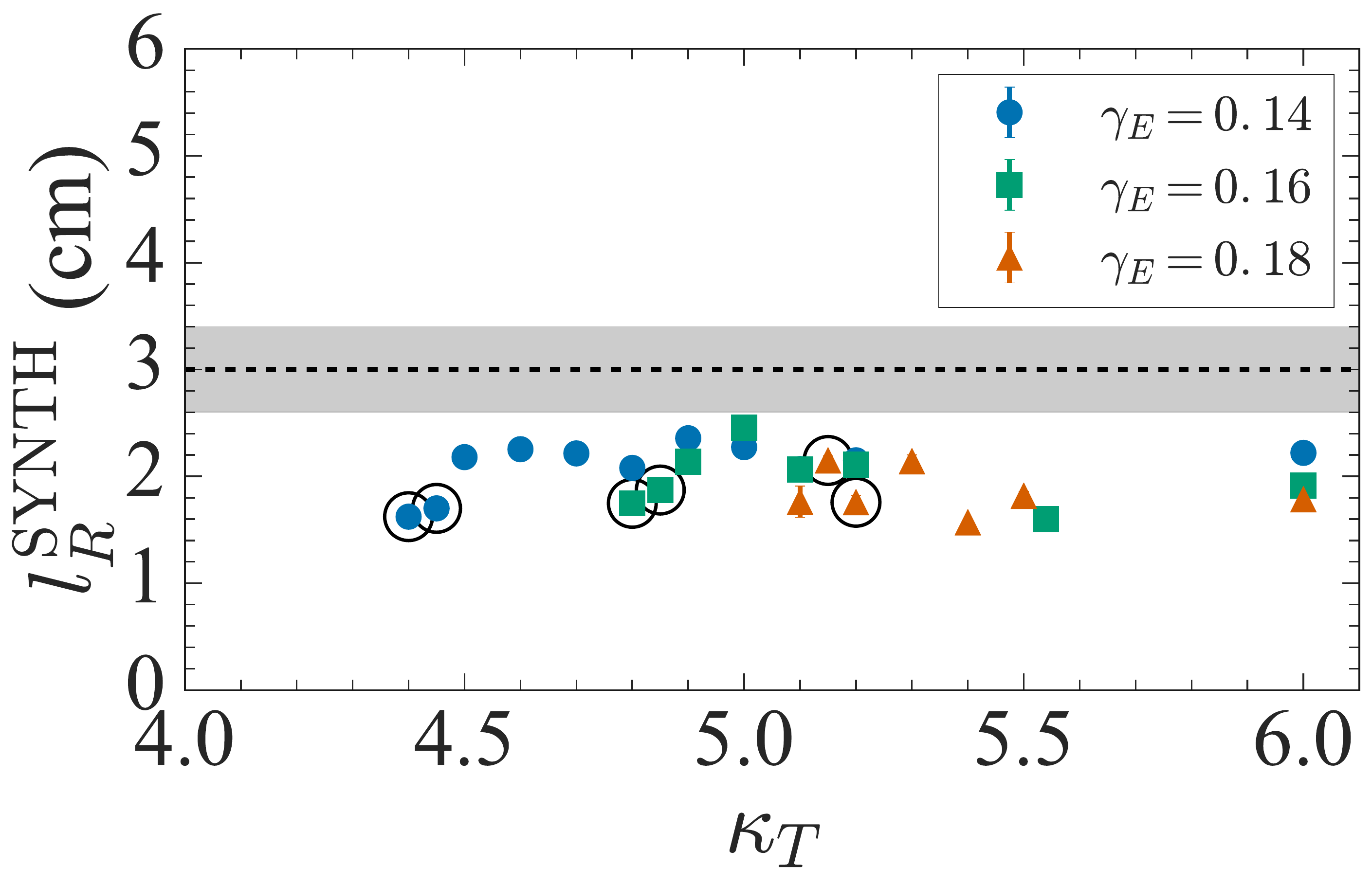}
      \caption{}
      \label{fig:lr_synth}
    \end{subfigure}
    \hfill
    \begin{subfigure}{0.49\linewidth}
      \includegraphics[width=\linewidth]{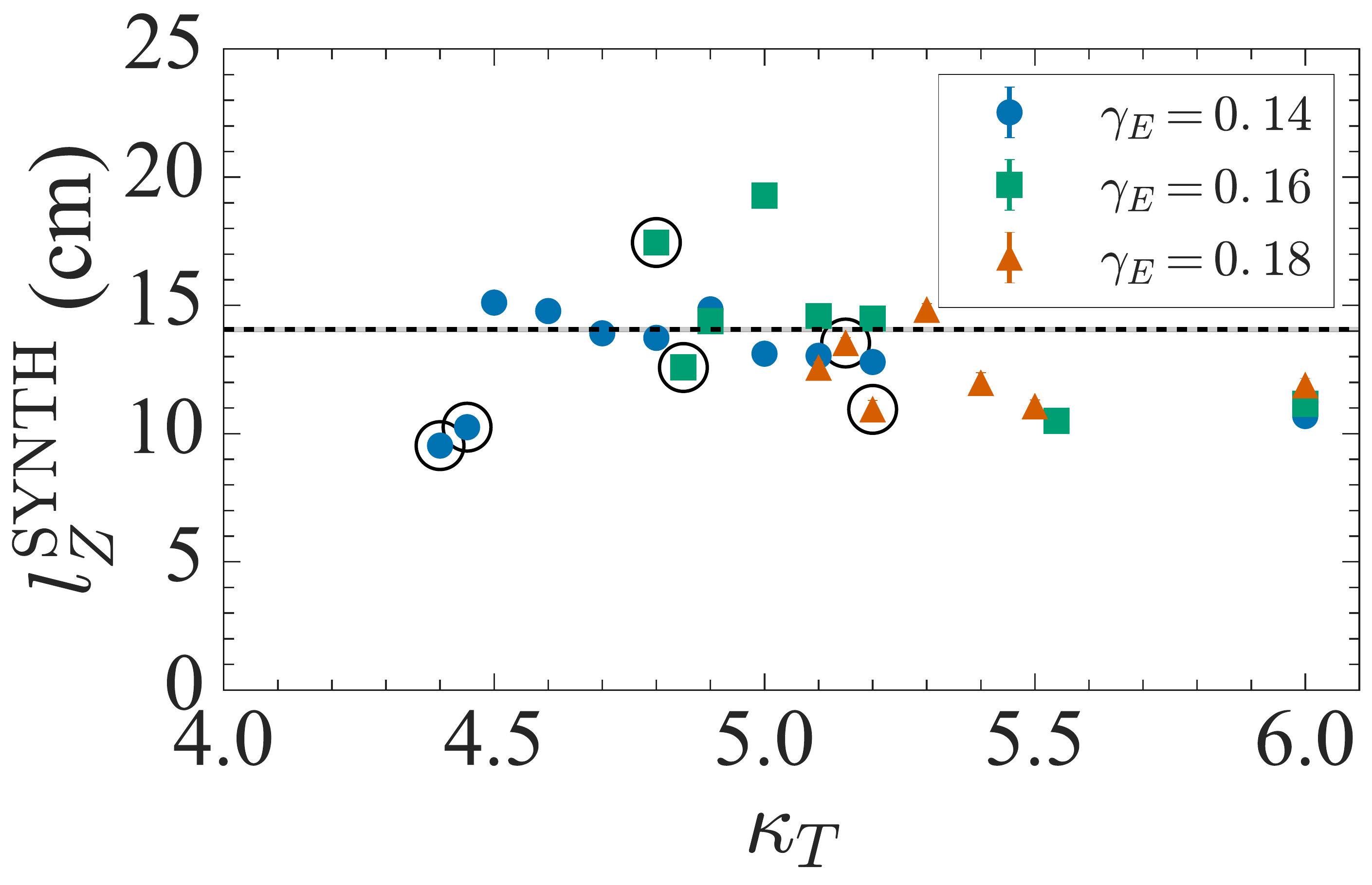}
      \caption{}
      \label{fig:lz_synth}
    \end{subfigure}
    \begin{subfigure}{0.49\linewidth}
      \includegraphics[width=\linewidth]{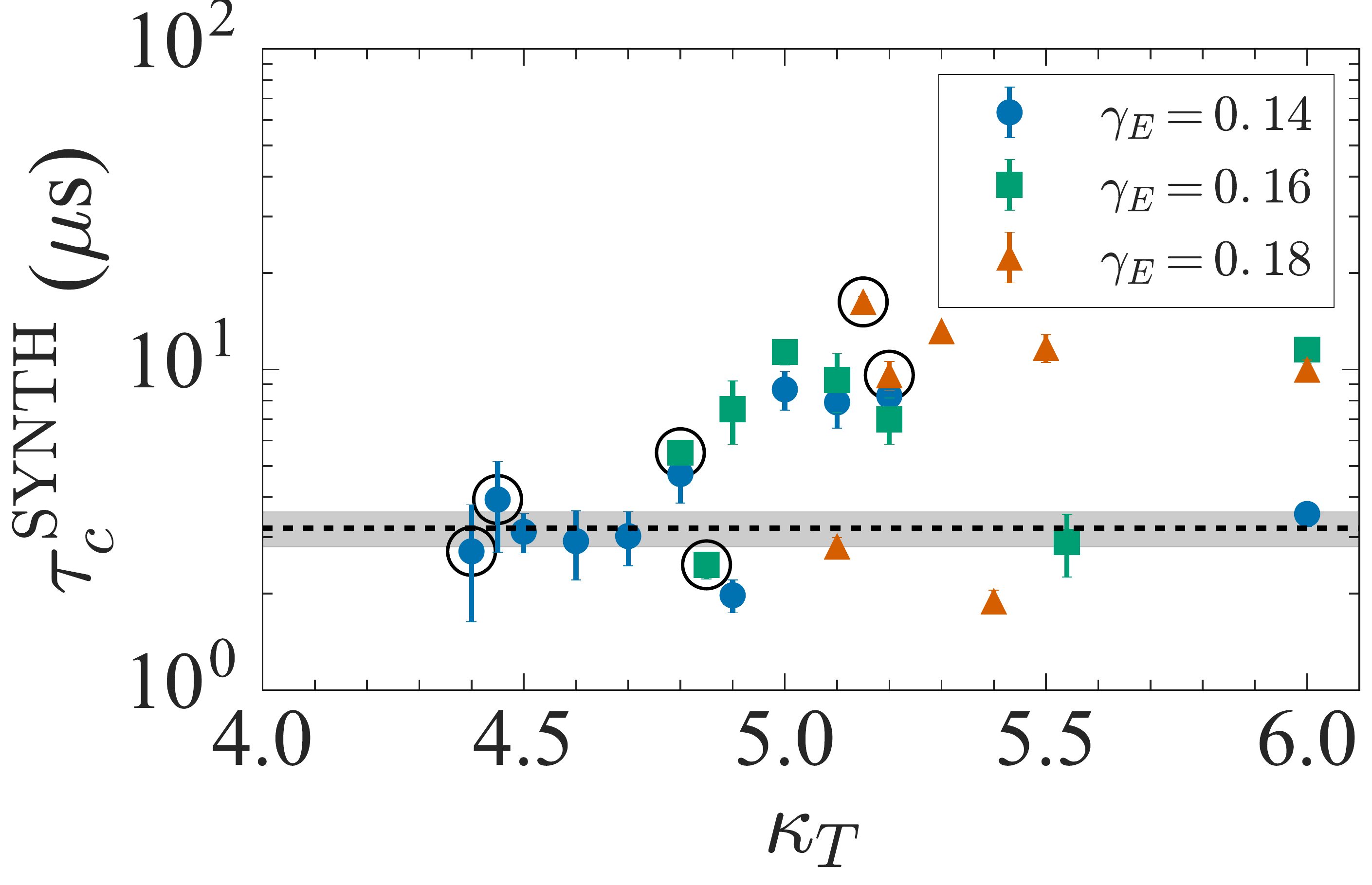}
      \caption{}
      \label{fig:tau_synth}
    \end{subfigure}
    \hfill
    \begin{subfigure}{0.49\linewidth}
      \includegraphics[width=\linewidth]{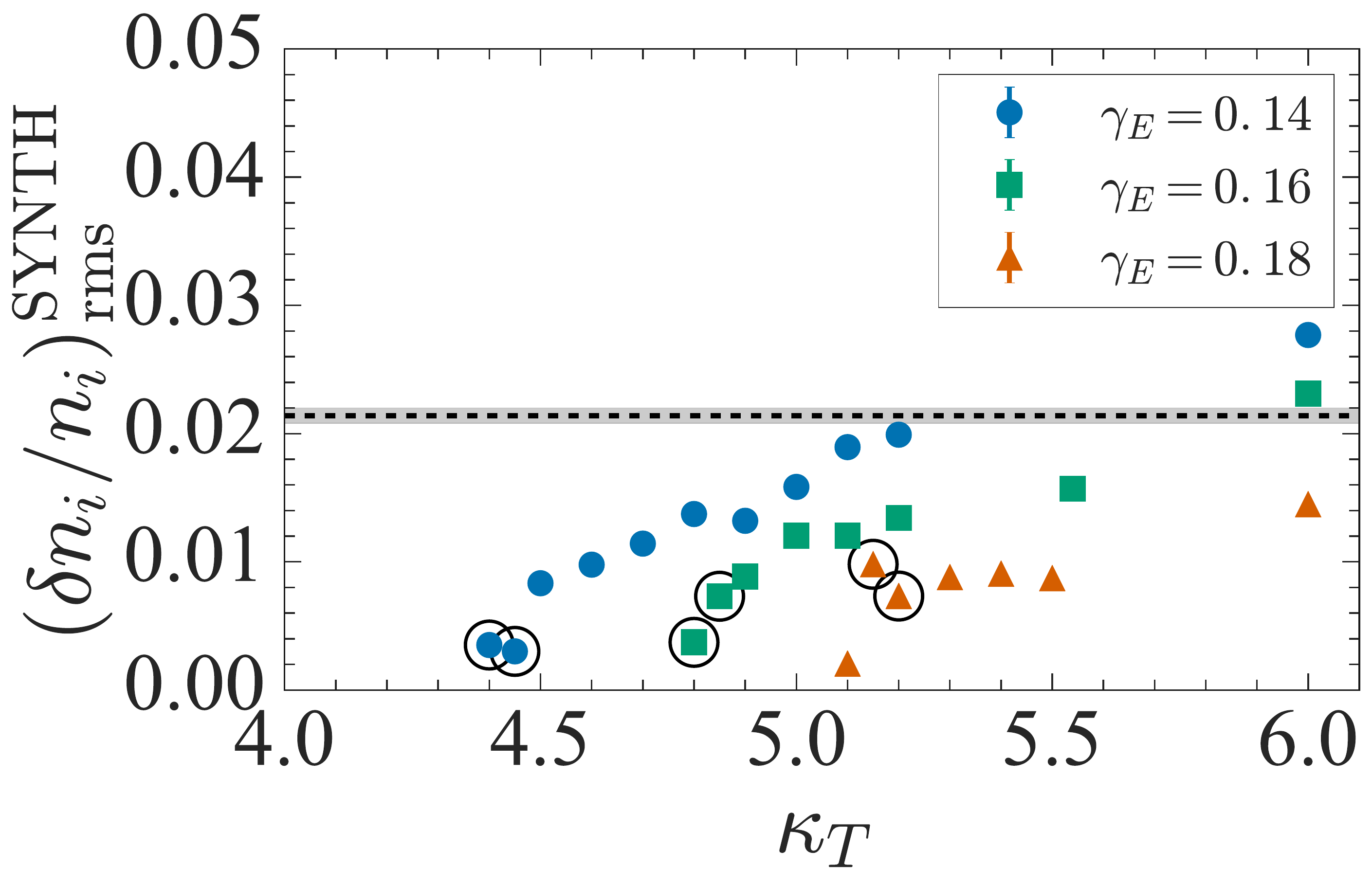}
      \caption{}
      \label{fig:n_synth}
    \end{subfigure}
    \caption[Correlation parameters of synthetic GS2 data]{
      Comparison of correlation parameters obtained via synthetic BES
      measurements of GS2-simulated density field:
      \subref*{fig:lr_synth} radial correlation length $l_R^{\mathrm{SYNTH}}$
      (Section~\ref{sec:radial_corr}),
      \subref*{fig:lz_synth} poloidal correlation length $l_Z^{\mathrm{SYNTH}}$
      (Section~\ref{sec:poloidal_corr}),
      \subref*{fig:tau_synth} correlation time $\tau_c^{\mathrm{SYNTH}}$
      (Section~\ref{sec:time_corr}), and
      \subref*{fig:n_synth} RMS fluctuation amplitude $\qty( \delta n_i /
      n_i)^{\,\mathrm{SYNTH}}_{\mathrm{rms}}$ (Section~\ref{sec:rms_density})
      as functions of $\kappa_T$ and for several values of $\gamma_E$ within
      experimental uncertainty. The circled points indicate the simulations
      match match the experimental heat flux, given in
      Table~\ref{tab:exp_match_sims}. The dashed lines indicate the
      experimental values and the shaded areas the associated error at $r =
      0.8$ obtained from interpolating between experimental measurements seen
      in \figref{exp_corr_results}, which correspond to the local equilibrium
      configuration studied in these simulations.
    }
    \label{fig:synth_corr_results}
  \end{figure}

  Examining \figref{lr_synth}, we see that the values of
  $l_R^{\,\mathrm{SYNTH}}$ are clustered around $2$~cm and below the
  experimental BES measurement of $3\pm0.4$~cm (see Section~\ref{sec:corr_exp}).
  According to the BES specifications~\cite{Field2009}, the approximate
  resolution limit in the radial and poloidal directions is $\sim2$~cm,
  the physical separation between BES channels. More recent work studying
  the measurement effect of the PSFs, concluded that the radial resolution limit
  can be between $2$ and $4$~cm depending on the orientation of the PSFs for a
  given configuration~\cite{Fox2016}. It is therefore likely that the results
  shown in \figref{lr_synth} simply confirm the radial resolution limit
  of the experimental analysis and the true value of $l_R$ may be lower than
  2~cm (as suggested by \figref{radial_fit}). We will confirm this in
  Section~\ref{sec:corr_gs2}, where we consider the correlation properties of
  the raw GS2 density fluctuations.

  Figures~\ref{fig:lz_synth}--\subref{fig:n_synth} give $l_Z^{\,\mathrm{SYNTH}}
  =$~$10$--$15$~cm, $\tau_c^{\,\mathrm{SYNTH}} =$~$2$--$15$~$\mu$s, and $\qty(
  \delta n_i / n_i)^{\,\mathrm{SYNTH}}_{\mathrm{rms}} \sim$~$0.005$--$0.03$. We
  see that these correlation parameters match experimental measurements for
  certain combinations of $\kappa_T$ and $\gamma_E$.  The values of
  $l_Z^{\,\mathrm{SYNTH}}$ are scattered around the experimental value
  $l_Z^{\,\mathrm{EXP}} = 14.06\pm0.09$~cm, showing no clear trend. While none
  of the cases that match the experimental heat flux (circled cases) match
  $l_Z^{\,\mathrm{EXP}}$, there are several simulations within the experimental
  uncertainty ranges of $\kappa_T$ and $\gamma_E$ that do match.
  Similarly, there are several values of $\tau_c^{\,\mathrm{SYNTH}}$ that match
  $\tau_c^{\,\mathrm{EXP}}$, including two cases that match the experimental
  level of heat flux. This is an important improvement over previous nonlinear
  gyrokinetic simulations of this MAST discharge~\cite{Field2014}, which
  overpredicted $\tau_c^{\,\mathrm{SYNTH}}$ by two orders of magnitude.
  Examining \figref{n_synth}, we see that $\qty( \delta n_i /
  n_i)^{\,\mathrm{SYNTH}}_{\mathrm{rms}}$ increases with increasing
  $\kappa_T$ or decreasing $\gamma_E$, and that increasing $\gamma_E$ leads
  to a increase in the value of $\kappa_T$ required to achieve the same $\qty(
  \delta n_i / n_i)^{\,\mathrm{SYNTH}}_{\mathrm{rms}}$. The latter is
  consistent with \figref{q_vs_tprim}, which showed that increasing $\gamma_E$
  shifted the nonlinear turbulence threshold to higher $\kappa_T$. While
  \figref{n_synth} shows that there is agreement between $\qty( \delta n_i /
  n_i)^{\,\mathrm{SYNTH}}_{\mathrm{rms}}$ and $\qty( \delta n_i /
  n_i)^{\,\mathrm{EXP}}_{\mathrm{rms}}$ at certain combinations of $(\kappa_T,
  \gamma_E)$, we see that the circled cases, representing simulations that
  match the experimental heat flux, have values of $\qty( \delta n_i /
  n_i)^{\,\mathrm{SYNTH}}_{\mathrm{rms}}$ well below $\qty( \delta n_i /
  n_i)^{\,\mathrm{EXP}}_{\mathrm{rms}}$. This may suggest that some effects are
  missing from the synthetic diagnostic procedure. For example, a more
  comprehensive analysis could be performed by translating both density
  \emph{and} temperature fluctuations to fluctuating emission
  intensity~\cite{Holland2009}. We note that this discrepancy between
  simulation and experimental density fluctuation measurements has been
  observed in previous BES diagnostic
  studies~\cite{Holland2009,Shafer2012,Gorler2014}, and so further work is
  clearly necessary.

  One phenomenon that was not present in our simulations but is present in the
  experiment is high-energy radiation (e.g., neutron, gamma ray, or hard X-ray)
  impinging on the BES detectors. These photons cause high-amplitude spikes in
  the time series, which are typically confined to a single detector channel
  and, therefore, uncorrelated with other channels. These radiation spikes then
  give rise to large auto-correlations at zero time delay, which are unrelated
  to the turbulent field that is being measured. A numerical ``spike filter''
  is normally used to remove radiation spikes by identifying changes above a
  certain threshold between one time point and the next, and replacing the
  high-intensity value with the value of a neighbouring point~\cite{Field2012,
  Fox2016a}. This ``spike filter'' is an important component of the
  experimental analysis of BES data and, while our simulations do not include
  such sources of radiation, we have included it in the analysis of our
  simulated density fluctuations for consistency with experimental analysis.
  For completeness, the results without the ``spike filter'' are given in
  Appendix~\ref{App:no_spike}. The results show little difference to those with
  the ``spike filter'' except for the value of $l_Z$. We found that in some
  cases, fast-moving structures in the poloidal direction (especially the
  long-lived structures found in our simulations close to the turbulence
  threshold) were removed by the ``spike'' filter and therefore did not
  contribute to the poloidal correlation function, resulting in a drop in
  $l_Z$. In particular, \figref{lz_bes_ns} in Appendix~\ref{App:no_spike} shows
  that $l_Z$ increased significantly in marginal cases compared to the results
  with the ``spike filter'', which may be dominated by coherent structures,
  since structures were no longer removed by the ``spike filter''.

  From the above results we can conclude that local gyrokinetic simulations are
  a reasonable approximation to the experimental turbulence. We showed that all
  correlation parameters apart from $l_R^{\,\mathrm{EXP}}$ show reasonable
  agreement with the experimental measurements within the
  experimental-uncertainty ranges. This shows that from the point of view of
  turbulence measured by the BES diagnostic, the experimental turbulence and the
  synthetic turbulence are comparable.

  Unlike the experiment, we have the raw density fluctuations, as calculated by
  GS2. In the next section we will repeat (and extend) the correlation analysis
  presented in this section for the raw density fluctuations.

\section{Correlation analysis of raw GS2 data}
  \label{sec:corr_gs2}

  Having considered the structure of turbulence processed through a synthetic
  BES diagnostic, we now want to investigate the raw GS2 density fluctuations,
  which will allow us to
  \begin{inparaenum}[(i)]
    \item study the (distorting) effect of the synthetic diagnostic, \item
    study the parallel structure using GS2 data along the field line, and \item
      consider our entire parameter scan to understand how the structure of
      turbulence in MAST might change with the equilibrium parameters
      $\kappa_T$ and $\gamma_E$.
  \end{inparaenum}
  This extends the previous analysis and comparison with simulations performed
  for this MAST discharge~\cite{Field2014}, which only considered for
  equilibrium parameters for a single equilibrium configuration and simulations
  with a synthetic diagnostic applied.

  \subsection{Correlation parameters within experimental uncertainty}
  We start by considering the correlation analysis results for simulations
  with values of $\kappa_T$ and $\gamma_E$  within the experimental
  uncertainty. The only operations applied to the raw density-fluctuation field
  output by GS2 are the transformation to the laboratory frame using
  equation~\eqref{lab_transform} and the transformation from the GS2 parallel
  coordinate $\theta$ to the real-space coordinate $\lambda$, as described in
  Appendix~\ref{App:real_space_transform}. Our correlation analysis is performed
  over a square $(R,Z)$-plane $20\times20$~cm$^2$ in size, located at the centre of
  our computational domain (see \figref{marginal_rz}). We do this to analyse a
  region of similar size to the region probed by the BES diagnostic and also to
  avoid the real-space remapping effect at the edges of the radial domain
  inherent to the GS2 implementation of flow shear (see
  Section~\ref{sec:flow_shear}).

  \subsubsection{Correlation parameters}
\begin{figure}[t]
    \centering
    \begin{subfigure}[t]{0.49\textwidth}
      \includegraphics[width=\linewidth]{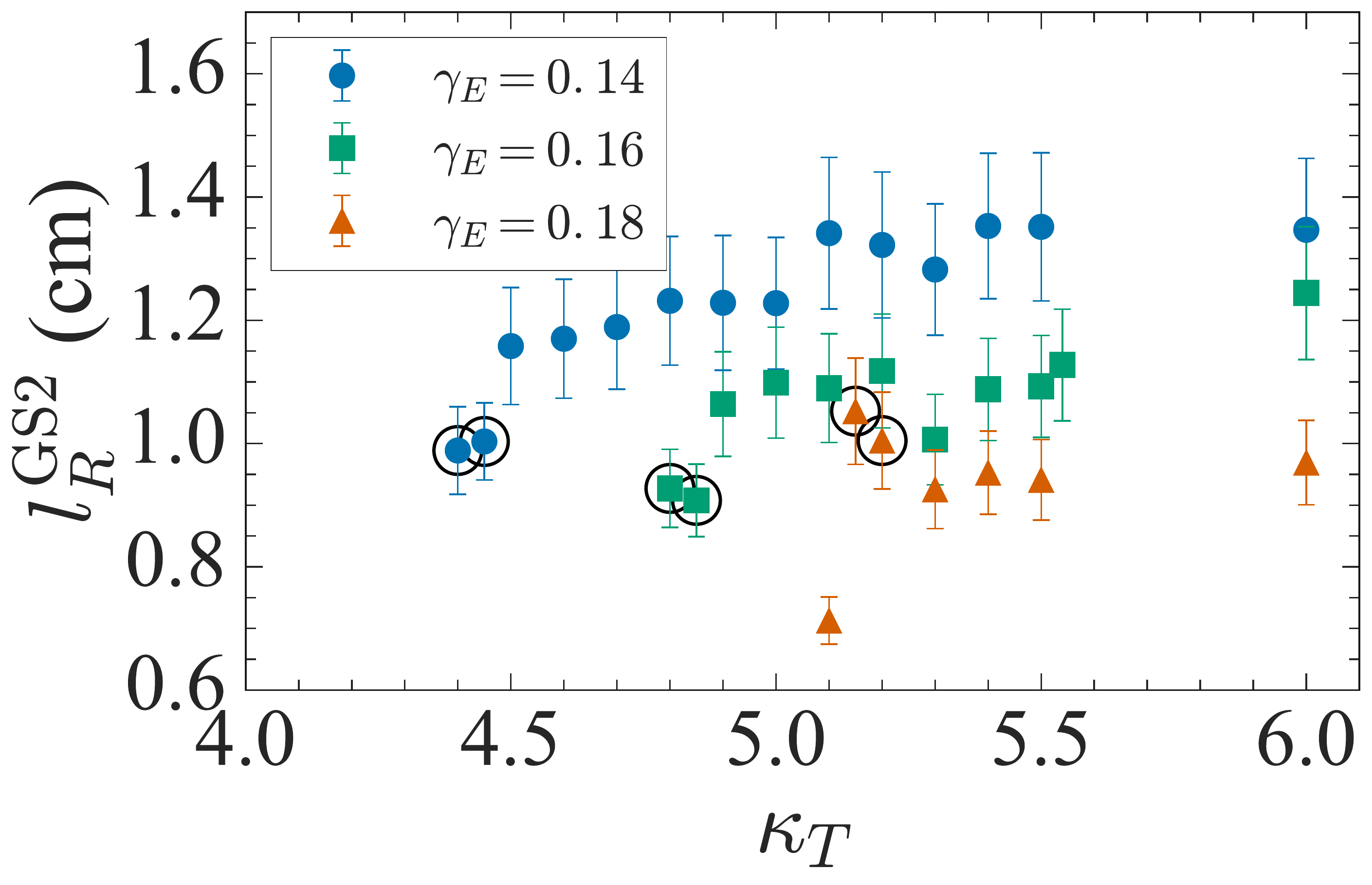}
      \caption{}
      \label{fig:lr_gs2}
    \end{subfigure}
    \begin{subfigure}[t]{0.49\textwidth}
      \includegraphics[width=\linewidth]{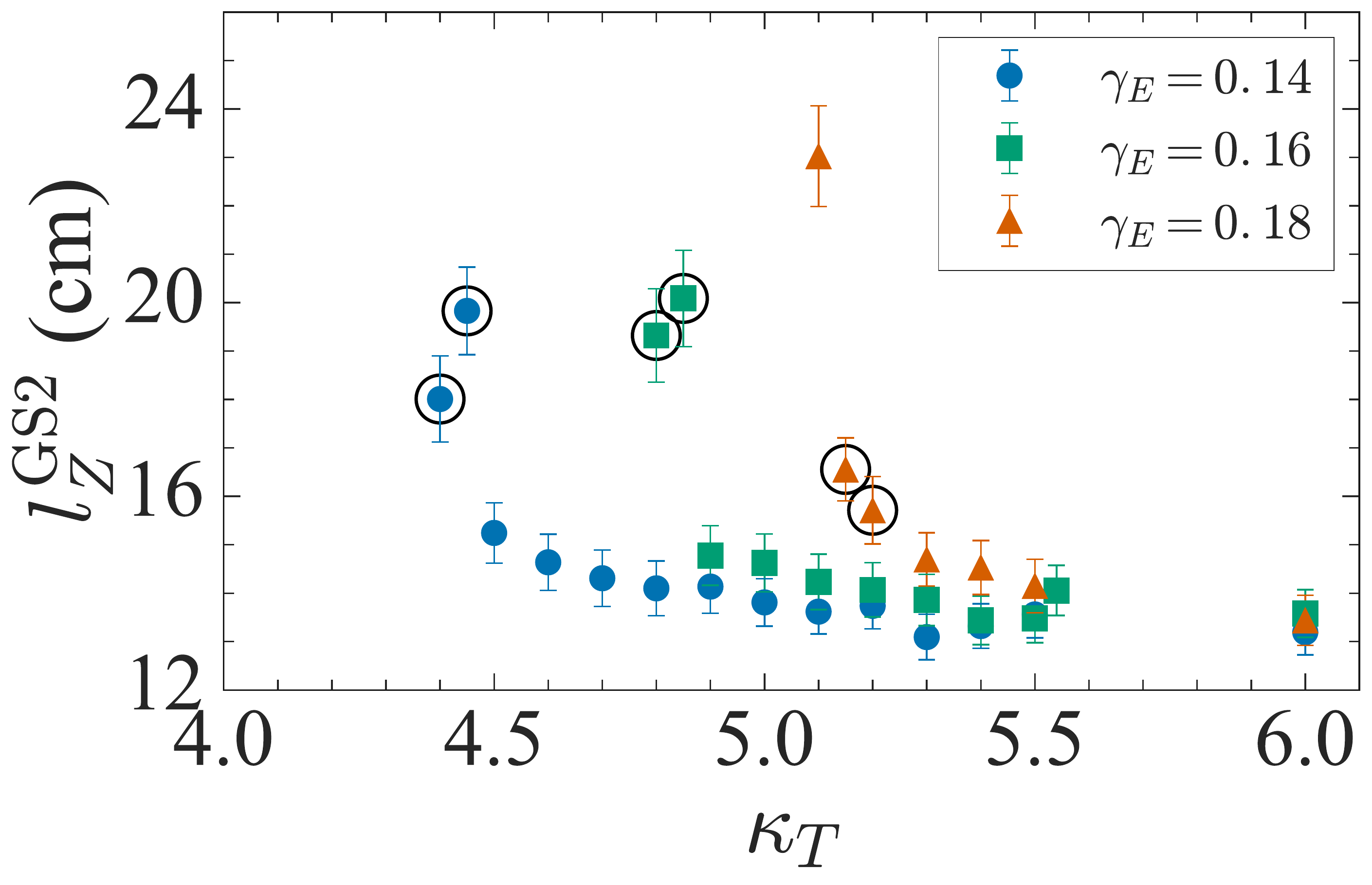}
      \caption{}
      \label{fig:lz_gs2_fixed}
    \end{subfigure}
    \\
    \begin{subfigure}[t]{0.49\textwidth}
      \includegraphics[width=\linewidth]{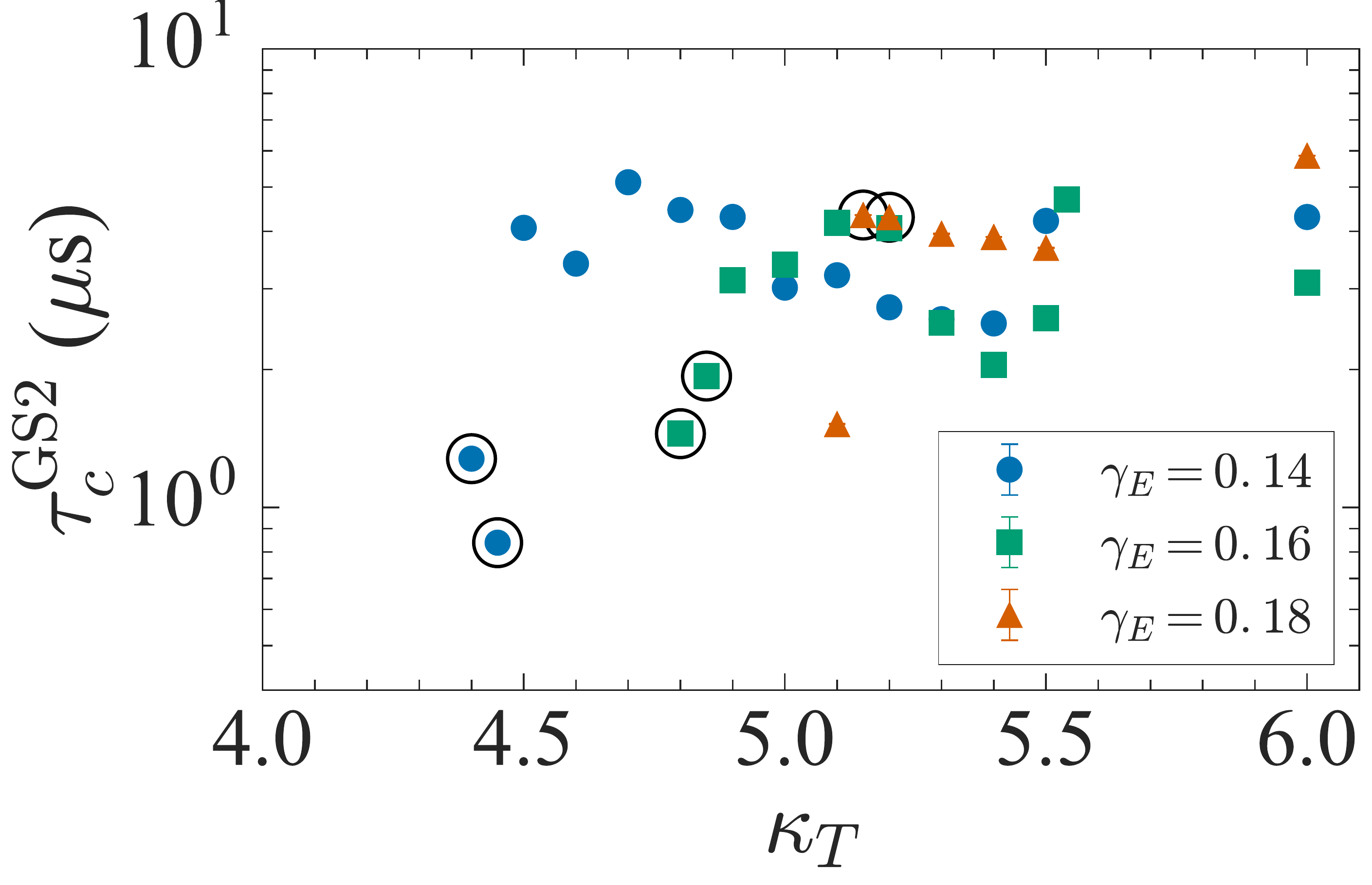}
      \caption{}
      \label{fig:tau_gs2}
    \end{subfigure}
    \begin{subfigure}[t]{0.49\textwidth}
      \includegraphics[width=\linewidth]{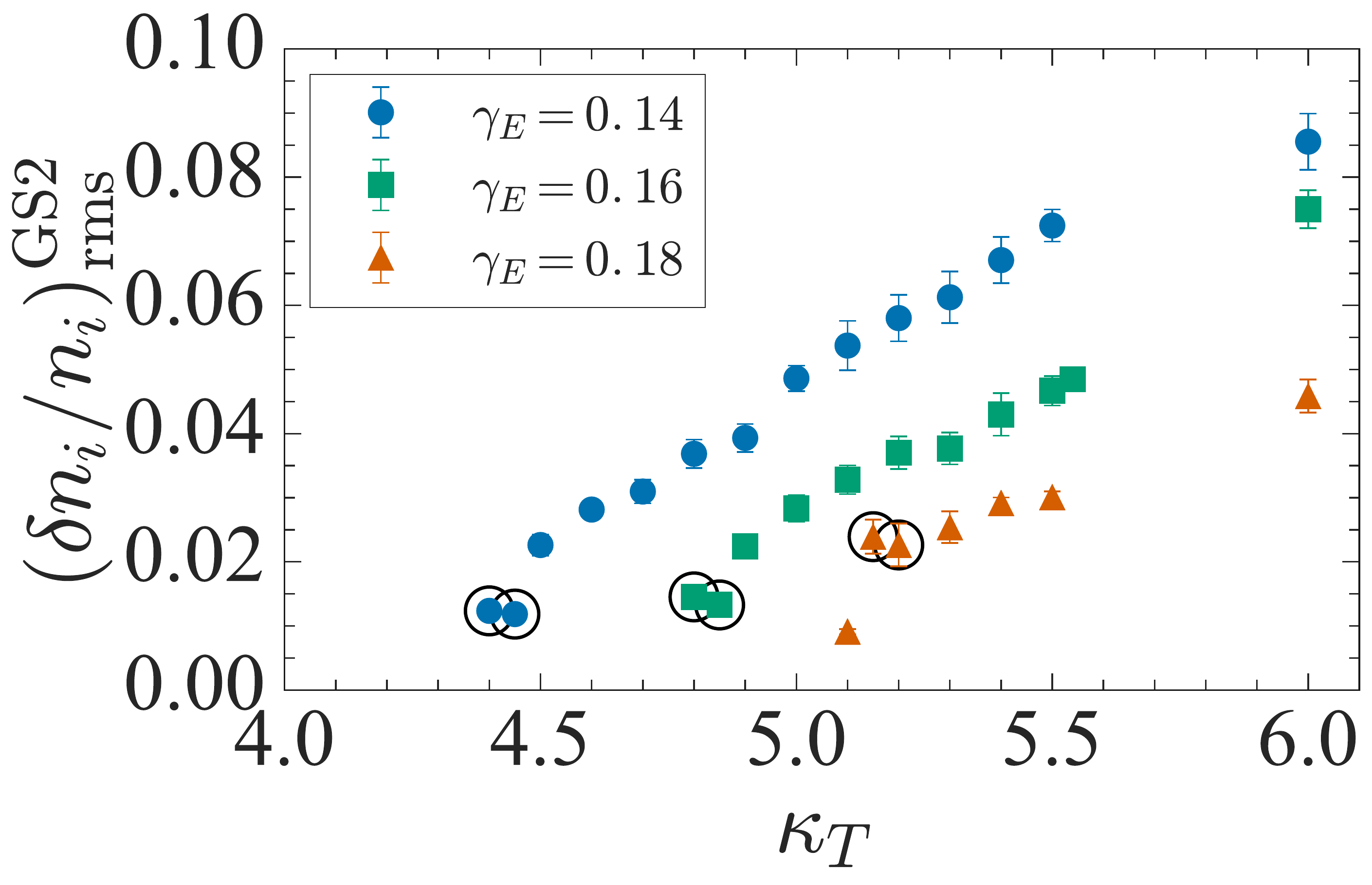}
      \caption{}
      \label{fig:n_gs2}
    \end{subfigure}
    \caption[Correlation parameters for raw GS2 density fluctuations]{
      Correlation parameters calculated for raw GS2 density fluctuations
      for $(\kappa_T, \gamma_E)$ within the region of experimental uncertainty
      indicated in \figref{contour_heatmap}:
      \subref*{fig:lr_gs2} radial correlation length $l_R^{\mathrm{GS2}}$
      (Section~\ref{sec:radial_corr}),
      \subref*{fig:lz_gs2_fixed} poloidal correlation length
      $l_Z^{\mathrm{GS2}}$ keeping $k_y$ fixed to $k_y = 2 \pi / l_Z$
      (Section~\ref{sec:poloidal_corr}),
      \subref*{fig:tau_gs2} correlation time $\tau_c^{\mathrm{GS2}}$
      (Section~\ref{sec:time_corr}), and
      \subref*{fig:n_gs2} RMS density fluctuations ${\qty(\delta n_i /
      n_i)}^{\mathrm{GS2}}_{\mathrm{rms}}$ (Section~\ref{sec:rms_density}).
    }
    \label{fig:gs2_corr_results1}
  \end{figure}

  \Figref{gs2_corr_results1} shows the radial correlation length
  $l_R^{\mathrm{GS2}}$, the poloidal correlation length  $l_Z^{\mathrm{GS2}}$,
  correlation time $\tau_c^{\mathrm{GS2}}$, and RMS density fluctuation
  ${\qty(\delta n_i / n_i)}^{\mathrm{GS2}}_{\mathrm{rms}}$ calculated for our GS2
  density-fluctuation field. The results shown in \figref{gs2_corr_results1}
  are for a range of values of $\kappa_T$ and for $\gamma_E = [0.14, 0.16,
  0.18]$, with circled points describing the simulations that match the
  experimental value of the heat flux. The results are as follows.

  We find that the radial correlation length is $l_R^{\mathrm{GS2}} \sim$
  $1$--$1.5$~cm, increasing with $\kappa_T$ and decreasing with $\gamma_E$.
  This suggests that $l_R^{\mathrm{GS2}}$ has a tendency to increase with
  $Q_i/Q_{\mathrm{gB}}$, as we will show explicitly later. In comparison with
  the synthetic diagnostic results shown in \figref{lr_synth}, where
  $l_R^{\mathrm{SYNTH}} \sim 2$~cm, the true radial correlation length of the
  turbulence $l_R^{\mathrm{GS2}}$ is below $2$~cm and, therefore, below the
  resolution threshold of the BES diagnostic (discussed in
  Section~\ref{sec:corr_synth}).

  \Figref{lz_gs2_fixed} shows that the poloidal
  correlation length is $l_Z^{\mathrm{GS2}} \sim$ $13$--$20$~cm, keeping the
  poloidal wavenumber $k_Z^{\mathrm{GS2}}$ fixed to $k_Z^{\mathrm{GS2}} = 2 \pi
  / l_Z^{\mathrm{GS2}}$ (giving $k_Z^{\mathrm{GS2}} \sim$~$30$--$50$~m$^{-1}$).
  In contrast to $l_R^{\mathrm{GS2}}$, we see that $l_Z^{\mathrm{GS2}}$
  decreases rapidly as $\kappa_T$ is increased from its value at the turbulence
  threshold.

  The correlation time [\figref{tau_gs2}] does not vary significantly with
  $\kappa_T$ or $\gamma_E$ and is in the range $\tau_c^{\mathrm{GS2}}\sim$
  $1$--$6$~$\mu$s.

  Finally, \figref{n_gs2} shows that ${\qty(\delta n_i /
  n_i)}^{\mathrm{GS2}}_{\mathrm{rms}} \sim$~$0.01$--$0.08$ and increases with
  increasing $\kappa_T$ or decreasing $\gamma_E$, i.e., has an upward tendency
  as heat flux increases.

  \subsubsection{Comparisons between experimental and GS2 correlation properties}
  We have presented the correlation parameters measured
  \begin{inparaenum}[(i)]
    \item by the BES diagnostic in Section~\ref{sec:corr_exp},
    \item from GS2 density fluctuations with the synthetic diagnostic applied
      in Section~\ref{sec:corr_synth}, and
    \item from the raw GS2 density fluctuations.
  \end{inparaenum}
  We show the results from all these analyses in Table~\ref{tab:corr_summary}.
  We can summarise the comparison between
  simulation results and experimental measurements as follows. Comparing the
  results of the correlation analysis of the GS2 density fluctuations with the
  experimental measurements, we see that the all the experimental values,
  except for the radial correlation length $l_R$, fall within the ranges found
  for the simulation results. This is particularly important in the case of
  $\tau_c$, which was significantly overestimated in the previous modelling
  effort for this MAST discharge~\cite{Field2014}. It is clear that the
  correlation parameters vary with the equilibrium parameters and there is no
  single simulation, i.e., no single combination of $(\kappa_T, \gamma_E)$, that
  perfectly matches the BES measurements in all four parameters (see
  \figref{gs2_corr_results1}), even for the correlation parameters where there
  is overlap between the experimental value and the simulation ranges.
  \begin{table}
    \centering
    \caption{Summary of results for the correlation parameters $l_R$, $l_Z$,
      $\tau_c$, and $(\delta n_i / n_i)_{\mathrm{rms}}$ from experimental BES
      measurements (EXP), from the correlation analysis of GS2 density
      fluctuations with synthetic diagnostic applied (SYNTH) using an identical
      correlation analysis to that used on the BES data, and from the
      correlation analysis of raw GS2 density fluctuations (GS2).
    }
    \begin{tabular}{c c c c}
      \toprule
      Parameter & EXP & SYNTH & GS2 \\
      \midrule
      $l_R$ (cm) & $3 \pm 0.4$ & 2 & 1--1.5 \\
      $l_Z$ (cm) & $14.06 \pm 0.09$ & 10--15 & 13--20 \\
      $\tau_c$ ($\mu$s) & $3.2 \pm 0.4$ & 2--15 & 1--6 \\
      $(\delta n_i / n_i)_{\mathrm{rms}}$ & $0.0214 \pm 0.0006$ &
        0.005--0.03 & 0.01--0.08 \\
      \bottomrule
    \end{tabular}
    \label{tab:corr_summary}
  \end{table}

  Considering the difference between the GS2 density fluctuations with and
  without the synthetic diagnostic gives us an indication of the effect of the
  PSFs on the measurement of turbulence correlation properties. Given that
  the value of $l_R$ measured from the raw GS2 density fluctuations is below
  the approximate resolution threshold, it is unclear what effect the PSFs have
  on the radial correlation length $l_R$. We see from
  Table~\ref{tab:corr_summary} that the ranges of values of the poloidal
  correlation length $l_Z$ are comparable in the SYNTH and GS2 cases. However,
  \figref{lz_synth} shows that, with the synthetic diagnostic applied, we do
  not see the clear trends versus $\kappa_T$ that we see in
  \figref{lz_gs2_fixed}. This may be due to the limited poloidal resolution,
  which can resolve the measured correlation lengths, but is not sensitive
  enough to recover the trend of decreasing $l_Z$ with $\kappa_T$ seen in
  \figref{lz_gs2_fixed}. The measurement of the correlation time $\tau_c$ is,
  again, less certain in the case of the correlation analysis of density
  fluctuations with a synthetic diagnostic applied, but there is reasonable
  agreement with the correlation time measured from the raw GS2 density
  fluctuations.  Finally, the application of the synthetic
  diagnostic leads to a reduction of roughly $50$\% of the RMS fluctuation
  amplitude, i.e., from ${(\delta n_i / n_i)}^{\mathrm{GS2}}_{\mathrm{rms}}
  \sim$~$0.01$--$0.08$ for the raw density fluctuations to ${(\delta
  n_i/n_i)}^{\mathrm{SYNTH}}_{\mathrm{rms}}~\sim$~$0.005$--$0.03$.  This
  observation is consistent with a recent detailed analysis of the effect of
  PSFs on the measurement of MAST turbulence using a subset of GS2 simulations
  found in this work~\cite{Fox2016}.

  \subsubsection{Poloidal and parallel correlation parameters}
  \label{sec:pol_par_corr}
  \begin{figure}[t]
    \centering
    \begin{subfigure}[t]{0.49\textwidth}
      \includegraphics[width=\linewidth]{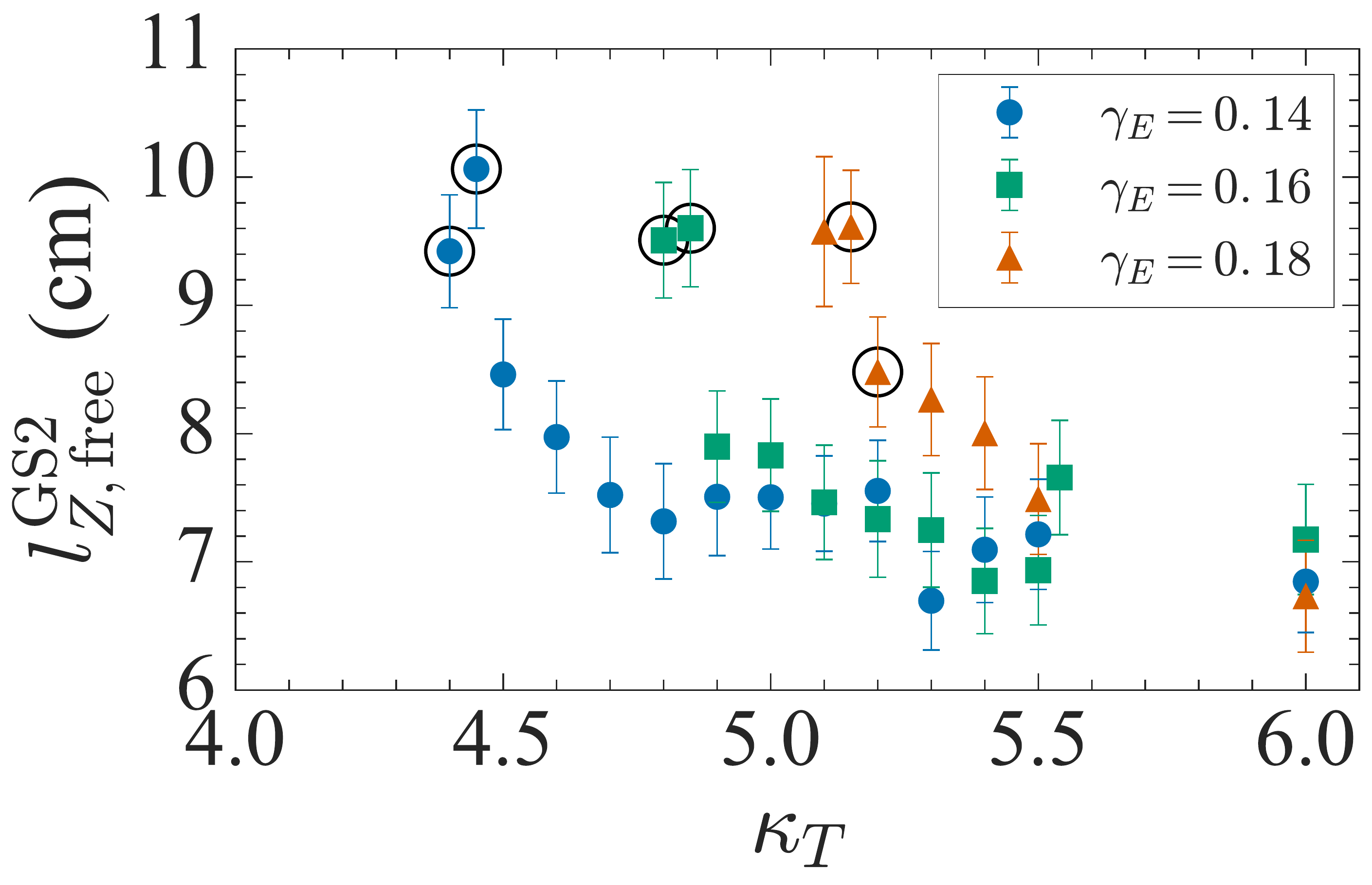}
      \caption{}
      \label{fig:lz_gs2_free}
    \end{subfigure}
    \begin{subfigure}[t]{0.49\textwidth}
      \includegraphics[width=\linewidth]{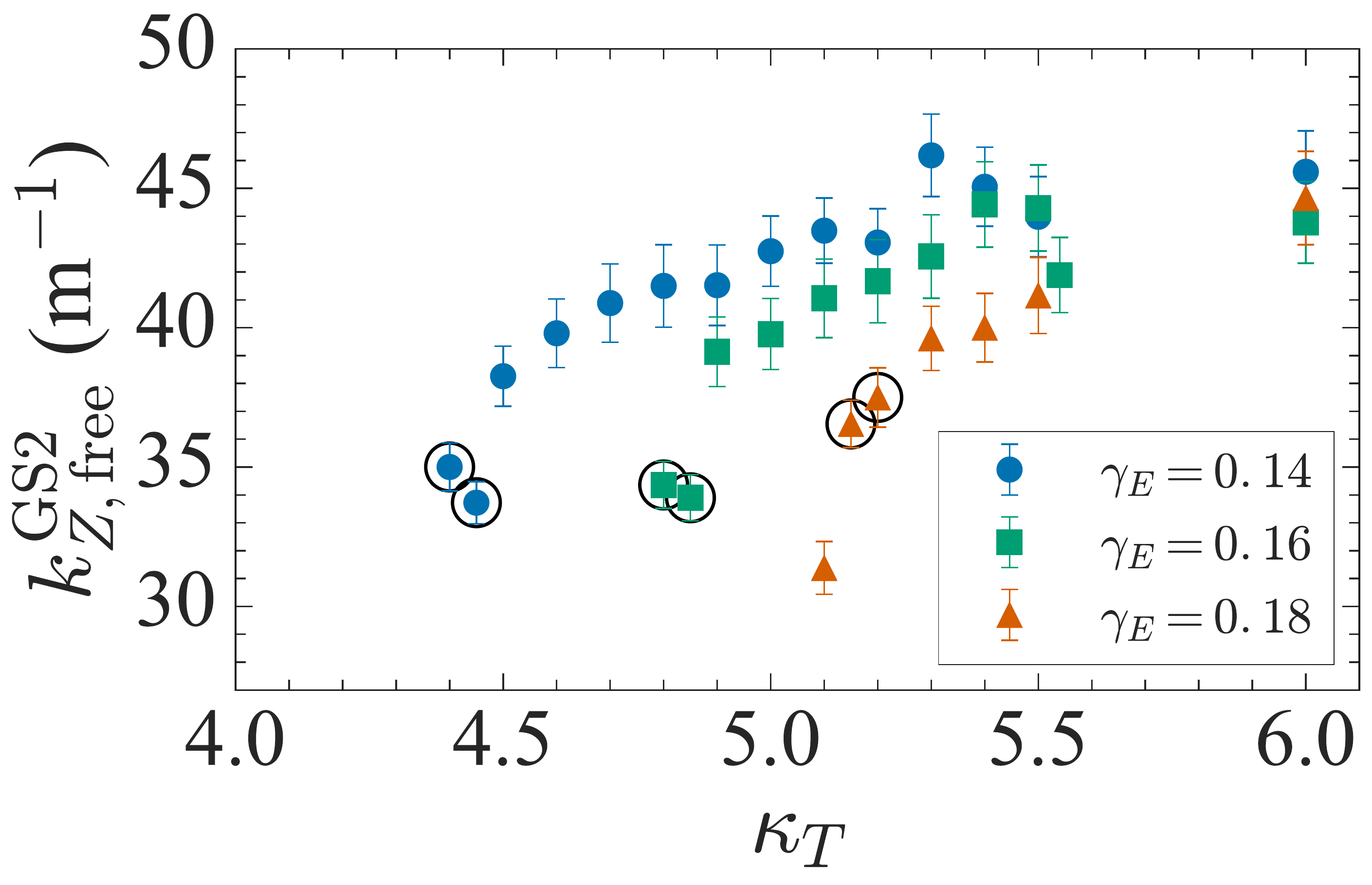}
      \caption{}
      \label{fig:kz_gs2}
    \end{subfigure}
    \\
    \begin{subfigure}[t]{0.49\textwidth}
      \includegraphics[width=\linewidth]{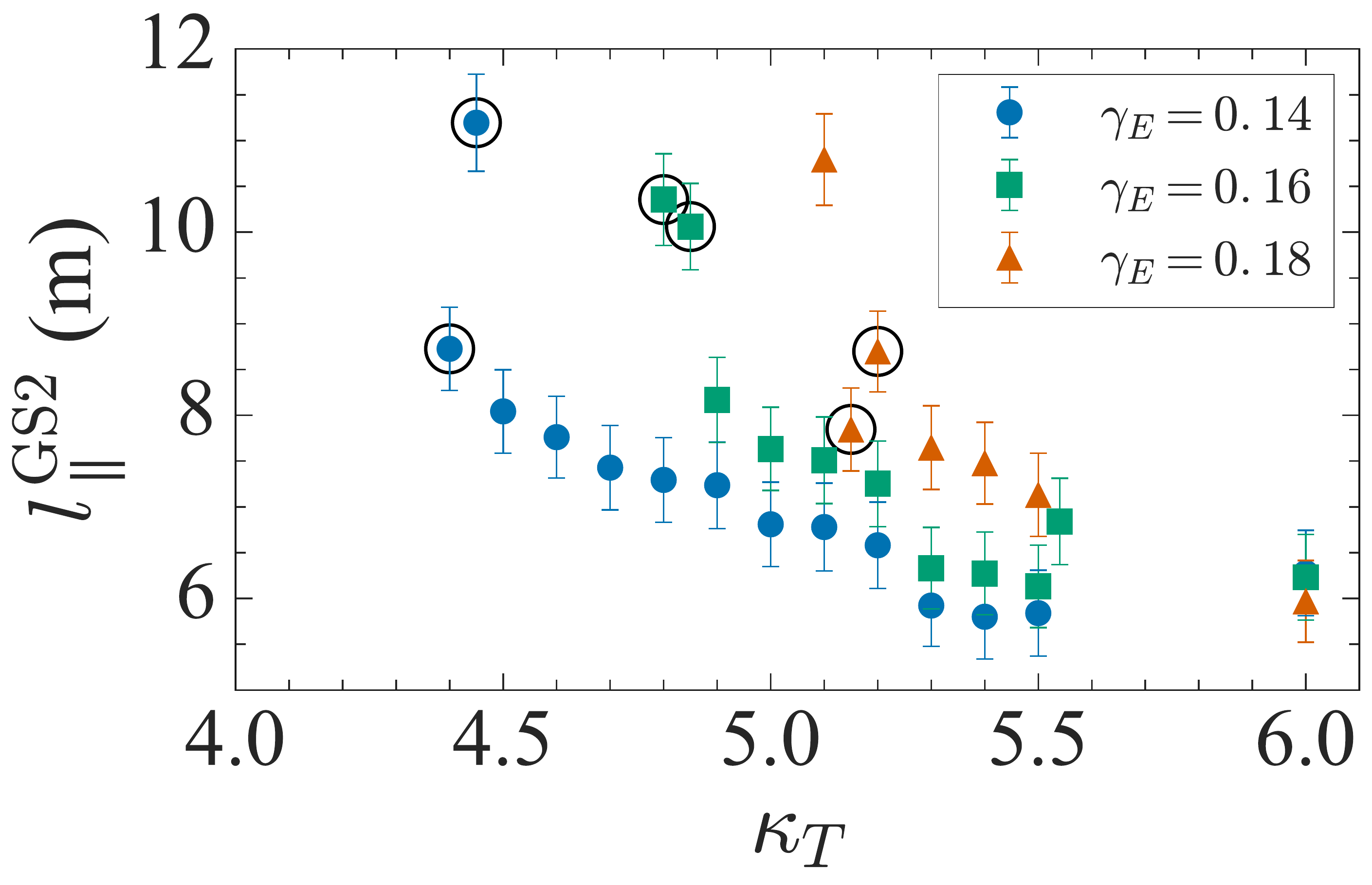}
      \caption{}
      \label{fig:lpar_gs2}
    \end{subfigure}
    \begin{subfigure}[t]{0.49\textwidth}
      \includegraphics[width=\linewidth]{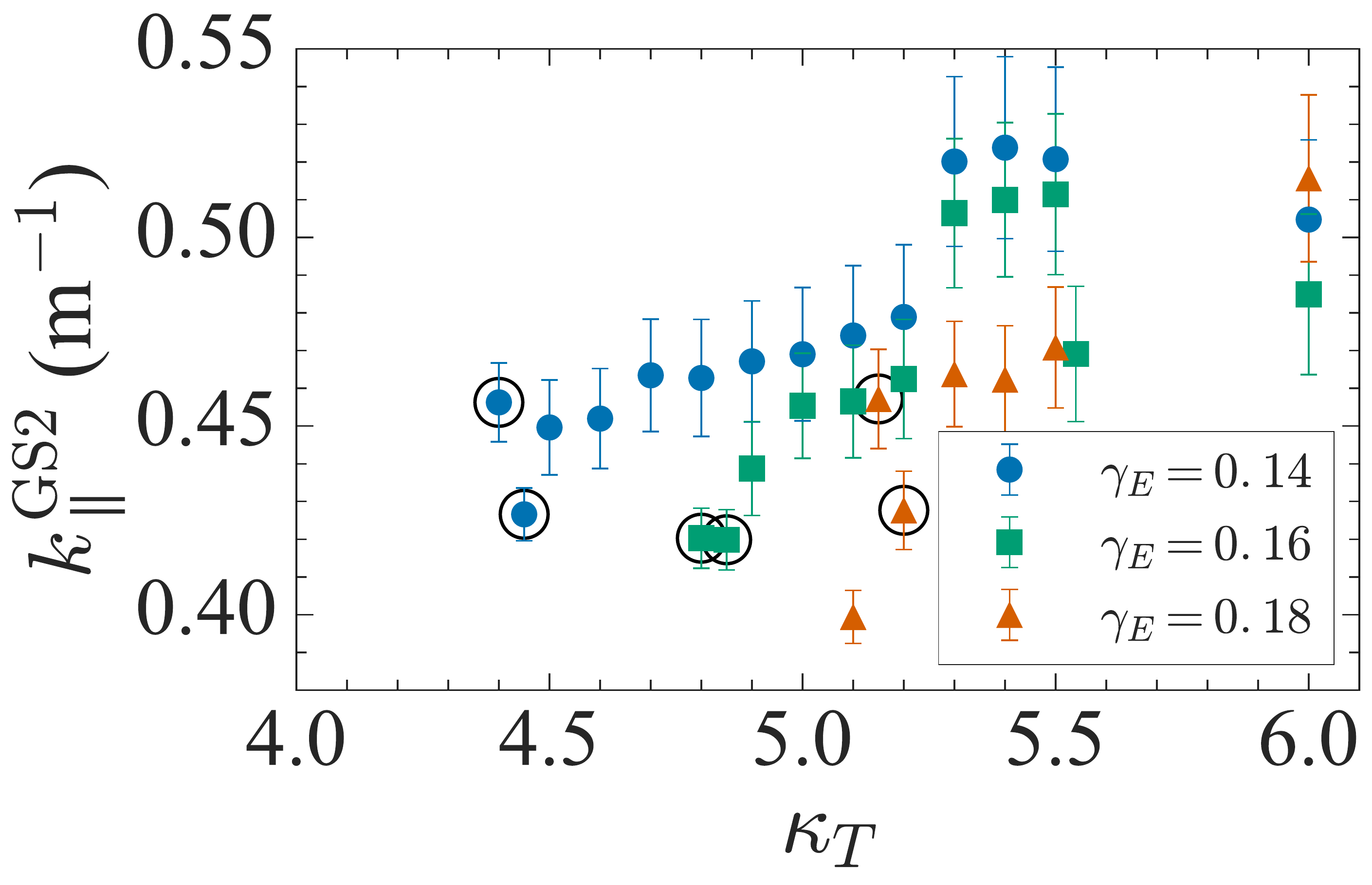}
      \caption{}
      \label{fig:kpar_gs2}
    \end{subfigure}
    \caption[Correlation parameters for raw GS2 density fluctuations (cont'd)]{
      Correlation parameters calculated for raw GS2 density fluctuations
      for $(\kappa_T, \gamma_E)$ within the region of experimental uncertainty
      indicated in \figref{contour_heatmap}:
      \subref*{fig:lz_gs2_free} poloidal correlation length
      $l_{Z,\mathrm{free}}^{\mathrm{GS2}}$ with $k_y$ as a free fitting
      parameter,
      \subref*{fig:kz_gs2} poloidal wavenumber
      $k_{Z,\mathrm{free}}^{\mathrm{GS2}}$ (Section~\ref{sec:poloidal_corr}),
      \subref*{fig:lpar_gs2} parallel correlation length
      $l_{\parallel}^{\mathrm{GS2}}$, and
      \subref*{fig:kpar_gs2} parallel wavenumber $k_{\parallel}^{\mathrm{GS2}}$
      (Section~\ref{sec:par_corr}).
    }
    \label{fig:gs2_corr_results2}
  \end{figure}
  We now consider two further diagnostics, which were not available to us
  experimentally: the poloidal and parallel correlation lengths and wavenumbers
  calculated as independent fitting parameters to the corresponding correlation
  functions (see Sections~\ref{sec:poloidal_corr} and~\ref{sec:par_corr}).  As
  explained in Section~\ref{sec:poloidal_corr}, the higher poloidal resolution
  of GS2 data compared to the experimental BES measurements allows us to fit
  the poloidal correlation function with $l_Z$ and $k_Z$ as independent fitting
  parameters. In addition, GS2 predicts density fluctuations in the parallel
  direction allowing us to calculate parallel correlation functions.

  Figures~\ref{fig:lz_gs2_free} and \subref{fig:kz_gs2} show the
  result of such fitting: $l_{Z,\mathrm{free}}^{\mathrm{GS2}}$ and
  $k_Z^{\mathrm{GS2}}$ versus $\kappa_T$. As already anticipated by
  \figref{poloidal_fit}, we see a roughly $50$\% decrease in
  $l_{Z,\mathrm{free}}^{\mathrm{GS2}}$ compared to $l_Z^{\mathrm{GS2}}$
  [\figref{lz_gs2_fixed}], from $13$--$20$~cm to $7$--$10$~cm, again decreasing
  as $\kappa_T$ increases or $\gamma_E$ decreases.  The value of
  $k_{Z,\mathrm{free}}^{\mathrm{GS2}}$ is in the range $35$--$45$~m$^{-1}$ --
  comparable to one obtained via fitting the procedure where $k_Z = 2\pi/l_Z$.
  Regardless of the fitting method, \figref{lz_gs2_fixed} and
  \figref{lz_gs2_free} show a similar dependence of $l_Z$ on $\kappa_T$ and
  $\gamma_E$.

  Currently the BES diagnostic on MAST is not capable of determining both $l_Z$
  and $k_Z$, but these estimates may be used for future comparisons between
  experimental measurements and numerical results if higher-resolution BES
  measurements become available. Similarly there is currently no diagnostic on
  MAST capable of measuring the parallel correlation length, but our
  estimates may guide future attempts at designing diagnostics to measure it.

  The results of the parallel correlation analysis, given in
  \figref{lpar_gs2} and \subref{fig:kpar_gs2}, are the values
  $l_\parallel^{\mathrm{GS2}}$ and $k_\parallel^{\mathrm{GS2}}$ versus
  $\kappa_T$. We see that $l_\parallel^{\mathrm{GS2}} \sim$~$6$--$12$~m and
  decreases with increasing $\kappa_T$ and decreasing $\gamma_E$. Based on this
  measurement of the parallel correlation length, it is clear that the
  turbulence is highly anisotropic, i.e., $l_\parallel \gg l_\perp$, as it is
  expected to be~\cite{Abel2013}.

  Using the measurement of $l_\parallel^{\mathrm{GS2}}$, we can return to, and
  confirm, the assumption upon which the calculation of $\tau_c$ depends. In
  Section~\ref{sec:time_corr}, we assumed that reliably estimating the
  correlation time depends on the temporal decorrelation dominating over
  effects due to the finite parallel correlation length
  [see~\eqref{time_assumption}]. Using the value of $l_\parallel$ above, we can
  estimate that $l_\parallel \cos \vartheta / u_\phi \sim$ $80$--$160$~$\mu$s,
  where we have used $R = 1.32$ m, $\omega = 4.71 \times 10^4$
  $\mathrm{s}^{-1}$, and $\vartheta \approx 0.6$. This confirms that $\tau_c$
  is smaller than $l_\parallel \cos \vartheta / u_\phi$ by more than an order
  of magnitude and that the time correlation analysis is valid in this MAST
  configuration.

  \subsubsection{Comparison between linear and nonlinear time scales}
  \label{sec:time_scales}
  With the knowledge of the correlation parameters, we can return to the
  comparison of the transient-growth time $t_0$ and nonlinear time
  $\tau_{\mathrm{NL}}$ discussed in section~\ref{sec:subcritical}.  In
  particular, we want to determine one of the two conditions for the onset of
  subcritical turbulence [equation~\eqref{schek_t0}] proposed
  in Ref.~\cite{Schekochihin2012}. We also follow Ref.~\cite{Field2014} and compare
  $\tau_{\mathrm{NL}}$ with the correlation time of the turbulence $\tau_c$ and
  compare with the corresponding experimental results.

\begin{figure}[t]
    \centering
    \begin{subfigure}{0.49\linewidth}
      \includegraphics[width=\linewidth]{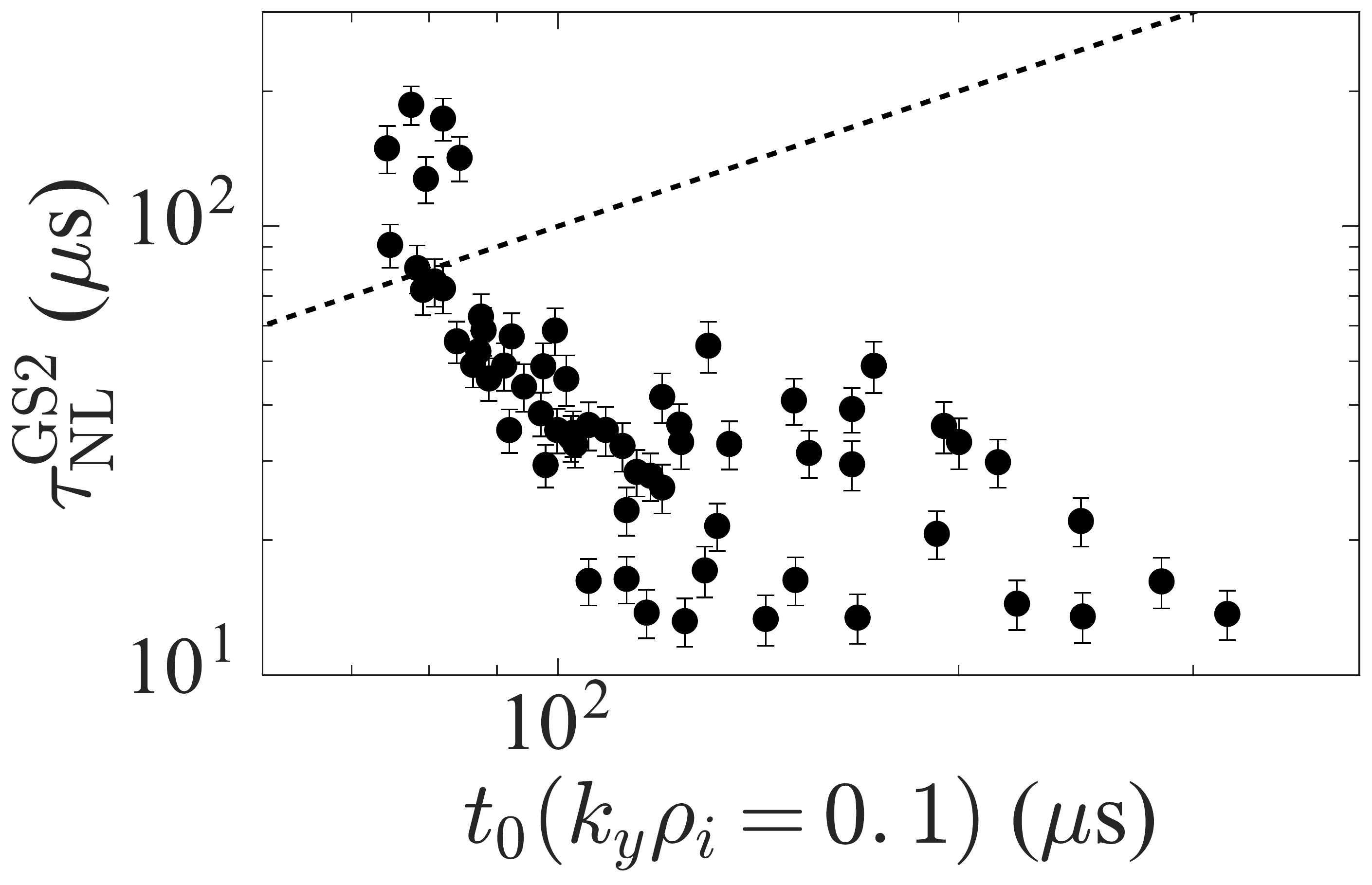}
      \caption{}
      \label{fig:tnl_t0}
    \end{subfigure}
    \hfill
    \begin{subfigure}{0.49\linewidth}
      \includegraphics[width=\linewidth]{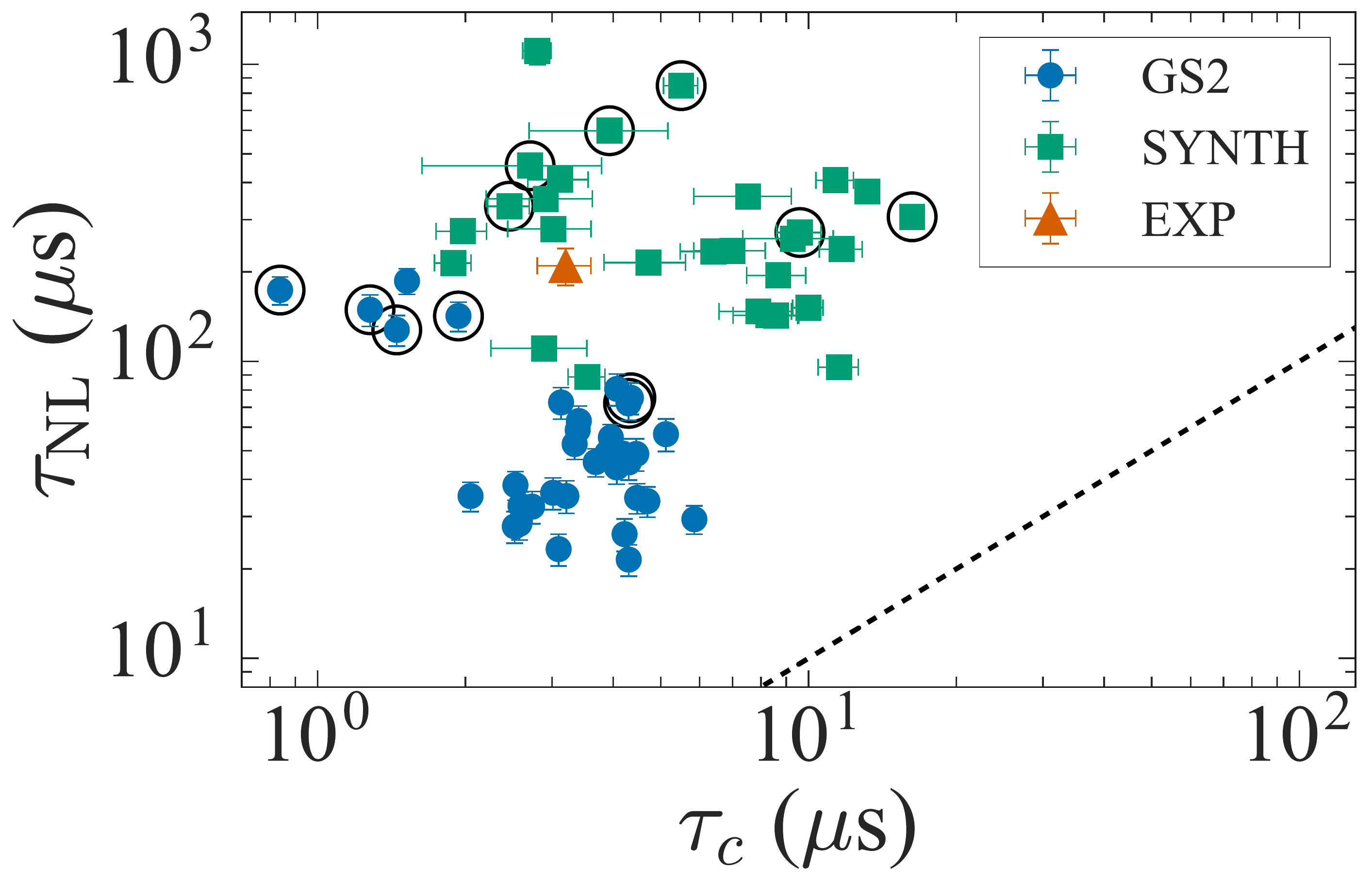}
      \caption{}
      \label{fig:tnl_tauc}
    \end{subfigure}
    \caption[Comparison of linear and nonlinear time scales]{
      \subref*{fig:tnl_t0} Nonlinear interaction time of the raw density
      fluctuations $\tau_{\mathrm{NL}}^{\mathrm{GS2}}$, calculated
      using~\eqref{tau_nl}, versus the transient-growth time $t_0$.
      We have taken $t_0$ at $k_y \rho_i = 0.1$, where $t_0$ is largest (see
      \figref{N_and_t0_16}). We show all simulations in our parameter scan with
      $\gamma_E>0$.
      \subref*{fig:tnl_tauc} $\tau_{\mathrm{NL}}$ versus $\tau_c$ for
      the correlation parameters calculated from the raw GS2 density
      fluctuations (GS2), from density fluctuations with the synthetic
      diagnostic applied (SYNTH), and from experimental measurements (EXP).
      The cases shown are for values of $(\kappa_T, \gamma_E)$ within
      experimental uncertainty and the circled simulations indicate the
      simulations that match the experimental heat flux. The dashed lines in
      each plot indicate where the time scales are equal.
    }
    \label{fig:time_scales}
  \end{figure}

  The non-zonal nonlinear interaction time is estimated to be~\cite{Ghim2013}:
  \begin{equation}
    \tau_{\mathrm{NL}}^{-1} =
    \frac{v_{\mathrm{th}i} \rho_i}{l_R l_Z} \frac{T_e}{T_i}
      \qty(\frac{\delta n_i}{n_i})_{\mathrm{rms}},
    \label{tau_nl}
  \end{equation}
  where we have assumed $l_Z \approx l_y$ (where $l_y$ is the correlation
  length in the binormal direction as defined in~\cite{Ghim2013}) because $l_Z =
  l_y \cos \vartheta$, where $\vartheta$ is the magnetic field pitch-angle (see
  \figref{pitch_angle}), and $\cos \vartheta \sim 1$ for this magnetic
  equilibrium.  The transient-growth time $t_0$ was calculated from linear
  simulations and plotted in
  \figref{N_and_t0_16}, showing that, at ion scales, the longest transient
  growth occurred at $k_y \rho_i \sim 0.1$. \Figref{tnl_t0} shows
  $\tau_{\mathrm{NL}}^{\mathrm{GS2}}$ versus $t_0$ (at $k_y \rho_i = 0.1$) for
  all simulations with $\gamma_E > 0$, where the dashed line indicates
  $\tau_{\mathrm{NL}}^{\mathrm{GS2}} = t_0$. We see that the majority of
  simulations are below the line defined by $\tau_{\mathrm{NL}}^{\mathrm{GS2}}
  = t_0$, showing that the condition for the onset of turbulence given
  by~\eqref{schek_t0} is approximately true, i.e., that $t_0 \gtrsim
  \tau_{\mathrm{NL}}$.

  Ref.~\cite{Field2014} compares $\tau_{\mathrm{NL}}$ with the
  turbulence correlation time $\tau_c$, both calculated from experimental
  measurements, and provides another possible point of comparison using the
  results from our correlation analysis of raw GS2 density fluctuations.
  \Figref{tnl_tauc} shows $\tau_{\mathrm{NL}}$ versus $\tau_c$ for
  nonlinear simulations with values of $(\kappa_T, \gamma_E)$ within
  experimental uncertainty. The values of $\tau_{\mathrm{NL}}$ were calculated
  from correlation parameters of raw GS2 density fluctuations (GS2), from
  correlation parameters calculated from GS2 density fluctuations with a
  synthetic diagnostic applied (SYNTH), and from the experimental BES
  measurements at $r=0.8$ (EXP). The dashed line indicates a line defined by
  $\tau_{\mathrm{NL}} = \tau_c$. First, we see that $\tau_{\mathrm{NL}} >
  \tau_c$ for both the GS2 and SYNTH cases, consistent with the experimental
  value: the red triangle at approximately $(\tau_{\mathrm{NL}}, \tau_c) =
  (3, 2 \times 10^2)$. Secondly, we see that $\tau_{\mathrm{NL}}$ for the raw
  GS2 density fluctuations tends to be below the experimental value, whereas
  the SYNTH cases are comparable. The results shown in \figref{tnl_tauc} are
  consistent with the experimental results in~\cite{Field2014} that showed
  $\tau_{\mathrm{NL}} > \tau_c$ for this and other experimental cases, and so
  gives us further confidence in the ability of GS2 to predict the properties
  of turbulence in MAST. However, we can also conclude from \figref{tnl_tauc}
  that $\tau_{\mathrm{NL}} \gg \tau_c$ in all cases, with $\tau_{\mathrm{NL}}$
  being up to three orders of magnitude larger in some cases. The value of
  $\tau_c$ is measured from the turbulence itself, and so \figref{tnl_tauc}
  suggests that the estimate of $\tau_{\mathrm{NL}}$ \eqref{tau_nl} can
  significantly overestimate the actual interaction time, given that it does
  not make sense to consider the interaction of eddies (over a time scale
  $\tau_{\mathrm{NL}}$) that have already decorrelated (over a much shorted
  time scale $\tau_c$).

  \subsection{Correlation parameters versus $Q_i/Q_{\mathrm{gB}}$}

  The correlation analysis results in Figures~\ref{fig:gs2_corr_results1}
  and~\ref{fig:gs2_corr_results2}, in particular $l_Z^{\mathrm{GS2}}$, $(
  \delta n_i / n_i)^{\mathrm{GS2}}_{\mathrm{rms}}$, and
  $l_\parallel^{\mathrm{GS2}}$, show similar trends versus $\kappa_T$ for
  different values of $\gamma_E$. As we showed in \figref{contour_heatmap},
  increasing $\kappa_T$ or decreasing $\gamma_E$ effectively amounts to
  controlling the distance from the turbulence threshold. Furthermore, our
  investigations of the transition to turbulence (see~\cite{VanWyk2016} and
  Section~\ref{sec:subcritical}) and the effect of flow shear on its
  structure~\cite{Fox2016a} suggest that the key determining factor is the
  distance from the threshold. This is most conveniently parametrised by the
  ion heat flux $Q_i/Q_{\mathrm{gB}}$. Here we describe the results of our
  correlation analysis of raw GS2 density fluctuations as a function of this
  parameter.
  \begin{figure}[t]
    \centering
    \begin{subfigure}[t]{0.49\textwidth}
      \includegraphics[width=\linewidth]{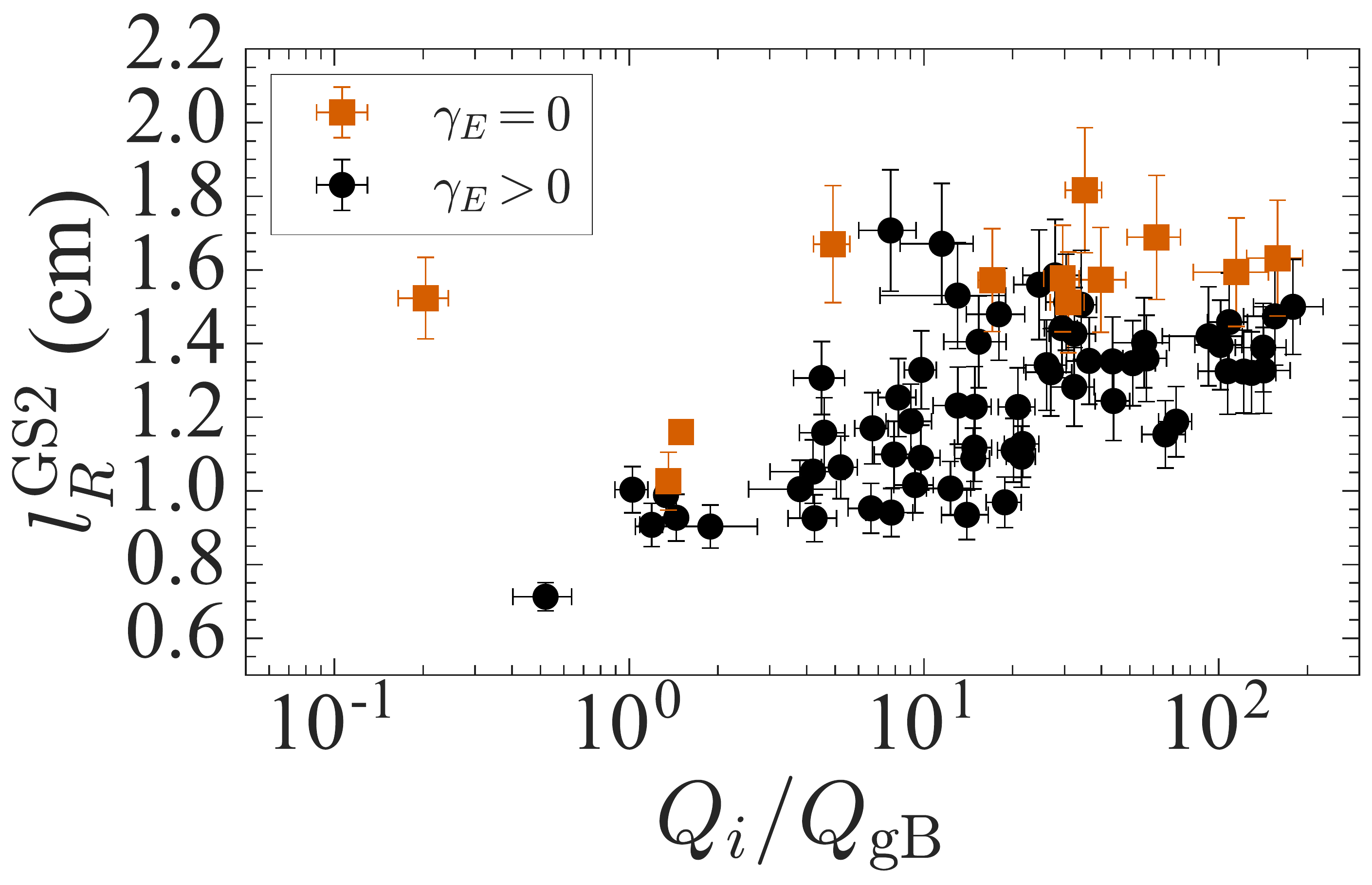}
      \caption{}
      \label{fig:lr_q}
    \end{subfigure}
    \begin{subfigure}[t]{0.49\textwidth}
      \includegraphics[width=\linewidth]{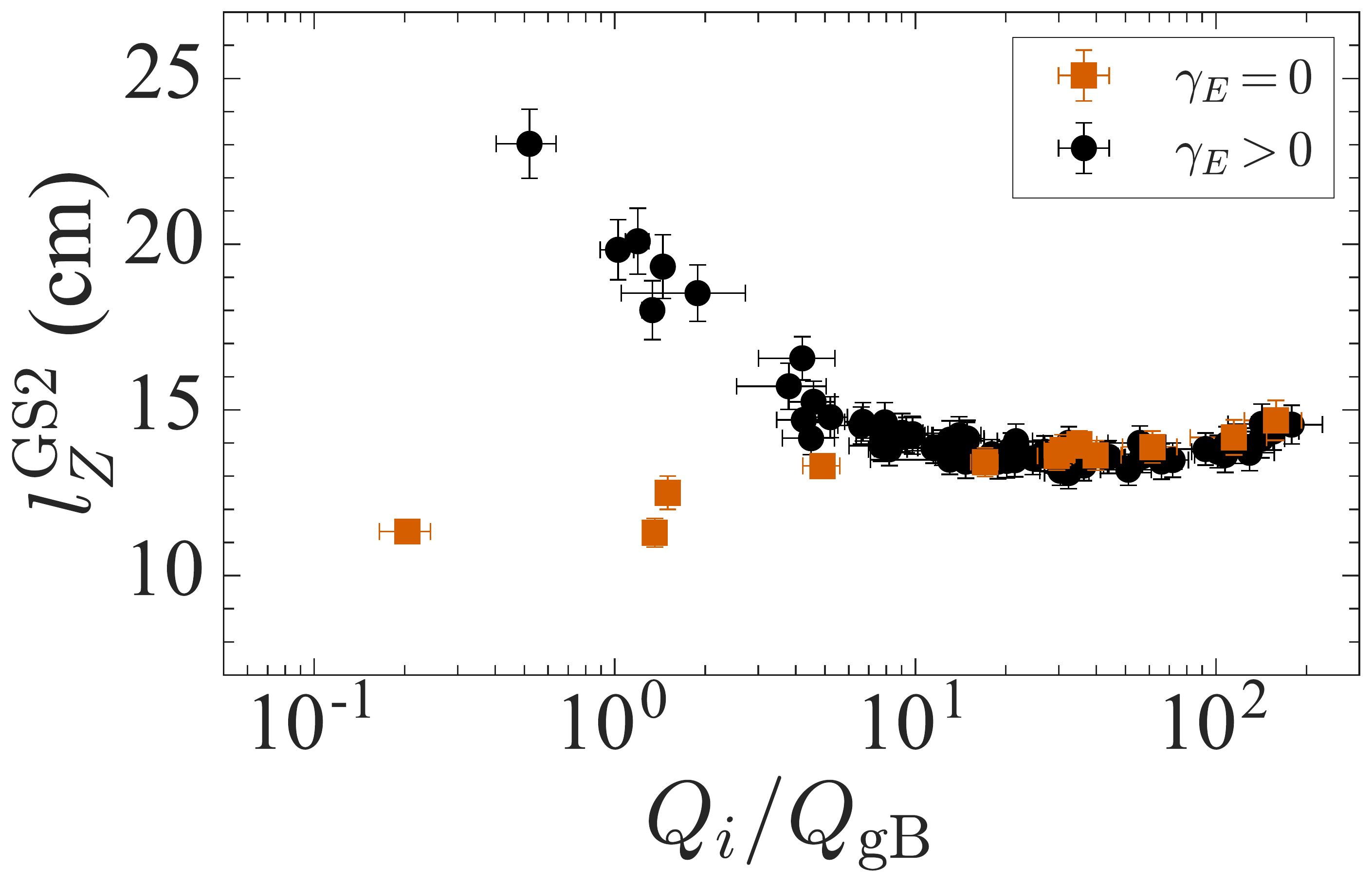}
      \caption{}
      \label{fig:lz_q}
    \end{subfigure}
    \\
    \begin{subfigure}[t]{0.49\textwidth}
      \includegraphics[width=\linewidth]{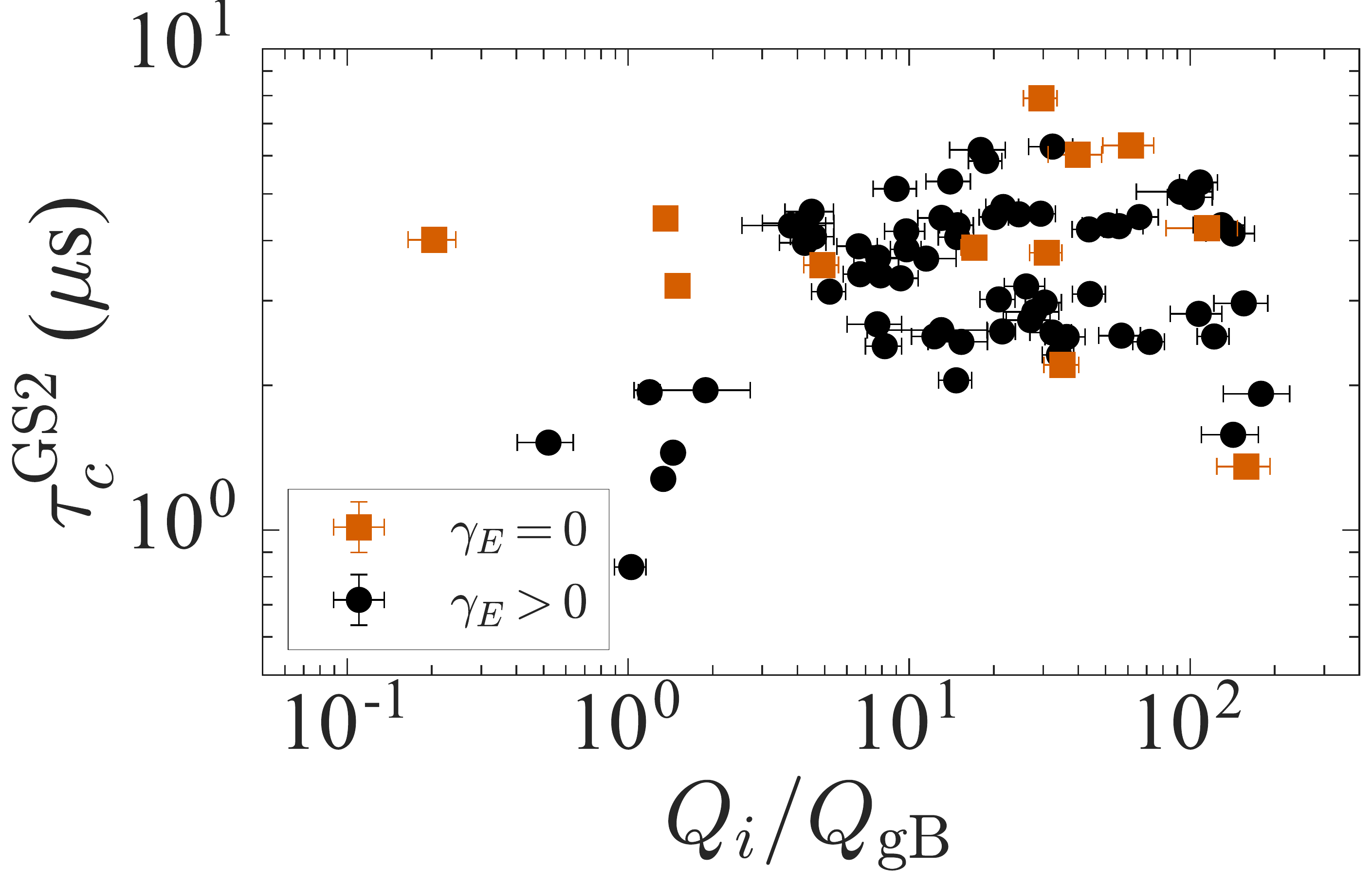}
      \caption{}
      \label{fig:tau_q}
    \end{subfigure}
    \begin{subfigure}[t]{0.49\textwidth}
      \includegraphics[width=\linewidth]{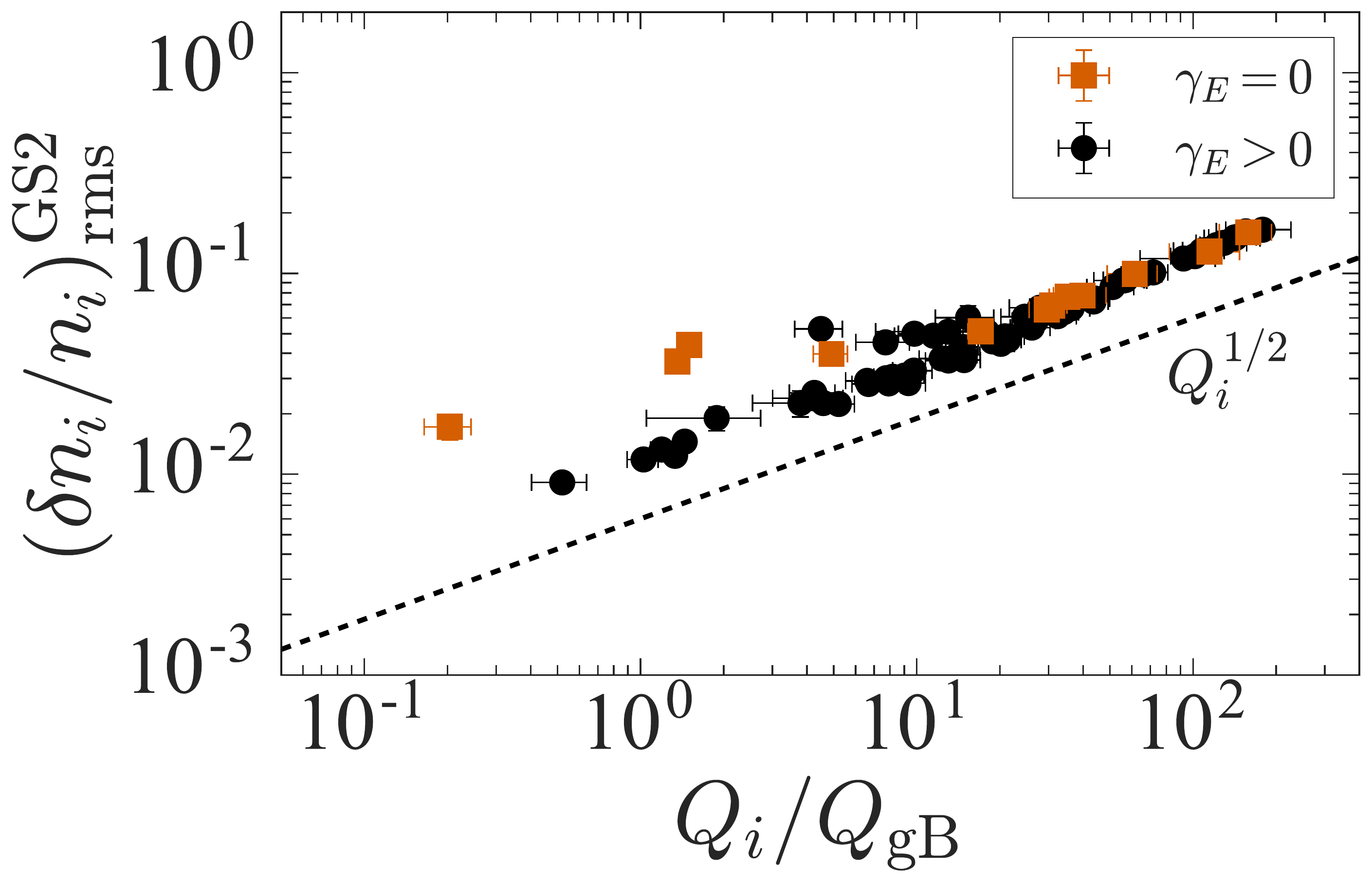}
      \caption{}
      \label{fig:n_q}
    \end{subfigure}
    \caption[Correlation parameters for GS2 density fluctuations versus
             $Q_i/Q_{\mathrm{gB}}$]{
      Correlation parameters calculated for raw GS2 density fluctuations
      for the entire parameter scan as a function of $Q_i/Q_{\mathrm{gB}}$:
      \subref*{fig:lr_gs2} radial correlation length $l_R^{\mathrm{GS2}}$
      (Section~\ref{sec:radial_corr}),
      \subref*{fig:lz_gs2_fixed} poloidal correlation length
      $l_Z^{\mathrm{GS2}}$ keeping $k_y$ fixed to $k_y = 2 \pi / l_Z$
      (Section~\ref{sec:poloidal_corr}),
      \subref*{fig:tau_gs2} correlation time $\tau_c^{\mathrm{GS2}}$
      (Section~\ref{sec:time_corr}), and
      \subref*{fig:n_gs2} RMS density fluctuations $(\delta n_i /
      n_i)^{\mathrm{GS2}}_{\mathrm{rms}}$ (Section~\ref{sec:rms_density}),
      where the dashed line indicates the scaling \eqref{q_scaling}.
    }
    \label{fig:gs2_q_scatter1}
  \end{figure}
  \begin{figure}[t]
    \centering
    \begin{subfigure}[t]{0.49\textwidth}
      \includegraphics[width=\linewidth]{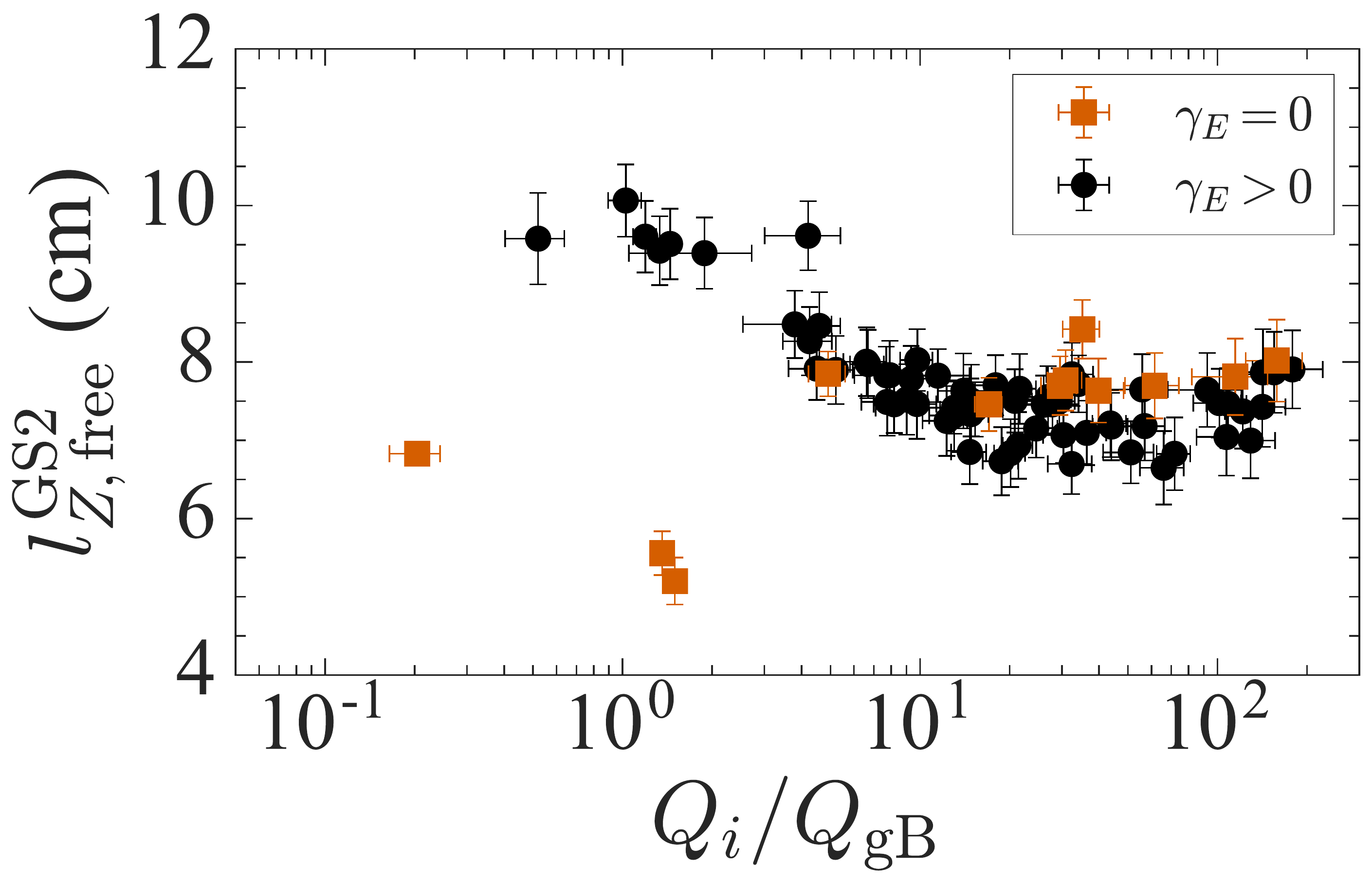}
      \caption{}
      \label{fig:lz_free_q}
    \end{subfigure}
    \begin{subfigure}[t]{0.49\textwidth}
      \includegraphics[width=\linewidth]{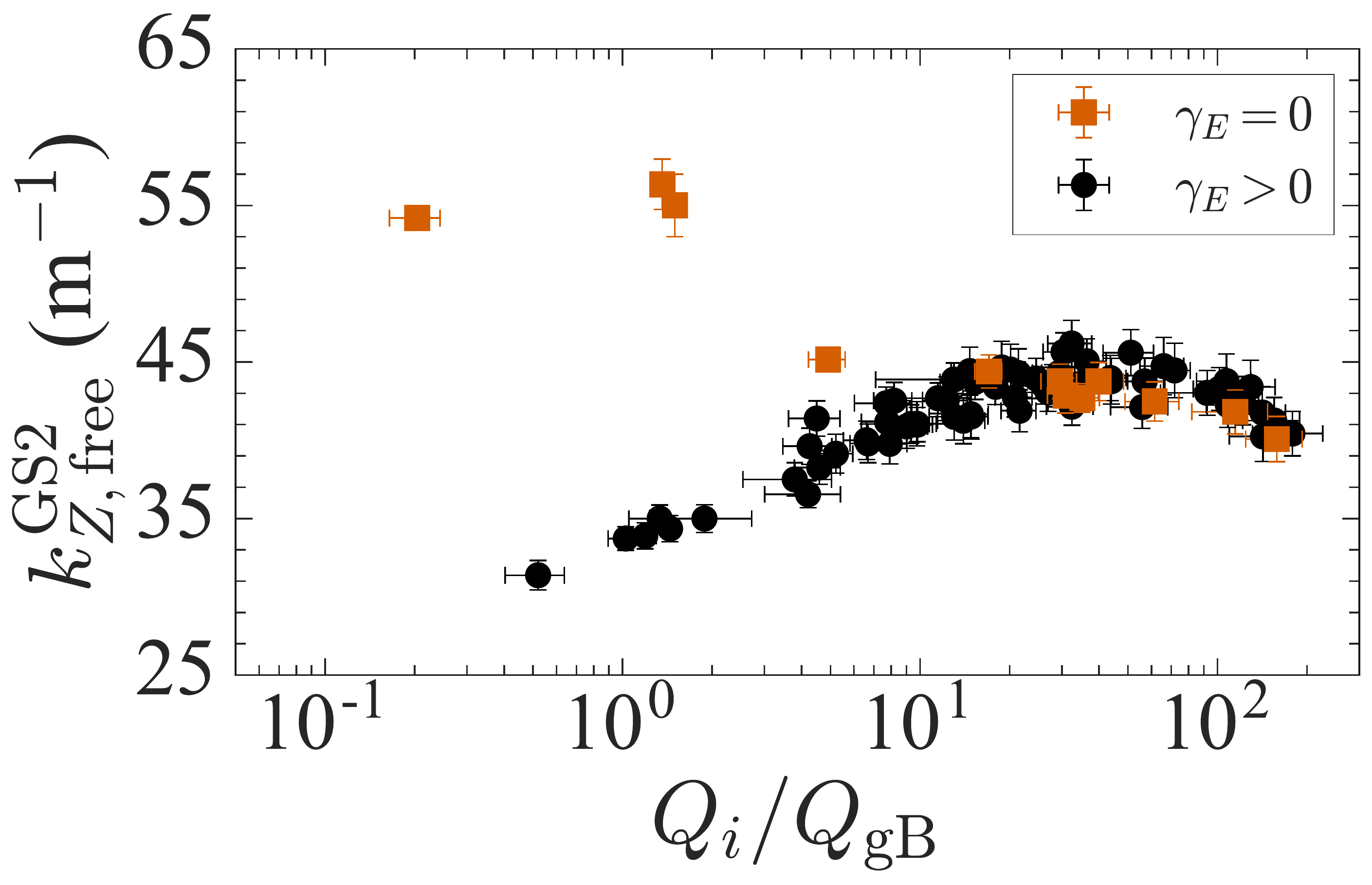}
      \caption{}
      \label{fig:kz_free_q}
    \end{subfigure}
    \\
    \begin{subfigure}[t]{0.49\textwidth}
      \includegraphics[width=\linewidth]{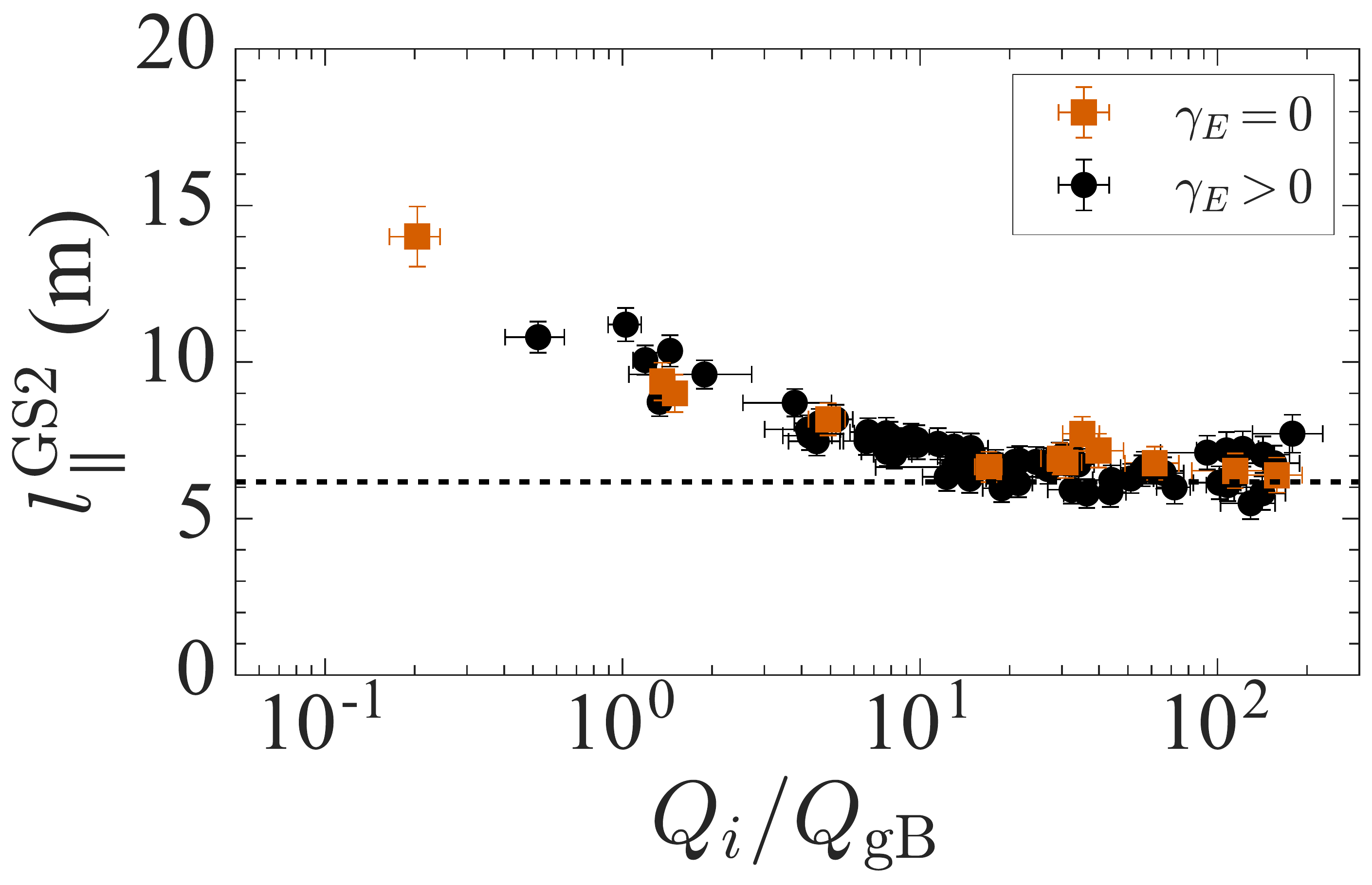}
      \caption{}
      \label{fig:lpar_q}
    \end{subfigure}
    \begin{subfigure}[t]{0.49\textwidth}
      \includegraphics[width=\linewidth]{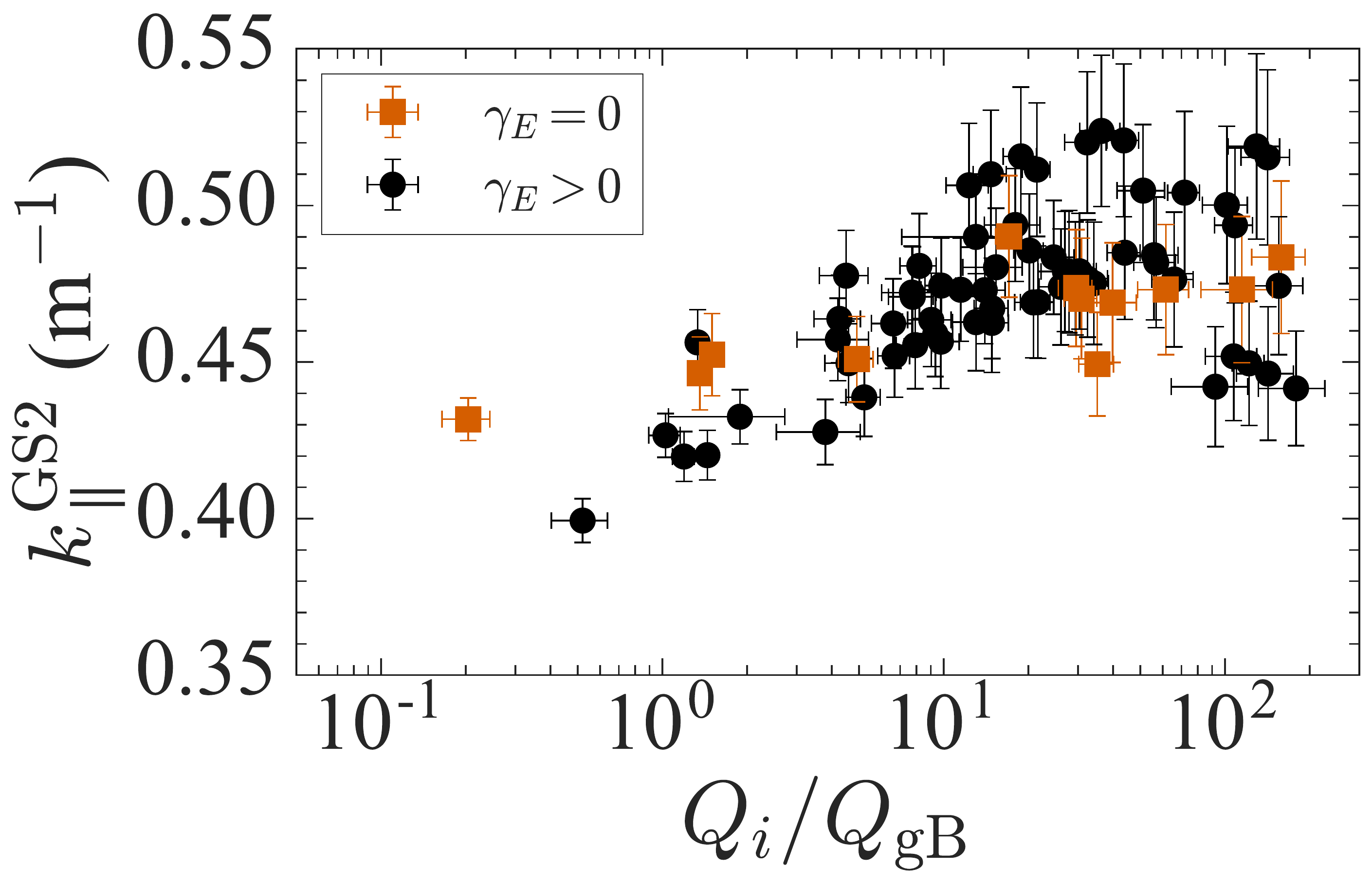}
      \caption{}
      \label{fig:kpar_q}
    \end{subfigure}
    \caption[Correlation parameters for GS2 density fluctuations versus
             $Q_i/Q_{\mathrm{gB}}$ (cont'd)]{
      Correlation parameters calculated for raw GS2 density fluctuations
      for the entire parameter scan as a function of $Q_i/Q_{\mathrm{gB}}$:
      \subref*{fig:lz_gs2_free} poloidal correlation length
      $l_{Z,\mathrm{free}}^{\mathrm{GS2}}$ with $k_y$ as a free fitting
      parameter, and
      \subref*{fig:kz_gs2} poloidal wavenumber
      $k_{Z,\mathrm{free}}^{\mathrm{GS2}}$ (Section~\ref{sec:poloidal_corr}),
      \subref*{fig:lpar_gs2} parallel correlation length
      $l_{\parallel}^{\mathrm{GS2}}$, and
      \subref*{fig:kpar_gs2} parallel wavenumber
      $k_{\parallel}^{\mathrm{GS2}}$ (Section~\ref{sec:par_corr}). The dashed
      line in \subref*{fig:lpar_gs2} indicates a line of $l_\parallel \sim qR$
      (see main text).
    }
    \label{fig:gs2_q_scatter2}
  \end{figure}

  Figures~\ref{fig:gs2_q_scatter1} and~\ref{fig:gs2_q_scatter2} show the
  correlation parameters from Figures~\ref{fig:gs2_corr_results1}
  and~\ref{fig:gs2_corr_results2} as functions of $Q_i/Q_{\mathrm{gB}}$ for our
  entire parameter scan, including $\gamma_E = 0$.  These figures clearly show
  that it is distance from threshold that determines the structure of
  turbulence and characterise this structure for realistic MAST configuration
  and for a large range of $Q_i/Q_{\mathrm{gB}}$. We start by discussing the
  $\gamma_E>0$ cases, which we can characterise as follows.

  We see a roughly monotonic increase in the radial correlation length
  $l_R^{\mathrm{GS2}}$ [\figref{lr_q}], which is consistent with an increasing
  $Q_i/Q_{\mathrm{gB}}$ because the formation of larger radial structures is
  one way the turbulence can transport heat more effectively.

  \Figref{lz_q} [along with figures~\ref{fig:lz_free_q}
  and~\subref{fig:kz_free_q}] shows the poloidal correlation length
  $l_Z^{\mathrm{GS2}}$ decreasing (and the corresponding wavenumber
  $k_Z^{\mathrm{GS2}}$ increasing) with increasing $Q_i/Q_{\mathrm{gB}}$.
  Again, this is consistent with an increasing $Q_i/Q_{\mathrm{gB}}$, where
  structures which are poloidally thin (large $k_Z$) are the most efficient at
  transporting heat out of the plasma, according to~\eqref{q_scaling} (given
  $k_Z \propto k_y$). However, an increase in amplitude may also lead to
  increased heat transport and so radially elongated and poloidally thin
  turbulent structures are not necessarily expected for turbulence in general.
  \Figref{lz_q} shows that $l_Z^{\mathrm{GS2}}$ decreases to roughly $14$~cm
  for $Q_i/Q_{\mathrm{gB}} \sim O(10)$ and possibly starts \emph{increasing}
  again for $Q_i/Q_{\mathrm{gB}} \sim O(100)$. Theoretical and numerical
  estimates  of the scaling of $l_Z$ far from the turbulence threshold
  suggested that $l_Z \sim q \kappa_T$~\cite{Barnes2011}. While our data shows
  that the value of $l_Z^{\mathrm{GS2}}$ increases at large
  $Q_i/Q_{\mathrm{gB}}$, further simulations at higher $\kappa_T$ are necessary
  to confirm whether our simulations adhere to this scaling.

  The RMS density fluctuations $(\delta n_i /
  n_i)^{\mathrm{GS2}}_{\mathrm{rms}}$ [\figref{n_q}] increase as
  $(Q_i/Q_{\mathrm{gB}})^{1/2}$ far from threshold, as expected from the
  scaling \eqref{q_scaling}. However, in contrast to the results in
  \figref{amplitude}, we do not see a flattening of $( \delta n_i /
  n_i)^{\mathrm{GS2}}_{\mathrm{rms}}$ at low $Q_i/Q_{\mathrm{gB}}$ (as in
  \figref{amplitude}, where we plotted the \emph{maximum} amplitude), for
  $\gamma_E>0$ simulations. This is due to the relatively little volume taken
  up by the coherent structures and, hence, their small contribution to the RMS
  value. We verified this by calculating the RMS density fluctuations while
  excluding varying amounts of the turbulence structures (near the threshold)
  and found that the RMS value did not change very much, showing that for the
  cases near the threshold the RMS value is dominated by the low-amplitude
  density fluctuations.

  Finally, we see that the parallel correlation length
  $l_\parallel^{\mathrm{GS2}}$ [\figref{lpar_q}] decreases as the system is
  taken away from the turbulence threshold.  Estimates of $l_\parallel$ for
  strongly driven ITG turbulence~\cite{Barnes2011} suggested that $l_\parallel$
  should be proportional to the connection length, i.e., $l_\parallel \sim \pi qR$.
  This estimate is indicated by the dashed line in \figref{lpar_q}, and we see
  that, indeed, $l_\parallel$ is of the order of the connection length.

  We have highlighted cases for which $\gamma_E=0$ (red) and $\gamma_E>0$
  (black) in \figsref{gs2_q_scatter1}{gs2_q_scatter2} to highlight two
  important features of sheared versus unsheared turbulence previously
  discussed in Section~\ref{sec:struc_analysis}. First, close to the turbulence
  threshold, the cases with $\gamma_E=0$, represent a different regime of
  turbulence to those cases with $\gamma_E>0$. In particular,
  $l_Z^{\mathrm{GS2}}$ shown in \figref{lz_q} [as well as
  Figures~\ref{fig:lz_free_q} and~\subref{fig:kz_free_q}], shows an \emph{increasing}
  trend for cases with $\gamma_E=0$: from $\sim 10$~cm near the turbulence
  threshold to $\sim 15$~cm far away from it, whereas cases with $\gamma_E>0$
  \emph{decrease} from $\sim 23$~cm near marginality to $\sim 15$~cm far away
  from it. This represents a different dependence on $Q_i/Q_{\mathrm{gB}}$ as
  well as showing a significantly lower value of $l_Z^{\mathrm{GS2}}$ at
  experimentally relevant $Q_i/Q_{\mathrm{gB}}$ ($=2 \pm 1$).
  \Figref{tau_q} shows that $\tau_c^{\mathrm{GS2}}$ predicted by
  $\gamma_E=0$ simulations stays roughly constant over a large range of
  $Q_i/Q_{\mathrm{gB}}$ whereas for $\gamma_E>0$ simulations,
  $\tau_c^{\mathrm{GS2}}$ diminishes rapidly for small $Q_i/Q_{\mathrm{gB}}$.
  Secondly, we see that far from the threshold, the
  $\gamma_E=0$ and $\gamma_E>0$ cases for \emph{all} correlation parameters
  show the same dependence on $Q_i/Q_{\mathrm{gB}}$. This shows that far from
  the threshold there is little difference between sheared and unsheared (by a
  background flow) turbulence. This result is consistent with the results in
  Section~\ref{sec:zf_shear}, further confirming the conclusions reached in
  Section~\ref{sec:zf_shear}: close to the turbulence threshold the background
  flow shear has a significant effect on the turbulence leading to reduced heat
  transport (as shown in Chapter~\ref{sec:nl}), whereas far from the threshold
  the turbulence is much like conventional ITG-driven turbulence in the absence
  of flow shear. This has been studied in related work~\cite{Fox2016a}, which
  attempted to argue a similar case in terms of symmetry breaking of
  fluctuation spectra close to the threshold in the presence of flow shear.
  Far from the threshold, however, the symmetry is effectively restored, and
  resembles turbulence in the absence of flow shear.
\section{Summary}

  In this chapter, we made quantitative comparisons between our GS2 simulations
  and the experimental measurements from the BES diagnostic. We first presented
  an overview of the correlation techniques in Section~\ref{sec:corr_overview},
  developed in Ref.~\cite{Ghim2013}, to measure the properties of turbulence
  from density fluctuations and extended the correlation analysis to the
  parallel direction, in which it is not currently possible to measure density
  fluctuations in order to calculate correlation lengths. The results from
  BES diagnostic measurements~\cite{Field2014} were presented in
  Section~\ref{sec:corr_exp}.

  In Section~\ref{sec:corr_synth}, we presented the first of our two
  correlation analyses, which looked strictly at simulations with equilibrium
  parameters within the experimental uncertainty ranges, we applied a synthetic
  diagnostic to the GS2 density-fluctuation fields before performing a
  correlation analysis exactly like the one used on experimental data. We
  showed reasonable agreement between our simulations and the BES measurements in the
  poloidal correlation length and correlation time (a major improvement
  compared to previous attempts at measuring this quantity). We also found that
  the radial correlation length was likely below the resolution threshold of
  the BES diagnostic. We showed agreement for the RMS density fluctuation amplitude
  within the experimental uncertainties of $\kappa_T$ and $\gamma_E$; however,
  this was at values of the equilibrium parameters far from those found to be
  relevant to the experiment, i.e., far from the turbulence threshold.

  In Section~\ref{sec:corr_gs2}, we performed a correlation analysis of the raw
  GS2 density fluctuations. We first presented the results within the
  experimental-uncertainty ranges of $\kappa_T$ and $\gamma_E$ and showed the
  following. We confirmed that the radial correlation tended to
  be below the resolution threshold of the BES diagnostic and showed that the
  poloidal correlation length and correlation times were comparable to both the
  results with a synthetic diagnostic applied and the experimental results. The
  effect of the synthetic diagnostic and associated PSFs was to reduce the
  measured density fluctuation amplitude compared to the raw GS2 density
  fluctuations. We compared the results from our two correlation analyses and
  experimental measurements and showed reasonable agreement across all the
  correlation properties of turbulence.

  Calculating the nonlinear decorrelation time, we confirmed in
  Section~\ref{sec:time_scales} that the onset of subcritical turbulence
  requires that the transient-growth time be approximately greater than the
  nonlinear interaction time in a given simulation. Furthermore, we showed that
  nonlinear interaction time tends to be much greater than the correlation
  times -- in agreement with the experimental results in Ref.~\cite{Field2014}.

  Finally, we showed that the correlation
  properties of the turbulence in our simulations are effectively determined by
  how far the system is from the turbulence threshold; quantified by the ion
  heat flux $Q_i/Q_{\mathrm{gB}}$. This was consistent with the results shown
  in Sections~\ref{sec:subcritical} and~\ref{sec:zf_shear}, which showed that
  the number of structures, their maximum amplitude, and the relative
  importance of zonal flows were also effectively functions of
  $Q_i/Q_{\mathrm{gB}}$. Presenting the data in this way highlighted two
  important properties of the turbulence:
  \begin{inparaenum}[(i)]
    \item close to the turbulence threshold, the background flow shear has a
      significant effect on the properties, and
    \item far from the threshold, the properties of sheared and unsheared
      turbulence were similar.
  \end{inparaenum}

\chapter{Conclusions}
\label{sec:conclusion}

We have simulated the conditions inside MAST discharge \#27274 using local
gyrokinetic simulations and performed a systematic parameter scan in the
ion-temperature-gradient length scale $\kappa_T$ and the flow shear $\gamma_E$. We
have demonstrated in Section~\ref{sec:heat_flux} that, within experimental
uncertainty, simulations reproduce the experimental ion heat flux and that the
experimentally measured equilibrium gradients lie close to the turbulence
threshold inferred from the simulations (see \figref{contour_heatmap}).
Importantly, this is one of the first numerical demonstrations that a MAST
plasma is close to the turbulence threshold.  The parameter scan performed in
this work has clearly shown that $\kappa_T$ and $\gamma_E$ are useful control
parameters, in agreement with several previous experimental and numerical
studies~\cite{Dimits1996, Mantica2009, Ritz1990, Burrell1997}.

We have shown in Section~\ref{sec:subcritical}, that the system is subcritical
for $\gamma_E>0$, i.e., finite initial perturbations, which we assume are
generated by the experiment, are required in order to achieve a saturated
nonlinear state. Subcriticality is a defining feature of this system: for
$\gamma_E>0$, even the largest values of $\kappa_T$ that we considered required
large initial perturbations to ignite turbulence. Using linear and nonlinear
simulations, we have estimated the conditions necessary for the onset of
subcritical turbulence: we require that maximum transient-amplification factor
be $N_{\gamma,\max} \gtrsim 0.4$ (see \figref{max_trans_amp}), and that the
transient-growth time $t_0$ be approximately greater than the nonlinear
interaction time, i.e., $t_0 \gtrsim \tau_{\mathrm{NL}}$
(Section~\ref{sec:time_scales}).  These conditions were comparable to those in
previous work for simpler systems~\cite{Schekochihin2012}.
Furthermore, we have showed that the linear dynamics do not show significant
changes as the turbulence threshold is passed, and so nonlinear simulations are
essential in predicting the exact onset of subcritical turbulence.

Our simulations have shown that, near the turbulence threshold, a previously
unreported turbulent state exists in which fluctuation energy is concentrated into
a few coherent, long-lived structures, which have a finite minimum amplitude
(Section~\ref{sec:coherent_strucs}). We have argued that this phenomenon is due
to the subcriticality of the system, which cannot support arbitrarily
small-amplitude perturbations (as in supercritical turbulence).  We have investigated
the changes in the nature of these nonlinear structures by tracking the maximum
fluctuation amplitude (Section~\ref{sec:max_amp}) and the number of structures
(Section~\ref{sec:struc_count}) as we changed our equilibrium parameters, and
have shown the following. Near the turbulence threshold, the system is
comprised of just a few finite-amplitude structures. As the system is taken away from
the turbulence threshold, the number of these structures increases (at
constant amplitude). Upon increasing in number sufficiently to fill the spatial
simulation domain, they begin to increase in amplitude (at a roughly constant
number of structures) (see \Figsref{amplitude}{nblobs}). Interestingly, the
evolution of our system as the system is taken away from the turbulence
threshold is reminiscent of the transition to subcritical turbulence via
localised structures in pipe flows~\cite{Barkley2015}.  We have further shown
that, in contrast to conventional ITG-driven turbulence regulated by
zonal flows~\cite{Dimits2000} (and their associated shear), in our system,
close to the turbulence threshold, the shear due to the mean toroidal flow
dominates over the shear due to the zonal flows.  We have shown that
the experimental gradients lie close to the threshold, meaning that it is
essential to include the background flow shear in simulations of MAST plasmas.
Only reasonably far from the turbulence threshold does the effect of the zonal
shear and the flow shear due to the background flow become comparable (see
\figref{zf_shear}), and further still the turbulence becomes similar to
ITG-driven turbulence in the absence of background flow shear.

We have made quantitative comparisons between density fluctuations in our
simulations and those measured by the MAST BES diagnostic~\cite{Field2009,
Field2012} (Section~\ref{sec:struc_of_turb}). A correlation
analysis~\cite{Ghim2012} was previously performed on the measurements of
density fluctuations from the BES diagnostic~\cite{Field2014}
(Section~\ref{sec:corr_exp}), giving the following properties of the
turbulence: the radial correlation length $l_R$, the poloidal correlation
length $l_Z$, and the correlation time $\tau_c$. We have performed two types of
correlation analysis on our simulated density fluctuations: one after applying
a synthetic BES diagnostic (Section~\ref{sec:corr_synth}), and one directly on
the raw GS2-generated density fluctuations (Section~\ref{sec:corr_gs2}). We
have compared these results to experimental measurements and achieved reasonable
agreement of the correlation lengths, time, and amplitude measurements, except
for the radial correlation length, which was predicted by us to be lower than
the resolution limit of the BES diagnostic.  Notably, the simulated and
experimentally measured correlation times were in good agreement, unlike in
previous global, gyrokinetic simulations of the same MAST
discharge~\cite{Field2014}.

Finally, we have shown that the nature of the turbulence is effectively a
function of the distance from the turbulence threshold [for example, see
Figures~\ref{fig:amplitude}, \ref{fig:nblobs}, \ref{fig:zf_shear_q_scatter},
\ref{fig:gs2_q_scatter1}, and \ref{fig:gs2_q_scatter2}]. We have quantified
this distance from threshold via the ion heat flux $Q_i/Q_{\mathrm{gB}}$, and
have shown that it is this quantity, rather than the specific values of the
equilibrium parameters $\kappa_T$ and $\gamma_E$, that determines the
properties of the turbulence. Throughout this work, we have presented our data
as functions of the distance from threshold to highlight the two distinct
turbulence regimes that we have identified. Close to the threshold, where
coherent structures dominate the dynamics, and far from the threshold, where
the turbulence appears to be similar to conventional strongly driven ITG
turbulence in the absence of flow shear. It is important to note that the
experiment is located at the boundary of these two regimes, in parameter
space, and may suggest that this boundary is most relevant to the experiment,
as opposed to the boundary separating the non-turbulent and turbulent states
--- the so-called ``zero-turbulence manifold''~\cite{Highcock2012}.

Using the local gyrokinetic code GS2, we have been able to reproduce both the
experimental heat flux and the quantitative measurements of turbulence
obtained using the BES diagnostic. This has given us confidence in
our simulations and has allowed us to trust some conclusions from them that do
not (yet) have direct experimental backing. More broadly, we have gained
confidence in the future use of local gyrokinetic simulations in predicting
turbulence and transport in high-aspect-ratio spherical tokamaks such as MAST.

\section{Future directions}

The most interesting experimental question that has arisen from this study is
about the existence of the long-lived, coherent structures near the turbulence
threshold, which support heat fluxes that are experimentally relevant.  Given
that these structures occur at ion scales, the BES diagnostic is well-suited
for detecting them.  However, as we have found in this investigation, the
``spike filter'', which plays an important role in cleaning experimental data
of high-energy radiation, may complicate the detection of these structures,
since it may erroneously remove long-lived, poloidally fast-moving structures.
Currently, the ``spike filter'' is a simple and efficient algorithm to remove
any spike in the emission above a certain threshold; however, future work might
involve more carefully filtering out only high-energy radiation and ensuring
that high-intensity emission that is correlated in time or across detectors
(such as that produced by a fast moving structure) is not overlooked. It might
also be possible to investigate the existence of structures statistically.
Recent work on this question has provided some tentative but encouraging
indications that a regime dominated by isolated structures might manifest
itself in experimentally observed skewed probability distributions of density
fluctuations \cite{Fox2016a}. Clearly, further more extensive analysis of MAST
BES measurements is needed.

In addition to detecting the coherent structures in experiments, it may be
useful to attempt to formulate an analytical description of their structure and
behaviour. Our simulations were of a realistic experimental configuration;
however, it may be possible to observe them in simpler systems and in this way
identify the key physical mechanisms that give rise to them. Our
simulations have identified the flow shear as a key physical mechanism and that
the relevant part of parameter space where the structures are found, is close
to the turbulence threshold. However, open questions remain regarding, for
example, the importance of the MAST geometry, the influence of dissipation
mechanisms such as collisions, and the role played by electron-scale
turbulence.

In this work we have identified two regimes of turbulence: a
coherent-structure-dominated regime and a more conventional ITG-turbulence
regime. Future studies could attempt to more precisely identify the criteria
that define the boundary between the two regimes, since it may be this boundary
that is most relevant to experiments, as is the case for the system we have
investigated.

Finally, we may ask: how universal are the turbulence regimes that we have
identified? First, with respect to other fusion devices and secondly, with
respect to other subcritical systems. We have shown in
Section~\ref{sec:subcritical} that even turbulence that has reached a saturated
state may still be quenched at a seemingly unpredictable time.  Previous work
on subcritical systems in neutral fluid flow down a
pipe~\cite{Hof2006,Avila2011} and Keplerian magnetorotational accretion
flows~\cite{Rempel2010} have shown (using large numbers of experiments and/or
numerical simulations) that subcritical turbulence has a finite life time and
is a statistical property of the system that depends on how far the system is
from the turbulence threshold, much like the ion heat flux in our study. Most
recently, it has been shown, for neutral fluid flow down a pipe, that
subcritical turbulence has a finite life time \emph{regardless} of how far the
system is from the turbulence threshold.  Currently, our simulations are much
too expensive to carry out the number of simulations required to determine the
turbulence life time as in the above studies. However, it would be an exciting
demonstration of the universality of subcritical turbulence if the turbulence
life time could be shown to behave similarly in tokamak plasmas.

\newpage
\appendix

\chapter{Linear and nonlinear effect of hyperviscosity}
\label{App:hyperviscosity}

\begin{figure}[t]
    \centering
    \begin{subfigure}{0.49\linewidth}
      \includegraphics[width=\linewidth]{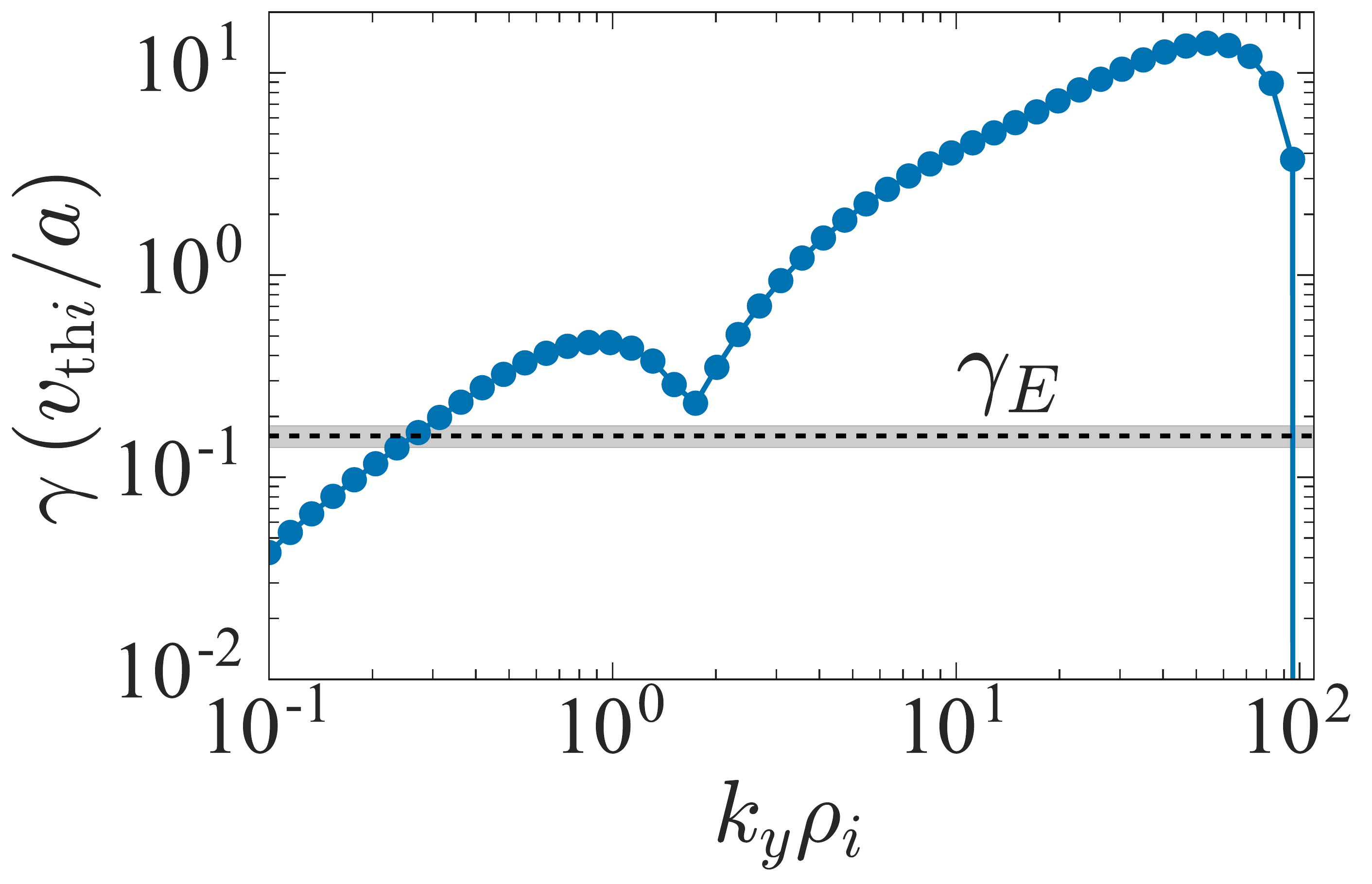}
      \caption{}
      \label{fig:gamma_vs_ky}
    \end{subfigure}
    \begin{subfigure}{0.49\linewidth}
      \includegraphics[width=\linewidth]{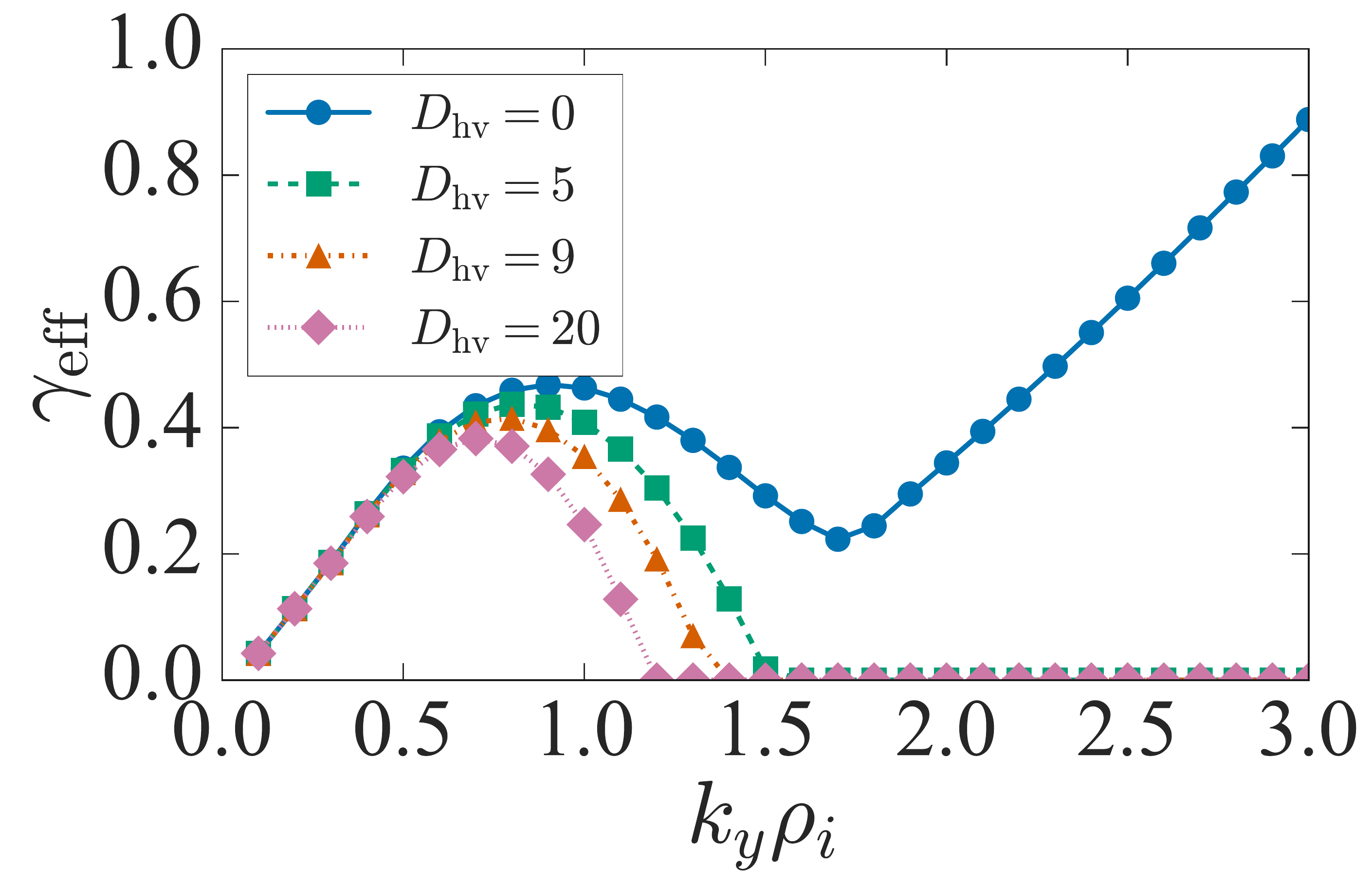}
      \caption{}
      \label{fig:growth_with_hyp}
    \end{subfigure}
    \caption[Linear effect of hyperviscosity]{
      \subref*{fig:gamma_vs_ky} Linear growth rate $\gamma$ as a function of $k_y
      \rho_i$ covering both ion and electron scales for $\kappa_T = 5.1$ and
      $\gamma_E=0$.  There is no clear scale separation between the ion and
      electron dynamics.
      \subref*{fig:growth_with_hyp} Effective linear growth
      rates $\gamma_{\mathrm{eff}}$ versus $k_y \rho_i$ at $k_x \rho_i = 0$ for
      a range of different values of $D_{\mathrm{hv}}$, calculated from
      \eqref{growth_rate_with_hyp}.  We have used $k_{y,\max} \rho_i = 3.1$
      because this was the maximum value resolved in our nonlinear simulations.
    }
  \end{figure}

  For the MAST configuration that we investigated, hyperviscosity was a key
  requirement in order for us to be able to run ion-scale-only simulations to
  saturation. To demonstrate the need for hyperviscosity, we start by
  considering the linear growth rate $\gamma$ (calculated with zero flow shear,
  $\gamma_E=0$) over a range of $k_y \rho_i$ that covers both ion ($k_y \rho_i
  \sim 1$) and electron scales ($k_y \rho_i \gtrsim 10$). This is shown in
  \figref{gamma_vs_ky}. We see that there is no clear scale separation between
  ion- and electron-scale instabilities  and, therefore, it is problematic to
  choose a maximum value of $k_y \rho_i$ at which our nonlinear simulations
  could naturally be cut off.  \Figref{gamma_vs_ky} suggests that multiscale
  simulations, covering both ion and electron scales, are required as any
  intermediate cut-off scale would lead to finite growth at the smallest
  resolved scales.

  Using equation~\eqref{hypervisc} we can determine the effect of different
  levels of hyperviscosity on linear growth rate (in the absence of flow shear)
  without running additional linear simulations.  Hyperviscosity is implemented
  as a wave-number-dependent factor applied to the distribution function at
  every time step, with the result that, in the presence of hyperviscosity,
  a perturbed quantity like $\varphi$ evolves in a linear simulations in time as
  \begin{equation}
    \varphi(t) \sim \exp \left[\left( \gamma -
      D_{\mathrm{hv}} \frac{k_\perp^4}{k_{\perp,\max}^4}\right)t\right],
    \label{phi_with_hyp}
  \end{equation}
  where $D_{\mathrm{hv}}$ is a constant
  coefficient controlling the strength of the hyperviscosity (denoted by
  \texttt{d\_hypervisc} in GS2), $k_\perp^2 = k_x^2 + k_y^2$, $k_{\perp,\max}$
  is the largest perpendicular wavenumber resolved in the simulation.
  Hence, the effective growth rate is given by
  \begin{equation}
    \gamma_{\mathrm{eff}} = \gamma -
                            D_{\mathrm{hv}} \frac{k_\perp^4}{k_{\perp,\max}^4}.
    \label{growth_rate_with_hyp}
  \end{equation}
  \Figref{growth_with_hyp} shows the effective linear growth rate, calculated
  using \eqref{growth_rate_with_hyp} as a function of $k_y \rho_i$ for $k_x
  \rho_i = 0$ for a range of values of $D_{\mathrm{hv}}$.  We have used
  $k_{y,\max} \rho_i \approx 3$, which was the maximum resolved wavenumber in
  our nonlinear simulations. The $D_{\mathrm{hv}} =0$ curve shows the need for
  hyperviscosity in our nonlinear simulations: there is no clear scale
  separation between ion ($k_y \rho_i \sim 1$) and electron scales ($k_y \rho_i
  \gtrsim 2$). Therefore, a purely ion-scale nonlinear simulation would have
  strongly growing electron modes at the smallest simulated scales, but
  wouldn't resolve the electron dissipation scale at $k_y \rho_i \sim 60$.
  Hence, hyperviscosity provides the damping needed to run an ion-scale
  simulation and stop an unphysical build up of free energy at the smallest
  scales.  In our nonlinear simulations we settle on the value $D_{\mathrm{hv}}
  = 9$ and prove later that it does not affect the transport properties.

\begin{figure}[t]
    \centering
    \begin{subfigure}[t]{0.49\textwidth}
      \includegraphics[width=\linewidth]{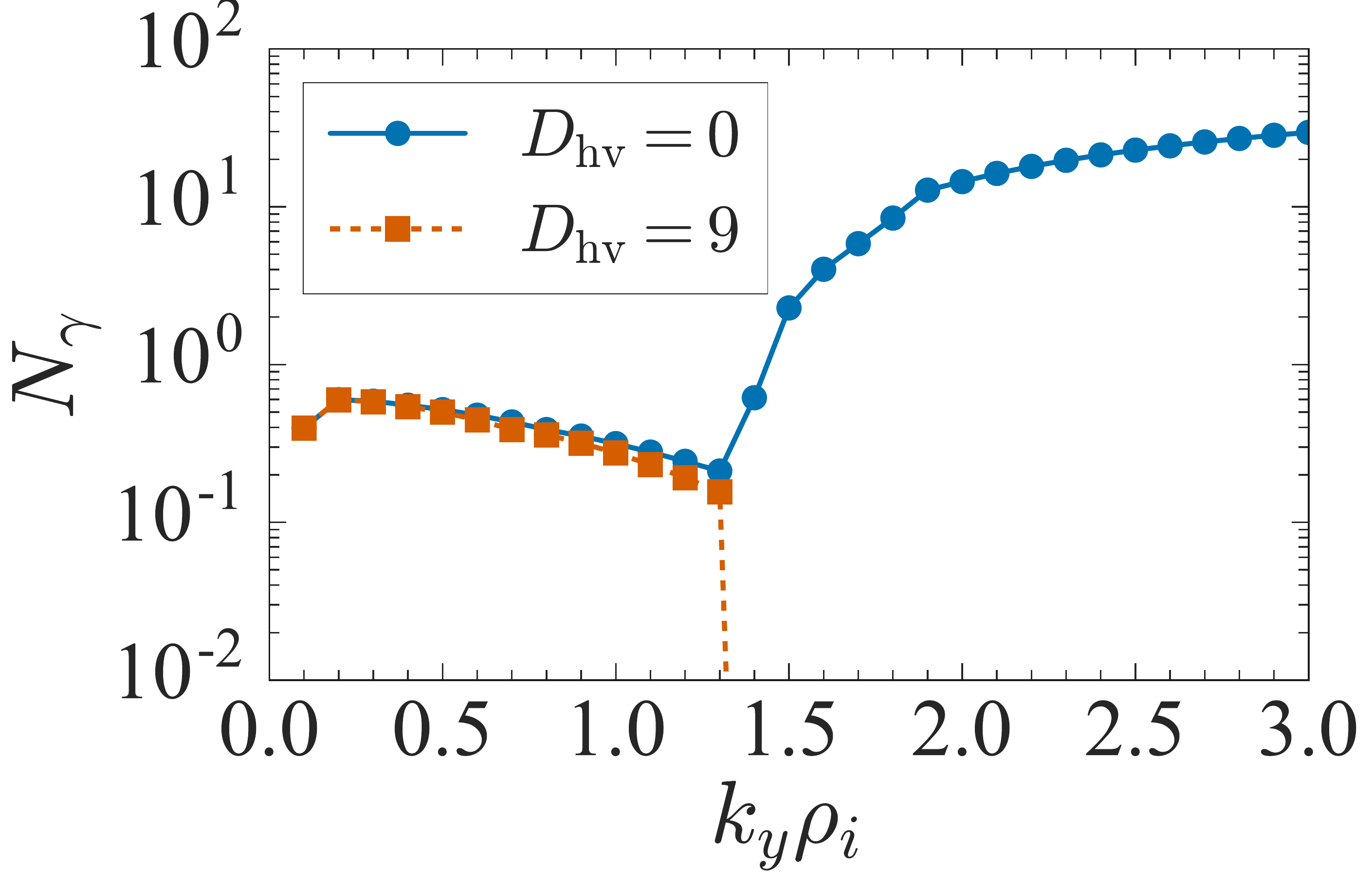}
      \caption{}
      \label{fig:N_with_hyp}
    \end{subfigure}
    \hfill
    \begin{subfigure}[t]{0.49\textwidth}
      \includegraphics[width=\linewidth]{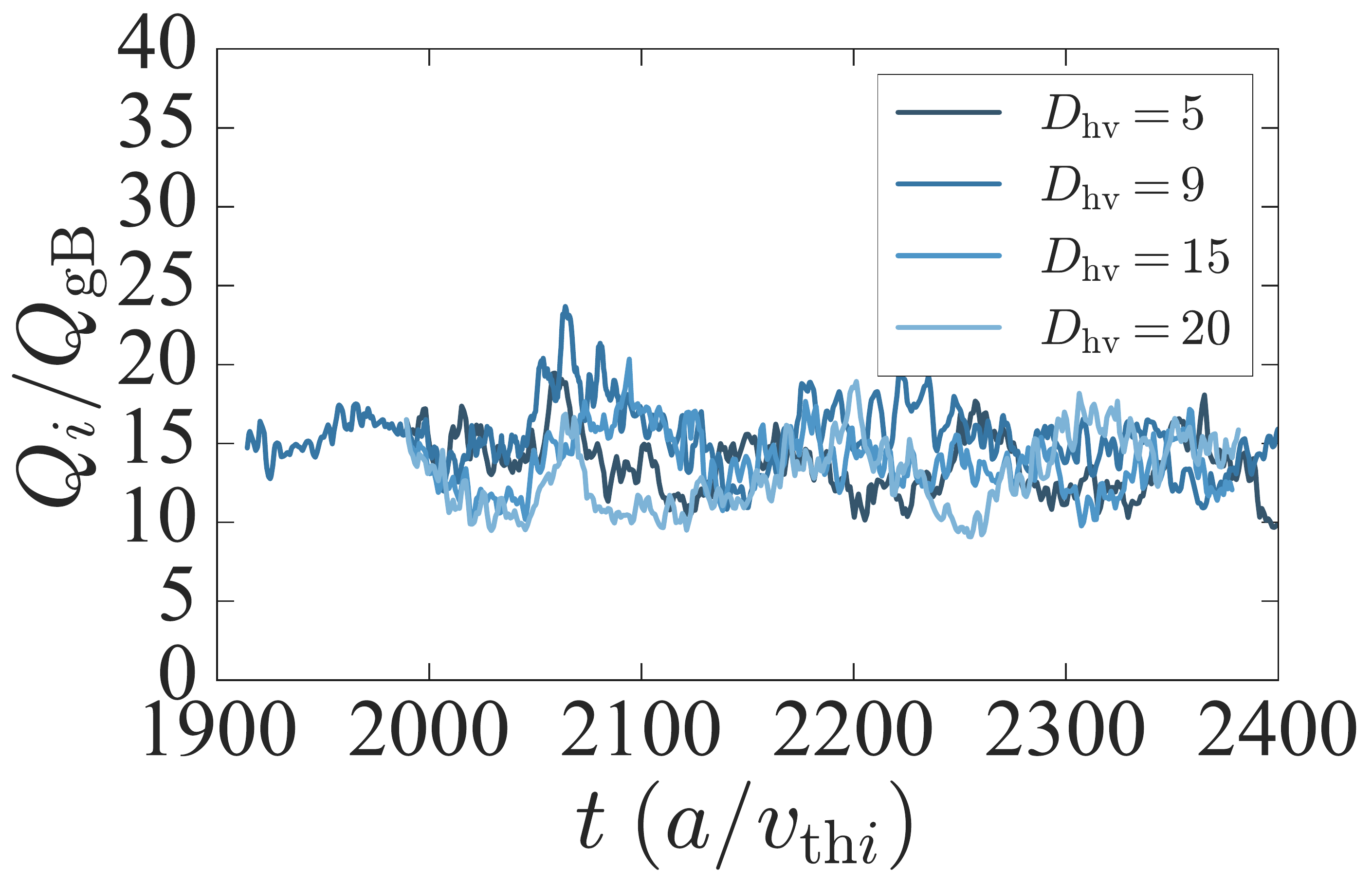}
      \caption{}
      \label{fig:hyp_scan}
    \end{subfigure}
    \caption[Effect of hyperviscosity on $N_\gamma$ and $Q_i/Q_{\mathrm{gB}}$]{
      \subref*{fig:N_with_hyp} Transient-amplification factor $N_\gamma$ (at
      $(\kappa_T,\gamma_E) = (5.1, 0.16)$ and $k_x \rho_i = 0$) for a range of
      $k_y \rho_i$ identical to that in our nonlinear simulations with
      $D_\mathrm{hv} = 0$ (blue line) and $D_\mathrm{hv} = 9$ (red line). While
      ion-scale transient growth is unaffected electron-scale modes are
      suppressed by the hyperviscosity.
      \subref*{fig:hyp_scan} Ion heat flux $Q_i/Q_{\mathrm{gB}}$ as a function
      of time for nonlinear simulations with $(\kappa_T,\gamma_E) = (5.1,0.16)$
      with $D_\mathrm{hv} =$ $5$, $15$, and $20$.
    }
    \label{fig:hypervisc}
  \end{figure}

  In the presence of flow shear, the picture is made more complicated by the
  fact that the system is subcritical; however, we are still able to study the
  effect of hyperviscosity. Setting $\gamma_E>0$, and calculating
  the transient-amplification factor $N_\gamma$, instead of $\gamma$, leads to a
  similar conclusion as for $\gamma_E=0$ simulations without hyperviscosity:
  there is no clear maximum value of $k_y \rho_i$ that would ensure there is
  no growth at the smallest scales, as shown by the blue line in
  \figref{N_with_hyp} (with $\gamma_E = 0.16$). The red line in
  \Figref{N_with_hyp} shows the effect of hyperviscosity on $N_\gamma$ [at
  $(\kappa_T,\gamma_E) = (5.1, 0.16)$ and $k_x \rho_i = 0$] for a value of
  $k_{\perp, \max}$ equal to that in our nonlinear simulations. We see that
  ion-scale transient growth is not strongly affected by the hyperviscosity
  while electron-scale transient growth is effectively damped (mainly due to
  their long transient growth time), i.e, $N_\gamma$
  goes to zero. This allowed us to choose a cut-off scale for our nonlinear
  simulations at $k_y \rho_i \sim O(1)$ and focus our attention at ion scales
  while still simulating electrons via a kinetic equation and including their
  effect on the ions.

  The key requirement when artificially removing energy from the system, as
  hyperviscosity does, is that the nonlinear saturated state should not depend
  strongly on the value of $D_\mathrm{hv}$.  \Figref{hyp_scan} shows four
  nonlinear simulations at $(\kappa_T,\gamma_E) = (5.1, 0.16)$ with different
  levels of hyperviscosity. The simulation at $D_\mathrm{hv} = 9$ was run until
  saturation and then restarted three times with different values of
  $D_\mathrm{hv}$: $D_\mathrm{hv} =$ $5$, $15$, and $20$.  \Figref{hyp_scan}
  shows that these level of $D_\mathrm{hv}$ do not affect the level of
  transport strongly while allowing our simulations to saturate.  Based on
  \figref{hyp_scan}, we have used $D_\mathrm{hv} = 9$ for all of our nonlinear
  simulations.

  In conclusion, using hyperviscosity we were able to damp high wavenumber
  dynamics and allowed us to run ion-scale-only simulations, with a cut-off
  scale around $k_y \rho_i \sim 3$. As a consequence of being limited to ion
  scales only, our simulations will miss the effects of turbulence at electron
  scales, as well as possible cross-scale coupling effects between electron and
  ion scales. Previous realistic multiscale
  studies~\cite{Howard2014,Howard2014a} have shown that these effects may
  increase the level of turbulence via the stabilisation of zonal flows by
  electron scale turbulence. However, for the purposes of this work we will
  assume that we are capturing the majority of the physics at ion scales, and
  are not introducing any artificial effects through our high-wavenumber
  cut-off.

\chapter{Resolving the effect of flow shear}
\label{App:res_flow_shear}

  In this appendix, we estimate the conditions that need to be satisfied in
  order to resolve the effect of flow shear using the results from  nonlinear
  simulations in the absence of flow shear.

  In Section~\ref{sec:flow_shear}, we showed that flow shear is implemented in
  GS2 by allowing the radial wavenumber $k_x$ to vary with time according
  to~\eqref{kx_time}, and by ``shifting'' the fluctuation fields along the $k_x$
  dimension. The frequency at which GS2 shifts the fluctuation fields in the
  $k_x$ dimension depends on the value of the radial grid spacing
  $\Delta k_x$, $\gamma_E$, and the poloidal wavenumber $k_y$. From
  \eqref{kx_time}, the time taken before the exact shift is $\Delta k_x/2$
  (at which points GS2 shifts the fluctuation fields by $\Delta k_x$ as
  explained in Section~\ref{sec:flow_shear}) is
  \begin{equation}
    \tau_{\mathrm{shift}} = \frac{\Delta k_x}{2 \gamma_E k_y}.
    \label{tau_shift}
  \end{equation}
  In order for the effect of flow shear to be considered ``resolved'', this
  shifting operation should occur at least once during the lifetime of an
  eddy, otherwise turbulence will interact and decorrelate as though the
  simulation were shearless. The turbulence decorrelation time
  $\tau_{\mathrm{NL}}$ is estimated from the correlation properties of
  turbulence via~\eqref{tau_nl}, and the condition for flow shear to be
  resolved is, therefore,
  \begin{equation}
    \tau_{\mathrm{shift}} \lesssim \tau_{\mathrm{NL}}.
    \label{shift_cond}
  \end{equation}

\begin{figure}[t]
    \centering
    \includegraphics[width=0.6\linewidth]{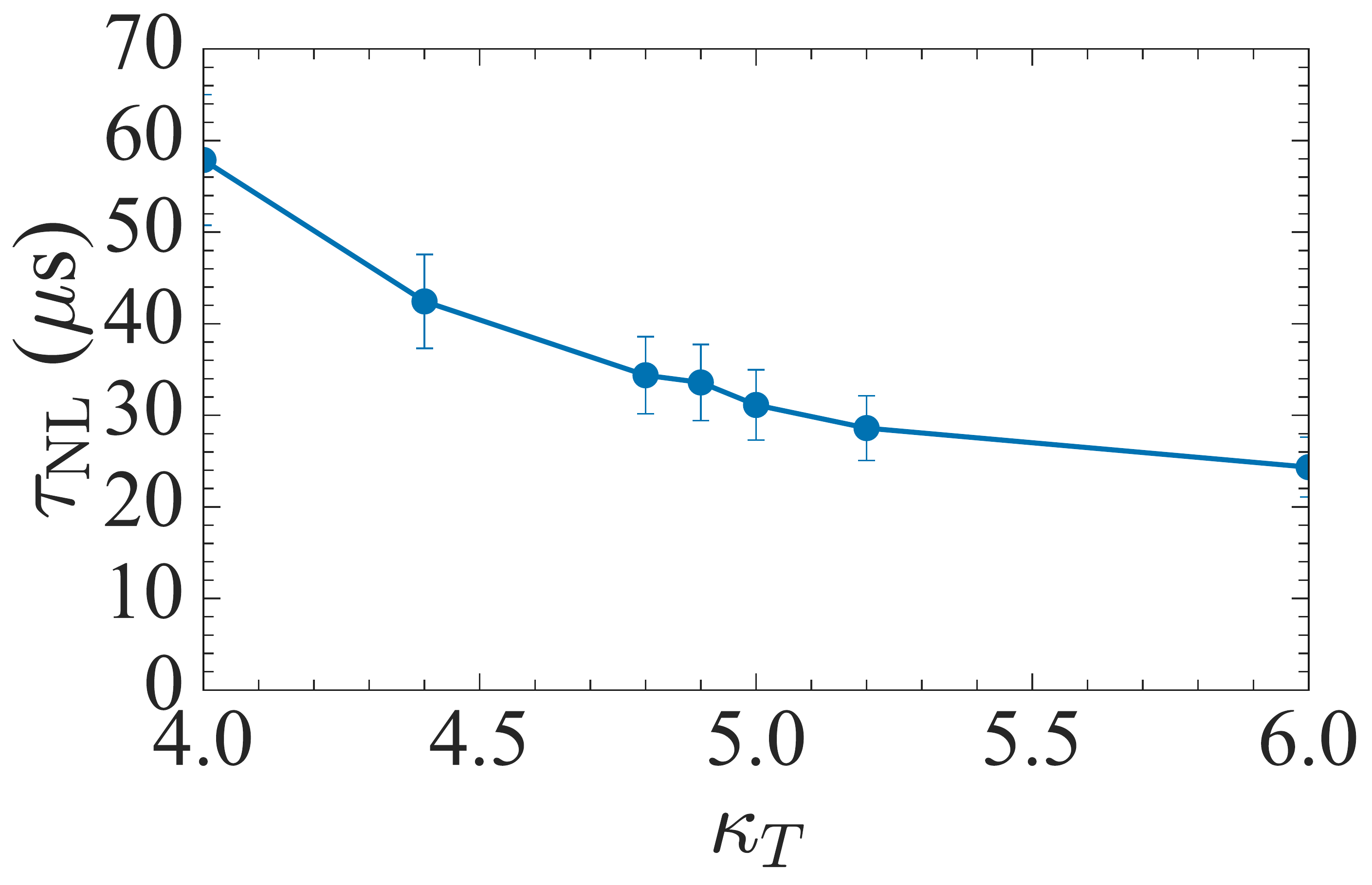}
    \caption[Nonlinear decorrelation time versus ITG]{
      Nonlinear decorrelation time $\tau_\mathrm{NL}$, calculated using
      \eqref{tau_nl}, as a function of $\kappa_T$ for simulations with
      $\gamma_E=0$.
    }
    \label{fig:taunl_tprim}
  \end{figure}

  To estimate the value of $\tau_\mathrm{NL}$ relevant to our parameter scan,
  we performed a series of nonlinear simulations at a range of different values
  of ion temperature gradient $\kappa_T$ in the absence of flow shear. The
  results are shown in \figref{taunl_tprim}, and we see that at the
  experimental value $\kappa_T=5.1$, $\tau_{\mathrm{NL}} \sim 30$~$\mu$s. We
  now want to find the approximate value of $\gamma_E$ that ensures
  \eqref{shift_cond} is satisfied, given the value of $\tau_{\mathrm{NL}}$
  above.  Returning to \eqref{tau_shift}, the radial grid spacing we employed
  in our nonlinear simulations was $\Delta k_x \approx 0.03$, and the most
  important scales in the system is $k_y \rho_i \sim 0.2$
  [see \figref{N_with_hyp}]. Using \eqref{tau_shift}, the value of $\gamma_E$
  that satisfies \eqref{shift_cond} is $\gamma_E \approx 0.08$, where values
  less than this satisfy \eqref{shift_cond} less well. Therefore, we have taken
  this to be the minimum value of flow shear for our parameter scan in this
  work.

\chapter{Linear simulations with $\gamma_E = 0$}
\label{App:linear_sims}

  In Section~\ref{sec:subcritical}, we showed that, in the presence of flow
  shear, the turbulence is subcritical. This means that one cannot easily
  define a linear growth rate for $\gamma_E>0$ simulations; however, it is
  still useful to consider the linear physics in the absence of flow shear to
  investigate which scales are important. Here, we look at the linear growth
  rates and frequencies for simulations with adiabatic and kinetic electron
  species.

\begin{figure}[t]
    \centering
    \begin{subfigure}{0.49\linewidth}
      \includegraphics[width=\linewidth]{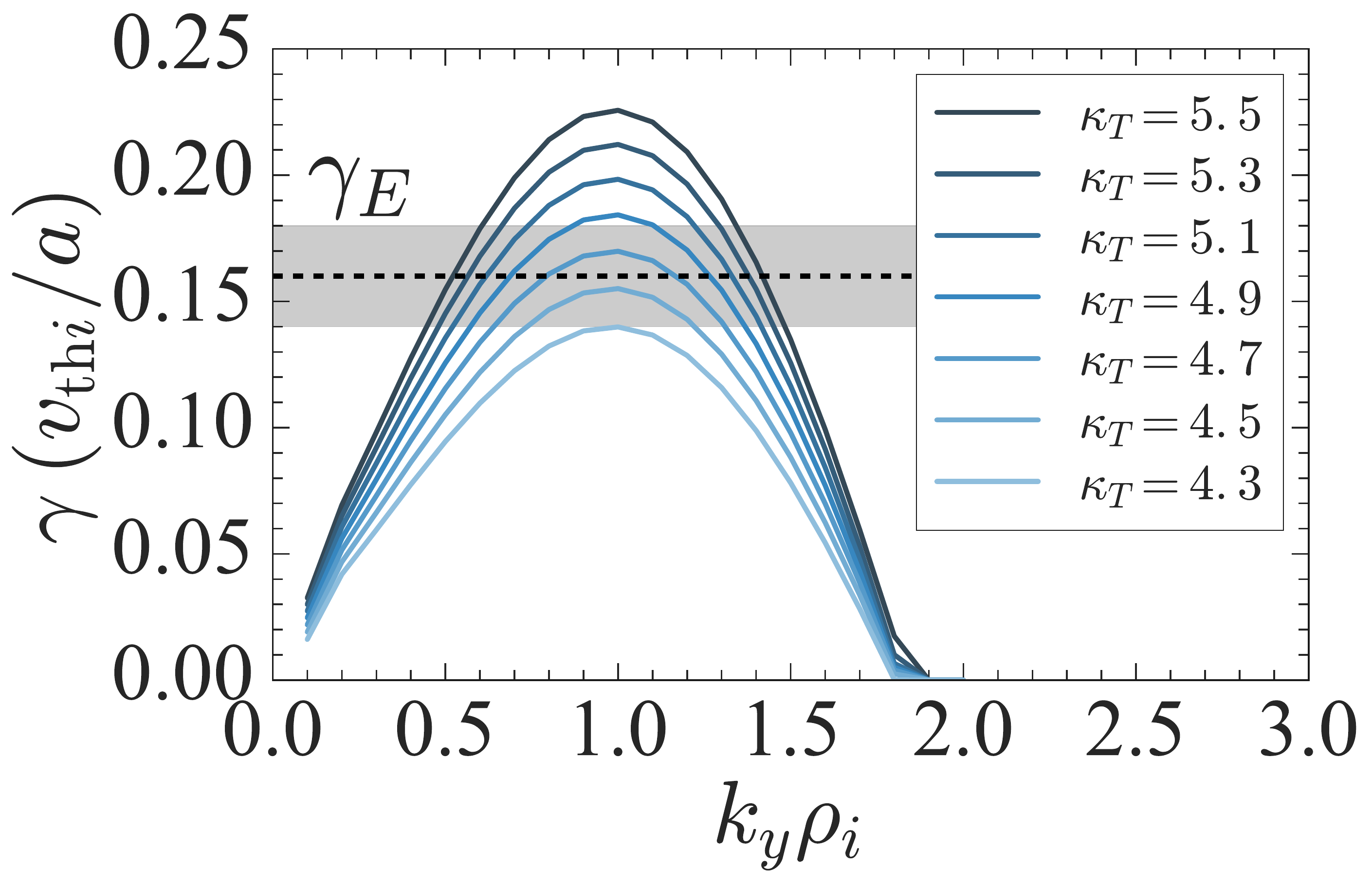}
      \caption{}
      \label{fig:gamma_nspec_1}
    \end{subfigure}
    \begin{subfigure}{0.49\linewidth}
      \includegraphics[width=\linewidth]{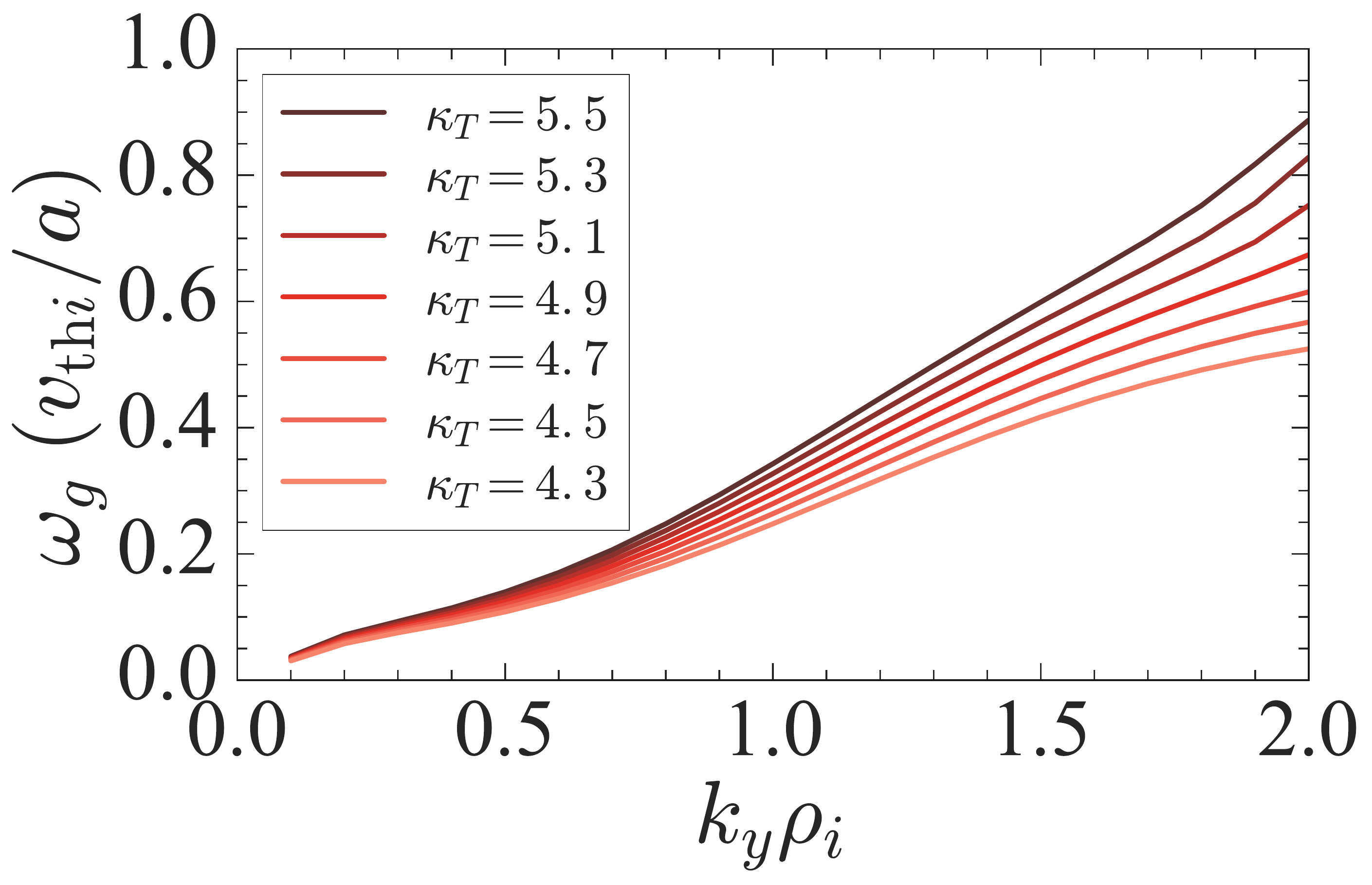}
      \caption{}
      \label{fig:omega_nspec_1}
    \end{subfigure}
    \caption[Linear growth rates and frequencies for one species, adiabatic
    electron simulations]{
      \subref*{fig:gamma_nspec_1} Linear growth rate $\gamma$ and
      \subref*{fig:omega_nspec_1} real frequency $\omega_g$ versus $k_y$ for
      simulations with a single kinetic ion species and adiabatic electrons.
      For these linear simulations plots, $k_x \rho_i = 0$. The shaded region
      shows the experimental level of flow shear $\gamma_E = 0.16 \pm 0.02$.
    }
    \label{fig:linear_nspec_1}
  \end{figure}

  In the absence of flow shear, $\varphi$ will evolve in time according to
  $\varphi \sim e^{\gamma t}$, where $\gamma$ is the linear growth rate.  We
  start by looking at $\gamma$ and real frequency $\omega_g$ versus $k_y$
  for simulations with kinetic ions and adiabatic electrons for a range of ion
  temperature gradient length scales $\kappa_T$ as shown in
  Figure~\ref{fig:linear_nspec_1}. The dashed line indicates the experimental
  value of flow shear $\gamma_E = 0.16 \pm 0.02$. We see that the flow shear is
  comparable to the maximum linear growth rate, i.e., $\gamma_E \sim
  \gamma_{\max}$. Previous numerical studies with adiabatic electrons and flow
  shear~\cite{Waltz1994} have defined the so-called ``Waltz Rule'', which
  states that ion-scale turbulence tends to be quenched when $\gamma_{\max}
  \sim \gamma_E$.  Indeed, nonlinear simulations of our system with adiabatic
  electrons and flow shear show that steady-state turbulence cannot be achieved
  for any $\kappa_T$ within the experimental error range, in agreement with the
  above quenching rule.
  \begin{figure}[t]
    \centering
    \begin{subfigure}{0.49\linewidth}
      \includegraphics[width=\linewidth]{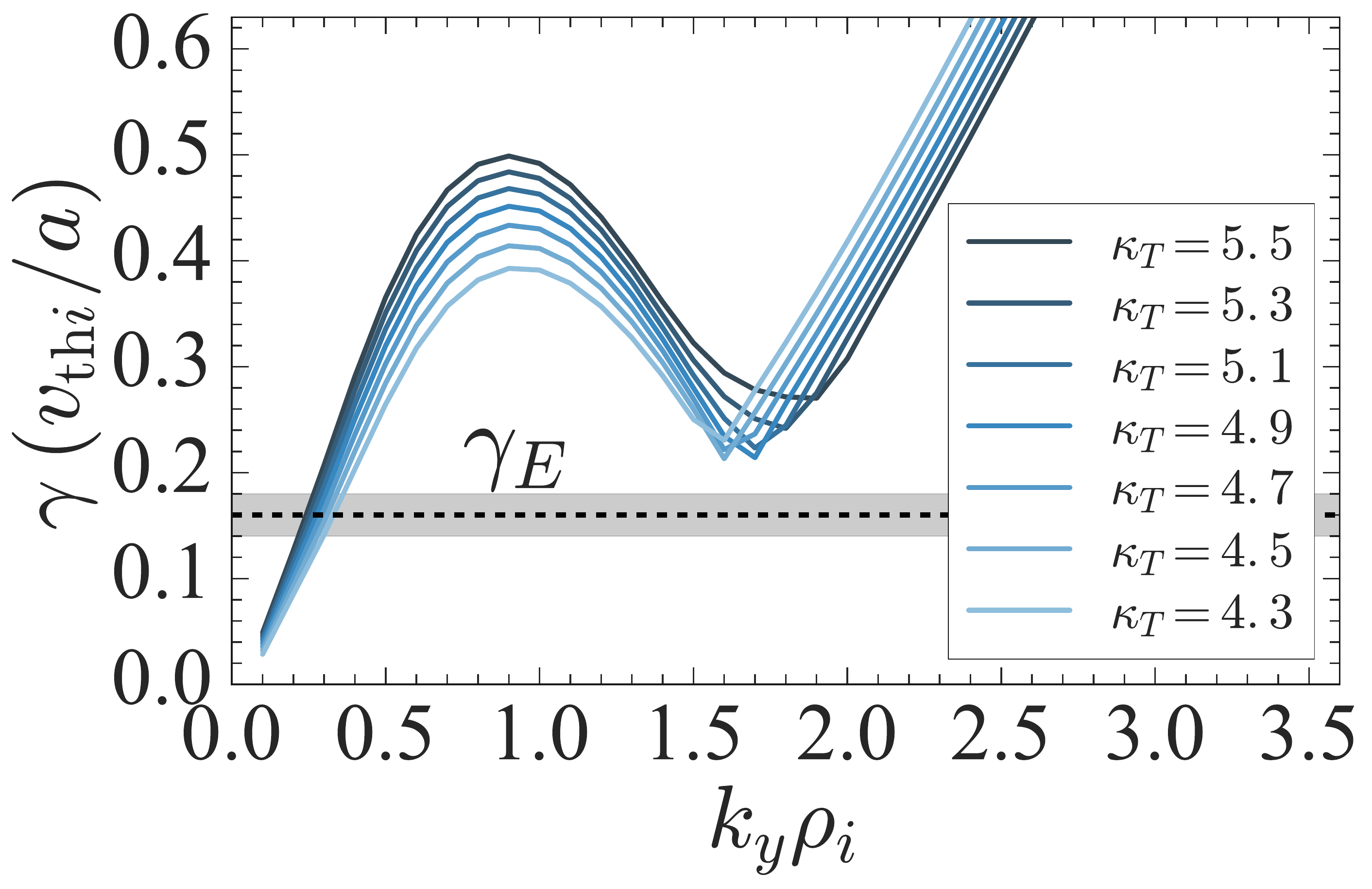}
      \caption{}
      \label{fig:gamma_nspec_2}
    \end{subfigure}
    \begin{subfigure}{0.49\linewidth}
      \includegraphics[width=\linewidth]{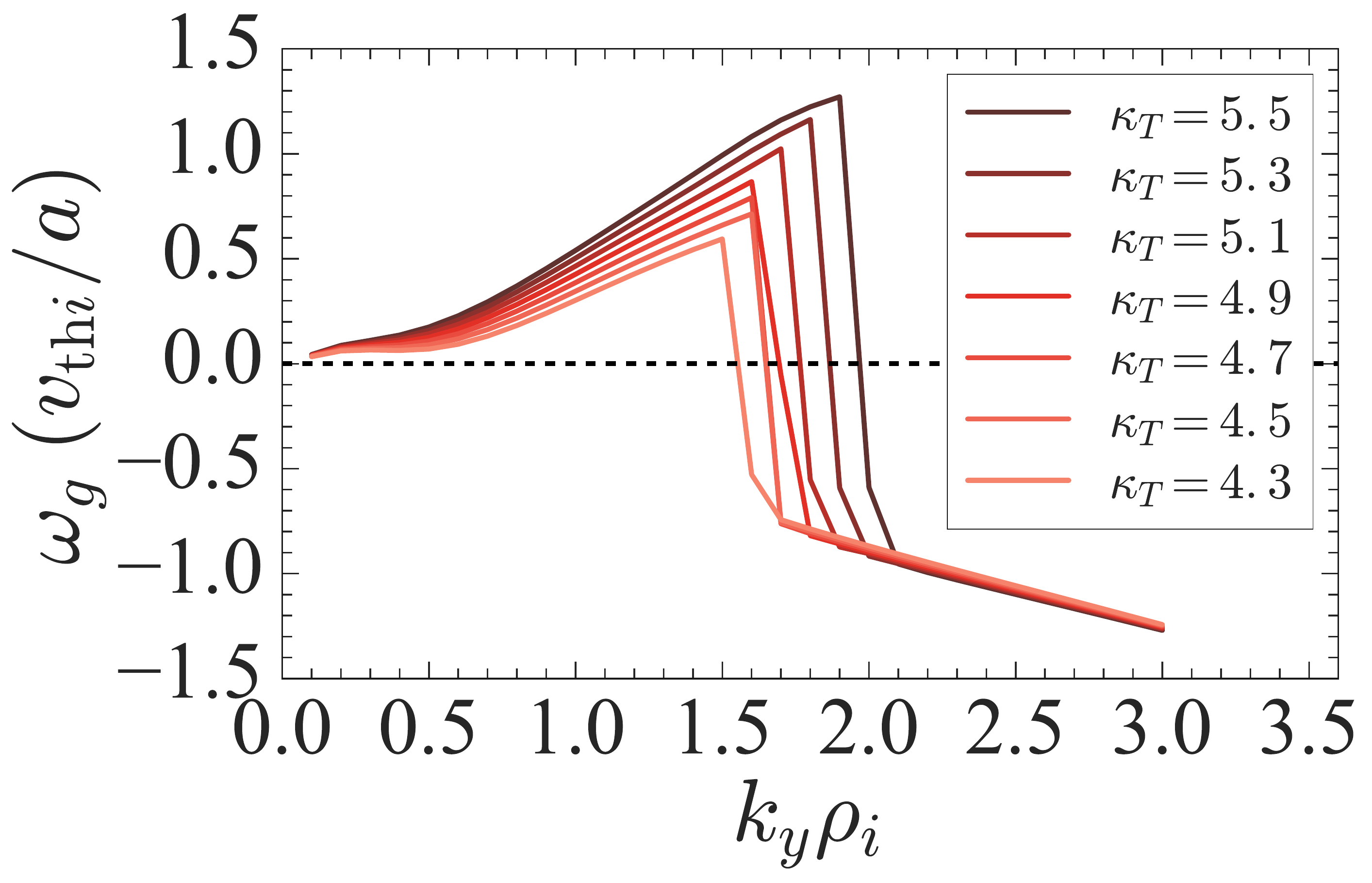}
      \caption{}
      \label{fig:omega_nspec_2}
    \end{subfigure}
    \caption[Linear growth rates and frequencies for two-species simulations]{
      \subref*{fig:gamma_nspec_2} Linear growth rate $\gamma$ and
      \subref*{fig:omega_nspec_2} real frequency $\omega_g$ versus $k_y$ for
      simulations with a kinetic ion and electron species. For these linear
      simulations plots, $k_x \rho_i = 0$. The shaded region shows the
      experimental level of flow shear $\gamma_E = 0.16 \pm 0.02$.
    }
    \label{fig:linear_nspec_2}
  \end{figure}

  Including a kinetic electron species, leads to much stronger linear growth as
  shown in \figref{linear_nspec_2}, which again shows $\gamma$ and
  $\omega_g$ as a function of $k_y \rho_i$. We focus here on the dynamics at
  ion scales ($k_y \rho_i \sim 1$), given that the hyperviscosity we apply in
  our nonlinear simulations acts predominantly on the electron scales (see
  Appendix~\ref{sec:hyperviscosity}). \Figref{max_growth_rate} shows the maximum
  growth rate at ion scales as a function of $\kappa_T$ with $\kappa_T=4.8$.
  The horizontal dashed line indicates $\gamma_E = 0.16 \pm 0.02$ and the
  vertical dashed line indicates $\kappa_T=4.8$, which was the value of
  $\kappa_T$ at which turbulence was quenched in our nonlinear simulations at
  this flow shear [see \Figsref{contour_heatmap}{value_heatmap}]. We see that
  the maximum growth rate at ion scales is clearly much larger than $\gamma_E$,
  and that $\gamma_E/\gamma_{\max} \sim 1/3$ at $\kappa_T=4.8$. Previous
  numerical investigations with kinetic electrons investigating the quenching
  of turbulence with flow shear estimated that~\cite{Kinsey2007}:
  $\gamma_E/\gamma_{\max} = 0.71 (\kappa/1.5) / (A/3)^{0.6}$, where
  $A$ aspect ratio and $\kappa$ is the flux surface elongation. For the flux
  surface we are considering, $A \sim 1.5$ and $\kappa=1.46$ (see
  Table~\ref{tab:sim_params}), giving $\gamma_E/\gamma_{\max} \sim 1$, similar
  to the quench condition for adiabatic electrons. We see that in our nonlinear
  simulations, turbulence is quenched for a much lower ratio of
  $\gamma_E/\gamma_{\max}$ suggesting that, for the system we are
  investigating, flow shear is more effective than expected at quenching
  ion-scale turbulence, at least compared to the estimates
  in~\cite{Kinsey2007}.
  \begin{figure}[t]
    \centering
    \includegraphics[width=0.6\linewidth]{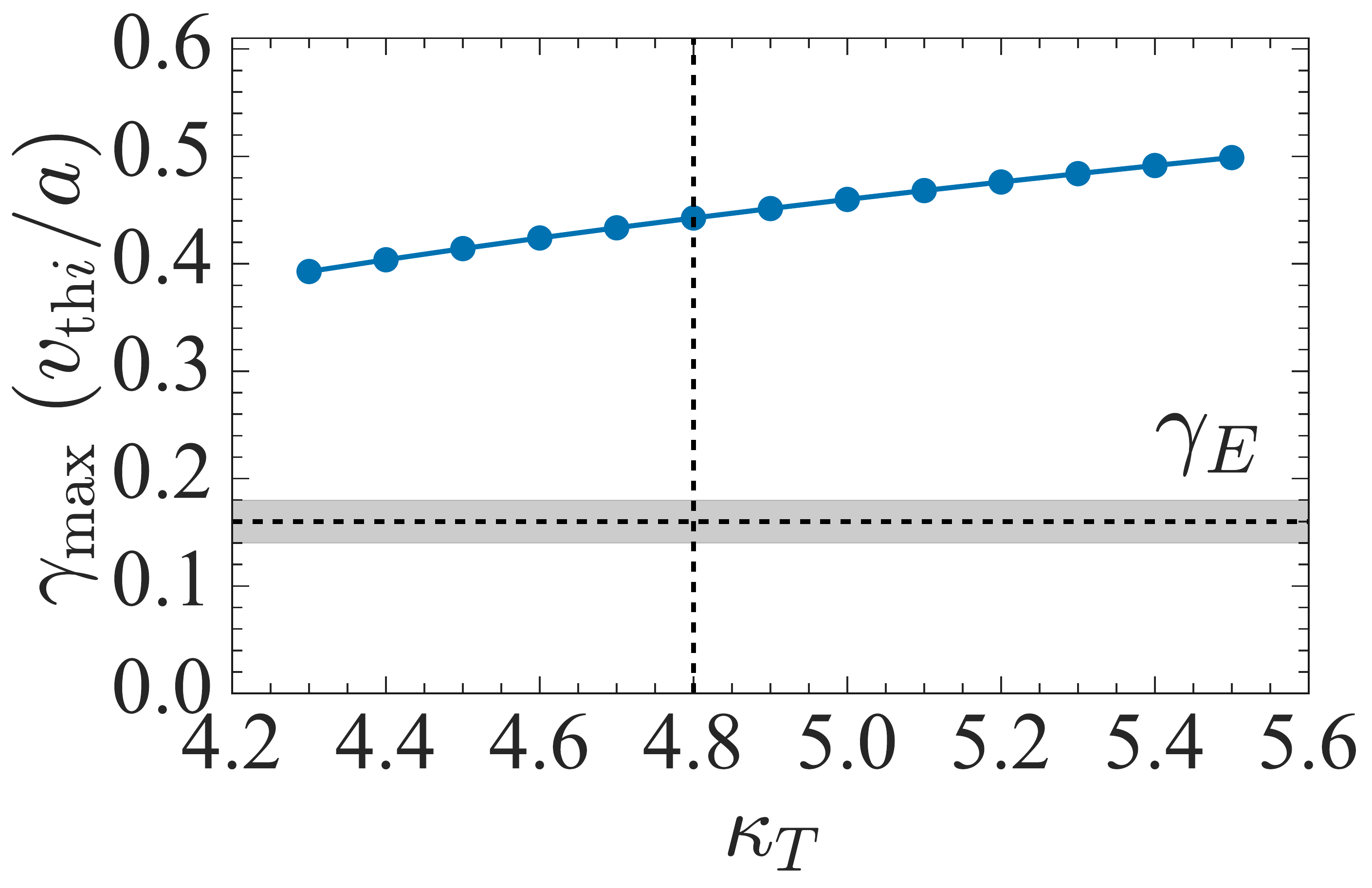}
    \caption[Maximum growth rate]{Maximum linear growth rate $\gamma_{\max}$ as
    a function of $\kappa_T$. The dashed line and shaded area indicate
    $\gamma_E = 0.16 \pm 0.02$.
    }
    \label{fig:max_growth_rate}
  \end{figure}

\chapter{Transforming to real space and laboratory frame}
  \label{App:real_space_transform}

  As explained in Section~\ref{sec:local_approx}, GS2 solves the gyrokinetic
  equation~\eqref{gk} in curvilinear coordinates~\cite{Beer1995a} in a domain
  known as a ``flux tube'', shown in~\figref{flux_tube}, that rotates with the
  plasma. In order to analyse the real-space structure of turbulence and
  compare with BES measurements, we need to transform our data from the
  rotating plasma frame to the laboratory frame and from flux-tube geometry to
  real-space geometry, i.e., from the GS2 coordinates $(x,y,\theta)$ to $(R, Z,
  \lambda)$ where $x$ and $y$ are the GS2 perpendicular coordinates, $\theta$ is the
  poloidal angle, $R$ is the major radius, $Z$ is poloidal height above the
  midplane of the machine, and $\lambda$ is the distance along the field line.

  \section{Laboratory frame}

  GS2 simulates the plasma in a frame rotating with the plasma [see
  equation~\eqref{rot_vel} and~\eqref{u_grad}], with toroidal
  rotation frequency $\omega_0$, whereas the BES diagnostic measures turbulence
  in the laboratory frame. In order to make realistic comparisons with BES
  measurements, we applied the following transformation to the GS2 distribution
  function to transform from the rotating to the laboratory
  frame~\cite{Holland2009}:
  \begin{equation}
    {\qty(\frac{\delta n_i}{n_i})}_{\mathrm{lab}}(t, k_x, k_y, \theta) =
    {\qty(\frac{\delta n_i}{n_i})}_{\mathrm{GS2}}(t, k_x, k_y, \theta)
                                        e^{- i n \omega_0 t},
    \label{lab_transform}
  \end{equation}
  where ${\qty(\delta n_i/n_i)}_{\mathrm{GS2}}(t, k_x, k_y, \theta)$ is
  the fluctuating density field calculated by GS2 in the rotating frame,
  ${\qty(\delta n_i/n_i)}_{\mathrm{lab}}(t, k_x, k_y, \theta)$ is the
  density field in the laboratory frame, and
  \begin{equation}
    n = k_y \rho_i \dv{\psi_N}{r} \frac{a}{\rho_i}
    \label{tor_mode_no}
  \end{equation}
  is the toroidal mode number of a given $k_y$ mode, $\psi_N$ is the normalised
  poloidal magnetic flux, $r=D/2a$ is the Miller~\cite{Miller1998} radial
  coordinate, $D$ is the diameter of the flux surface, $a$ is half of the
  diameter of the last closed flux surface (LCFS), and $\rho_i$ is the ion
  gyroradius.

  \section{Radial domain size}
  \label{sec:radial_domain}
  Here, we calculate the radial domain size $L_R$ at the outboard
  midplane from the radial domain size in GS2 coordinates $L_x$. We start by
  noting that gradients across the GS2 domain are held constant, meaning that
  \begin{equation}
    R'(\theta = 0) = \frac{1}{a} \dv{R(\theta = 0)}{r}
                   = \frac{1}{a} \frac{\Delta R(\theta = 0)}{\Delta r},
    \label{R_prime}
  \end{equation}
  where $R$ is the major radius, $R'(\theta)$ is the derivative of $R$ with
  respect to the poloidal angle $\theta$, and $\Delta R(\theta = 0) \equiv L_R$
  is the radial domain size. We calculate $\Delta r$ from the local GS2
  coordinate $x$ as follows. Using the Taylor expansion $r \approx r_0 +
  (\psi_N - \psi_{0N}) \eval{\dv*{r}{\psi_N}}_{r_0}$ and substituting
  into~\eqref{gs2_x} we get
  \begin{equation}
    x = (r - r_0) \frac{q_0}{r_0} \dv{\psi_N}{r} \frac{a}{\rho_i},
    \label{gs2_x_with_rho}
  \end{equation}
  where $r_0  = 0.8$ is the location of the flux surface we are investigating,
  and $q_0$ is the safety factor at $r = 0.8$.  The extent of the radial domain
  in the coordinate $x$ is then
  \begin{equation}
    \Delta x = \Delta r \frac{q}{r_0} \dv{\psi_N}{r} \frac{a}{\rho_i}.
    \label{delta_x}
  \end{equation}
  Using the following values from our simulations $\Delta x = 2 \pi /
  k_{x,\min} \rho_i \approx 200 \rho_i$, where $k_{x,\min}$ is the minimum
  resolved $k_x$ in our nonlinear simulations, $\qty(\dv*{\psi_N}{r})^{-1} =
  1.44$, and from the experiment [see Tables~\ref{tab:equil_params}
  and~\ref{tab:sim_params}] $a = 0.58$~m, $\rho_i = 6.08 \times10^{-3}$~m, $q_0
  = 2.31$, we calculate $\Delta r$ from equation~\eqref{delta_x} and substitute
  into equation~\eqref{R_prime} to find $\Delta R(\theta = 0) \equiv L_R
  \approx 65 \rho_i \approx 0.4$~m.  We note that while $x$ is a local
  coordinate and $R$ is a physical coordinate our simulations only describe the
  turbulence at $r=0.8$. Hence, our results are only comparable to experimental
  results at this radius.

  \section{Poloidal domain size}
  \label{sec:pol_domain}
\begin{figure}[t]
    \centering
    \includegraphics[width=0.6\linewidth]{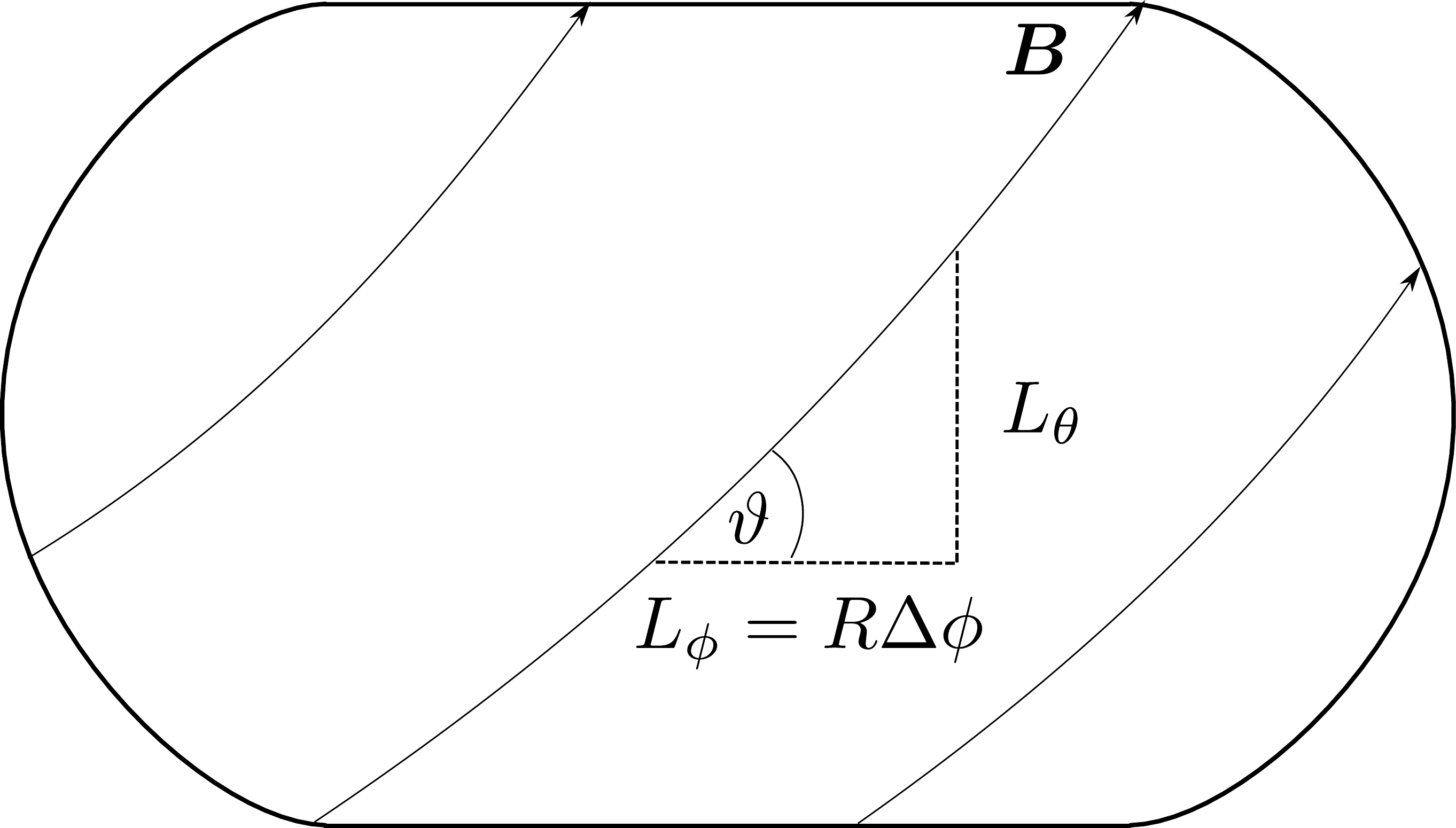}
    \caption[MAST magnetic-field pitch-angle]{
      Side view sketch of the MAST geometry shown in~\figref{flux_tube}.
      The magnetic field's pitch-angle, $\vartheta$ ($\approx 0.6$), relates the
      toroidal extent of the GS2 domain, $L_\phi$, with the poloidal extent,
      $L_\theta$, through $\tan \vartheta = L_\theta / L_\phi$.
    }
    \label{fig:pitch_angle}
  \end{figure}

  To calculate the poloidal domain size $L_Z$, we start by noting that, the GS2
  grid points lie on $(\phi, \psi)$ planes at constant values of $\theta$.
  Therefore, at $\theta = 0$, GS2 simulates turbulence on a radial-toroidal
  plane. The extent of the GS2 domain in toroidal angle $\phi$
  is~\cite{Beer1995a}
  \begin{equation}
    \Delta \phi = \frac{2 \pi}{n_0},
    \label{phi_box_size}
  \end{equation}
  where
  \begin{equation}
    n_0 = k_{y,\min} \rho_i \dv{\psi_N}{r} \frac{a}{\rho_i}
    \label{n0}
  \end{equation}
  is the minimum toroidal mode number simulated and $k_{y,\min} \rho_i$ is
  the smallest resolved $k_y$ mode in our nonlinear simulations. The toroidal
  extent of the domain is therefore given by $L_\phi = R \Delta \phi$, where
  $R$ is the major radius of the flux surface at the outboard midplane. We can
  relate $L_\phi$ to the poloidal extent of the GS2 domain, $L_\theta$, via the
  relation $\tan \vartheta = L_\theta / L_\phi$, where $\vartheta$ ($\approx
  0.6$) is the pitch-angle of the magnetic field, as shown
  in~\figref{pitch_angle}, for the flux surface $r=0.8$ at the outboard
  midplane.
  In our nonlinear simulations, $k_{y,\min} \rho_i = 0.1$, giving
  $n_0 \approx 7$ using \eqref{n0} and $L_\phi \approx 1.2$~m.
  Using the above relations we find that the poloidal projection of the plane
  at $\theta = 0$ is $L_\theta \approx 134 \rho_i \approx 0.81$~m.

\begin{figure}[t]
    \centering
    \includegraphics[width=0.6\linewidth]{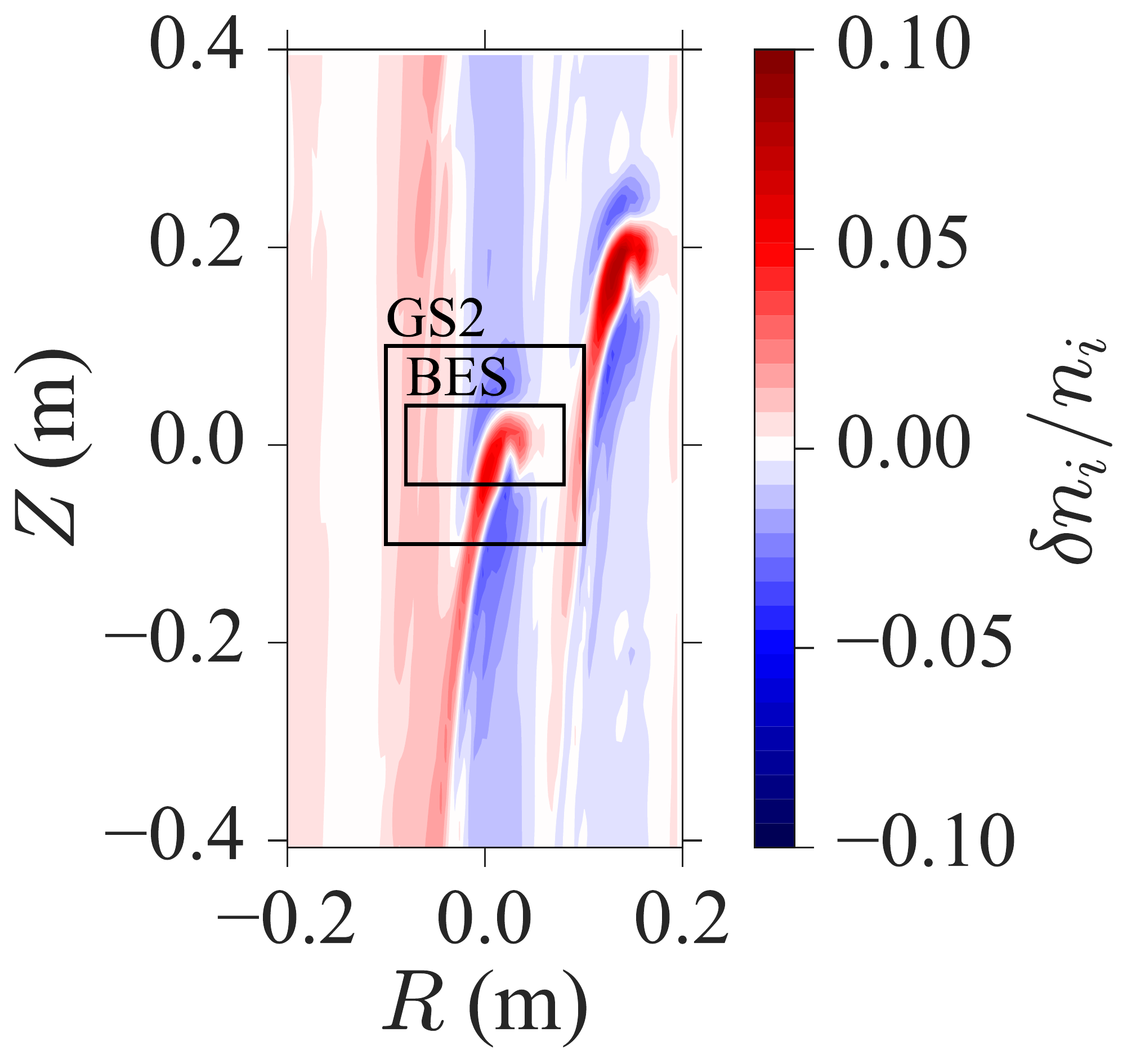}
    \caption[GS2 density fluctuations for a near-marginal case on an $(R,Z)$-plane]{
      Density-fluctuation field $\delta n_i/n_i$ for the same near-marginal
      shown in \Figref{marginal} for the equilibrium parameters
      $(\kappa_T,\gamma_E) = (4.8, 0.16)$ as a function $R$ and $Z$. The
      indicated domains are those used for the correlation analysis of raw GS2
      density fluctuations (GS2) and the approximate size of the BES viewing
      window (BES).
	}
    \label{fig:marginal_rz}
  \end{figure}

  Using the results from this section and Section~\ref{sec:radial_domain}, we
  can transform our density fluctuation fields at the outboard midplane to a
  radial-poloidal plane similar to the BES measurement window.  For example,
  \Figref{marginal_rz} shows the same plot as in \figref{marginal} at $\theta =
  0$ in terms of the real-space poloidal coordinates $R$ and $Z$.  Also
  indicated in \figref{marginal_rz} are the domains used for the correlation
  analysis of BES data and raw GS2 data, as used in Sections~\ref{sec:corr_exp}
  and~\ref{sec:corr_gs2}, respectively.

  \section{Parallel coordinate and domain size}

  Finally, we calculate the parallel distance along the magnetic field line at
  the centre of our flux tube. This procedure is non-trivial for a
  general geometry because a uniform grid in $\theta$ does not map to a uniform
  spatial grid along the field line as it would have done for circular flux
  surfaces. For our D-shaped geometry we want to find $\lambda(\theta)$, the
  distance along the field line parametrised by the poloidal angle $\theta$.
  The differential arc length of a line element along the field line in terms
  of $(R,Z,\phi)$ is
  \begin{equation}
    d \lambda^2 = dR^2 + dZ^2 + {(R d\phi)}^2,
    \label{line_element}
  \end{equation}
  where $R = R(\theta)$ and $Z = Z(\theta)$ are the coordinates of the magnetic
  field line at the centre of the flux tube.
  We can differentiate with respect to $\theta$ and integrate to get the arc
  length as a function of $\theta$:
  \begin{equation}
    \lambda(\theta) = \int_0^\theta d \theta' \sqrt{{\qty(\dv{R}{\theta'})}^2 +
                      {\qty(\dv{Z}{\theta'})}^2 +
                      {\qty(R \dv{\phi}{\theta'})}^2}.
    \label{l_theta}
  \end{equation}
  The quantities $R(\theta)$, $Z(\theta)$, $\dv*{\phi}{\theta}$ are obtained
  from GS2 and we then calculate their numerical derivatives with respect to
  $\theta$, and then the integral~\eqref{l_theta} to determine
  $\lambda(\theta)$. With the knowledge of the real-space parallel grid, we can
  calculate correlation lengths in the parallel direction.

\chapter{Synthetic correlation properties without the ``spike filter''}
  \label{App:no_spike}

\begin{figure}[t]
    \centering
    \begin{subfigure}[t]{0.49\textwidth}
      \includegraphics[width=\linewidth]{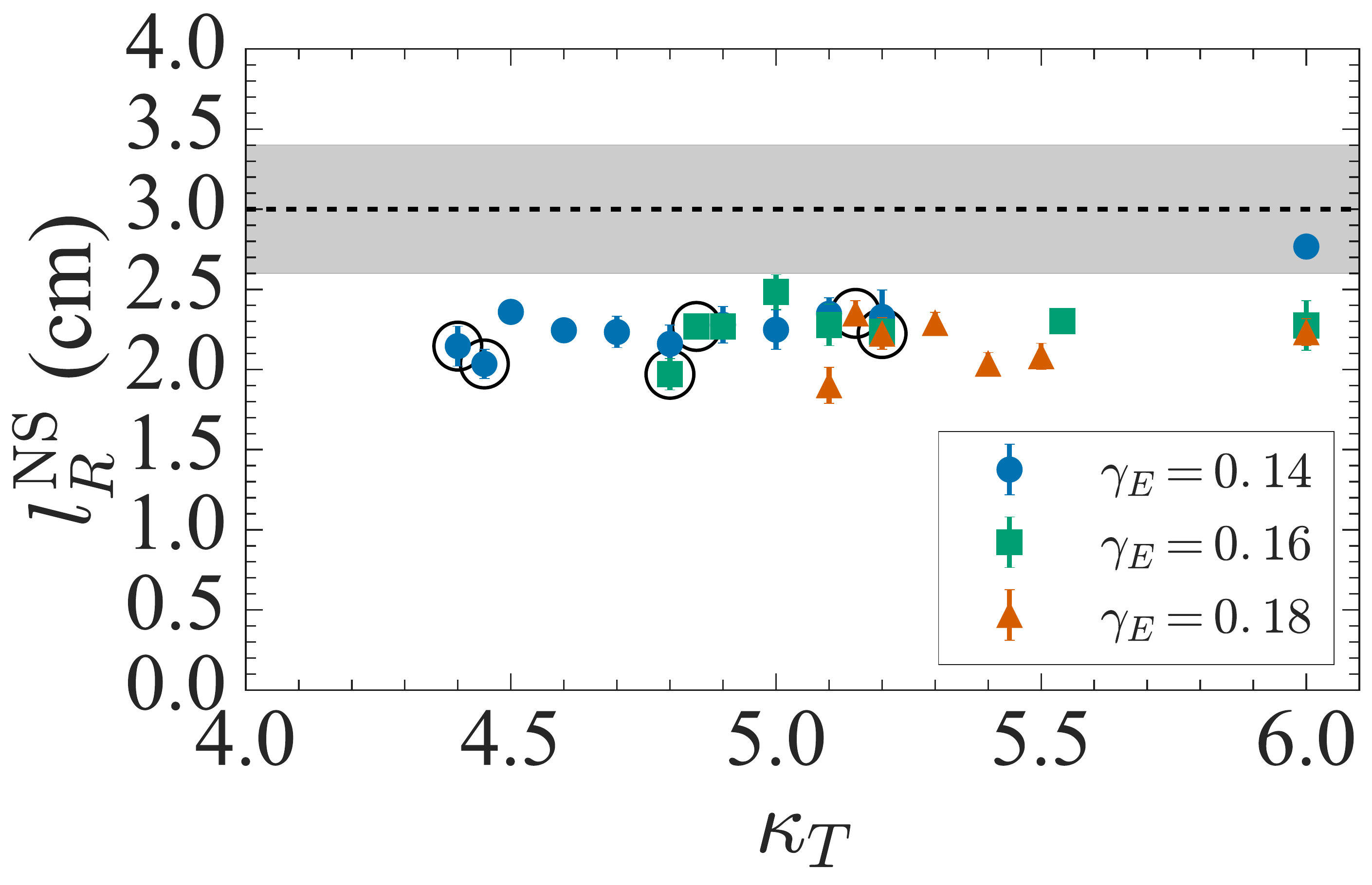}
      \caption{}
      \label{fig:lr_bes_ns}
    \end{subfigure}
    \hfill
    \begin{subfigure}[t]{0.49\textwidth}
      \includegraphics[width=\linewidth]{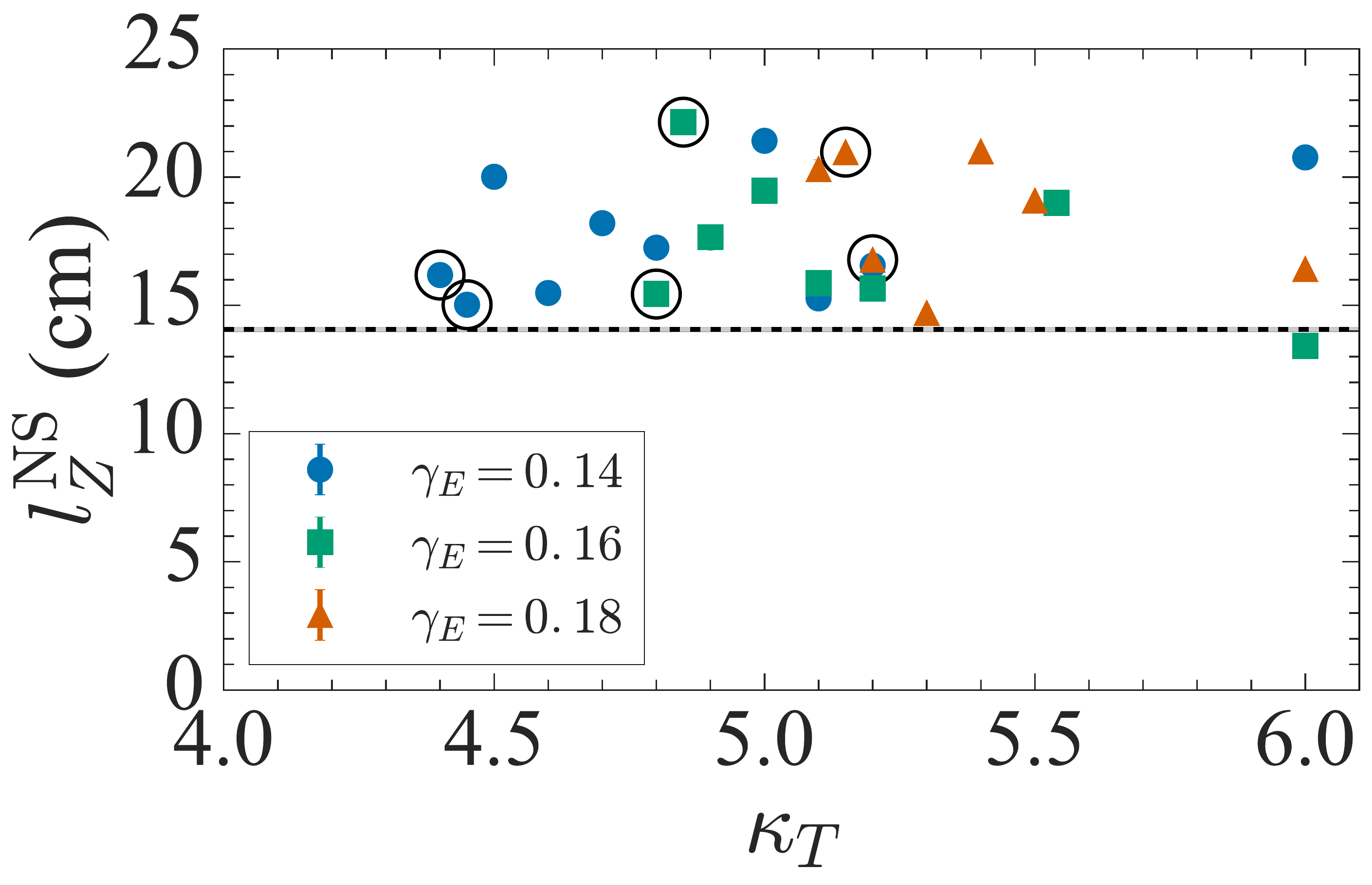}
      \caption{}
      \label{fig:lz_bes_ns}
    \end{subfigure}
    \\
    \begin{subfigure}[t]{0.49\textwidth}
      \includegraphics[width=\linewidth]{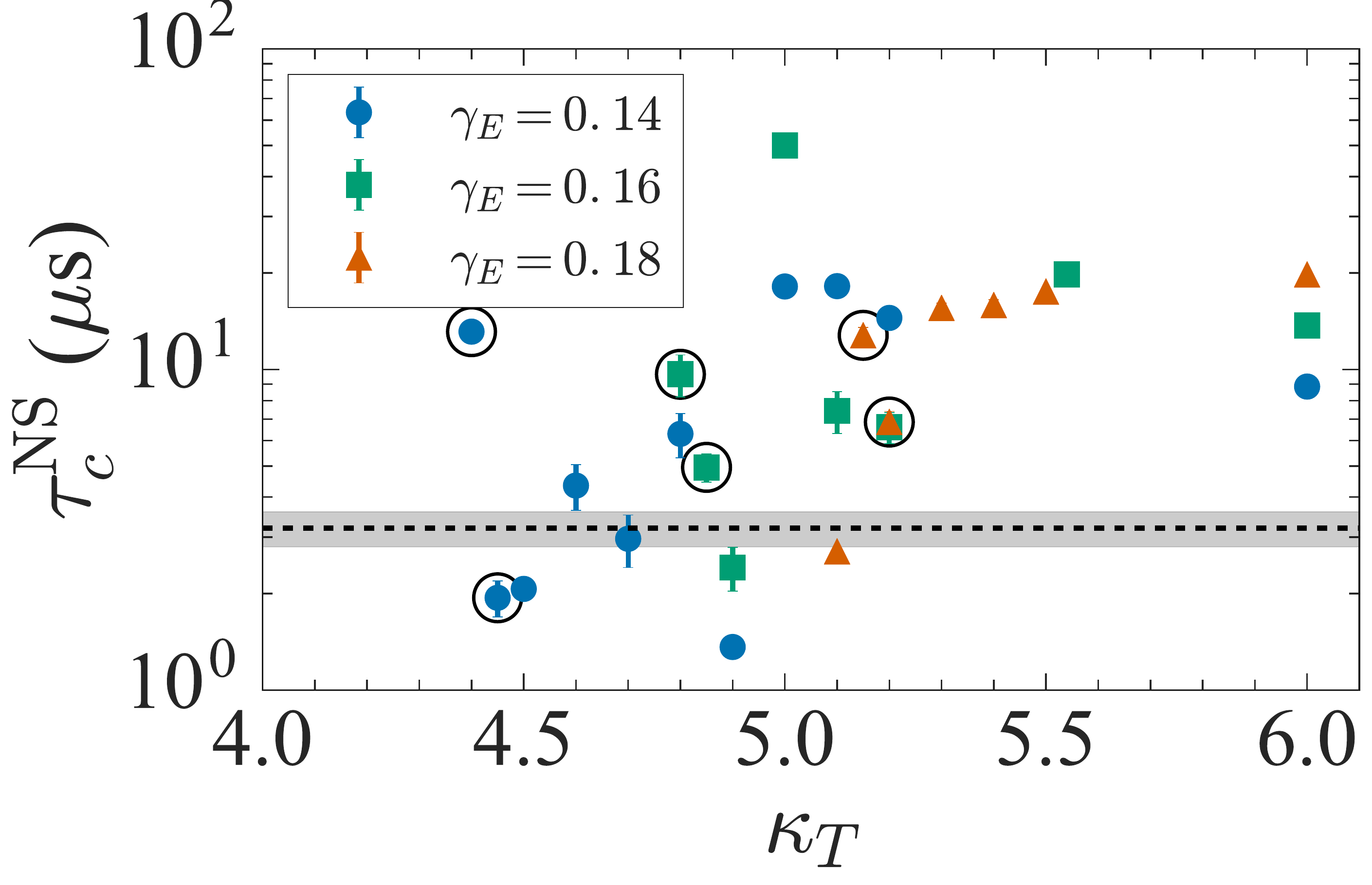}
      \caption{}
      \label{fig:tau_bes_ns}
    \end{subfigure}
    \hfill
    \begin{subfigure}[t]{0.49\textwidth}
      \includegraphics[width=\linewidth]{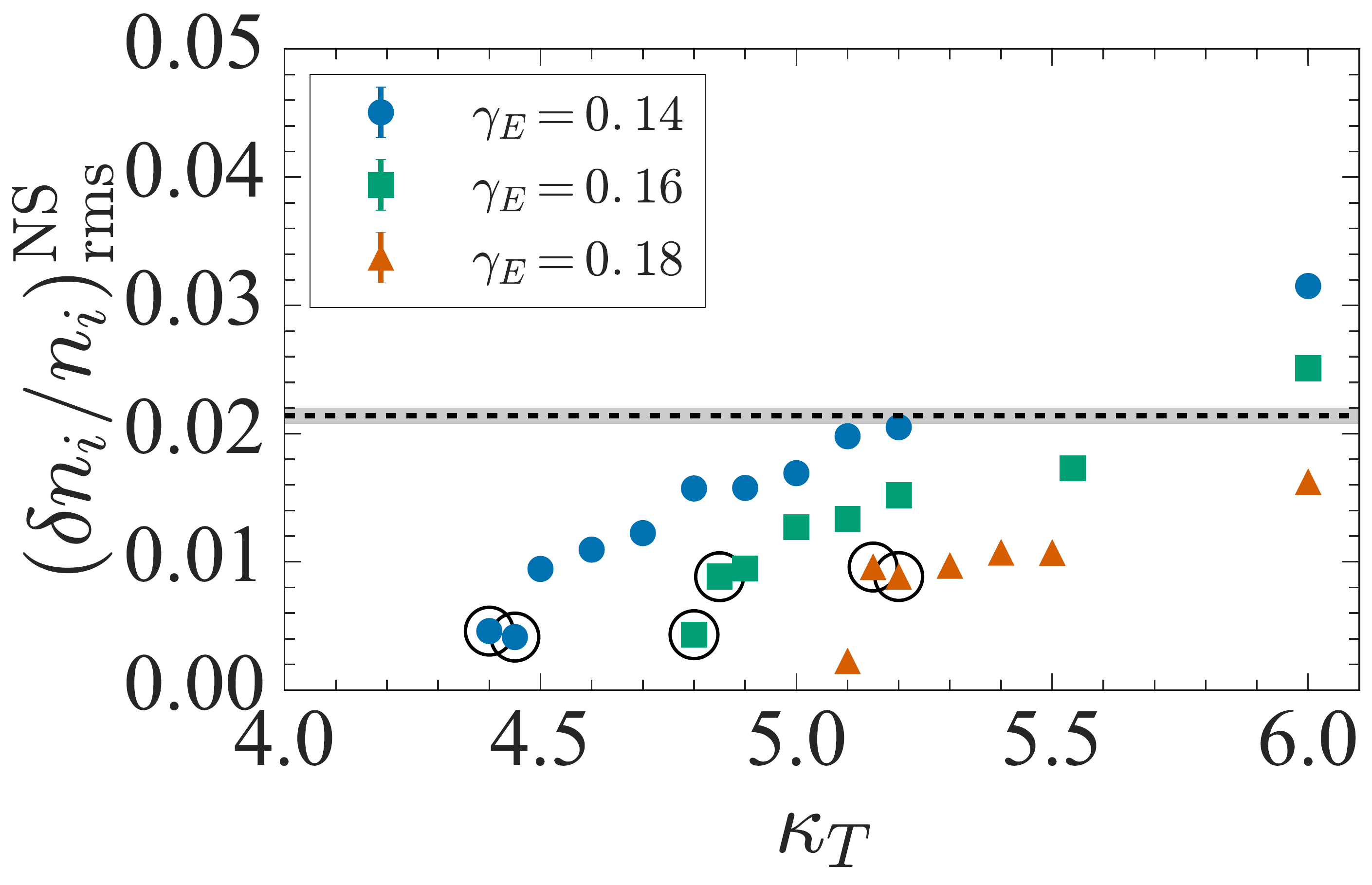}
      \caption{}
      \label{fig:n_bes_ns}
    \end{subfigure}
    \caption[GS2 correlation parameters with a synthetic diagnostic
             applied without a numerical ``spike filter'']{
      Correlation-analysis results calculated from the analysis of GS2
      fluctuation data (within the region of experimental uncertainty) after
      applying the synthetic diagnostic, but without the spike filter normally
      applied to experimental data:
      \subref*{fig:lr_bes_ns} radial correlation length $l^{\mathrm{NS}}_R$,
      \subref*{fig:lz_bes_ns} poloidal correlation length $l^{\mathrm{NS}}_Z$,
      \subref*{fig:tau_bes_ns} correlation time $\tau^{\mathrm{NS}}_c$, and
      \subref*{fig:n_bes_ns} RMS density fluctuation level
      ${(\delta n_i/n_i)}^{\mathrm{NS}}_{\mathrm{rms}}$. The simulations that
      matched the experimental heat flux are circled. The quantities plotted
      here are discussed in Section~\ref{sec:corr_overview}.
    }
    \label{fig:bes_ns}
  \end{figure}

  A key step in the analysis of experimental data involves the removal of
  high-energy radiation (e.g., neutron, gamma ray, or hard X-ray) impinging on
  the BES detector. This radiation manifests itself as
  delta-function-like spikes in time, typically only on a single BES channel.
  These are removed via a numerical ``spike filter''~\cite{Field2012,Fox2016a},
  which was included in the main analysis for consistency with experimental
  analysis.  Here, we show the results of a correlation analysis of GS2
  density fluctuations with the synthetic diagnostic applied, but without this
  ``spike filter''. \Figref{bes_ns} shows the correlation results for
  parameter values within the experimental uncertainty: the radial correlation
  length $l^{\mathrm{NS}}_R$ [\figref{lr_bes_ns}], the poloidal correlation
  length $l^{\mathrm{NS}}_Z$ [\figref{lz_bes_ns}], the correlation time
  $\tau^{\mathrm{NS}}_c$ [\figref{tau_bes_ns}], the RMS density fluctuation
  ${(\delta n_i/n_i)}^{\mathrm{NS}}_{\mathrm{rms}}$ [\figref{n_bes_ns}].

  Comparing these results to the results in Section~\ref{sec:corr_synth} with
  the ``spike filter'', we see that only the poloidal correlation length is
  affected: $l^{\mathrm{NS}}_Z$ is several centimetres lower with the ``spike
  filter'' compared to cases without it. We found that in some cases,
  fast-moving structures in the poloidal direction (especially the long-lived
  structures found in our near-marginal simulations) were removed by the
  ``spike filter'' and, therefore, would not affect to the poloidal correlation
  function, resulting in a drop in $l^{\mathrm{NS}}_Z$. In particular,
  \figref{lz_bes_ns} shows that $l^{\mathrm{NS}}_Z$ increased significantly in
  near-marginal simulations compared to the results with the ``spike filter'',
  suggesting that the coherent structures were no longer removed by the ``spike
  filter''. This observation may assist future attempts to observe
  experimentally the coherent structures predicted by our simulations.

\chapter{Example GS2 input file}
\label{App:gs2_input_file}

The following is an example GS2 input file used for this study (see
\url{http://gyrokinetics.sourceforge.net} on how to run the code with these
settings). A description of each of these variables can be found at
\url{http://gyrokinetics.sourceforge.net/wiki/index.php/Gs2_Input_Parameters}.

\begin{lstlisting}[language=Fortran]
! General parameters
&parameters
    beta = 0.0047310768
    zeff = 1.5899834
/

&kt_grids_knobs
    grid_option = "box"
/

! Resolution parameters
&kt_grids_box_parameters
    nx = 128
    ny = 96
    jtwist = 80
    y0 = 10.0
    x0 = 10.0
/

! Geometric parameters
&theta_grid_parameters
    ntheta = 20
    nperiod = 1
    shat = 3.9955695
    rhoc = 0.7966436
    qinp = 2.31493
    akappa = 1.45734
    akappri = 0.44877291
    tri = 0.205939
    tripri = 0.46296135
    shift = -0.30708938
    rmaj = 1.4891066
    r_geo = 1.6438144
/

&theta_grid_knobs
   equilibrium_option = "eik"
/

&theta_grid_eik_knobs
    itor = 1
    iflux = 0
    irho = 2
    local_eq = .true.
    bishop = 4
    s_hat_input = 3.9955695
    beta_prime_input = -0.1212404
    delrho = 0.001
/

! Velocity space grid parameters
&le_grids_knobs
    ngauss = 8
    negrid = 16
/

&dist_fn_knobs
    gridfac = 1.0
    boundary_option = "linked"
    g_exb = 0.16
    apfac = 1.0
    driftknob = 1.0
    opt_init_bc = .true.
    opt_source = .true.
/

&fields_knobs
    field_option = "local"
    field_subgath = .true.
    response_dir = "response"
    do_smart_update = .true.
    field_local_allreduce = .true.
    field_local_allreduce_sub = .true.
/

! Time parameters
&knobs
    fphi = 1.0
    faperp = 0.0
    delt = 0.01
    nstep = 100000
    avail_cpu_time = 21600
    margin_cpu_time = 600
/

&reinit_knobs
    delt_adj = 2.0
    delt_minimum = 1.0e-06
    delt_cushion = 10
/

&layouts_knobs
    layout = "xyles"
    unbalanced_xxf = .true.
    max_unbalanced_xxf = 0.5
    unbalanced_yxf = .true.
    max_unbalanced_yxf = 0.5
    intmom_sub = .true.
    intspec_sub = .true.
/

&collisions_knobs
    collision_model = "default"
/

&hyper_knobs
    hyper_option = "visc_only"
    const_amp = .true.
    d_hypervisc = 9
/

&nonlinear_terms_knobs
    nonlinear_mode = "on"
    flow_mode = "off"
    cfl = 0.5
/

&species_knobs
    nspec = 2
/

&species_parameters_1
    z = 1.0
    mass = 1.0
    dens = 1.0
    temp = 1.0
    tprim = 5.1
    fprim = 2.64278
    uprim = 0.0
    vnewk = 0.02098588
    type = "ion"
/

&dist_fn_species_knobs_1
    fexpr = 0.48
    bakdif = 0.05
/

&species_parameters_2
    z = -1.0
    mass = 0.0002723311
    dens = 1.0
    temp = 1.091722
    tprim = 5.773614
    fprim = 2.64278
    uprim = 0.0
    vnewk = 0.5900574
    type = "electron"
/

&dist_fn_species_knobs_2
    fexpr = 0.48
    bakdif = 0.05
/

! Initial conditions
&init_g_knobs
    phiinit = 1.0
    restart_file = "gs2.nc"
    ginit_option = "noise"
    restart_dir = "nc"
/

! Diagnostics
&gs2_diagnostics_knobs
    write_verr = .true.
    write_avg_moments = .true.
    write_eigenfunc = .true.
    write_final_fields = .true.
    write_final_moments = .true.
    nsave = 500
    nwrite = 100
    navg = 10
    omegatol = -0.001
    omegatinst = 500.0
    save_for_restart = .true.
    write_cross_phase = .true.
/
\end{lstlisting}

\backmatter

\printbibliography

\end{document}